ESA/SRE(2009)6
December 2009

# SPICA

## Revealing the origins of planets and galaxies

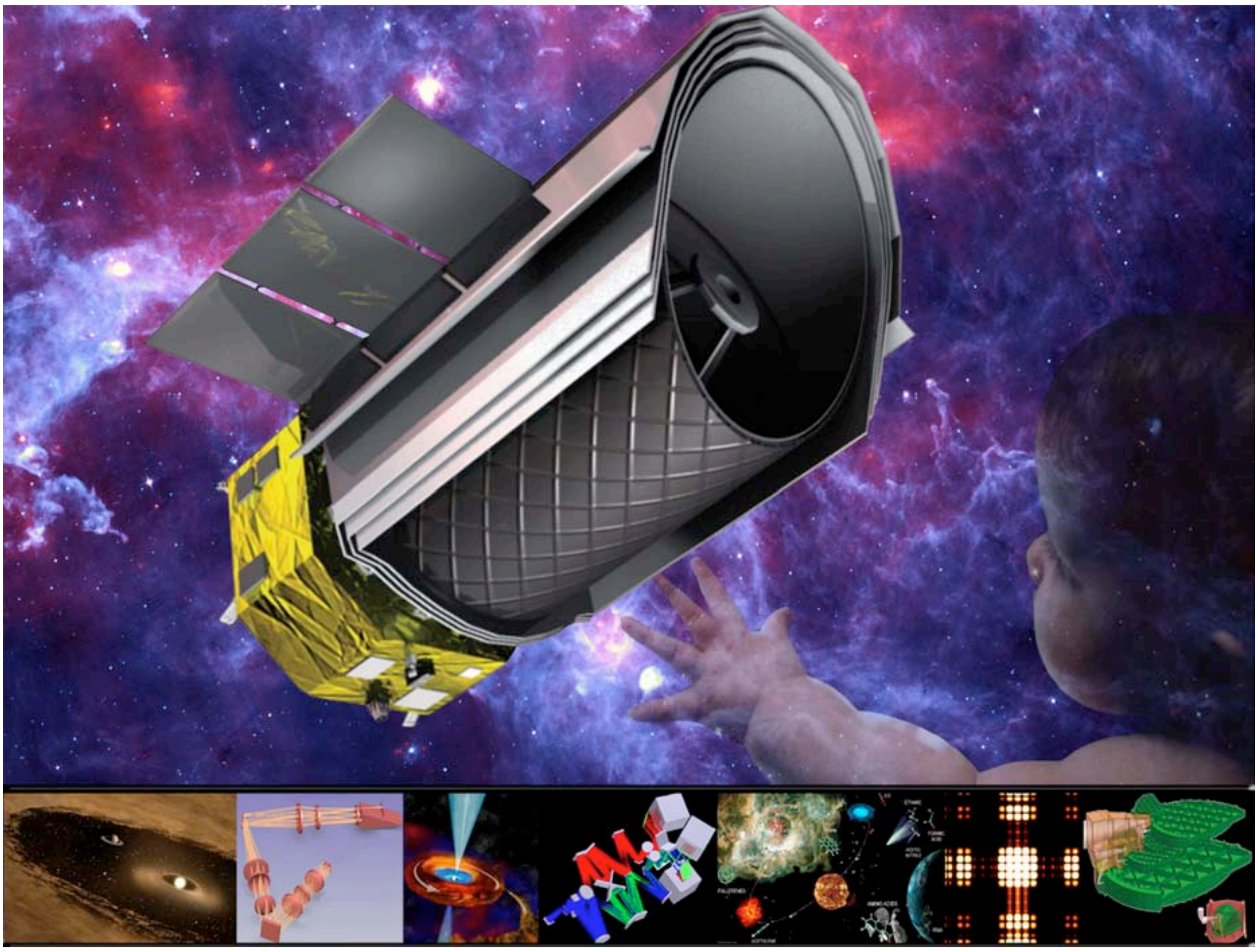

### Assessment Study Report

European Space Agency

# SPICA

JAXA Japan Aerospace Exploration Agency

esa

The University of Tokyo · Kobe University · Nagoya University · National Astronomical Observatory of Japan · Ibaraki University · Osaka University

| **SPICA – SPace Infrared telescope for Cosmology and Astrophysics** | |
|---|---|
| Science objectives | • Formation and evolution of planetary systems: Gas and dust in proto-planetary discs, including water, and their link to planetary formation; mineralogy of debris discs; gas exoplanets atmospheres; composition of Kuiper Belt objects.<br>• Life cycle of dust: Physics and chemistry of gas and dust in the Milky Way and in nearby galaxies; dust mineralogy; dust processing in supernova remnants and the origin of interstellar dust in the early Universe.<br>• Formation and evolution of galaxies: AGN/starburst connection over cosmic time and as a function of the environment; co-evolution of star formation and super-massive black holes; star-formation and mass assembly history of galaxies in relation with large scale structures; the nature of the Cosmic Infrared Background. |
| Wavelength range | Medium to far infrared (5–210 $\mu$m) |
| Telescope | Ritchey-Chrétien, 3.5 m diameter cryogenic infrared telescope, with diffraction limited performance at 5 $\mu$m (5 arcmin radius FOV). |
| Instruments | • MIRACLE (mid infrared camera and spectrophotometer)<br>• SAFARI (far infrared imaging spectrometer)<br>• MIRMES (mid infrared medium resolution spectrometer)<br>• MIRHES (mid infrared high resolution spectrometer)<br>• SCI (mid infrared coronagraph)<br>• BLISS (far infrared and sub-mm spectrometer; optional)<br>• FPC (dedicated Focal Plane Cameras for guidance)<br><br>On Board Science Data: 24 h * 4 Mb/s = 350 Gbit/day |
| Launcher and orbit | • H-IIB (JAXA) – direct launch in $L_2$ transfer trajectory.<br>• Launcher dispersion correction on first day, main trajectory manoeuvre at day L + 2 and touch up at L + 15, $L_2$ final insertion at L + 120; monthly orbit correction manoeuvres.<br>• Operational orbit is a large amplitude libration orbit around SE-$L_2$ (Halo orbit), with a semi-major axis amplitude of 750 000 km.<br>• Radiation environment: TID (2 mm Al shield) of 19.5 krad (5 yr), 26.8 krad (10 yr). |
| Mission phases and operations | • Launch by end of 2018<br>• LEOP duration ~ 3 days.<br>• Commissioning phase ~ 2 months. Cool-down ~ 168 days<br>• Guaranteed mission duration 3 yr (5 yr goal)<br>• MOC at JAXA (nominal ground station Usuda)<br>• Main SOC at JAXA, with additional SOC at ESA and SAFARI ICC in Europe. |
| Spacecraft | • Cryogenic PLM supported by octagonal SVM via a dedicated, low thermal conductivity, truss structure.<br>• Dimensions: ~ 4.5 m width, 7.5 m high. Total wet mass ~ 4000 kg<br>• Total maximum power of 2.4 kW (EOL)<br>• Total delta-V budget: 110 m $s^{-1}$<br>• Pointing requirements: Standard mode: APE < 0.135 arcsec, RPE < 0.075 arcsec/200 s; Coronagraph mode: APE < 0.03 arcsec, RPE < 0.03 arcsec/20 min.<br>• Attitude control system: 3-axis stabilised spacecraft, RW's. AOCS sensors complemented by dedicated focal plane cameras for fine guidance.<br>• Thermal Control System: Passive cooling via Sun-shield, 3 thermal shields, telescope shell and external telescope baffle. Active cooling via a redundant set of mechanical coolers based on 2-stage Stirling units and JT expansion units (4.5 and 1.7 K stage).<br>• Propulsion system: Monopropellant (blow down), 4x 23N and 8x 3N thrusters. Propellant: 220 kg.<br>• On Board Data Handling: Central DPU (common to PLM and SVM), SpaceWire, on board mass memory > 48 Gbyte.<br>• Telecommand, Telemetry & Communication: X-Band (HGA, MGA, and 3x LGA for down-link), S-band (2 LGA for up and down-link). |

Page intentionally left blank

# Foreword

In October 2007, the Space Science Advisory Committee selected SPICA (SPace Infrared telescope for Cosmology and Astrophysics) for an assessment study as a candidate M-class mission with the character of “mission of opportunity”. SPICA is a JAXA/ISAS led observatory that will operate in the mid and far infrared wavelength range with unprecedented sensitivity, thanks to the 3 m-class (3.5 m current baseline) cold telescope and advanced instruments. SPICA will carry mid infrared camera, spectrometers and coronagraph (built by JAXA institutes), a far infrared imager spectrometer (provided by an European/Canadian consortium), and a far infrared/sub-millimetre spectrometer (proposed by NASA). The collaboration between the far infrared instrument (SAFARI) European/Canadian consortium and JAXA, initiated more than a decade ago when the SPICA concept was outlined, led to the Cosmic-Vision proposal in which ESA is invited to become a mission partner. In particular, it is proposed that ESA procures the telescope assembly, a ground station, contributes to the science operations and manages the JAXA/SAFARI interfaces. From the scientific point of view, the participation in SPICA would enable European astronomers to greatly benefit from a first-class observatory with unique capabilities that will improve drastically our understanding of planetary systems formation and evolution, exoplanets and galaxy evolution.

As a first step of the assessment phase, the SPICA Telescope Science Study Team was organised to identify the science requirements associated with the telescope. Once the science and engineering requirements were established, an internal study on the telescope assembly was carried out between March and April 2008 at ESTEC, in close cooperation with JAXA. Following the release of the Invitation to Tender, two parallel competitive contracts of one year duration were awarded to EADS Astrium (F) and to Thales Alenia Space (F) for the industrial assessment study of the SPICA telescope. The two studies showed the overall feasibility of the telescope development given the heritage and expertise available in Europe.

The European instrument SAFARI was the subject of an internal ESA CDF study in May 2008, with a strong participation of the SAFARI specialists. The SAFARI Consortium has carried out the instrument assessment study until September 2009, including a Technology Readiness Review that was organised in collaboration with ESA. The study has led to a consolidation of the instrument design and a clear definition of the detector options with a plan for their selection.

The SPICA science case described in the document is the result of a combined effort from scientists from the SAFARI consortium, supported by astronomers from different disciplines, and the Japanese SPICA Task Force, which is in charge of the definition of the mission science objectives and requirements. A joint European/Japanese workshop took place in July 2009 in Oxford, where the science objectives were presented in detail and discussed in the light of their uniqueness and relevance.

At JAXA, SPICA has been in the Pre-project phase, which is approximately equivalent to ESA’s phase-A, since July 2008. It consists of two phases, the Concept Design Phase and System Definition Phase. The aim of the Concept Design Phase is to define the system requirements and obtain their approval in the System Requirements Review that will be held in the period December 2009-March 2010. After the System Definition Review in Autumn 2010, SPICA will go through the Project Approval Review by mid 2011, a management review by JAXA HeadQuarters, where the decision whether SPICA can proceed though phase-B to the end of mission will be made.

In this document we present the scientific objectives, science requirements and the description of the mission which is the result of the ESA and SAFARI assessment studies and of JAXA phase-A activities. Although we have put special emphasis on the elements of the European contribution, the mission as a whole is covered in the document. The reasons are two-fold. First, to provide the context for the European contribution and second, more importantly, to present the full SPICA observatory and its unprecedented capabilities as they will be available to the astronomical community.

Page intentionally left blank

# Authors

The SPICA Assessment Study Report has been prepared with inputs from the ESA SPICA Study Team, the SPICA Telescope Science Study Team, the JAXA/ISAS SPICA Team and the SAFARI Consortium. The authors, in alphabetical order, are:

| | |
|---|---|
| Marc Ferlet | Rutherford Appelton Laboratory (UK); *SPICA Telescope Science Study Team* |
| Norbert Geis | Max Planck Institute für extraterr. Physik (Germany); *SPICA Telescope Science Study Team* |
| Javier Goicoechea | Centro de Astrobiología, CSIC-INTA (Spain); *SAFARI Consortium, Galactic Science Coordinator* |
| Douglas K. Griffin | Rutherford Appelton Laboratory (UK); *SAFARI System Engineer* |
| Ana M. Heras | European Space Agency; *SPICA Study Scientist, Document Editor* |
| Kate Isaak | Cardiff University (UK); *SAFARI Consortium, Extragalactic Science Coordinator* |
| Takao Nakagawa | JAXA/ISAS; *JAXA SPICA Project Leader* |
| Takashi Onaka | University of Tokyo, Japan; *SPICA Telescope Science Study Team* |
| Nicola Rando | European Space Agency; *SPICA Study Manager* |
| Bruce Swinyard | Rutherford Appleton Laboratory (UK); *SAFARI PI, SPICA Telescope Science Study Team* |
| Nobuhiro Takahashi | JAXA; *SPICA System Engineer* |
| Sebastien Vives | Astronomy Observatory of Marseilles Provence (France); *SPICA Telescope Science Study Team* |

## Acknowledgements

The authors would like to acknowledge the essential contributions to the definition of the science objectives from: J.-C. Augereau (UJF, Fr), M. Barlow (UCL, UK), J.-P. Beaulieu (IAP, Fr), A. Belu (U. Nice, Fr), O. Berne (CAB/CSIC-INTA, E), B. Bézard (Obs. Paris, Fr), A. Boselli (LAM, Fr), V. Buat (LAM, Fr), D. Burgarella (LAM, Fr), J. Cernicharo (CAB/CSIC-INTA, E), D. Clements (Imperial College, UK), G. de Zotti (Oss. Astr. Padova, It), M. Delbo (UNS, Fr), Y. Doi (U. Tokyo, JP), S. Eales (U. Cardiff, UK), E. Egami (U. Arizona, USA), D. Elbaz (CEA, Fr), K. Enya (ISAS/JAXA, JP), A. Franceschini (U. Padova, It), H. Fraser (STRATH, UK), M. Fukagawa (Osaka U., JP), H. Gómez (Cardiff U., UK), J. Gómez-Elvira (CAB/CSIC-INTA, E), O. Groussin (LAM, Fr), C. Gruppioni (INAF, It), S. Hailey-Dunsheath (MPE, Ger), S. Hasegawa (ISAS/JAXA, JP), F. Helmich (SRON, NL), M. Honda (KAN-U, JP), Y. Itoh (Kobe U., JP), R. Ivison (U. Edinburgh, UK), H. Izumiura (NAO, JP), C. Joblin (CESR, Fr), D. Johnstone (NRC, CA), G. Joncas (ULAVAL, CA), A. Jones (IAS, Fr), I. Kamp (U. Groningen, NL), H. Kaneda (Nagoya U., JP), H. Kataza (ISAS/JAXA, JP), C. Kemper (Manchester U., UK), F. Kerschbaum (UNIVIE, AT), Y. Kitamura (ISAS/JAXA, JP), P. Lacerda (Belfast U., UK), F. Levrier (LRA-LERMA, Fr), D. Lutz (MPE, Ger), S. Madden (CEA, Fr), M. Magliocchetti (IFSI, It), J. Martín-Pintado (CAB/CSIC-INTA, E), H. Matsuhara (ISAS/JAXA, JP), R. Meijerinks (Sterrewacht Leiden, NL), S. Molinari (IFSI, It), R. Moreno (Obs. Paris, Fr), A. Moro-Martín (CAB/CSIC-INTA, E), P. Najarro (CAB/CSIC-INTA, E), N. Narita (NAO, JP), Y. Okamoto (Ibaraki U., JP), S. Oliver (U. Sussex, UK), T. Onaka (U. of Tokyo, JP), T. Ootsubo (ISAS/JAXA, JP), M. Page (MSSL, UK), E. Pantin (CEA, Fr), E. Pascale (U. Cardiff, UK), I. Pérez-Fournon (IAC, E), S. Pezzuto (IFSI, It), A. Poglitsch (MPE, Ger), E. Pointecouteau (U. Toulouse, Fr), C. Popescu (U. Lancaster, UK), F. Pozzi (U. Bologna,It), W. Raab (MPE, Ger), G. Raymond (Cardiff U., UK), D. Rigopoulou (Oxford U., UK), I. Roseboom (U. Sussex, UK), H. Rottgering (Leiden U., NL), F. Selsis (LAB, Fr), S. Serjeant (Open U., UK), A. Smith (U. Sussex, UK), M. Spaans (Kapteyn Astr. Inst., Neth), L. Spinoglio (IFSI, It), E. Sturm (MPE, Ger), M. Takami (Academia Sinica,Taiwan), T. Takeuchi (Nagoya U., JP), S. Takita (ISAS/JAXA, JP), M. Tamura (NAO, JP), G. Tinetti (UCL, UK), S. Tommasin (IFSI, It), J. Torres (CAB/CSIC-INTA, E), R. Tuffs (MPIK, Ger), M. Vaccari (Oss. Astr. Padova, It), P. van der Werf (Leiden Obs., NL), S. Viti (UCL, UK), C. Waelkens (KUL, Be), R. Waters (UVA, NL), M. Wyatt (CAM, UK), T. Yamashita (NAO, JP).

This report has been supported by the ESA Science Planning and Community Coordination Office: Marcello Coradini, Philippe Escoubet, Fabio Favata and Timo Prusti.

Page intentionally left blank

# Contents

# Executive Summary

Understanding of the origin and evolution of galaxies, stars, planets, our Earth and of life itself are fundamental objectives of Science in general and Astronomy in particular. Although impressive advances have been made in the last twenty years, our knowledge of how the Universe has come to look as it does today is far from complete. A full insight of the processes involved is only possible with observations in the long wavelength infrared waveband of the electromagnetic spectrum. It is in this range that astronomical objects emit most of their radiation as they form and evolve in regions where obscuration by dust prevents observations in the visible and near infrared.

Over the past quarter of a century successive space infrared observatories (IRAS, ISO, Spitzer and AKARI) have revolutionised our understanding of the evolution of stars and galaxies. Mid to far infrared observations have led to stunning discoveries such as the Ultra Luminous Infrared Galaxies (ULIRGS), the basic processes of star formation from "class 0" pre-stellar cores through to the clearing of the gaseous proto-planetary discs and the presence of dust excesses around main sequence stars. The Herschel Space Observatory launched this year will continue this work in the far infrared and sub-mm and JWST, due for launch in 2014, will provide a major boost in observing capability in the 2 – 28 $\mu$m range.

Previous infrared missions have been hampered by the requirement to cool the telescope and instruments to $< 5$ K using liquid cryogens. This has limited the size of the apertures to $< 1$ m and our view of the infrared Universe has been one of poor spatial resolution and limited sensitivity. The Herschel mission addresses the first of these by employing a 3.5 m mirror to dramatically increase the available spatial resolution but, because it is only cooled to 80 K, only offers a modest increase in sensitivity in the 55 – 210 $\mu$m range compared to previous facilities. JWST will provide a major increase in both spatial resolution and sensitivity but only up to 28 $\mu$m.

It is in this context that the JAXA led mission SPICA (SPace Infrared telescope for Cosmology and Astrophysics) is proposed. SPICA is an observatory that will provide imaging and spectroscopic capabilities in the 5 to 210 $\mu$m wavelength range with a 3.5 m telescope like Herschel, but now cooled to a temperature less than 6 K. In combination with a new generation of highly sensitive detectors, the low telescope temperature will allow us to achieve sky-limited sensitivity over the full 5 to 210 $\mu$m band for the first time. This unique capability means that SPICA will be between one and two orders of magnitude more sensitive than Herschel in the far infrared band. SPICA will cover the full 5 to 210 $\mu$m wavelength range, including the missing 28 $\mu$m to 55 $\mu$m octave which is out of the Herschel and JWST domains, with unprecedented sensitivity and spatial resolution. Furthermore, SPICA will be the only observatory of its era to bridge the wavelength gap between JWST and ALMA, and carry out unique science not achievable at sub-mm wavelengths with ALMA. In the mid infrared SPICA will be able to carry out medium and high-resolution spectroscopy ($R \sim 30\,000$), one order of magnitude higher than in JWST, and will include spatial high multiplexing imaging and medium spectral resolution capabilities. In addition, the characteristics of the SPICA monolithic telescope will provide unique and optimal conditions for mid infrared coronagraphy in imaging and, uniquely, spectroscopic mode.

**The science objectives of SPICA** are closely linked to three of the ESA Cosmic Vision Themes:

Theme 1 - What are the conditions for planet formation and the emergence of life?;

Theme 2 - How does the Solar System work?;

Theme 4 - How did the Universe originate and what is it made of?

With its powerful scientific capabilities, SPICA will provide unique and ground-breaking answers to these key questions, and it is with this goal that the SPICA science objectives have been defined:

**(i) The formation and evolution of planetary systems:** By accessing key spectral diagnostic lines, SPICA will provide a robust and multidisciplinary approach to determine the conditions for planetary systems formation. This will include the detection for the first time of the most relevant species and mineral components in the gas and dust of hundreds of transitional protoplanetary discs at the time when planets form. SPICA will also be able to trace the warm gas in the inner ($< 30$ AU) disc regions and by resolving the gas Keplerian rotation, will allow us to observe the evolution of disc structure due to planet formation.

SPICA will study debris discs and make the first unbiased survey of the presence of zodiacal clouds in thousands of exoplanetary systems around all stellar types. It will allow us to detect both the dust continuum emission and the brightest grain/ice bands as well as the brightest lines from any gas residual present in the disc. SPICA will have the unique capability to observe water ice in all environments, and thus fully explore its impact on planetary formation and evolution and in the emergence of habitable planets. In the closest debris discs, SPICA will spatially resolve the distribution of water ice and determine the position of the "snow line", which separates the inner disc region of terrestrial planet formation from that of the outer planets.

SPICA will drastically enhance our knowledge of the Solar System by making the first detailed characterisation of hundreds of Kuiper Belt Objects, and of different families of inner, hotter centaurs, comets and asteroids. SPICA will provide the means to quantify their composition and determine unambiguously their size distribution: critical observational evidence for the models of Solar System formation. No other planned or present facility will be able to carry out these observations.

SPICA will provide direct imaging and low resolution mid infrared spectroscopy of outer young giant exoplanets (e.g., at $\sim 9$ AU of a star at 10 pc), which will allow us for the first time to study the physics and composition of their atmospheres in a wavelength range particularly rich in spectral signatures (e.g., $H_2O$, $CH_4$, $O_3$, silicate clouds, $NH_3$, $CO_2$) and to compare it with the planets of our Solar System. In addition, mid infrared transit photometry and medium or high resolution spectroscopy of "hot Jupiters" will be routine with SPICA.

SPICA will also provide an unprecedentedly sensitive window into key aspects of the dust life-cycle both in the Milky Way and in nearby galaxies, from its formation in evolved stars, its evolution in the ISM, its processing in supernova-generated shock waves and massive stars, to its final incorporation into star forming cores and protoplanetary discs.

**(ii) The formation and evolution of galaxies**: SPICA observations will provide a unique insight into the basic questions about how galaxies form and evolve such as: What drives the evolution of the massive, dusty distant galaxy population, and what feedback/interplay exists between the physical processes of mass accretion and star formation? How and when do the normal, quiescent galaxies such as our own form, and how do they relate to (Ultra) Luminous Infrared Galaxies (ULIRGs)? How do galaxy evolution, star formation rate and AGN activity vary with environment and cosmological epoch?

Substantial progress in this area can only be made by making the transition from large-area photometric to large-area spectroscopic surveys in the mid to far infrared, which will be possible with SPICA. This is because the mid and far infrared region plays host to a unique suite of diagnostic lines to trace the accretion and star formation, and to probe the physical and chemical conditions in different regimes from AGN to star-forming regions. While the Herschel-PACS spectrometer will detect the brightest far infrared objects at $z \sim 1$, SPICA will be able to carry out blind spectroscopic surveys out to $z \sim 3$. This will lead to the first statistically unbiased determination of the co-evolution of star formation and mass accretion with cosmic time. Spectroscopic surveys will provide direct and unbiased information on the evolution of the large scale structure in the Universe from $z \sim 3$ and the unprecedented possibility to investigate the impact of environment on galaxy formation and evolution as a function of redshift.

The high sensitivity of SPICA will enable photometric surveys beyond $z \sim 4$ that will resolve more than 90% of the Cosmic Infrared Background (in comparison with 50% that Herschel will achieve). SPICA will also observe Milky Way type galaxies in the far infrared out to $z \sim 1$, where the cosmic star formation rate peaks. For the first time, we will be able to piece together the story of the evolution of our own galaxy and answer the question of whether we are in a "special place" in the cosmos.

The science objectives of SPICA have led to the identification of the **mission science requirements** and these, in turn, have informed the assessment study for the mission and the payload. This study has been carried out by JAXA, ESA and the consortium of the European far infrared instrument, SAFARI, for their respective contributed elements. In particular, **ESA's contribution** to the SPICA mission consists of the cyrogenic telescope assembly, a ground station, a collaboration in the science operations and the management of the SAFARI instrument interfaces to JAXA.

The 3.5 m **SPICA telescope** is the main ESA contribution to the mission. It needs to operate at a temperature of $< 6$ K in order for SPICA to achieve the unprecedented sky-limited sensitivity. The image quality requires diffraction limited performance at 5 $\mu$m with a wavefront error $< 350$ nm rms, and an unvignetted field of view larger than 12 arcmin. More stringent requirements on image quality, telescope pupil obscuration and telescope transmission are imposed by the SPICA coronagraph, and have been considered as a delta in the study phase. Two parallel competitive contracts were awarded to EADS Astrium (F) and Thales-Alenia (F) for the industrial assessment study. The results have confirmed the Ritchey-Chrétien configuration as the most suitable to meet the performance requirements. Both contractors have identified a design entirely based on ceramic material (SiC100 for Astrium and HB-Cesic for Thales) for the optical surfaces (M1 and M2), as well as for the overall telescope structure. Both assessment studies have shown that the mass, heat load, and wavefront error budgets are compliant with the JAXA allocations and the scientific requirements, including those from the coronagraph. The reference development schedule of the SPICA Telescope assumes an early freezing of all design requirements during the ESA Definition Phase (A/B1). In this scenario, the delivery of the telescope to JAXA could take place by Q4/2016, leading to a launch in 2018. Inclusion of the coronagraph requirements would extend polishing of an additional 6 months.

The **scientific instruments** on board SPICA are: a mid infrared camera and spectrophotometer, MIRACLE (5 – 38 $\mu$m, $6' \times 6'$ FOV), a mid infrared medium resolution spectrometer, MIRMES (10.32 – 36.04 $\mu$m, $R \sim 900 - 1500$), a mid infrared high resolution spectrometer, MIRHES (4 – 18 $\mu$m, $R \sim 20\,000 - 30\,000$), a mid infrared coronagraph, SCI (5(3.5) – 27 $\mu$m, $R \sim 20 - 200$, contrast $10^{-6}$, inner working angle $3.3\lambda/D_{tel}$), an optional far infrared spectrometer, BLISS (38 – 430 $\mu$m, $R \sim 700$), and a far infrared imaging spectrometer, SAFARI. **SAFARI**, which is to be built by a consortium of European institutes (with Canadian and Japanese participation), is an imaging Fourier Transform Spectrometer designed to provide continuous coverage in photometry and spectroscopy from 34 to 210 $\mu$m, with a field of view of $2' \times 2'$ and spectral resolution modes $R = 2000$ (at 100 $\mu$m), $R \sim$ few hundred and $20 < R < 50$. The spectral sensitivity is required to be $\sim 3 \times 10^{-19}$ W m$^{-2}$ at 48 $\mu$m (5$\sigma$, 1 hour). Four detector technologies are undergoing development to meet these requirements: TES bolometers, Ge:Ga photoconductors, Si-bolometers and KID detectors. Several of the detector technologies have already demonstrated that they should be capable of delivering the required sensitivity for SAFARI and all are undergoing a development programme aimed at proving their compatibility with the SPICA system - the final detector selection will be made by mid 2010.

SPICA will be launched with JAXA's H-IIB from the Tanegashima Space Centre and is planned as a nominal three year mission (goal 5 years) orbiting at $L_2$. A large amplitude halo orbit is baselined, with a period of around 180 days and a semi-major axis amplitude of about 750 000 km. The thermal environment required by the telescope and the instruments is maintained by a combination of passive cooling (via dedicated solar and thermal shields combined with radiators) and active cooling, using a number of mechanical coolers to provide base temperatures of 4.5 K and 1.7 K. The total mass of the spacecraft is about 4 tonnes and it has a width of 4.5 m and a height of 7.5 m.

SPICA is an **observatory** open to the world astronomical community. The "Guaranteed" time for the instrument builders will be 25% of the total operational time; and 60% will be opened to the scientific community ("Open time"), of which 22% will be reserved to astronomers from the ESA member states, 45% to scientists from Japan and other contributing countries and 33% to astronomers of any nationality. Proposals for "Legacy" and "normal" observing programmes will be selected by a single Time Allocation Committee through Announcements of Opportunity. A Science Advisory Committee will be established by JAXA, which will be chaired by the Japanese SPICA Project Scientist and in which the ESA Project Scientist will participate. In parallel, the ESA SPICA Science Team will be organised. JAXA will carry out the satellite and science **operations** in centres in Japan. A European SPICA Data Centre will be established, which will consist of the SAFARI Instrument Control Centre and the ESA SPICA Science Centre. The latter, located at ESAC, will be responsible for community support to the European astronomers for all observatory capabilities, and will be the interface between SAFARI and JAXA for operational and ground segment software development matters.

JAXA will be responsible for the overall SPICA project. ESA will be the prime project partner, and will be responsible for the delivery of the Telescope Assembly and for the formal delivery of the SAFARI instrument. The SAFARI is a nationally funded, Principal Investigator provided instrument, led by SRON (The Netherlands). A **SPICA Steering Group** will be established early in the programme. It will be chaired by a JAXA representative and will consist of members from JAXA, ESA and potentially other funding agencies. In 2010 the SPICA Steering Group will draft the content of a formal agreement between both agencies.

Page intentionally left blank

# Chapter 1

# Scientific Objectives

## 1.1 Introduction

A full understanding of the formation and evolution of galaxies, stars and planets can only be found through the investigation of the cold and obscured parts of the Universe where the basic processes of formation and evolution occur. The deep exploration of the cold Universe through high spatial resolution observations in the Far Infrared (FIR) and sub-mm started in earnest in 2009 with the launch of the Herschel Space Observatory. Herschel will open a new window on this almost unexplored Universe through photometric surveys of star formation in our own Galaxy and of distant galaxies, however its relatively warm telescope (between 82 and 90 K) greatly limits its sensitivity. A new mission reaching better sensitivity must therefore follow to continue the work started by Herschel and gain a deeper understanding of the physics of the objects discovered by Herschel: i.e., to spectroscopically measure star formation rates in the obscured extra-galactic sources at different cosmic epochs and explore the physics and chemistry of planet formation. To obtain this increase in sensitivity we need a cold (< 6 K) telescope of at least the same diameter as Herschel. The Japanese led SPace Infrared telescope for Cosmology and Astrophysics (SPICA) mission promises this in 2018 and Europe can play a vital role in its success.

Progress in mid- and far infrared astronomy has been slow because instruments and telescopes must be cooled to cryogenic temperatures to achieve high sensitivity and, for most of the frequency range, the observations can only be made from the space. Only four small space observatories have operated in the past quarter of a century (IRAS, ISO, Spitzer and AKARI) offering limited spatial resolution and sensitivity. Herschel, with a 3.5-m 80-K telescope, the largest telescope ever put in space, is providing greatly increased spatial resolution with modest increases in sensitivity over the 55 – 210 $\mu$m band, and JWST, with a ~ 6-m 45-K telescope, will provide a leap in both spatial resolution and sensitivity in the mid infrared (MIR) up to 28 $\mu$m. However, coverage of the full MIR/FIR band with high sensitivity and spatial resolution will still not have been achieved and it is here that SPICA will be a breakthrough mission. SPICA will have a similar size telescope to Herschel but cooled to < 6 K, thus removing its self-emission and allowing observations limited only by the astronomical background.

SPICA offers a sensitivity up to two orders of magnitude better than Herschel covering the mid-to-far-IR (the full 5 – 210 $\mu$m range, mostly unreachable from the ground) with imaging, spectroscopic and coronagraphic instruments (see Figure 1.1). Thanks to this tremendous increase in sensitivity, SPICA will make photometric images in a few seconds that would take hours for Herschel and will take a full 5 – 210 $\mu$m infrared spectrum of an object in one hour that would take several thousand hours for Herschel. We illustrate this major increase in sensitivity in Figure 1.2 (left panel) which shows the area covered by a full spatial and spectral survey in which we can detect *spectroscopically* all galaxies down to a luminosity of $\sim 10^{11}\,L_\odot$ at $z = 1$ and $\sim 10^{12}\,L_\odot$ at $z = 2$ in 900 hours. In theory, in about twice this time, the Herschel-PACS spectrometer would just be able to detect a single object over its full waveband to the same sensitivity. We can immediately see that this major increase in sensitivity, combined with a wide field of view and coverage of the full 5 – 210 $\mu$m waveband, will revolutionise our ability to spectroscopically explore the nature of the thousands of objects that Herschel,

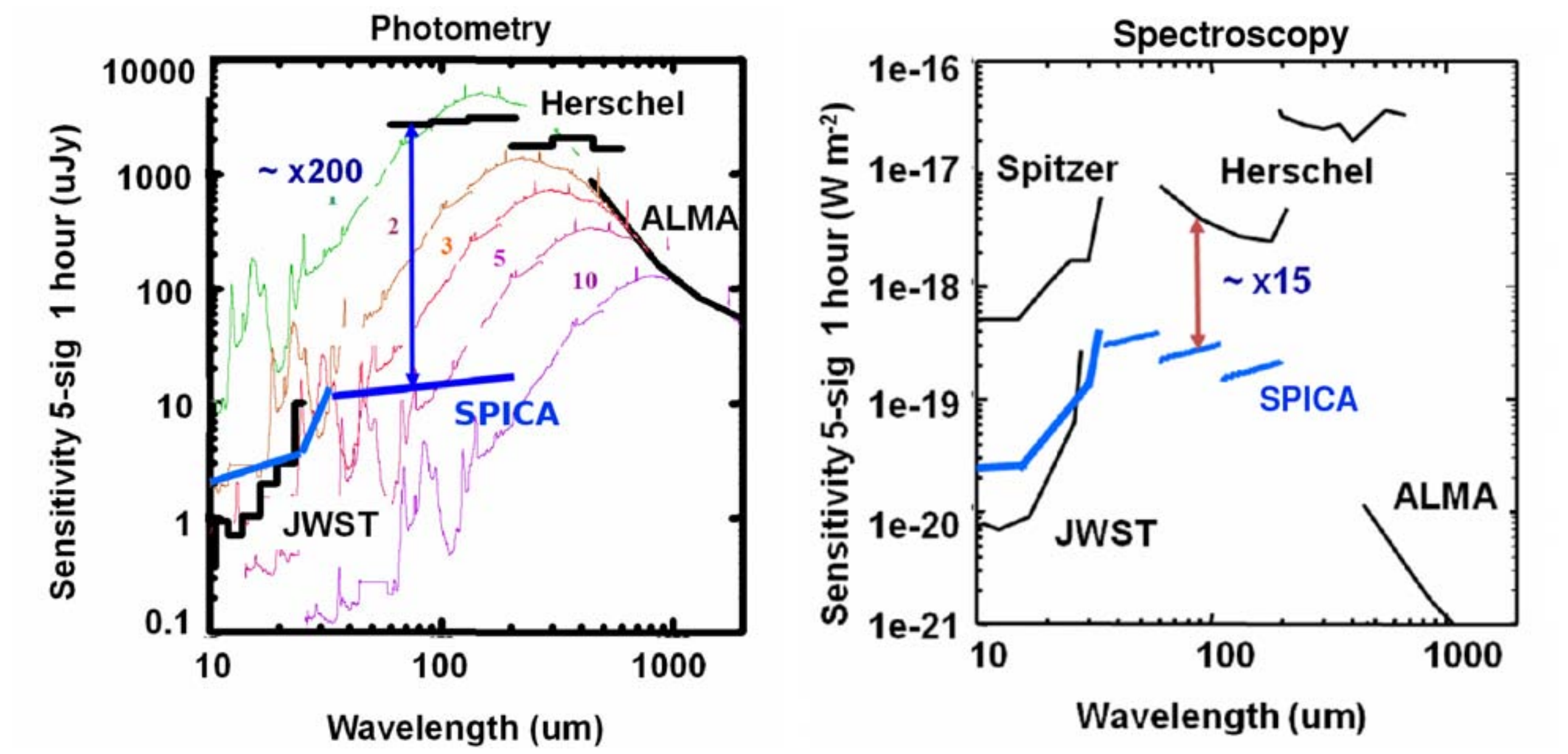

*Figure 1.1:* ***Left panel****: Photometric performance expected for SPICA (blue), compared to Herschel, ALMA and JWST (black), for a point source (in μJy for 5σ in 1 hour) using the goal sensitivity detectors on SPICA (NEP* $= 2 \times 10^{-19}$ *W Hz*$^{-1/2}$*). Note the ∼ 2 orders of magnitude increase in FIR photometric sensitivity compared to Herschel-PACS. The effects of confusion, and their mitigation, are discussed later. For illustrative purposes the SED of the starburst galaxy M82 as redshifted to the values indicated is shown in the background.* ***Right panel****: Spectroscopic performance expected for SPICA (blue) compared to predecessor and complementary facilities (black) for an unresolved line for a point source in W m*$^{-2}$ *for 5σ in 1 hour. For ALMA 100 km s*$^{-1}$ *resolution is assumed. The SPICA MIR sensitivities are scaled by telescope area from the JWST and Spitzer-IRS values respectively. Note that SPICA becomes more sensitive than JWST beyond 20 μm.*

JWST, and SPICA will discover in photometric surveys.

In the rest of this section we first place SPICA in context within the ESA Cosmic Vision science goals before discussing the critical importance of observing in the mid to far infrared and the diagnostic tools that are available. We then describe in detail how SPICA will shed light on the processes of planetary formation in the local Universe and the formation and evolution of galaxies in the more distant Universe.

### 1.1.1 SPICA in the Cosmic Vision

The ambition set out in the ESA Cosmic Vision (*ESA BR-247 2005*) is to seek "... the answers to profound questions about our existence, and our survival in a tumultuous cosmos". To do this European scientists have identified four grand themes in space science that will bring us closer to understanding how the Universe has come to look as it does and the place of our Earth within the Cosmos. Three of these themes directly require observations in the mid to far infrared and a space based far infrared observatory is identified as a key facility within the aegis of Cosmic Vision. We recapitulate those themes here and briefly state how SPICA will directly contribute to furthering our knowledge of the Universe:

**Theme 1: "What are the conditions for planet formation and the emergence of life?"** The Cosmic Vision calls for a mission that will "place the Solar System into the overall context of planetary formation, aiming at comparative planetology" and "Search for planets around stars other than the Sun". SPICA will have a state of the art MIR coronagraph that will allow imaging and spectroscopy of young massive planets for the first time.

There is also the call to, "Investigate star-formation areas, proto-stars and proto-planetary discs and find out what kinds of host stars, in which locations in the Galaxy, are the most favourable to the formation of planets" and "Investigate the conditions for star formation and evolution". SPICA will have an FIR spectrometer more than an order of magnitude more sensitive than any previous facility to probe further and into a wider range of objects and regions than ever before.

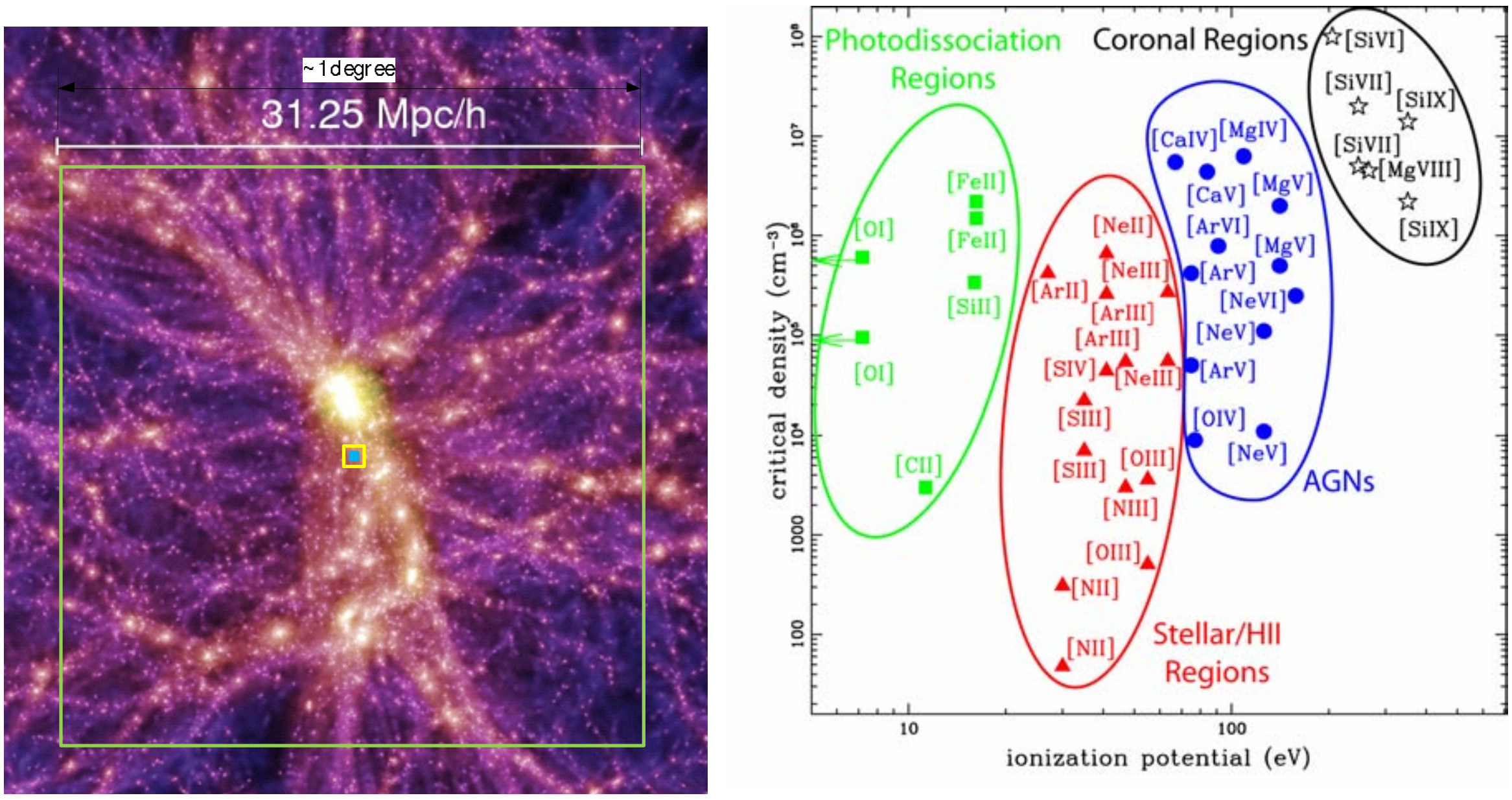

*Figure 1.2:* ***Left panel****: Spectroscopic mapping speeds of the SPICA far infrared instrument (SAFARI) and Herschel-PACS superposed on a realisation of the Millenium simulation at* $z \sim 1.4$ *(Springel et al. 2006). In the centre are the footprints of the instantaneous spectroscopic FOV of PACS (blue) and SAFARI (yellow). The large green box shows the area covered by SAFARI in a 900 hour spectral full wavelength spectral survey (~ 1 degree). In this survey SAFARI will see down to* $5 \times 10^{-19}$ $W\,m^{-2}$ *over the full 34 – 210* $\mu$*m band. PACS would require approximately twice this time just to cover a single pointing to this depth over its full waveband.* ***Right panel****: A selection of the fine-structure atomic and ionic lines accessible with SPICA, plotted as a function of critical density and ionisation potential. Using ratios between lines with different ionisation or critical density, we can trace out a wide range of different physical/excitation conditions.*

**Theme 2: "How does the Solar System work?"** This theme calls for the study of asteroids and comets as they are "the most primitive small bodies that can give clues to the chemical mixture and initial conditions from which the planets formed in the early solar nebula". SPICA's high sensitivity spectroscopy will allow the chemistry of these objects to be studied by remote sensing to an unprecedented level of detail, for the first time allowing us to link our own "debris" to that seen in the formation of planetary systems around other stars.

**Theme 4: "How did the Universe originate and what is it made of?"** Here SPICA is ideally suited to "…trace the subsequent co-evolution of galaxies and super-massive black holes" and can "Resolve the far infrared background into discrete sources, and the star-formation activity hidden by dust absorption". The combination of MIR and FIR spectroscopy on SPICA is also essential to "Trace the formation and evolution of the super-massive black holes at galactic centres – in relation to galaxy and star formation – and trace the life cycles of chemical elements through cosmic history".

In the following sections we illustrate in detail how observations with the full suite of instruments on SPICA are essential to answering the questions posed in these themes and show how a European involvement in SPICA is an essential part of Europe's Cosmic Vision for the 21$^{st}$ century. We first set the context of why observations in the mid to far infrared are critical to providing the observational evidence for galaxy, star and planetary formation and evolution.

### 1.1.2 A critical waveband

The composite spectrum shown in Figure 1.3 shows the importance of observations in the mid to far infrared. Here we see the spectral energy distribution (SED) from the X-rays to the radio of a typical galaxy undergoing a modest amount of star formation, showing the different physical components at work. The bulk of the radiation is emitted either in the optical (i.e., starlight) or in the mid to far infrared where the dust in the interstellar medium absorbs and re-radiates the starlight as grey body radiation. Superimposed on the contin-uum throughout the 5 – 210 $\mu$m range are fine structure lines from ions, atoms and molecular lines which, due to their low excitation potentials and because they are not readily absorbed, control the thermal energy balance in a large variety of physical conditions. The gas reservoirs cool through an extensive network of atomic, ionic and molecular lines, which include the fine-structure lines of carbon and oxygen, ionic lines of neon and sulphur as well as many rotational transitions of molecules such as hydrogen ($H_2$), water, hydroxyl and carbon monoxide. The lines can be strong, emitting up to a percent of the total far infrared luminosity. Through observations of combinations of lines it is possible to disentangle the complex dependencies and so to start to characterise the physical properties that parameterise star-formation such as the nature and strength of the interstellar radiation fields, chemical abundances, local physical temperatures and gas densities. As well as the low excitation/ionisation tracers measuring the star formation component in galaxies, the MIR contains important diagnostic lines that can reveal the presence of Active Galactic Nuclei (AGN), such as the [Ne v] and [O iv] emission lines. These lines are not severely affected by extinction, while optical and UV lines of forbidden and permitted transitions are heavily reddened and even X-rays are blocked by the large columns of hydrogen absorption in Compton thick AGN type 2 objects. Therefore observations in the mid-to-far infrared gives access to a large set of diagnostic line ratios to measure the physical conditions of astrophysical sources. Figure 1.2 (right panel) shows the mid- and far-IR lines from many different species, with a range of different ionisation potentials and excitation conditions. Used together, these transitions place constraints over a wide space of physical conditions and phases of the ISM, from the neutral atomic ISM, through the ionised ISM as seen in photo-dissociation regions and HII regions to the highly ionised AGN and "coronal" regions.

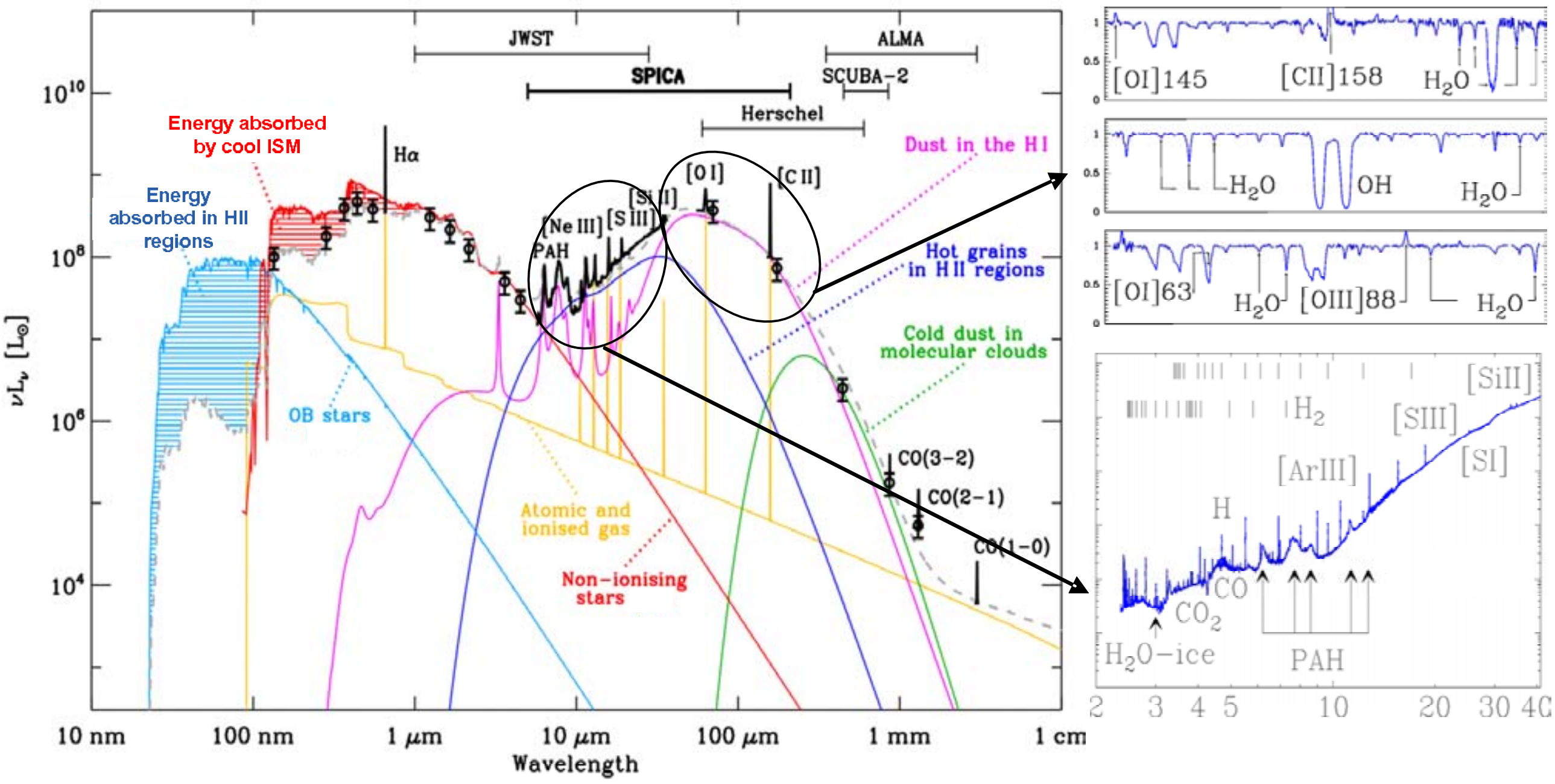

*Figure 1.3: A synthetic spectrum of a typical galaxy undergoing modest rates of star formation – similar to the Milky Way – showing the parts of the spectrum that will be covered by SPICA and contemporaneous facilities (cf. F. Galliano). The major importance and spectral richness of the mid to far infrared region is demonstrated by the detailed spectra from ISO shown to the right of the diagram (Rosenthal et al. 2000; Goicoechea et al. 2004; Polehampton et al. 2007).*

These powerful diagnostics can be used to study both star and planetary formation very locally, and star formation and the influence of AGN in the very distant Universe. For example, the wavelength region between 5 and 40 $\mu$m contains transitions of many ionised species that can be used to differentiate between the role of Active Galactic Nuclei (AGN) and starburst regions in the evolution of galaxies (see section 1.6.1) and, as these and other MIR lines are redshifted into the FIR, they can trace the star formation history of the Universe. Similarly, the most abundant "metals", C, N and O, have atomic and ionic fine structure emission lines in the FIR that allow direct, and unambiguous, determination of their relative abundance in distant galaxies. In the more local Universe, the MIR and FIR line ratios of these species can be used as direct probes of the temper-ature, density and UV fields present in star formation regions, Planetary Nebulae, Supernova Remnants and proto-planetary discs. In addition to the atomic and ionic species, important molecular reservoirs of H, C, N and O, such as $H_2$, OH, $H_2O$, CO, $HC_xN$ and carbon clusters ($C_x$), have emission features throughout the mid to far infrared, which are uniquely diagnostic of the physical conditions of the sources where they arise and, in the case of water, are critically significant in any discussion about how and where the conditions for life have emerged. The $H_2$, OH, $H_2O$ and high-J lines of CO can only be observed from space and require high sensitivity coupled with sufficient spatial and spectral resolution to distinguish the different environments within complex emission regions. Importantly for cosmogonological studies, the deuterated species of hydrogen bearing molecules (HD, HDO, OD, etc.) also have transitions through the mid and far infrared that can only be adequately accessed from space. Observation of these molecules in a variety of objects and comparison with the terrestrial deuterium ratio gives a direct insight into how the Earth formed and the origins of the oceans.

Dust emission and absorption dominates the continuum spectrum in the MIR/FIR range with the most prominent features of MIR continua spectra being the UV excited emission bands of PolyAromatic Hydrocarbon molecules (PAHs) at 6.2, 7.7, 8.6 and 11.3 $\mu$m. These PAH molecules appear to be omnipresent in all phases of dust evolution both in our own galaxy and the most distant galaxies yet seen using MIR spectroscopy. One of the most important features of PAHs is that they are excited by absorption of single UV photons; consequently, and in contrast to emission from larger grains, their output does not depend on their distribution around the heating source. Thus they are relatively insensitive to radiation from older stellar populations, making them good tracers of the UV field and so of star formation as a function of redshift. Also present in the 5 –100 $\mu$m range are diagnostic spectral features arising from the minerals that make up the larger interstellar dust grains, as well as solid state features from the ices that condense onto them under the right physical conditions. Particularly prominent in the MIR range are the “silicate” feature at $\sim$ 9.7 $\mu$m – often seen in absorption in extragalactic spectra – and another at 20 – 24 $\mu$m. Some solid state features, for instance that at 69 $\mu$m from forsterite, can also be used as a direct measurement of the dust temperature. All these features are broad, faint and difficult to detect with the current or near future generation of facilities, which are either unsuitable for the task or do not have the full uninterrupted wavelength coverage required. Only the exquisite, and unrivalled, sensitivity of SPICA in the full mid to far infrared range will allow astronomers to trace the full story of dust evolution in our own and distant galaxies.

This overview of the power of the mid and far infrared “toolbox” is of necessity brief and incomplete; in the remainder of this chapter we will give specific examples of how these tools can be used to answer some of the most vital questions about how and when planets, stars and galaxies came to form and their subsequent evolution into the Universe we see today.

## 1.2 Planetary systems formation and evolution

Modern astrophysics is just beginning to provide answers to one of the most basic questions about our place in the Universe: Are Solar Systems like our own common among the millions of stars in the Milky Way and, if so, what implications does this have for the occurrence of exoplanets that might give rise to life? The most straightforward method we have to start answering these questions is to compare our current knowledge of our own Solar system evolution with observations of planets and planetary disc systems around stars other than the Sun. Observations of the most primitive bodies in the Solar System are also critical to infer the physical and chemical conditions in the early solar nebula as well as to provide clues about the water abundance, the most obvious solvent for life, its distribution, and its transfer from the outer Kuiper Belt regions to the inner terrestrial planet regions. The basic building blocks of an extra-solar system, the gas and dust, emit predominantly at mid infrared and far infrared wavelengths, and thus this is the critical domain in which to unveil the processes that transform the interstellar gas and dust into stars and into planetary systems.

We discuss below how only the very sensitive observations over the entire mid infrared to far infrared (MIR and FIR) wavelength range provided by SPICA will give access to key spectral diagnostics and provide a robust and multidisciplinary approach to determine the conditions for planetary systems formation. Such a comprehensive study will include the first detailed characterisation of hundreds of pristine bodies in our own

Solar System, the detection for the first time of the most relevant chemical species and mineral components of hundreds of transitional protoplanetary discs at the time when planets form, the spatial resolution of the water *snow line* in the closest and more evolved planetesimals debris discs, the first unbiased survey of the presence of zodiacal clouds in thousands of exoplanetary systems around all stellar-types and the first direct determination of the chemical composition of outer exoplanet atmospheres. These and other challenging science goals require more than an order of magnitude sensitivity improvement in the FIR compared to Herschel and higher spectral resolution than provided by JWST in the mid infrared. Additionally, direct spectroscopic coronagraphy of exoplanets and planetary discs in the critical mid infrared domain is not planned in any space telescope other than SPICA.

### 1.2.1 Protoplanetary and debris discs

All planets are thought to form in the accretion discs that develop during the collapse and infall of massive dusty and molecular cocoons ($\gtrsim$ 10 000 AU) where stars are born. However, we still have a very incomplete understanding of the physical and chemical conditions in such discs, how they evolve when dusty bodies grow and collide, their mineral content, how they clear as a function of time and, ultimately, how planets as diverse as the Earth or hot-Jupiters form around different types of stars and at different places of the galaxy.

Circumstellar discs are classified into 3 classes according to their evolutionary stage. **Primordial protoplanetary discs**, which are rich in atomic and molecular gas, and are composed of relatively unprocessed interstellar material left over from the star formation process. Such young discs are very optically thick in dust, with high radial midplane optical depths in the visual ($\tau_V >> 1$). These protoplanetary discs start to become optically thin ($\tau_V \simeq 1$) in a few million years after formation (Haisch et al. 2001) and evolve into **transitional discs** as their inner regions begin to clear at $\gtrsim$ 10 Myr. This is the critical intermediate stage when planetary formation is believed to take place, with dust particles colliding and growing to form larger bodies reducing the disc opacity. Spitzer has shown that their outer regions (beyond $\sim 10 - 20$ AU) can remain intact for longer, and thus residual gas can exist in the disc and play an important role in its evolution. In spite of its importance, the gas content in transitional discs at the first stages of planet formation is very poorly constrained, and many clues on these early stages of planetary formation can be provided by spectral line observations. Discs with ages above $\gtrsim$ 10 Myr are thought to be practically devoid of gas (Duvert et al. 2000) and the dust in these older debris discs is generally not primordial but continuously generated “debris” from planetesimal and rocky body collisions. The smallest dust grains have, at this stage, either been dispersed or have coagulated into larger grains and the disc becomes very optically thin ($\tau_V << 1$). **Debris discs** are thus younger and more massive analogs of our own asteroid (hot inner disc, $T_d \simeq 200$ K) and Kuiper belts (cool outer disc, $T_d \simeq 60$ K) so their study is vital to place the Solar System in a broader context.

#### The gas content in transitional, planet forming discs

The physical and chemical conditions in young protoplanetary discs (*primordial* and *transitional*) set the boundary conditions for planet formation, and an understanding of the formation and evolution of such discs will finally link star formation and planetary science. Planets are believed to form in transitional protoplanetary discs, which emit most of their energy in the MIR/FIR region (Boss 2008). Although the dust is relatively easily detected by photometric observations in the far infrared range, very little is known about the gas phase. It seems clear that an abrupt transition from massive optically thick discs to tenuous debris discs occurs at $\gtrsim$ 10 Myr (Meyer et al. 2008), when the majority of at least the giant, gaseous planets (e.g., Jupiters and “hot Jupiters”) must have formed. The very fact that these planets are largely gaseous means they must be formed before the gaseous disc dissipated, making the study of the gas in discs essential to understand how and where they formed. For instance, the recently detected massive “hot Jupiters” orbiting close to the parent star are very unlikely to have formed in situ, but rather they must have migrated inwards from outer disc regions beyond the *snow line*. Likewise it seems clear that Neptune and Uranus must have formed in a region closer to the Sun and migrated outwards. The mechanism for either of these scenarios is not certain.

According to our current theories, gas giant planets are thought to form in discs either via accretion of gas onto rocky/icy cores of a few earth masses (Lissauer 1993; Pollack et al. 1996; Kornet et al. 2002) or by gravitational instability in the disc that triggers the formation of overdense clumps that afterwards compress to form giant planets (Boss 2003). In the latter scenario, gas giants form quickly and the gas may dissipate early ($\lesssim 10$ Myr), whereas longer gas disc lifetimes ($\gtrsim 10$ Myr) may leave enough time for building a rock/ice core and ease the subsequent accretion of large amounts of gas. Therefore, observations of the amount of gas in transitional discs around a large sample of stars can discriminate between the two most accepted planet forming theories. The residual gas content in the innermost regions ($< 2$ AU for a Sun-like star but much larger for more massive stars) at the time terrestrial rocky protoplanets assemble will determine their final mass, chemical content and orbit eccentricity, and therefore its possible habitability (Agnor & Ward 2002).

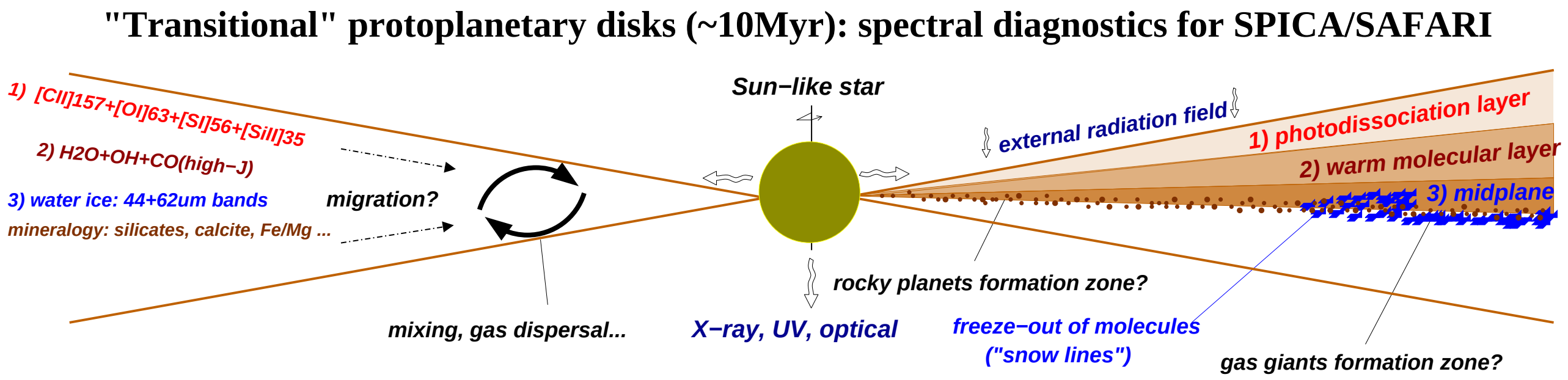

*Figure 1.4: Diagram showing where radiation arises from a protoplanetary disc (at the time when planets assemble) and why the FIR (i.e., SAFARI) and the millimetre (e.g., ALMA) are essential to understanding the full picture of planetary formation and the primordial chemistry that leads to the emergence of life.*

Protoplanetary disc models predict a “flared” disc structure (see Figure 1.4) which allows the disc to capture a significant portion of the stellar UV and X-ray radiation even at large radii (Qi et al. 2006), boosting the MIR/FIR dust thermal emission and the MIR/FIR lines emission from gas phase ions, atoms and molecules. Theoretical models of protoplanetary discs recognise the importance of these X-UV irradiated surface layers, which support active photochemistry and are responsible for most of the MIR/FIR line emission. They are, in many ways, similar to the well studied interstellar photodissociation regions (PDRs). State-of-the-art 2D and 3D disc/PDR models are nowadays able to simulate the disc dynamics, thermodynamics, chemistry and radiative transfer (Gorti & Hollenbach 2004; Aikawa et al. 2002; Gorti & Hollenbach 2008; Woitke et al. 2009a,b; Cernicharo et al. 2009) and they guide us in the interpretation of the observed disc emission (generally spatially unresolved). In particular, the FIR fine structure lines of the most abundant elements and metals (O, C, S, Si...) together with the FIR rotational line emission of light hydrides ($H_2O$, OH...) are predicted to be the strongest gas coolants (i.e., the brightest lines) of the warm disc, especially close to the star. In particular the FIR [Si II] 34 $\mu$m, [O I] 63 $\mu$m, [S I] 56 $\mu$m and [C II] 158 $\mu$m fine structure lines are likely the most intense lines emitted in the disc, and thus the best diagnostic of the gas content in discs. **By detecting these FIR lines we can directly probe a wide range of physical and chemical conditions (e.g., those associated with the UV-illuminated disc regions) that are very difficult, if not impossible, to trace at other wavelengths** (e.g., by ALMA). Herschel can only search for the strongest FIR lines (above a few $10^{-18}$ W m$^{-2}$) toward the brightest, closest and most massive young protoplanetary discs. In fact, only a few discs will be fully surveyed in the FIR domain, i.e., those with strong dust continuum which unfortunately hampers line detection at low spectral resolution. In particular, Herschel does not reach the sensitivity to either detect the gas emission in the much more tenuous transitional discs or detect the predicted line emission of less massive, and thus more numerous, discs around cooler stars.

***SPICA studies of the gas dispersal and of the chemical complexity in protoplanetary discs:***
In order to shed some light on the gas dispersal time scales, and thus on the formation of gaseous Jovian-type planets, high sensitivity infrared to sub-mm spectroscopic observations over large statistical stellar samples tracking all relevant disc evolutionary stages and stellar types are clearly needed. To date, studies of the gas content are biased to young and massive protoplanetary discs (probably not the most representative) through observations with ground-based (sub)mm interferometers and optical/near-IR telescopes. Sub-mm observations allow one to trace the outer cooler disc extending over a few hundred AU where most molecular species start to freeze-out onto dust grains (Dutrey et al. 2007). Recently the very inner regions of protoplanetary discs have been probed by means of optical/IR observations (Najita et al. 2007). Spitzer and ground-based telescopes have just started to show the potential diagnostic power of MIR/FIR spectroscopy in a few “template” discs. This has allowed the exploration of the gas content and composition at intermediate radii ($1-30$ AU), i.e., the crucial region for the formation of planets. Recent MIR detections toward young discs include atomic lines such as [Ne II], [Fe II], etc. (Lahuis et al. 2007), molecules like $H_2$, $H_2O$, OH, HCN, $C_2H_2$ and $CO_2$ (Carr & Najita 2008; Salyk et al. 2008), and complex organics like PAHs (Geers et al. 2006; Habart et al. 2006).

Molecular hydrogen ($H_2$) is the most abundant gas species in a primordial protoplanetary disc ($\sim 90\%$ of the initial mass). Due to the relatively poor line sensitivity that can be achieved with even the largest ground-based mid infrared telescopes (a few $10^{-17}$ W m$^{-2}$), $H_2$ has been detected only toward a few protoplanetary discs so far (Bitner et al. 2008). JWST will improve this situation enormously and many more discs will have been observed by the time SPICA is active. To complete and complement the JWST capability, SPICA’s mid infrared high resolution echelle spectrometer (MIRHES) will be able to detect the brightest $H_2$ line (the $v$=0-0 S(1) line at $\sim 17\ \mu$m) with $\sim 10$ km s$^{-1}$ resolution with much higher sensitivities than those achieved from the ground and with an order of magnitude higher spectral resolution than JWST. Such a high spectral resolution will be enough to start resolving the gas Keplerian rotation, trace the warm gas in the inner ($< 30$ AU) disc regions and be sensitive to $H_2$ masses of the order of $\sim 0.1\ M_{earth}$. However, given the peculiar excitation conditions of $H_2$, its emission does not sample the whole circumstellar disc. Instead, the deuterated isotopologue of molecular hydrogen, HD (with the lowest energy lines at 112, 56 and 37 $\mu$m), although a few thousand times less abundant, can serve as a proxy of $H_2$ column density. The D/H abundance ratio of different species can also be used as a diagnostic of the interstellar origin of the young disc material.

In terms of chemistry, the protoplanetary disc is the major reservoir of key species with prebiotical relevance, such as oxygen, ammonia ($NH_3$), methane ($CH_4$) or water ($H_2O$) to be found later in (exo)planets, asteroids and comets. But how the presence and distribution of these species relates to the formation of planets, and most particularly, rocky planets with substantial amounts of water present within the so called habitable zone, remains open to speculation without a substantial increase in observational evidence. Water is an obvious ingredient for life, and thus it is very important that we understand how water transfers from protostellar clouds and primordial protoplanetary discs to more evolved asteroids, comets and planets like our own. Ultimately one has to understand how the water we see today in our oceans was delivered to the Earth.

At a distance of $\sim 50$ pc from the Sun, a given gas line will produce a line flux of $\sim 10^{-19}(L_{line}/10^{-8}\ L_\odot)$ W m$^{-2}$ where $L_{line}$ is the radiated power in the line expressed in solar luminosities. SAFARI will be able to detect gas lines with luminosities $\gtrsim 10^{-8}\ L_\odot$ in nearby, isolated discs, or $\gtrsim 10^{-7}\ L_\odot$ in more distant discs in the closest star forming regions at $\sim 150$ pc (see Figure 1.5). State-of-the-art models of transitional discs around Sun-like stars and with gas masses as low as $\sim 0.001\ M_{Jup}$ ($\sim 0.3\ M_{earth}$) predict luminosities around $10^{-8}\ L_\odot$ for [O I]145 $\mu$m, [S I]56 $\mu$m and $H_2O$, OH or CO high-$J$ emission lines (Gorti & Hollenbach 2004). The [O I]63 $\mu$m line is expected to be the brightest line ($10^{-6} - 10^{-7}\ L_\odot$), even for modest gas masses, and therefore it is one of the most robust gas tracers in transitional discs. The exciting prospect of detecting this small amount of gas in a statistically significant sample of discs, e.g., toward the 6 closest ($\lesssim 150$ pc) young stellar clusters with ages of $\simeq 1-30$ Myr, Taurus, Upper Sco, TW Hya, Tuc Hor, Beta Pic and Eta Cha, will provide a crucial test on the lifetime of gas and its dissipation timescales in planet forming discs and therefore on the exoplanet formation theory itself.

Nevertheless, the angular resolution of SPICA ($\sim 4''$ at the [O I]63 $\mu$m line) can not compete with the resolution of telescopes with much larger collecting area (e.g., ALMA) and thus most protoplanetary discs will be spatially unresolved (a young disc like TW Hya, at a distance of $\sim 50$ pc, has a size of $\sim 10''$ on the sky). However, SAFARI has access to the critical FIR domain that can not be observed with ALMA or JWST, thus enabling the detection of unique diagnostics of the disk (dust, ice, atomic and molecular features). SAFARI will also permit us to: (1) Carry out ***deep searches of gas towards large samples of discs*** (including more evolved debris discs) by e.g., searching for the bright [O I]63 $\mu$m line and (2) take advantage of the broadband coverage of SAFARI-FTS to obtain ***unbiased FIR spectral line surveys***. Even if the line emission can not be fully spatially resolved, SPICA observations will allow us to detect the presence of water vapour and the

**Expected gas line emission in transitional protoplanetary disks with SPICA/SAFARI**

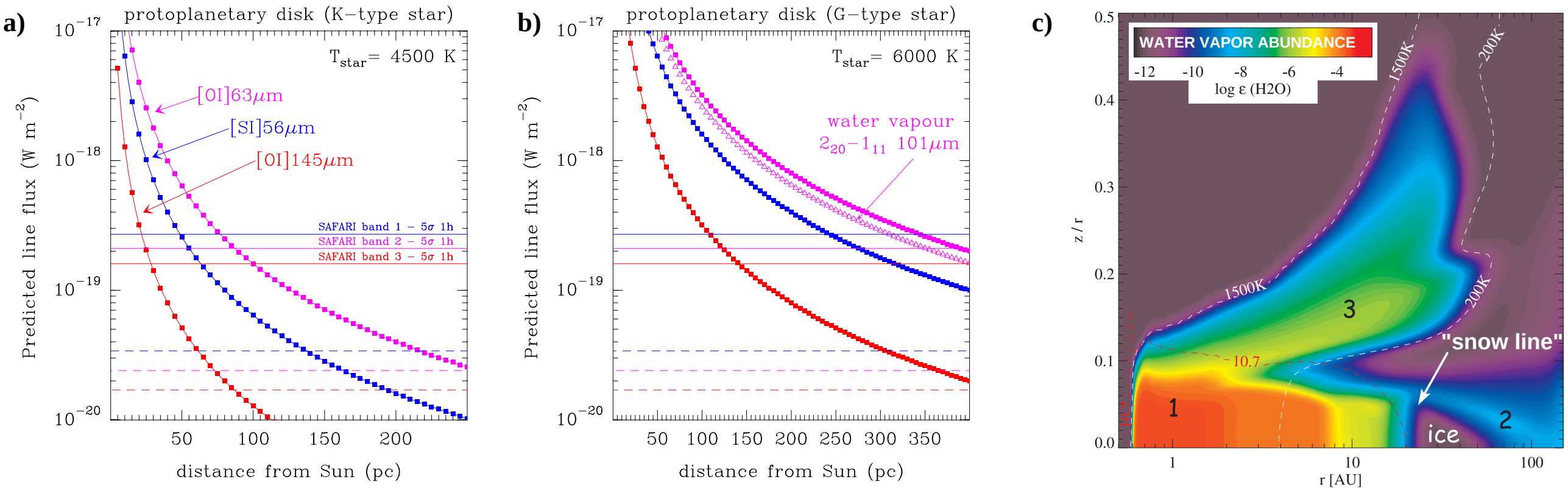

*Figure 1.5: Predicted fluxes of key FIR cooling lines in transitional protoplanetary discs (~ 10 Myr) around* ***(a)*** *a cool K star and* ***(b)*** *a G solar-type as a function of distance. The horizontal lines indicate 5σ-1hr detection limits of SAFARI. Dashed lines correspond to sensitivities using narrow band filters on a restricted set of line wavelengths. The total gas mass in the disc is low ~$10^{-2}$ $M_{\rm Jup}$. SPICA will have the sensitivity to observe the gas content of transitional discs in the closest star forming regions (~ 150 pc) at the time planets are thought to assemble (adapted from Gorti & Hollenbach 2004).* ***(c)*** *Water abundance model for a young Herbig/Ae protoplanetary disc (Woitke et al. 2009a,b). Three different regions with high water vapour abundance can be distinguished and be traced by different MIR/FIR $H_2O$ lines with MIRHES and SAFARI: 1) the inner water reservoir where rocky planets form, 2) the distant water belt, and 3) the hot water layer. The "snow line" refers to the radii where water is expected in ice phase (and can be directly traced by the ~44 μm and ~62 μm water ice bands with SAFARI). Gas giant exoplanets are thought to form beyond there, and migrate inwards.*

basic building blocks of the UV illuminated-disc chemistry ($C^+$, $CH^+$, CH, O, OH, $Si^+$, S, HD, ...) for the first time in hundreds of *transitional* discs and, together with disc/PDR models, relate their presence with the main parameters of each planetary system (stellar mass, temperature, age ...).

In summary, SPICA spectrometers will provide continuous coverage from ~5 to 210 μm with medium to high spectral resolution and high sensitivity to detect the gas line emission ($H_2$, [O I], $H_2O$, HD...) from hundreds of tenuous planet forming discs and thus to infer their gas masses, chemical composition, etc. Additionally, the MIR coronagraph (SCI) will cover the ~ 5 − 27 μm range continuously, providing high stellar suppression within a small inner working angle (IWA ≃ 1″). Instantaneous images of entire planetary systems with SCI will reveal their morphology in great detail and also will allow us to carry out MIR spectroscopy.

### Planetesimals, dust and mineralogy in debris discs

Dust is a key building block of rocky planets and the cores of giant gas planets. Dust appears to be present at all stages of planetary system formation with the mass ratio of gas to dust, the amount of dust present, its temperature and its distribution evolving rapidly through the process of planetary formation (Tanaka et al. 2005; Su et al. 2005). The processing and evolution of dust from protoplanetary discs to evolved solar systems like ours is key to understanding the formation and mineralogy of rocky, Earth-like, planets. For instance, it appears that in discs around intermediate mass, pre-main sequence stars, the dust in the inner regions of the disc (≲ 2 AU) can be more evolved. That is, it shows signatures of grain growth and crystallisation whereas the dust in the outer region is often seen to be pristine and similar to the interstellar dust (Natta et al. 2006). This implies a strong radial dependence of the dust processing whereby the inner regions are dominated by the stellar radiation field, leading to grain heating and crystallisation, and by higher densities, leading to coagulation, whilst the outer regions remain largely unprocessed and carry the signature of the pre-stellar nebula from which the star formed. And yet in our own Solar System we see crystalline silicates present in comets that clearly originate from regions far from the zones where this processing must have occurred. Only with detailed mapping of the MIR/FIR mineral and ice band emission/absorption of our own and distant circumstellar material, will we be able to understand the evolutionary track that leads to this situation. Observations using

SPICA spectrometers will allow us to carry out mineral and ice spectral studies (started with ISO and followed by Spitzer and AKARI) with unprecedented sensitivity and angular resolution.

The final stage of planetary system formation, the formation of small planetesimals that sweep up much of the disc's material, together with the formation of larger planetary bodies via collisions of planetesimals, often appears to result in an almost gas-free, "second generation" dusty disc, the debris disc. The dust in these discs is produced by mutual collisions of planetesimals in the final stages of planetary formation and the "heavy bombardment" phase (> 300 Myr) evidenced in the impact craters seen in all rocky Solar System planets. Debris discs can survive over billions of years. This points towards the presence of large reservoirs of colliding asteroids and evaporating comet-like bodies. Moreover, since colliding planetesimals need to be present to replenish the dust grains in the disc, detecting a debris disc is a strong signature of an emerging planetary system and is indicative of the presence of analogous asteroid and Kuiper belts (Wyatt 2008), or of regions where the formation of Earth-like or Pluto-like planets is ongoing (Kenyon & Bromley 2008). Indeed, the very recent direct detection of exoplanets towards several stars that host bright debris discs (Kalas et al. 2008; Marois et al. 2008; Lagrange et al. 2009) have quantitatively confirmed that studies of debris discs are critical to advance in our understanding of the formation and diversity of extra-solar planetary systems.

The dust emission in circumstellar discs produces an "excess" of FIR continuum emission that becomes apparent and reaches its maximum in the SAFARI wavelength range, which is not seen in the shorter wavelength mid infrared range covered by JWST (see Figure 1.6a). The exact wavelength location of the disc dust emission peak depends on the grains temperature, size distribution and composition. Around ~ 300 debris discs have been photometrically discovered so far with ISO and Spitzer. Although there are very few debris disc detections around stars later than K2, this is likely an observational bias because these discs would have been too cold and faint to be detected. This large discovery space of discs around cool stars is yet to be explored. The most recent FIR photometric census suggests that about 10% of lower-mass, Sun-like stars, are surrounded by debris discs, at least down the limiting fluxes observables with Spitzer (Meyer et al. 2008).

The study of debris discs is an emerging field that is currently limited by the available FIR sensitivity which makes it impossible to **(1)** perform large statistical surveys over meaningful star volumes; **(2)** probe more distant discs and/or less massive dusty discs down around all types of stars including young M dwarfs and cooler substellar objects. In particular, the sensitivity required to detect the dust disc over stellar luminosity ratio ($L_{dust}/L_{star}$) in a planetary system analogous to the Kuiper Belt disc ($L_{dust}/L_{star} \simeq 10^{-7}$) or Asteroid Belt ($L_{dust}/L_{star} \simeq 10^{-8}$) in our Solar System remains far below the far infrared capabilities provided by IRAS, ISO, AKARI, Spitzer and Herschel (see Figure 1.6b). Note that the flux of a Kuiper Belt disc with a mass of ~0.1 $M_{earth}$ at 30 pc corresponds to ~ 4 mJy at 70 $\mu$m (Meyer et al. 2008). In other words, due to sensitivity constraints, previous and current FIR telescopes are probably only showing us the "tip of the iceberg" as far as the detection and characterisation of extra-solar planetary discs is concerned.

SAFARI offers two powerful methods to detect and characterise the dust emission from debris discs **(1)** fast and ultra sensitive photometric searches in the 48 $\mu$m, 85 $\mu$m and 160 $\mu$m bands simultaneously reaching the $\lesssim$ 100 $\mu$Jy level sensitivity for 5$\sigma$ in 1 min and **(2)** very sensitive spectral characterisation at modest spectral resolution ($R \sim 100$) over the full 34 – 210 $\mu$m waveband with ~ 1 mJy level sensitivity for 5$\sigma$ in 1 hour. Both modes take advantage of a large instantaneous FOV for efficient wide area surveys of multiple objects. Figure 1.6c shows the minimum detectable fractional luminosities ($L_{dust}/L_{star}$) for debris discs at different distances as observed with Spitzer-MIPS at 70 $\mu$m, Herschel-PACS at 70 $\mu$m and 100 $\mu$m, and SPICA/SAFARI at 48 $\mu$m and 85 $\mu$m. For solar-type stars, we see that **Herschel will not detect dust at the Kuiper or Asteroid belt levels even for the closest discs**, and debris disc detections at 70 $\mu$m have generally been limited to ~ 100 times the luminosity of the dust in the Kuiper belt. Hundreds of nearby main sequence stars, ~ 1000 stars at $d < 25$ pc, without known far infrared excesses will be targets for SAFARI to search for the faintest dust excesses and the presence of dusty belts indicating the presence of hitherto undetected planetary systems.

The ability of SAFARI to survey and detect large numbers of FIR excesses will represent a revolution in debris discs studies because of the large number stars that harbour debris discs that may be detected. Models and observations indicate that debris discs might be very common and there are ~ $10^5$ F0 – K2 stars within ~ 150 pc, a number to be compared to the ~ 190 debris discs around these type of stars known to date. SAFARI will increase the number of debris discs detections around solar-type stars by 3 orders of magnitude, allowing realistic statistics of the disc frequencies and properties as a function of stellar type, age and environment. In particular, **SAFARI will be the first instrument to detect Kuiper Belt analogues out to** ~ 150 **pc** (see Figure 1.6c). As discussed before, this is the distance to the closest star forming regions, meaning that **SAFARI will be particularly adept at studying discs during the brief epoch at $\gtrsim$ 10 Myr when the transition from protoplanetary to debris disc occurs**, a transition which is as yet poorly understood but which is of prime importance due to its curtailment and direct link with planet formation.

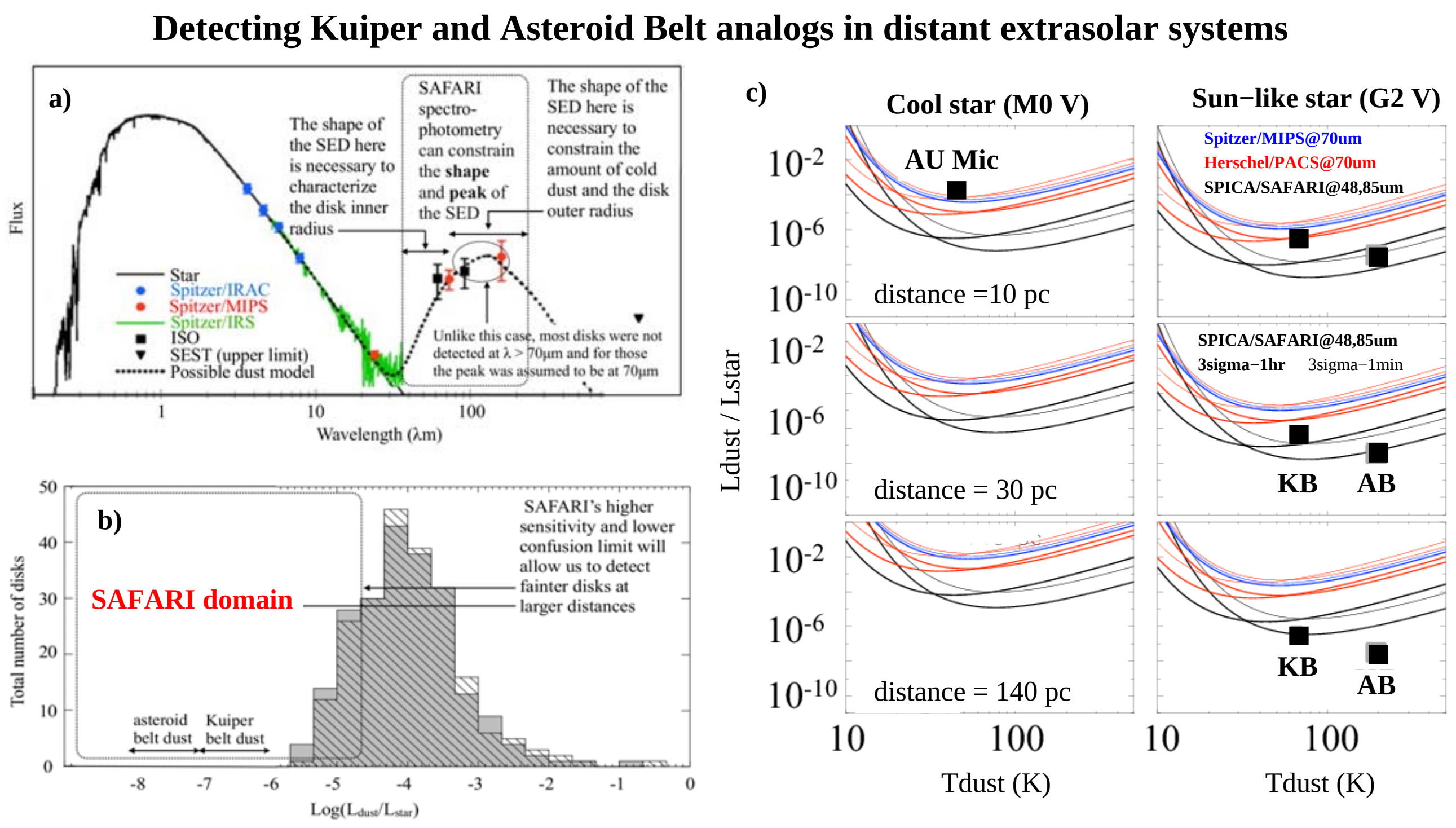

*Figure 1.6: **(a)** SED of HD105 star. The lack of excess emission at MIR wavelengths is due to the depletion of dust at distances <35 AU from the central star. The FIR wavelengths covered by SAFARI are thus critical (adapted from Meyer et al. 2008). **(b)** Distribution of fractional luminosities, $L_{dust}/L_{star}$, for the number of debris discs known to date. The characteristic values for the Kuiper and Asteroid belts (KB and AB) are shown at the left (Moro-Martín 2009). **(c)** Minimum detectable fractional luminosities for discs at different distances around stars of different types: cool M0 V (left) and solar G2 V (right) – with characteristic values for the KB and AB shown as black squares. The different colours correspond to different far infrared instruments (3σ-1hr): Spitzer-MIPS at 70 μm (blue); Herschel-PACS at 70 and 100 μm (red); and SPICA/SAFARI at 48 and 85 μm (black). The thick/thin lines for SAFARI refer to 1hr/1min integrations (Moro-Martín 2009).*

SAFARI will be so sensitive that it will be perfectly adapted to detect the faintest and maybe the most profuse planet forming systems in the Galaxy, those around cool stars and brown dwarfs (BDs) (Rebolo et al. 1995; Teixeira et al. 2009). Note that ~77% of stars in the local neighbourhood are cool M-stars. In fact there is observational evidence that at least 3% harbour planets and that "super-earths" are more common than giants. Figure 1.6 shows that debris discs orders of magnitude fainter than that around AU-Mic (at 9.9 pc) will be detectable, and there are ~160 M-type stars within ~10 pc. These studies will open new frontiers in planetary studies: What kind of planets could eventually form around such cool stars? Will they be habitable? The detection of the faint far infrared excesses around cool stars and brown dwarfs is a great challenge for current instrumentation, with disc fluxes around mJy or less. However, SAFARI will routinely observe them and massively increase our knowledge of how these enigmatic objects are formed.

***Very sensitive spectrophotometric characterisation of circumstellar discs:***
Despite the poorer angular resolution of SPICA compared to the future facilities (e.g., ALMA or SKA), much will be learnt by going beyond photometric detections and carrying out extensive MIR/FIR spectroscopy studies of the strongest dust/ice features in a large sample of debris discs. SAFARI will provide the continuous FIR spectral energy distribution (SED) and be able to detect both the dust continuum emission and the brightest grain/ice bands as well as the brightest lines from any gas residual present in the disc. The ability of FIR spectroscopy to determine the mineral makeup of dusty discs around young stars is illustrated by the ISO spectrum of HD142527 shown in Figure 1.7a. SPICA will be hundreds of times more sensitive than ISO and so will see not only the young and massive opaque circumstellar discs like this, but will be able to trace the mineralogy of dust within discs at all stages of planetary system formation. SPICA will not only determine the detailed mineralogy of discs, but will also trace the variation grain size distribution and temperature, which are both expected to evolve with disc age leading to a variation in the disc SED. Models suggest that there is little predicted evolution of the mid infrared SED with disc age and a possible ambiguity in the mid infrared intensity between age and disc structure which is removed with observations in the far infrared (Tanaka et al. 2005). A key step forward needed to cope with all model complexities (Dullemond et al. 2001) and to fully characterise the nature of dusty discs is to obtain full SED wavelength sampling (at least in the critical far infrared domain) in order to accurately determine the emission peak, slope, and dust spectral features, and so be able to constrain grain sizes and dust opacities, critical ingredients to determine the disc mass. Either in photometric or in spectroscopy modes, **SAFARI will be the true *"hunter of debris discs"*** of the next decade.

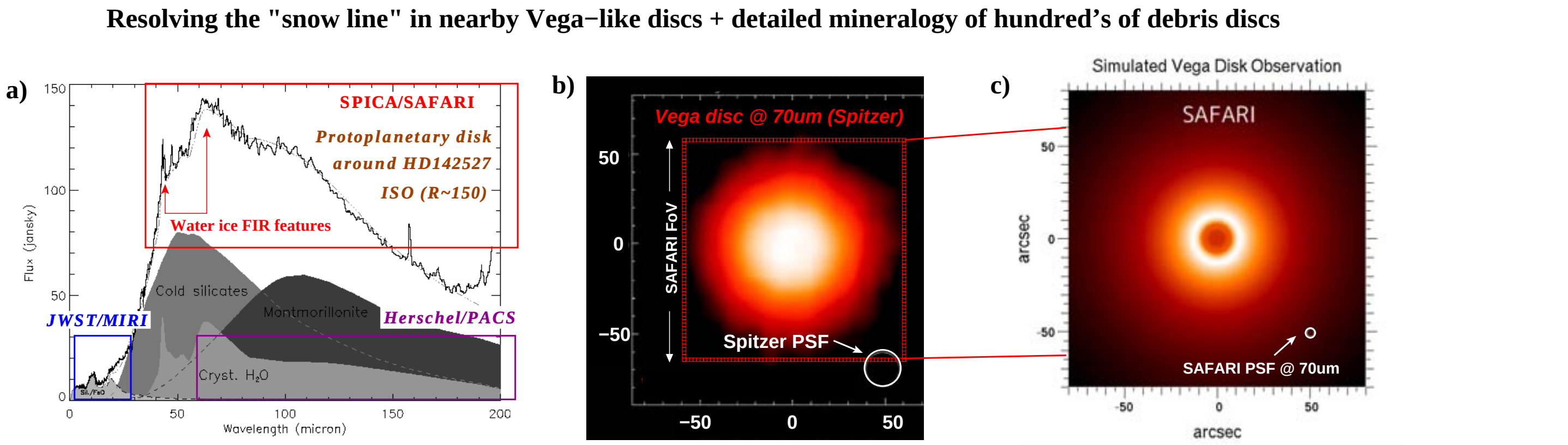

*Figure 1.7: **(a)** ISO spectrum of the disk around HD142527 (Malfait et al. 1999). Note that amorphous water ice only shows bands at* ~ 44 *μm in the FIR (Moore & Hudson 1992). Such features can not be accessed with Herschel or JWST. SPICA will take the equivalent spectra of objects at flux levels less than ~ 10 mJy per minute. **(b)** Image of Vega debris disk at 70 μm with Spitzer (Su et al. 2005). SAFARI's large FOV and smaller pixel size (*~ 1.8″ *at ~44 μm, red squares) will provide very detailed spectroscopic images of nearby disks. **(c)** Simulated image with SAFARI at 70 μm assuming an inner hole in the disk with a radius of* ~ 11″.

*Water Ice in discs:*
Below gas temperatures of ~ 150 K water vapour freezes-out onto dust grains and the main form of water in the cold circumstellar disc midplane and at large disc radii will be ice. The physical location of the point at which water freezes out determines the position of the so called "*snow line*", i.e., the water ice sublimation front, which separates the inner disc region of terrestrial, rocky, planet formation from that of the outer giant planets (Nagasawa et al. 2007; Woitke et al. 2009a,b) (see model predictions in Figure 1.5c). Grains covered by water icy mantles can play a significant role in planetary formation, enabling the formation of planetesimals and the core of gas giants protoplanets beyond the *snow line*. Observations of the Solar System's asteroid belt suggest that our *snow line* occurred near a disc radius of ~ 2.7 AU (Lecar et al. 2006). In the outer reaches of our own Solar System, i.e., beyond the snow line, most of the satellites and small bodies contain a significant fraction of water ice; in the case of comets this fraction is as high as 80% and the presence of water in the upper atmospheres of the four gas giants is thought to be highly influenced by cometary impacts such as that of Shoemaker-Levy 9 on Jupiter. It is possible that it is during the later phases of planetary formation that the atmospheres, and indeed the oceans, of the rocky planets were formed from water ice contained in the comets and asteroids that bombarded the inner Solar System. For more distant exoplanetary systems, very little is known as (1) we first need to detect the presence of water-ice and infer its abundance in a large sample of protoplanetary systems (spanning a broad range of host star types and ages) and (2) even for the closest discs, the exact location of the *snow line* is very hard to resolve spatially with current instrumentation.

In the far infrared there is a powerful tool for the detection of water ice and determination of the amorphous/crystalline nature: Namely the transverse optical mode at $\sim 44\,\mu$m both from crystalline and amorphous water ice and the longitudinal acoustic mode at $\sim 62\,\mu$m arising only from crystalline water ice (Warren 1984; Moore & Hudson 1992; Maldoni et al. 1999). In contrast to the mid infrared $\sim 6.1\,\mu$m water ice feature (that has a stronger band strength and that SPICA MIR spectrometers will observe with sub arcsec angular resolution), the difference between the amorphous and crystalline phase is very well defined in the FIR (see Figure 1.7a) and, again unlike the MIR features, the FIR bands are not confused with other solid state features of less abundant species. In optically thin discs it is extremely difficult to use MIR absorption to trace water ice and the material is too cold to emit in the NIR/MIR bands. Hence, these strong FIR features are robust probes of **(1)** the presence/absence of water ice, even in cold or heavily obscured or cold regions without a MIR background, and **(2)** the amorphous/crystalline state which provides clues on the formation history of water ice. Note that JWST cannot access these FIR ice solid-state bands (Moore & Hudson 1994).

First observed in emission towards the Frosty Leo Nebula (Omont et al. 1990), water ice has been detected in young protoplanetary discs in a few bright sources either in the FIR using the ISO-LWS (Dartois et al. 1998; Malfait et al. 1999) or MIR (Pontoppidan et al. 2005; Terada et al. 2007). The FIR features were also observed using ISO-LWS in comets within our own Solar System (Lellouch et al. 1998). Since the bands change shape i.e., they narrow for crystalline ice and the peak shifts in wavelength with the temperature, relatively high spectral resolution is needed to extract all the available information from the ice band profiles. In its highest resolution mode SAFARI provides $R \sim 4500$ at $\sim 44\,\mu$m which is appropriate for very detailed ice spectroscopy studies. Indeed **SPICA is the only planned mission that will allow water ice to be observed in all environments** and fully explore its impact on planetary formation and evolution and the emergence of habitable planets. In many cases, the absorption from other features (e.g., water ice at $\sim 6.1\,\mu$m, but also methanol, carbon dioxide, formaldehyde and methane ice) will be studied also with complementary MIR spectra taken with SCI and MIRMES. In combination with SAFARI observations, such complete MIR/FIR ice spectra will accurately constrain disc models, determine for how long ice mantles survive under the irradiation of different stellar UV fields (Grigorieva et al. 2007) and determine the thermal history of the ice crystallinity.

***Spatially resolved debris discs (resolving the snow line in discs?):***
Images of the few spatially resolved debris discs, either seen in reflection or directly in the FIR or sub-mm, can show gaps and ring-like structures indicating the presence of planets which “shepherd” the dust. This has been most readily observed so far using ground based near-IR scattered emission and thermal MIR emission (e.g., Subaru, Gemini; Fukagawa et al. 2006; Fujiwara et al. 2006) and sub-mm and mm telescopes (JCMT, CSO and IRAM). However, these facilities cannot detect specific dust grain spectral features, nor the FIR water ice bands. ALMA will undoubtedly add a great deal to the subject by detecting the optically thin sub-mm continuum emission of the cold dust, but will not be very efficient in mapping the very extended emission of the closest discs which can extend to a few arcmin in size. Spitzer has revealed that the size of some disc systems, such the Vega debris disc, is surprisingly larger in the FIR (e.g., at the emission peak) than in the sub-mm. Obviously this is telling us something about the disc nature and it is clear that only through multi-wavelength observations that we can fully understand the complete picture of the formation and evolution of these discs.

SPICA will greatly increase the number of debris discs that can be spatially resolved (only $\sim 20$ currently), allowing us to break the degeneracy in the spectral energy distribution between dust location and mineral composition. For instance, at the 48 $\mu$m photometric band the expected angular resolution is $3.5''$ allowing a 100 AU disc to be just resolvable at $\sim 28$ pc. The number of stars within that distance is approximately $\sim 210$ (A), $\sim 1100$ (F0-K2) and $\sim 3600$ (K2-M). In the closest ones, we will be able to indirectly infer from observations of clumpy/warped/asymmetric dust structures that planets are gravitationally perturbing the planetesimals and dust (Wyatt et al. 1999; Wyatt 2003; Kalas et al. 2005). In nearby debris discs ($d < 10$ pc) such as those around Vega or Formalhaut ($\sim 100$ A-type stars), the spatial resolution of SPICA at the 44 $\mu$m feature ($\sim$25 AU) will be sufficient to spatially resolve the distribution of water ice – the *snow line* (Figure 1.7). In all these “Vega-like” systems the expected snow line is pushed away to $\sim 22$ to 44 AU, or even larger radii (Ida & Lin 2005; Grigorieva et al. 2007; Kennedy & Kenyon 2008).

Herschel will only be able to observe the ∼ 62 $\mu$m water ice band, i.e., only regions where crystalline water is present in large quantities. Even then, observing a broad indistinct spectroscopic feature with the narrow band Herschel-PACS grating spectrometer will be difficult and the sensitivity will be very much worse than SAFARI with inferior spatial resolution. SAFARI will be so sensitive that it will not only be able to produce fully sampled spectroscopic images of the ∼ 44 $\mu$m and ∼ 62 $\mu$m water ice bands, but also of any “secondary” or “residual” gas content (e.g., by detecting FIR OH and [O I] lines) produced by the photoevaporation of ice grain mantles, outgassing of comets or collisional evaporation. Only with SPICA will we be able to confirm these predictions and provide the critical tests on which to base future models of the formation of planetary systems. All in all, the study of discs with SPICA spectrometers will shed light on the diversity of planetary systems, the link between circumstellar discs and planets and the link between extra-solar planetary systems and our own.

### 1.2.2 The inner and outer Solar System “our own debris disc”

The observation of circumstellar discs around distant stars at different evolutionary stages provide clues on how our planetary system was formed, from which materials, and how they were processed. However, in order to understand the observed diversity of extra-solar planetary systems, we also need to explain how our own Solar System emerged. Clearly both approaches complement each other and represent intimately and increasingly interconnected research fields. The study of the Solar System extends from the traditional investigation of the planets and their rings and moons, to the most recent characterisation of the different populations of primitive leftovers of their construction (comets, asteroids, Kuiper belt bodies, etc.). Finally, investigations of how life came to exist on Earth are fundamental and becoming of great general interest.

Our current view of the Solar System’s early evolution is based on the “Nice Model” (Gomes et al. 2005). This model argues that, after a relatively slow evolution, the orbits of Saturn and Jupiter crossed their mutual 2:1 mean motion resonance about 700 Myr after formation, which caused a violent destabilisation of the orbits of planetesimals throughout the disc. This event populated the Kuiper belt (> 40 AU) with different families of leftover bodies and delivered pristine, icy planetesimals to the inner Solar System. This model agrees with the geochemical evidence (e.g., in the Moon) of a very peaked cratering rate at that time, and is generally known as the late heavy bombardment period (or LHB). Because there is also evidence of planet migration and planetesimal belts in extra-solar systems, a natural question arises: Are LHB-type events common in other planetary systems? From a broader astrobiological point of view, those primitive bodies coming from the outer Solar System (with little or no chemical processing) could have delivered significant amounts of volatiles and chemical species to the inner rocky planets (e.g., water and organic matter) that are relevant for the habitability of such planets. The study of the enigmatic nature of the outer Solar System is very challenging because **(1)** at such large distances from the Sun, rocky and icy bodies are cold, below ≃ 60 K, therefore their thermal emission peak occurs at FIR wavelengths that cannot be observed from the ground and **(2)** the FIR fluxes are very weak (a few mJy and below). As we shall describe hereafter, the very sensitive instruments on board SPICA, together with their broadband spectroscopic capabilities, will provide a new perspective of the Solar System’s outermost belts, the regions that hide a record of the earliest phases of the solar nebula.

#### Studying the Kuiper Belt object by object

Kuiper Belt Objects (KBOs) refer to the physico-chemically unaltered population of bodies beyond Neptune’s orbit. Since the discovery of the first KBO (Jewitt & Luu 1993), more than 1200 objects have been detected so far in the outer Solar System (> 30 – 40 AU), including planetary-sized objects such as Eris (Bertoldi et al. 2006). Several thousands of KBOs are expected to exist, especially at high ecliptic latitudes. Unlike asteroids in the inner Solar System, KBOs must have formed relatively slowly and they are thought to be composed of pristine, almost unprocessed chemical material. The detailed study of their physical parameters (temperature, size distribution and albedo) and chemical characteristics (mineral/ice content) is thus fundamental to consistently link the history of our system with that of distant extra-solar systems. Unfortunately, most of the KBO attributes and chemical composition are almost unknown. Together with the inner asteroid belt, these remnant

**Detection and Characterization of outer Solar System Objects with SPICA/SAFARI**

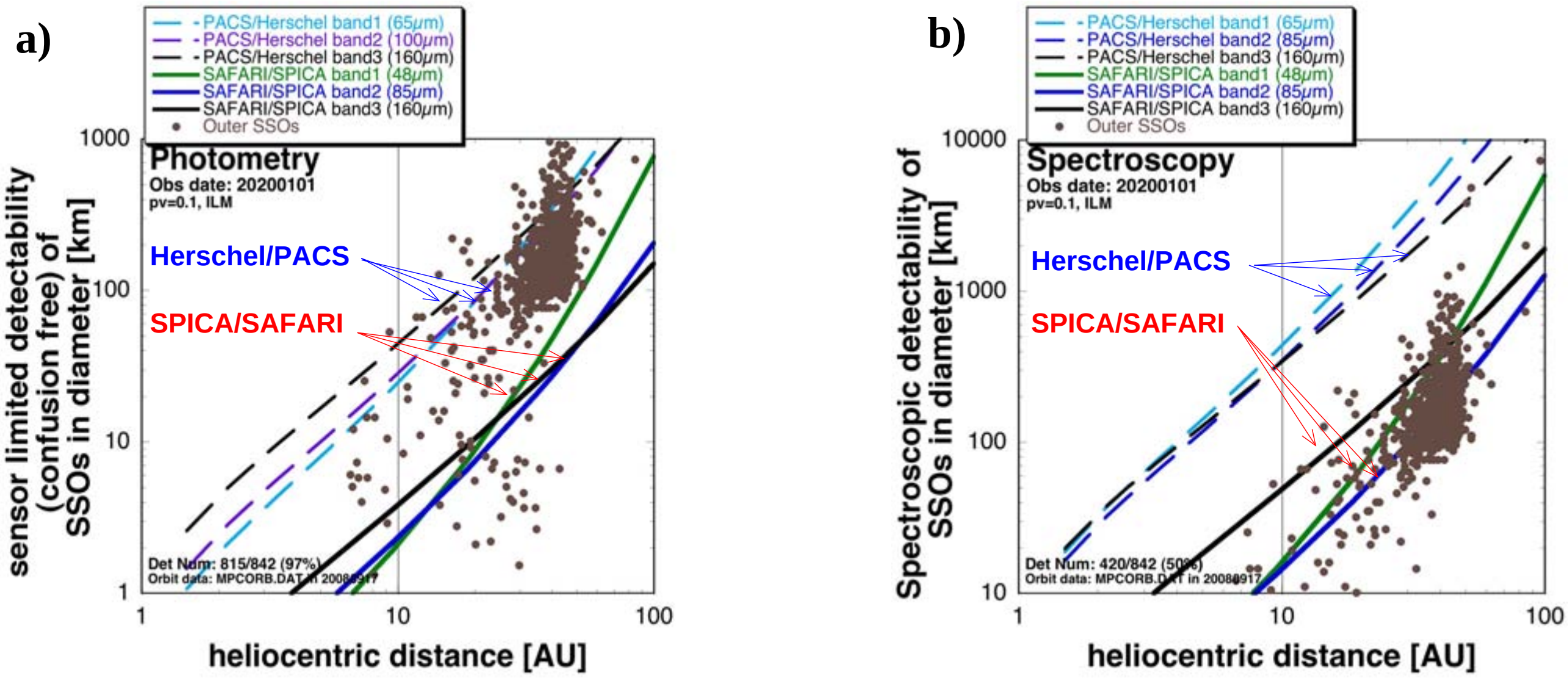

*Figure 1.8:* ***(a)*** *Diameters (in km) of known outer Solar System objects (SSOs), that emit at FIR wavelengths, as a function of their heliocentric distance. The coloured curves display detectability estimations for the different FIR photometric bands of Herschel-PACS (dashed) and SPICA/SAFARI (continuous). Predictions take into account that SSOs are moving objects so that they can be detected below the FIR confusion limit by pair-subtracting within a reasonable time interval.* ***(b)*** *Same but for spectroscopy, which will allow us to detect the mineral/ice content of the most primitive SSOs for the first time (updated from Hasegawa 2000).*

planetesimal belts are analogues of the cold debris discs observed as FIR photometric excess towards more distant stars. In this sense, the outer Solar System provides the closest "template" to study the composition, processing and transport of minerals, ices and organic matter by ***studying debris disc bodies "one by one"***.

Spitzer-MIPS has detected a few KBOs photometrically at 24 and 70 $\mu$m by observing from minutes to hours per target. Measurements of the far infrared thermal emission of KBOs (where the bulk of the KBOs' energy is radiated) reveal the thermal properties of the near-surface layers (Stansberry et al. 2006). Complementary observations in the visible are needed to establish the position accurately, determine the objects albedo, and, in combination with the FIR observations, determine the object size and mass. About $\sim 30$ KBOs have known albedos; low albedo values (such as those in comets) are presumed to be darkened by the presence of organic matter while high albedos are thought to be associated with ice mantles. Surprisingly, all inner and cold KBOs targets detected in the Spitzer sample (~20) have much higher albedos (Brucker et al. 2009) than previously assumed (Jewitt & Luu 1993). The underestimation of the KBO albedos leads to a significant overestimation of their mass. This is an unexpected result, and it is clear that future FIR studies on the size and mass distribution of the outer Solar System will need to provide a much more robust confirmation by observing a large sample of thousands of KBOs.

***SAFARI photometry – FIR detection of all known KBOs:*** While Herschel will require $\sim 300$ hr to detect photometrically $\sim 10\%$ of the currently known KBOs (those with diameters larger than $> 250$ km at rate of $\sim 1$ per hr), SAFARI will detect almost all known KBOs (those with diameters $> 100$ km) in only $\sim 50$ hr at a rate of 1 object per minute. SAFARI's ultra-deep FIR photometric surveys will provide the crucial observational input to determine their sizes and albedos. SAFARI will also detect bodies as small as $\sim 10$ km diameter, thus including the new KBOs to be discovered from now to the launch of SPICA, the number which is expected to increase by a factor $\sim 2$, to $\sim 2500$ by $\sim 2018$. To be more precise, Figure 1.8 shows the diameters (in km) of a population of known KBOs as a function of their heliocentric distance. We plot on the same figures the detection limit predictions for the different photometric bands of Herschel-PACS and SAFARI showing the large number of KBOs that could be detected photometrically in the FIR in short time exposures by SAFARI. The ability to constrain the size and albedo of such a huge number of KBOs is a unique science driver for SAFARI.

***SAFARI spectroscopy: What are the most primitive objects in the Solar System made of?*** Neither Spitzer nor Herschel have the required sensitivity to detect spectral features from KBOs in the FIR as the expected flux at 70 $\mu$m is below a few mJy and below a few $\mu$Jy at 24 $\mu$m (Brucker et al. 2009). SAFARI will be the first FIR spectrometer to cover simultaneously the $\sim 34 - 210$ $\mu$m spectrum of KBOs with the required sensitivity, thus opening the outer Solar System studies to FIR spectroscopy. SAFARI low-resolution spectra ($R \sim 100$) will be a **new powerful tool** to constrain KBOs thermal models, because it does not just add a few photometric measurements, but fully samples the peak and shape of the thermal emission. Even more importantly, with spectroscopy we will be able investigate the presence of the main minerals and ices in a statistically significant sample of targets.

The broad wavelength range and high sensitivity of SAFARI will allow us to detect both the water ice features at $\sim 44$ and $\sim 62$ $\mu$m and the broadband FIR emission from Mg-rich crystalline silicates (e.g., forsterite, clino-enstatite and diopside; Bowey et al. 2002) and other Fe-rich minerals. As a result, we will be able to determine the mineral composition of KBOs and link it to the mineralogy of extra-solar planetary systems. SAFARI promises even more ambitious discoveries such as the detection of the possible tenuous gas atmospheres of some KBOs, plausibly formed through cryo-volcanism as in Enceladus, by looking for the lowest-energy-level transitions of water vapour and the FIR lines of elements such as neutral sulphur or oxygen. For the most distant/cooler/faintest KBOs, the FIR features will be the best and only way to characterise the outer Solar System composition.

**The inner Solar System, its size distribution and its chemical composition:**

SPICA will not only characterise the composition of the cold and outer KBOs, but also of the different families of inner, hotter centaurs, comets and asteroids, thus probing the primary/secondary dust processing in the Solar System. A full determination of their surface composition requires direct spectroscopic studies over the entire critical MIR/FIR range, where specific minerals and ices can be identified unambiguously. The similarity between interstellar and cometary ices (Ehrenfreund et al. 1997; Crovisier et al. 1997; Bockelée-Morvan et al. 2000) may not directly prove that comets accrete unprocessed material, but rather suggest that similar chemical processes were at work in the early Solar Nebula and in interstellar clouds (Cernicharo & Crovisier 2005). However, abundant water, ammonia and methanol ice mantles in Solar System objects could well be the seeds of more complex biogenic molecules, such as the aminoacids, that do form when interstellar ice analogues are irradiated with UV radiation (Muñoz Caro et al. 2002). Some of the amino acids identified in the laboratory are also found in meteorites which suggests that prebiotic molecules could have been delivered to the early Earth by cometary dust, meteorites or interplanetary dust particles. Spectroscopic studies with SPICA in the critical MIR/FIR domain will clearly advance our understanding of the chemical complexity in planetary systems by direct searches of the “ingredients for life”.

In summary, SPICA will provide for the first time the means to quantify the composition of hundreds of Solar System objects. Sensitive FIR photometric and spectroscopic observations will give the first unambiguous determination of their size distribution (and thus mass) and composition providing critical observational evidence for models of Solar System formation.

## 1.3 Exoplanet characterisation in the Infrared

The first detection of an exoplanet around a solar-type star (Mayor & Queloz 1995) has been one of the most relevant discoveries of the last century. Indeed, the existence of extra-solar planetary systems has global implications that go beyond astrophysics, and captures the interest of the general public. Extra-solar systems may include a variety of planets even richer than the selection we have in our own Solar System: gas-giants, icy bodies as well as less massive rocky planets with their oceans, moons and rings (Gillon et al. 2007; Fortney et al. 2008). Almost 15 years after this outstanding discovery, we know that extra-solar planetary systems clearly exist wide in range of masses and orbital parameters, and new exoplanets (EPs) are found at surprisingly in-creasing rates. At the time of writing, 373 EPs have been detected by different methods (radial velocity, transits, microlensing, imaging and timing). In terms of individual EP detections, most of them (~ 294) have been found through radial velocity searches, and refer to massive (out to a few Jupiter masses), hot ($T_{\rm eff} \sim 1000 - 1500$ K) gas planets, commonly named as "hot Jupiters", which orbit very close (<< 1 AU) to Sun-like stars (G-type) and thus have short periods ($P < 10$ days). A few lower-mass planets ($M < 10$ $M_{\rm earth}$), so called "super-Earths", have also been discovered, which suggests that rocky planets relatively similar to the Earth may not be uncommon. Planets with larger orbital axes, e.g., comparable to Jupiter's orbit, will be detected in the future through ground-based direct detection capabilities (e.g., Subaru-HiCIAO, VLT-SPHERE, Gemini-GPI) or astrometry methods (e.g., with Gaia). In fact, a few outer EPs (~ 20 to 120 AU) were directly imaged in 2008 for the first time: a planet around Formalhaut at ~ 7.7 pc (Kalas et al. 2008) and multiple planets around HR8799 at ~ 40 pc (Marois et al. 2008).

A further step forward in our understanding of these intriguing objects requires not just the detection of their presence (with many specialized ground- and space-based telescopes such as MEarth, COROT, Kepler, etc., designed to carry out extensive photometric searches of the sky), but also a detailed characterisation of their main physical properties and chemical conditions (temperature, densities, climate, atmospheric composition and biosignatures). Surprisingly, it has been possible to infer from Hubble and Spitzer observations some atmospheric components in a few EPs: sodium (Charbonneau et al. 2002; Redfield et al. 2008), methane (Swain et al. 2008b), CO and $CO_2$ (Swain et al. 2009) and even water vapour (Barman 2007; Tinetti et al. 2007). All these recent works show that, with more sensitive space instrumentation, a firm characterisation of a representative sample of EPs would be possible. Two different observational approaches can be used to characterise both *outer* and *inner* EPs: direct detection and transit techniques respectively. ***Direct detection***, which refers to observations where the star and the planet can be spatially separated on the sky with coronagraphs. Due to the limited size of current telescopes, this technique is only able to image EPs at large orbital distances (~ 50 – 100 AU). Besides, the huge contrast between the host star and the planet flux (e.g., the Sun is $10^9$ times brighter than the Earth in the visible) makes the direct detection of an Earth-like planet at ~ 1 AU a distant goal for future instrumentation. Fortunately, EPs orbiting very close to the host star ($\lesssim 0.05$ AU) and with a favourable inclination (almost edge-on systems) can be indirectly studied via the ***transit technique*** by which the dimming of the starlight as the EP transits the star is followed (Seager & Sasselov 2000). Because of the precise timing of a transit event, this technique allows high-contrast observations if very sensitive, low background and stable observing conditions are available (i.e., from space). Although not specifically designed for EP research, transit observations with Hubble and Spitzer have revolutionised and redefined the field of EP characterisation over the past 5 years (Barman 2007; Beaulieu et al. 2008; Charbonneau et al. 2002; Deming et al. 2005, 2006, 2007; Marley et al. 2007; Demory et al. 2007; Gillon et al. 2007; Grillmair et al. 2007; Harrington et al. 2006, 2007; Knutson et al. 2007; Richardson et al. 2006, 2007; Machalek et al. 2008; Swain et al. 2008b,a; Tinetti et al. 2007). Collectively, this work has conclusively established that the detailed characterisation of EP atmospheres is feasible and today we can discuss the observational signatures including weather and atmospheric chemistry (e.g., Selsis et al. 2002; Tinetti & Beaulieu 2009).

### 1.3.1 New frontiers in exoplanet research

The mid infrared region (from ~ 3 to ~ 30 $\mu$m) is especially important in the study of planetary atmospheres as it spans both the peak of thermal emission from the majority of EPs so far discovered, and is particularly rich in molecular features that can uniquely identify the composition of planetary atmospheres and trace the fingerprints of primitive biological activity. In the coming decades many space and ground based facilities are planned that are designed to search for EPs on all scales from massive, young "hot Jupiters", through large rocky "super-Earths". Few of the planned facilities, however, will have the ability to characterise the atmospheres which they discover through the application of infrared spectroscopy.

The observation of EPs at IR wavelengths offers several advantages compared to traditional studies in the visible domain. First, the star-to-planet flux contrast is much lower than in the visible (~ $10^3$ for a typical "hot Jupiter" around a Sun-like star) and second, transiting planets around stars much cooler than the Sun have to be observed in the IR where their emission peaks. Spitzer has measured EP photometric transits out to 24 $\mu$m demonstrating that the light-curve is simpler ("box-like") than in the visible domain due to the negligible role of stellar limb-darkening effects. This allows a robust and precise determination of the EP radius as a function of wavelength and provides further strong constraints to the atmospheric properties. SPICA will achieve a precision similar or better than that reachable with Spitzer instruments, enabling us to detect the atmospheric features of hot-Jupiters and Neptunes. Transit observations with the Spitzer-IRS spectrometer have been used to extract the absolute intrinsic spectrum of HD209458b hot Jupiter around a Sun-like star – with the resultant spectrum in physical units (e.g., in $\mu$Jy) as opposed to the typical relative contrast measurements (Swain et al. 2008a) (see Figure 1.9). Infrared observations have allowed to characterise temperature-pressure profiles, chemistry and circulation patterns of a select subset of massive, close-in hot Jupiters along with several less massive EPs: the hot Saturn HD14926b and the cooler Neptune-mass planet GJ 436b.

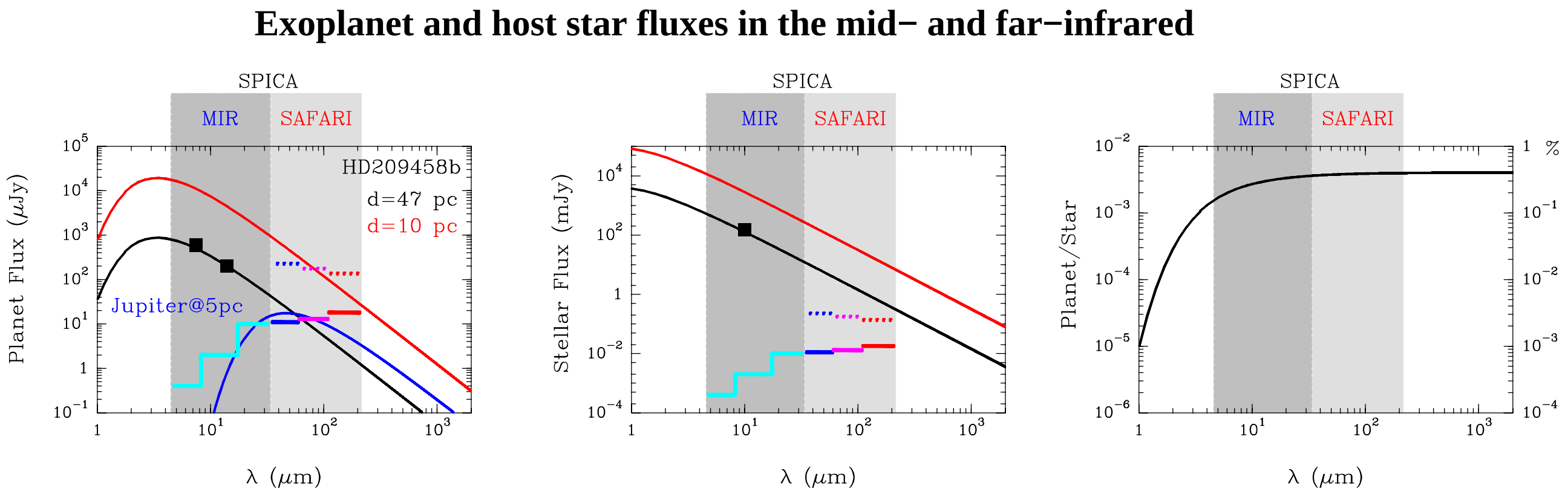

*Figure 1.9: **Left panel**: Fit to HD 209458b "hot Jupiter" ($T_{\rm eff} \simeq 1000$ K) MIR fluxes inferred from a secondary transit with Spitzer (Swain et al. 2008a) around a G0 star ($d \sim 47$ pc, in black) and interpolation to $d = 10$ pc (red). For comparison, the emission of a cooler Jupiter-like planet at 5 pc is shown in blue (reflected emission neglected). The thick lines are the 5$\sigma$-1hr photometric sensitivities of SPICA MIR instruments (cyan) and SAFARI (blue, magenta and red). Dashed lines show sensitivities in spectrophotometric mode ($R \simeq 25$). SPICA will observe similar inner "hot Jupiter" transits routinely and will potentially extract their IR spectrum (rich in $H_2O$, $O_3$, $CH_4$, $NH_3$ and HD features as in Solar System planets). **Middle panel**: Flux from the host star at different distances. **Right panel**: Increasing planet-to-star contrast at long wavelengths (Goicoechea et al. 2008).*

By operating at long wavelengths, SPICA will be well adapted to carry out transit observations of rocky planets closely orbiting around the numerous cool M stars ($\sim$ 75 stars at $d < 5$ pc and $\sim$ 6500 within a $d < 25$ pc volume!). Owing to their low mass ($\sim 0.09 - 0.68\ M_\odot$) and low effective temperature (2500 – 3800 K), M stars are better studied at infrared wavelengths. The location of the habitable zone (i.e., the distance from the host star to a planet where liquid water could exist) scales with $\sim L_{star}^{1/2}$, and thus M stars have habitable zones much closer ($\sim$ 0.02 – 0.5 AU) than Sun-like stars. In addition, the light curve of an earth-like EP around a M star is about as deep as for a Jupiter-like planet around a G star. The characterisation of rocky "super-Earths" with volatile-rich, maybe prebiotic atmospheres (e.g., with traces of $O_3$, $O_2$, $CO_2$, $CH_4$ or $H_2O$) orbiting within the habitable zone of M stars, in spite of being challenging (several transit measurements needed) will have a tremendous impact as these cool EPs are much more interesting from an astrobiological point of view. Searches for bright M star transit candidates with modest ground-based telescopes are currently underway (e.g., MEarth Project; Nutzman & Charbonneau 2008), and more targets will be searched by new radial velocity surveys. We anticipate that several of these will be good targets for detailed characterisation with SPICA.

### 1.3.2 Mid-infrared coronagraphy: "Direct" imaging and spectroscopy of EPs

Although the transiting technique is a powerful method to characterise "hot-Jupiters" spectroscopically, it is limited to EPs orbiting very close to the host star and to planetary systems with a favourable viewing geometry. High–contrast "direct" observations with coronagraphs are needed to study outer and cooler planets. The projected coronagraph on SPICA (SCI) is a unique instrument that provides several advantages over JWST coronagraphs. First, the monolithic mirror of SPICA will be better optimised for coronagraphy than the seg-mented mirror of JWST due to its much simpler and clean PSF. The segmented geometry of JWST also requires complex Lyot stops for the suppression of the light diffracted by each mirror segment, reducing the throughput and requiring a high degree of alignment stability. Secondly, the SPICA telescope itself will be actively cooled down to ∼ 6 K (compared to passive cooling down to 45 K for JWST); it is, therefore, further optimised for MIR/FIR astronomy. From the scientific point of view, **the main difference from JWST is the possibility of undertaking "direct" spectroscopy with SPICA/SCI in the critical MIR domain** (using a grism/prism providing $R \sim 20 - 200$) in addition to imaging. This spectral capability in a continuous wavelength domain rich in chemical signatures (∼ 3.5 – 27 $\mu$m) represents a unique science possibility of SPICA (Abe et al. 2007) compared to JWST, which will just have quadrant phase and Lyot photometric coronagraphs in the MIR and will provide lower contrast than SCI. As we have seen earlier, exo-giant-planets (EGPs) are thought to form beyond the "snow line". Indeed, several of them are known from Doppler shift measurements or direct imaging in the visible (Wetherill & Stewart 1989; Kalas et al. 2008; Marois et al. 2008) and more will be detected in the near future (with large optical telescopes equipped with coronagraphs, JWST and SPICA itself). This key population of outer and young EGPs can not be studied through transit experiments but will be targets for SCI direct characterisation. In fact, in the next decade ***SCI will be the only instrument available to characterise outer and cool EPs through direct spectroscopy in the MIR***. Therefore, SPICA will add greatly to EP research (and much earlier than the future Terrestrial Planet Finder type missions) by imaging young EPs directly and by recording their MIR spectra, thus constraining their temperature and atmospheric composition. The SCI simpler binary mask type approach will achieve a contrast of $10^{-6}$ at the equivalent to ∼ 9 AU (∼ Saturn's orbit) at ∼ 5 $\mu$m for a star at 10 pc (IWA∼1″). At this wavelength we probe the younger end of the planet age range (∼100 Myr to 1 Gyr, see Figure 1.10a). SCI is expected to observe ∼ 200 targets and produce an "spectral atlas" of several outer EPs, therefore completing the discovery and characterisation space of other telescopes and methods. According to their expected flux at long wavelengths ($\lambda > 5\,\mu$m), the main targets for SCI spectroscopy will be ∼ 2 – 3 $M_{Jup}$ planets around solar-type stars and lower mass planets, ∼1 $M_{Jup}$, around M dwarfs. We anticipate ∼ 200 targets for direct spectroscopy, for which ∼ 1 hr integration per EP will be required. In photometric mode, SCI will perform imaging surveys integrating a few minutes per target (both young M stars: ∼300 Myr, < 50 pc; and FGK stars: ∼1 Gyr, < 25 pc).

The most significant atmospheric features expected in the EP spectra can be summarised as follows: (1) The ∼ 4 – 5 $\mu$m "emission bump" due to an opacity window in EGPs and cooler objects with temperatures between 100 and 1000 K; (2) Molecular vibration bands of $H_2O$ (∼ 6 – 8 $\mu$m), $CH_4$ (∼ 7.7 $\mu$m) , $O_3$ (∼ 9.6 $\mu$m), silicate clouds (∼ 10 $\mu$m), $NH_3$ (∼ 10.7 $\mu$m), $CO_2$ (∼ 15 $\mu$m) and many other trace species such as hydrocarbons and more exotic nitrogen/sulphur-bearing molecules. If detected, the relative abundance of all these species could be compared among different EPs and with Solar System planets and bodies. Note that giant planets like Saturn in the Solar System show strong $NH_3$, $PH_3$ and $H_2O$ features around $\sim 5\ \mu$m (de Graauw et al. 1997). SPICA will also have access to the PAHs band features which are excellent tracers of UV-radiation fields; (3) He-$H_2$ and $H_2$-$H_2$ collision induced absorption band features as tracer of the He/H relative abundance; (4) Features from deuterated molecular species to distinguish cool brown dwarfs from "real" EP (e.g., $CH_3D$ at $\sim 8.6\ \mu$m) and non-equilibrium species (e.g., $PH_3$ at $\sim 8.9$ and $\sim 10.1\ \mu$m). Were the spectral coverage to be extended down to 3.5 $\mu$m then $H_3^+$ ($\sim 4\ \mu$m) would also be accessible, and with it the key to understanding the cooling processes in the upper atmosphere (Koskinen et al. 2007).

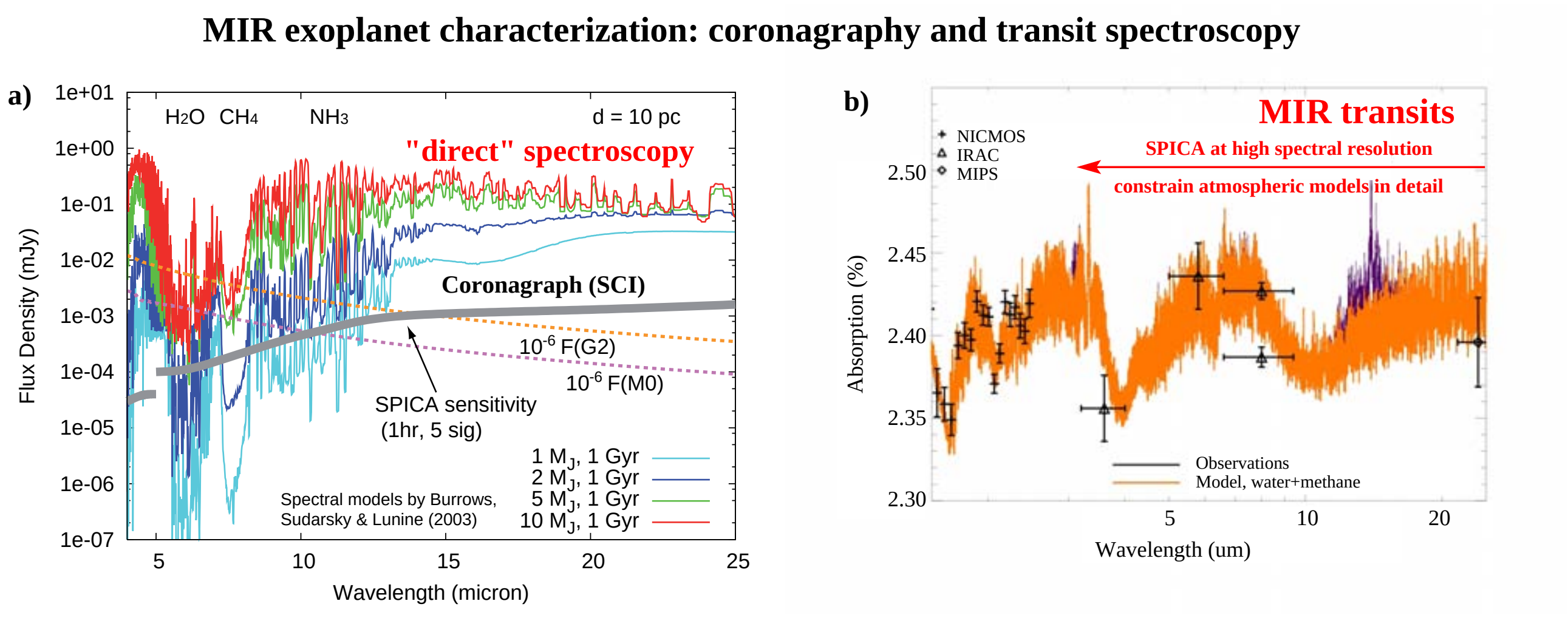

*Figure 1.10:* ***(a)*** *Simulated spectra for a range of exoplanet masses at an age of 1 Gyr and a distance of 10 pc (from Burrows et al. (2003) models). SCI is the only planned coronagraph that will carry out MIR "direct" spectroscopy of outer (>10 AU) and young EPs.* ***(b)*** *Hubble and Spitzer primary transit of HD 189733b hot Jupiter (photometry) and $H_2O$ and $CH_4$ models (by G. Tinetti). Transit observations with SPICA spectrometers will characterise the atmospheric composition of many EPs at the required higher spectral resolution.*

### 1.3.3 Mid-IR "transit" photometry and spectroscopy of exoplanets

The SPICA MIRACLE, MIRMES and MIRHES instruments will cover the mid infrared range with low, medium, and high spectral resolution (out to $R \sim 30\,000$). Following the unexpected but successful observations of "hot-Jupiters" EPs made by Spitzer, these instruments will be used to study primary and secondary transits of EPs orbiting close to the star. SPICA's spectrometers will be used to perform (1) multi-wavelength transit photometry of hot Jupiters routinely and (2) carry out "transmission" and "occultation" spectroscopy of an appropriate sample of bright EPs. Hot-Jupiters in particular, are bright and the possibility to carry out detailed atmospheric spectroscopy will allow us to study exo-atmospheres in great detail. This will help to develop the techniques and models that will be needed to interpret the spectra of habitable exo-Earths in the future. We estimate that the $\sim 24\ \mu$m thermal emission of gas giant planets similar to HD 209458b could be extracted from secondary transit observations around stars as far as $\sim 150$ pc (a few hundred star targets) as the contrast requirement is relatively modest ($\sim 0.1\%$). We anticipate that several giant EPs, with a great diversity of mass, semi-major axis, eccentricities, etc. will be available to SPICA for detailed mid infrared characterisation by $\sim 2018$. With even higher photometric precision ($\sim 0.01\%$), SPICA will have access to more challenging goals such as inferring potential Saturn-like rings around EPs (Barnes & Fortney 2004).

In the case of EPs around cooler stars, a super-Earth ($2-3R_{\rm earth}$; $T_p \sim 300$ K) orbiting around the habitable zone of a cool M8 star ($T \sim 2500$ K) will produce an intrinsic flux of $\sim 25\ \mu$Jy at MIR wavelengths, roughly a few times higher than the projected $1\sigma$-1min photometric sensitivity of SPICA's MIR instruments (but very hard to achieve with ground-based MIR instruments on "Extremely Large Telescope"-type observatories). These photometric flux levels are accessible for SPICA secondary transit studies with a contrast of $\sim 0.1\%$. Mid infrared transmission spectroscopy will probe the low-pressure layers of EP atmospheres where non-equilibrium chemical conditions occur. The first optical transmission spectrum taken from space used a medium resolution spectrometer ($R \sim 5500$; Charbonneau et al. 2002). More recently, high resolution ($R \sim 60\,000$) ground-based optical transmission spectra have been used to resolve EP spectral features (Redfield et al. 2008). Either at modest or high spectral resolution, SPICA mid infrared spectra of transiting EPs will provide a unique opportunity to determine the properties of their atmospheres through detailed analysis of band profiles, e.g., to extract molecular abundances and isotopic ratios accurately. In summary, SPICA will be an important intermediate milestone in exoplanet research, both in science achievements (direct spectroscopy of outer EPs for the first time in the MIR, atmospheric characterisation of transiting EPs at long wavelengths...) and in the required technological developments for future longer-term missions (e.g., coronagraphic techniques, cryogenic systems...).

## 1.4 The life cycle of gas and dust

SPICA will provide an unprecedented window into key aspects of the dust life-cycle both in the Milky Way and in nearby galaxies: from the dust formation in evolved stars, its evolution in the ISM, its processing in supernova-generated shock waves and massive stars (winds, HII regions, etc.), to its final incorporation into star forming cores and protoplanetary discs. Dust grains are a key player in determining the energy budget of the interstellar and circumstellar media because of their fundamental role in reprocessing stellar UV/visible photons into IR to sub-mm radiation, and in heating the gas via photo-electron emission from small grains and PAHs. SPICA will be the first space telescope since ISO (launched in 1995) covering in spectroscopy the uninterrupted MIR/FIR domain where astro-mineralogy studies can be carried out. ISO caused a revolution in the field of astromineralogy (Henning 2003) and showed that much can be learnt by following the evolution of the grains composition. The critical keys to unlock our understanding of the dust evolutionary cycle are MIR/FIR spectroscopy and the FIR dust SEDs. The former gives direct access to the poorly known ice and grain composition and its evolution along the dust life-cycle, and the latter enables us to determine the dust temperature uniquely.

In addition to dust, other very important gas-phase species that are hard, often impossible, to detect from the ground have their spectral signatures in the MIR/FIR: ions, atoms, light molecules and heavier organic species. Their associated MIR/FIR spectral features provide means to derive physical conditions in environments that are difficult to probe at other wavelengths (e.g., in UV/X-ray illuminated PDR/XDRs, shocked regions, galactic nuclei, etc.). They provide clues on the elemental abundances (C, O, S, D, Si, Fe ...) and also provide deep insights into the gas/dust chemical interplay: ice formation/evaporation, deuterium enrichment, grain growth and metal depletion (Okada et al. 2008). In this context, SAFARI will provide far infrared spectroscopic images of the regions that are too obscured for JWST to examine in the mid infrared or too warm/extended to be efficiently traced by ALMA and SKA interferometers. Thanks to its superb sensitivity, SPICA will extend our knowledge of the physics and chemistry of the gas and dust in our galaxy (e.g., acquired with Herschel) to similar detailed studies in nearby galaxies. In the following we list several fields where we anticipate that SPICA will play a unique role. This is not an exhaustive list, and probably does not do justice to the great discovery space available to SPICA (see section 1.8):

***Chemical complexity beyond our galaxy:*** Spectral observations from UV to cm wavelengths reveal a high degree of chemical complexity in our Galaxy that was until recently not expected. Such complexity is demonstrated by the diversity of detected species, that range from simple light hydrides, molecules carrying heavy metals, alcohols (methanol, ethanol,...), and a collection of many organic families (acetylenic chains, methane or benzene; Cernicharo et al. 2001) that lack permanent electric dipole (i.e., do not have radio spectrum to be observed with ALMA) but show MIR/FIR features. Either in the gas phase or as ice mantles, those species can be precursors of more complex molecules (sugars, amino acids, etc. Muñoz Caro et al. 2002). SPICA spectrometers will open new windows to probe the chemical complexity in the universe.

***Evolved stars (dust factories):*** The principal objects injecting dust into the ambient ISM in our Galaxy are the evolved stars, mainly red giants and AGB stars that predominantly emit at MIR/FIR wavelengths. SAFARI will be specially adapted to study the inner parts of their circumstellar envelopes (CSE), and to study the mass-loss history and mineralogy of a large sample of targets in very different environments including the closest galaxies. SAFARI observations will include (1) detailed spectroscopic studies of unresolved CSEs, (2) point source photometry of unresolved AGB populations in nearby galaxies, and (3) spectroscopic images of spatially resolved CSEs and their faint interaction with the ISM. For nearby objects, SAFARI will provide spectroscopic images of faint but very extended "detached" shells around evolved stars revealing their physical and chemical structure in great detail (for a prototype shell like TT Cyg, $\sim 5''$ corresponds to only $\sim$100 years of expansion). Photometric measurements will constrain the mass-loss rates in AGB stars out to the Large Magellanic Cloud (LMC; $d \sim 50$ kpc) in modest integration times. For example, the expected flux at 100 $\mu$m of a star like R Cas at the distance of the LMC is $\sim 200\,\mu$Jy, while detached shells like TT Cyg will have fluxes of $\sim 500\,\mu$Jy, requiring an integration time of a few minutes with SAFARI (see Figure 1.1).

***Supernova remnants (dust processing):*** Stars much more massive than our Sun end their lives with a violent explosion where massive flows are blown outward and shock waves propagate through the surrounding medium. When supernova (SNe) remnants encounter molecular clouds, they drive slower shock waves that compress and heat the molecular material and result in strong MIR/FIR line emission (Neufeld et al. 2007). At the same time, dust grains are processed (eroded, fragmented, amorphised ...), their physical state changes, and large quantities of refractory elements are sputtered back to the gas phase (Si, Fe, Mg, etc.). Dust grains may condense after the shock passage, and therefore SNe may contribute to the formation of dust. In fact, high-z galaxies seems to contain dust that can only have formed in SNe ejecta because the galaxies are too young for AGB stars to have formed dust. Thus, dust condensation must be efficient, requiring about 1 $M_\odot$ of dust per SNe. However, the efficiency of SNe dust formation in our own Galaxy is still debated and typically seems to be inefficient, for example, SNe 1987a only formed about $10^{-4}$ $M_\odot$ of dust. SPICA spectroscopic images will allow us to infer the physical conditions prevailing in the shocked gas while directly tracing the dust emission to the faint levels required to measure the dust production, and even its composition.

***ISM (The impact of low metallicity):*** The ISM is the repository of the metals ejected by dying stars as well as the site for the birth of the next generation. Winds of low and high mass stars and SNe explosions continually raise the metallicity of the ISM where recycling is ongoing. In local metal-poor ISM, representative of the early epochs of star formation, the spectral fingerprints of both dust and gas differ markedly from those of gas-rich starburst galaxies. The conditions for star formation as well as the effects of star formation on the ISM in metal-poor ISM can give us insight on the process and the effects of cosmic enrichment. Thus the study of the lifecycle of dust and gas in galaxies is incomplete without the dimension of metallicity. Large numbers of low-metallicity, actively starforming dwarf galaxies are known in the local Universe, and these provide ideal laboratories in which to study the interplay between star formation and the ISM, in environments that are reminiscent of conditions likely to be found in very young galaxies in the distant and youthful Universe. Recent surveys with ISO and Spitzer have observed the dust and MIR properties of less than 75 dwarf galaxies: only a handful of the most metal-poor (1/20 of solar) were detected (Madden et al. 2006; Engelbracht et al. 2006), however. Spectroscopic surveys of the major FIR cooling lines are planned with Herschel: Just detecting the [C II]158 $\mu$m line in the lowest metallicity galaxy will be a major breakthrough. To determine the properties of the dense and diffuse components of the ISM in these unexplored systems we need both the molecular and atomic as well as FIR fine structure diagnostics. The factor of 100 gain in sensitivity of SAFARI will bring FIR spectroscopy of this enigmatic population well within our reach, and provide an important benchmark against which to compare studies of distant objects. While the more massive galaxies have been the targets of the majority of high redshift surveys, recent spectroscopic surveys of the low-mass population at $z \sim 1$ (Davies et al. 2009) are beginning to put dwarf galaxies into a clearer perspective in galaxy evolution, favouring the downsizing scenario of galaxy evolution (Cowie et al. 1996). A SAFARI survey of these targets would be the next step in understanding the evolution of metals in the gas and dust, and how these properties affect the star formation density as a function of redshift. The importance of low metallicity at high redshift is highlighted in recent studies of gamma ray burst hosts (Chen et al. 2009), Lyman-Break galaxies (Mannucci et al. 2009) and damped Ly-$\alpha$ systems (Pettini et al. 2003, 2008), which all bear the tell-tale hallmarks of low metallicity dwarf galaxies.

## 1.5 Galaxy evolution: the coevolution of stars and supermassive black holes

The study of galaxies in their cosmological context has undergone important changes during the past years, with the accumulation of evidence that the standard picture based on a naive interpretation of the hierarchical scenario of galaxy formation is not consistent with observations. This may be summarised using simple key-words such as the “downsizing” or “bimodality” of galaxies, however it certainly reflects a major limitation of our theoretical framework that only observations which can be directly compared to theory will overcome.

When trying to define the observations required, astronomers have been confronted with two major limitations – firstly, the increasing importance of obscuration acting both on newly formed stars and supermassive black holes (SMBH) in their growing phase, and secondly the so-called confusion limit. The confusion limit is a technical limitation that cannot be avoided when observing at long wavelengths. Due to large beam sizes and high projected source densities, the very same galaxies we wish to study make the Universe opaque. These limitations call for a new generation of infrared instrumentation such as SPICA, as other existing or scheduled observatories will all fail to address them. To overcome these two sources of blindness – i.e., intrinsic dust absorption and projected overlapping of galaxies – there is an optimal definition of a telescope diameter for a given wavelength range. Herschel is limited by confusion in five of its six broadband filters, i.e., above 100 $\mu$m, and as a result will not resolve more than half of the infrared background light resulting from the cumulative dust obscured star formation and accretion in galaxies over the Hubble time. SPICA, like Herschel, will have a 3.5 m telescope due to the limitations imposed by the launch vehicle lift capability and payload shroud. In (Figure 1.11a) we show the detection limit versus wavelength at which a 3.5 m aperture becomes limited by confusion between sources and the amount of the observable Universe that will be resolved at each wavelength. We can see two things from these plots: (1) the optimal wavelength at which to observe the extragalactic background with a 3.5 m telescope is around 70 $\mu$m and (2) that Herschel-PACS will never reach the confusion limit at this wavelength but SPICA SAFARI will reach it easily. The extremely high spatial resolution of ALMA, on the other hand, is not suited to the study of galaxies in their cosmological context, since it will only follow-up targeted galaxies or produce pencil beam surveys. These are, by definition, not suited to developing an understanding of the anti-hierarchical behaviour of galaxies, since they will suffer from cosmic variance and limited statistics. Submillimetre observations from ground-based facilities are also limited by confusion to a few mJy, hence to the very brightest tip of the galaxy population (i.e., confined to galaxies more luminous than a hundred times $L_*$ galaxies, the average galaxy luminosity).

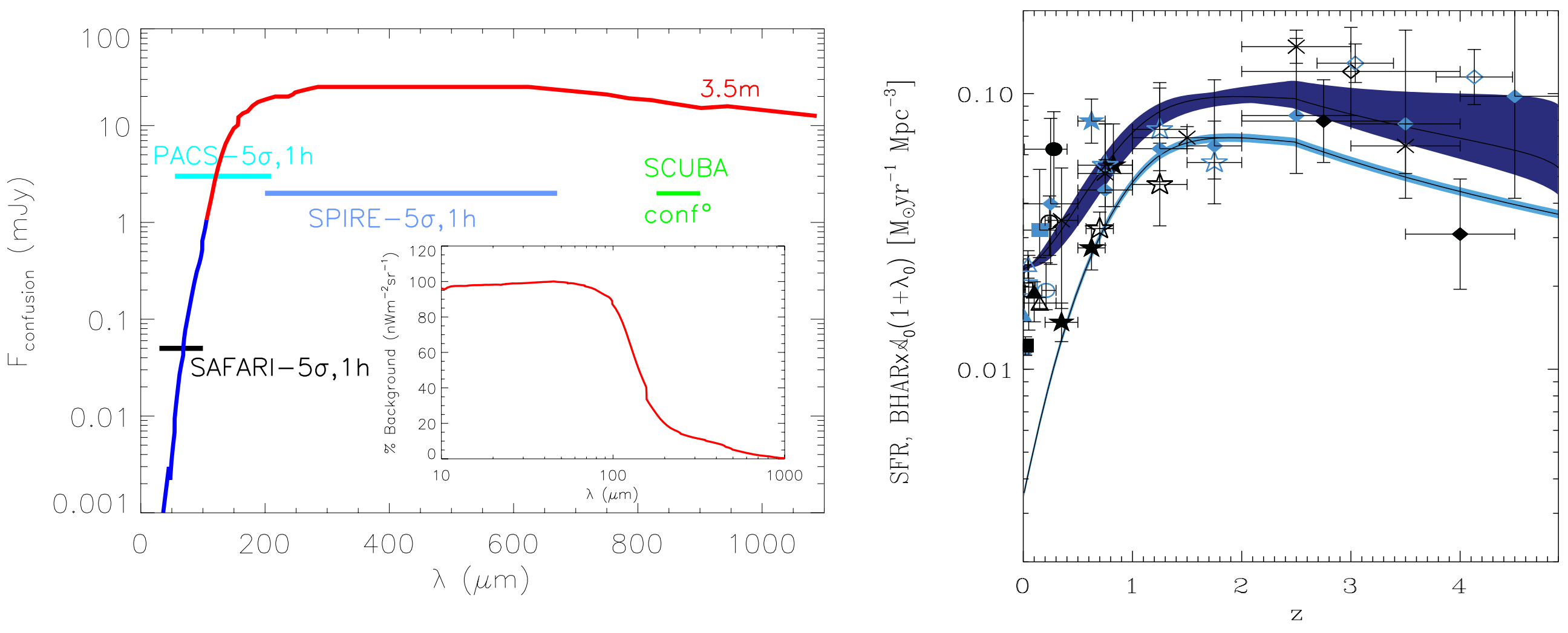

*Figure 1.11:* ***(a) Left panel****: A plot of the confusion limit in the FIR/sub-mm, with the wavelength coverage and sensitivity of selected instruments overlaid. The inset plot gives the % of the background at a given wavelength that is resolved by 3.5 metre-class telescope, as a function of wavelength – from this one can see that below 80 $\mu$m one can resolve almost all of the background;* ***(b) Right Panel****: A composite plot of the evolution of the star formation rate (SFR) density and black hole accretion rate, adapted from Merloni et al. (2006). The points represent values derived from observations (listed in Merloni et al. 2004) and the dark-blue shaded region the best fit model to the data. The light blue locus is a model of the black-hole accretion rate (BHAR, Merloni et al. 2006). Both the SFR and BHAR peak between z* $\sim 1-2$.

SAFARI onboard SPICA will survey the Universe in the most efficient waveband with which to study galaxy formation with a 3.5 m class telescope in space, and will therefore:

- Unveil the population of heavily obscured active nuclei responsible for the missing half of the cosmic X-ray background (CXB) at its peak around 30 keV (Comastri et al. 1995; Gilli et al. 2001, 2007).

- Provide, for the first time, a complete census on the star formation history of galaxies over the last 90% of the age of the Universe, by measuring dust obscured star formation down to the limits at which galaxies become optically thin, and redshifted ultra-violet light becomes a robust tracer of star formation.

- By combining the points above we will be able to follow the relative pace at which stars and black holes formed, to pinpoint the major events of both processes, and therefore to understand the connection between the two major sources of light in the Universe after the Big Bang, nucleosynthesis and accretion.

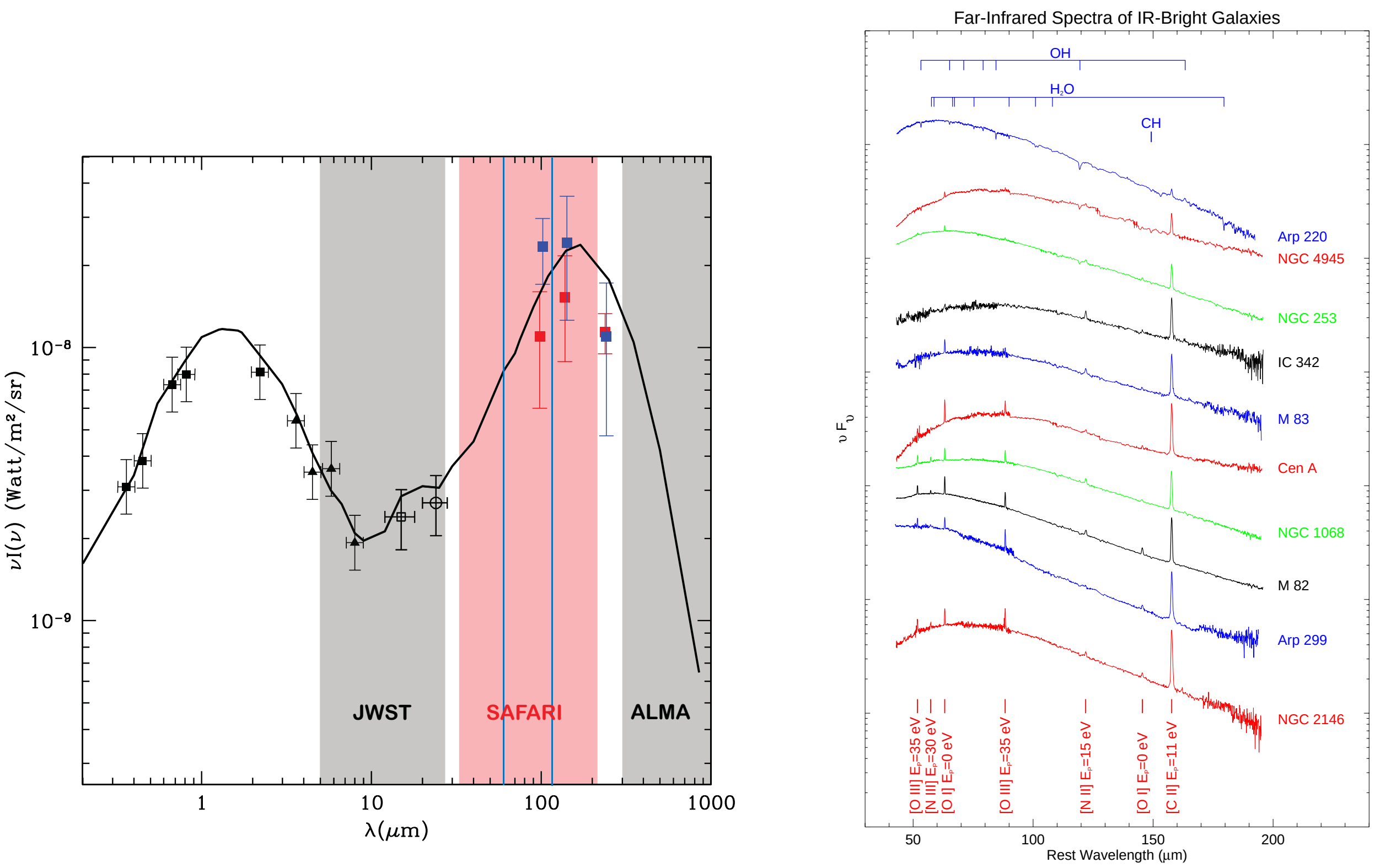

*Figure 1.12: **(a) Left Panel**: Recent constraints placed on the cosmic infrared background (CIRB) by Spitzer show that the integrated extragalactic background flux density at MIR/FIR wavelengths is comparable to that found in the cosmic optical background (COB: UV-visible-near-IR). The CIRB peak is a measure of the re-radiated emission from light processed by dust over cosmic time, whilst the COB measures the light emitted by stars and AGN. It is clear that a significant piece of the galaxy evolution puzzle is missed if we neglect the IR. The solid line traces out the CIRB spectrum that is predicted by the galaxy evolution model of Franceschini et al. (2009). References for the observational data points can also be found in this paper. **(b) Right panel**: The full ISO-LWS spectra of ten local galaxies, illustrating the range of emission line strengths observed. These differences will be explored in detail by Herschel, and will provide a zero-redshift benchmark against which to compare distant galaxies. The spectra have been offset vertically and plotted as a sequence (bottom-to-top) of increasing IR luminosity (Fischer et al. 1999).*

- Resolve the individual sources that make up the cosmic infrared background (CIRB).

### 1.5.1 Major recent discoveries and open problems

Two important discoveries in recent years have changed our perspective of galaxy evolution: the first is the strong observed correlation in the local Universe between the mass of the black hole at the centres of massive galaxies and the key properties of the host galaxy, and the second the existence of a dusty population of galaxies that emits a significant fraction of their bolometric luminosity in the FIR/submillimetre (sub-mm). The strong correlation observed in the local Universe between the masses of central black holes and the luminosities, dynamical masses, and velocity dispersions of their host spheroids (e.g., Magorrian et al. 1998; Ferrarese & Merritt 2000; Ferrarese & Ford 2005; Shankar et al. 2009) implies that the processes of black hole growth (through gravitational accretion) and bulge formation (through star formation (SF)) are intimately linked. Optical studies of the local massive galaxy population show that most, if not all, galaxy spheroids host massive relic black-holes (Richstone et al. 1998) which, in turn, suggests that a mass-accreting AGN (active galactic nucleus) phase is one through which all massive galaxies pass. The origin of the correlation cannot be explained by observations of the local Universe, where the two processes appear to be almost independent of one another, and so must be seeded at much earlier epochs. The evolution of the two processes with time can be seen in Figure 1.11b. Global accretion power, measured using X-rays (Hasinger et al. 2005) and the star formation power, measured by H$\alpha$ and rest-frame UV observations (Shim et al. 2009), were both ~20 times higher at $z = 1 - 1.5$ than today. That both peak at around $z \sim 1 - 2$ strongly suggests the co-evolution of SMBHs and the star formation rate (SFR) (e.g., Marconi et al. 2004; Merloni et al. 2004). To understand and unravel the relationship between the blackhole growth and bulge formation we need to track back the evolution of black hole growth and star formation over cosmic time, and determine whether there is a causal link between the growth of the central black hole and its host spheroid. If there is such a link, how and when is it established, and which, if either, process drives or regulates the other?

We know from galactic and extragalactic studies that the processes of star formation and early stellar evolution are typically deeply enshrouded in dust. This is also the case for active galactic nuclei (AGN), which are often optically obscured both locally (e.g., Goulding & Alexander 2009) and at high-z (Hernán-Caballero et al. 2009). Galaxies therefore pass through the most active periods of their lives deeply obscured by dust. The second important discovery – of a population of distant, dusty IR-bright galaxies that were missed by optical surveys – is a direct manifestation of this. Luminous and ultra-luminous infrared galaxies (ULIRGs) are rare in the local Universe, but become much more common (factor of 1000) at high-redshift. Evidence for this first came from the identification of the so-called sub-mm galaxies (SMGs) (Smail et al. 1997; Hughes et al. 1998) which emit a significant fraction of their rest-frame bolometric luminosity in the FIR/sub-mm, and was subsequently confirmed with Spitzer. The integrated luminosity of the (U)LIRG population accounts for the majority of the CIRB (Dole et al. 2006), the infrared component of the extragalactic background light, yet this population is insignificant in the optical. Shown in Figure 1.12a is a plot of the extragalactic background light from optical to millimetre wavelengths. The two peaks in the energy density – the peak in the CIRB which includes reprocessed light emitted by dust, and the peak in the optical (the cosmic optical background, COB), which is made up of contributions from stellar processes – are of approximately equal height, and underline the importance of the IR. Without a detailed understanding of the IR-bright, dust-obscured galaxy population we miss a substantial piece in the galaxy evolution puzzle.

The discoveries discussed above shed new light onto one of the most topical problems of cosmology: the origin and shaping of the galaxy mass and luminosity functions. On the one hand, the very luminous phases of galaxy evolution revealed by FIR/sub-mm observations may explain the formation of spheroidal galaxies and galaxy bulges, with their massive stellar populations. On the other, the nuclear activity from gravitational accretion, very often found concomitant with the ultra-luminous IR phase, is believed to play a key role in shaping the galaxy stellar mass and luminosity functions. The highly energetic feedback from the central AGN could be responsible for regulating the cooling and collapse of primordial gas into the forming galaxy, essentially stopping star formation above a given threshold for the SMBH and host galaxy mass. This feedback might be primarily responsible for the well-known exponential shape of the universal mass and luminosity functions of galaxies, and its fundamental divergence from the power-law shape of the parent dark-matter halo mass functions. Unfortunately, this will remain a mere conjecture without suitable instrumentation to penetrate the dust opaque media wherein this activity occurs.

**What are the key questions to be answered?** Our current picture of galaxy evolution raises many unanswered questions as we attempt to reconcile observational data with models:

- What drives the evolution of the massive, dusty distant galaxy population, and what feedback/interplay exists between the AGN and star-forming phase?
- What physical processes are responsible for shaping the mass and luminosity functions of galaxies, the universal form of which both differ considerably from that of dark-matter halos?
- How do galaxy evolution, star formation rate and AGN activity vary with environment and cosmological epoch?
- How and when do the normal, quiescent galaxies such as our own form, and how do they relate to (U)LIRGs?

Two key requirements need to be met to address these questions: first, we need access to the MIR/FIR to overcome the obscuring effects of dust and to fingerprint and track both star formation and AGN activity through cosmic time. Second, we need an imaging capability combined with a spectroscopic capability to undertake the large-scale surveys required to locate and study these dusty sources during their evolution. These two requirements translate into high sensitivity, large instantaneous spectral coverage and spectroscopic imaging capabilities. SPICA provides all of these.

In the remainder of this section we demonstrate how these instrumental features will address the science questions posed above: in subsection 1.5.2 we make the case for an MIR/FIR spectroscopic imaging capability, and highlight how this waveband can be used to identify and investigate the physical processes driving energy production in individual galaxies; in subsections 1.6.1 and 1.6.2 we illustrate how the deep FIR spectroscopic and photometric imaging capabilities of SPICA will revolutionise our picture and understanding of galaxy evolution.

## 1.5.2 MIR/FIR imaging spectroscopy is the key observational technique

Hot and young stars and black hole accretion discs show strong differences in the shape of their ionising continuum. However the far UV continuum, dominating the total bolometric output luminosity in both processes, is in general not observable directly, due to absorption by HI. The best tracers and indeed discriminators of accretion and star formation are therefore emission lines from the photo-ionised gas. In both stabursts and AGN a fairly constant fraction (10 – 20%) of the ionising continuum gets absorbed by gas and then re-radiated as line emission, making spectroscopy a powerful diagnostic tool. Detecting the exact fraction of ionising radiation absorbed by gas surrounding the power source is however not crucial if one uses emission line ratios, which are therefore the most efficient observable probes of energy production mechanisms. The principle spectroscopic tracers of star formation are in the rest-frame optical, however, these can be rendered useless by extinction in dusty objects. Emission lines in the near-, mid- and far-IR are much less affected by dust extinction, and thus provide powerful diagnostics of obscured regions dominated by star formation and AGN activity. MIR/FIR spectroscopy is therefore essential to tracing the energetics and evolution of these regions because it is only with MIR/FIR spectroscopy that we can push past the effects of dust obscuration to determine the physical processes powering the individual galaxies. A wide field imaging capability with broad instantaneous spectral coverage provides the key to surveying large areas of the sky, so tracing galaxy evolution in large numbers of objects in an unbiased way.

By the launch of SPICA, deep cosmological surveys undertaken by ISO, AKARI and Spitzer along with those planned for Herschel and SCUBA-2 will have produced catalogues containing the fluxes of many tens of thousands of faint MIR/FIR/sub-mm sources. Photometric surveys can be used to determine source counts and establish the contribution of the observed populations at the given wavelength to the extragalactic background. However, to derive source luminosities, a reliable estimate of the source redshift is required implying a great amount of multiwavelength photometric data and the use of galaxy templates derived from local observations. Even more importantly, *photometric data alone do not allow us to unambiguously differentiate between AGN and star formation activity.*

Substantial progress in studying galaxy evolution therefore can only be made by making the transition from photometric to spectroscopic surveys as it is only with direct MIR/FIR spectroscopy that we can unambiguously characterise the detected sources. This will not only allow for a direct census of galaxies to be made, but will also measure the AGN and starburst contributions to the bolometric luminosity of all detected galaxies over a wide range of cosmological epochs. *And this will all happen in a single observation*. The *enormous* advantage between the photometric surveys that Herschel will perform and those that SPICA-SAFARI will make is precisely this: the capability of obtaining **unbiased spectroscopic data in a rest-frame spectral range that have been shown in the local Universe to be mostly unaffected by extinction and to contain strong, and unambiguous signatures of both AGN and star formation emission.**

#### The diagnostic power of the rest-frame MIR and FIR

Pioneering spectroscopy of a handful of nearby star forming galaxies and AGN with ISO-LWS (Fischer et al. 1999; Braine & Hughes 1999) has demonstrated the importance and diagnostic power of the rest-frame FIR.

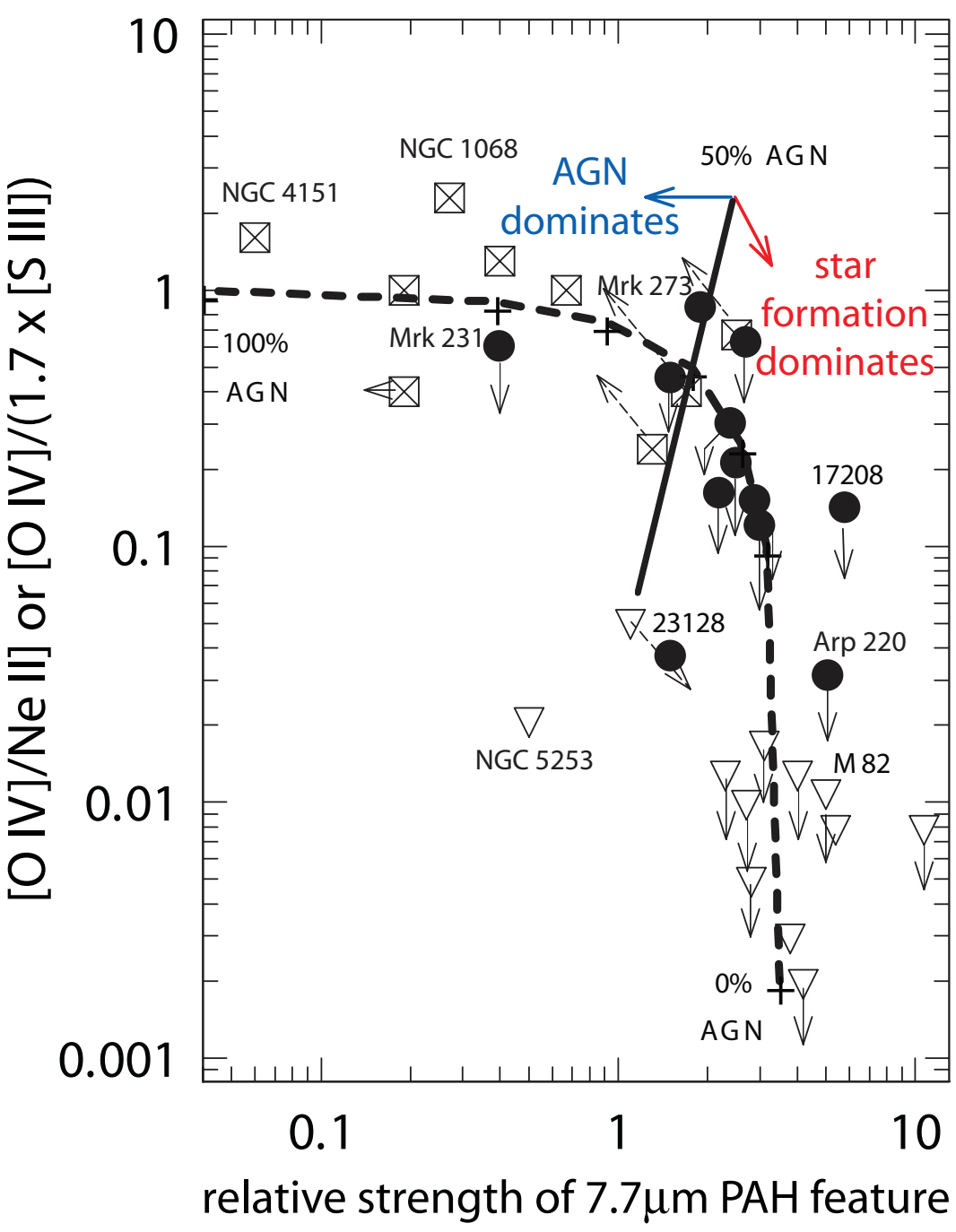

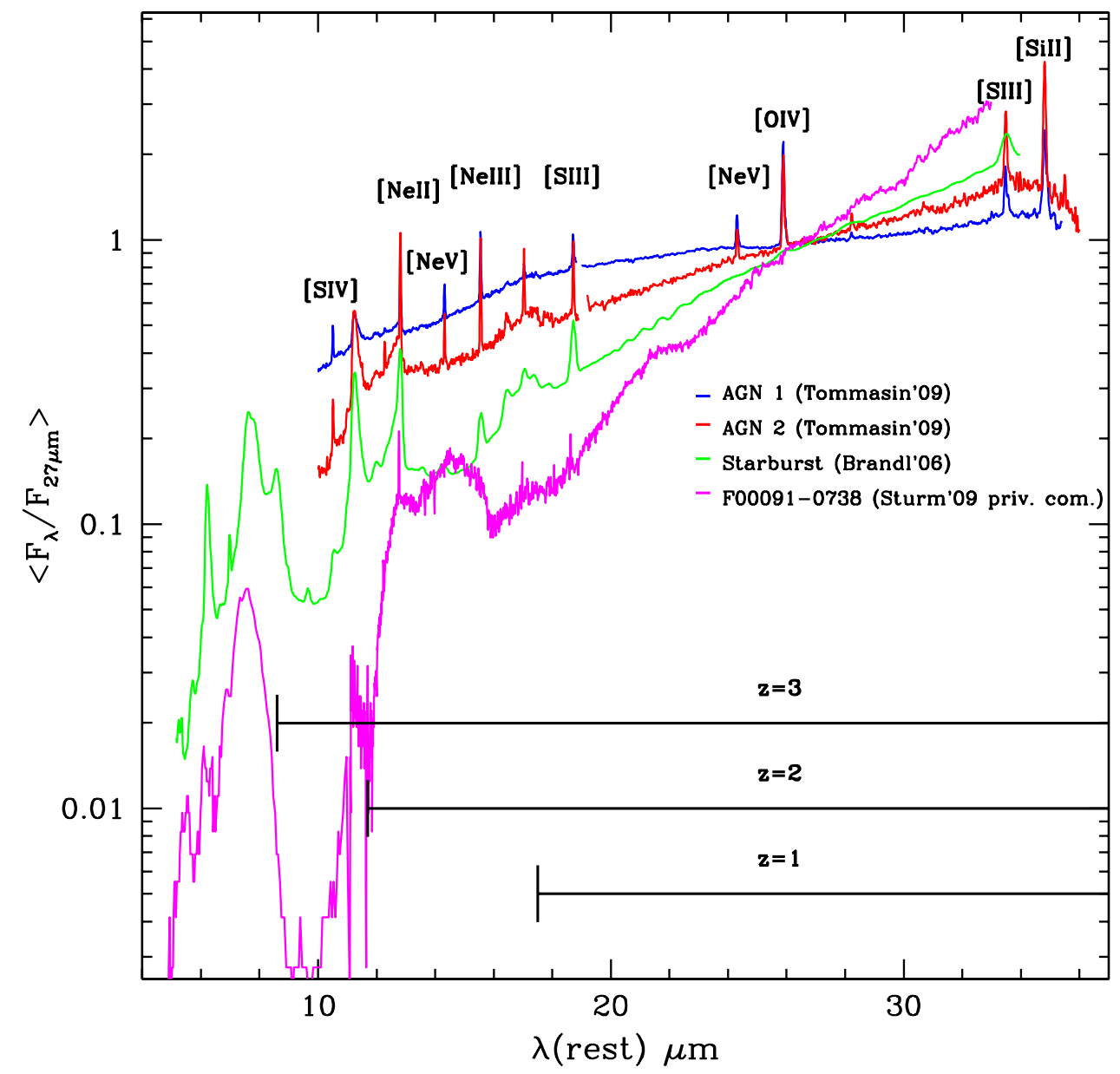

*Figure 1.13: **(a) Left panel**: An example of how combinations of MIR line emission/feature emission can be used to determine the relative importance of AGN- and starburst-activity in IR-luminous galaxies in the local Universe (Genzel et al. 1998). This and other line ratio/feature combinations that will be formed from SPICA spectra will provide the key to identifying and separating AGN and stabursts in distant galaxies; **(b) Right panel**: An illustration of the range of MIR diagnostic lines and features that can be seen in Seyfert type-1/type-2 AGN (Tommasin et al. 2009), a starburst galaxy (Brandl et al. 2006) and a heavily obscured ULIRG (Sturm, priv. comm.). Many of the strongest of these lines will be detectable with SAFARI in a single 1-hr observation. Superposed onto the plot is the fraction of the waveband that will be accessible to SAFARI at z ∼ 1, z ∼ 2 and z ∼ 3. The spectra have been normalised at ∼ 27 μm.*

Shown in Figure 1.12b are ten different spectra that have been observed: from strong atomic/ionic fine-structure line emission originated in photon dominated regions (PDR) and HII regions, to spectra showing strong molecular and even atomic absorption lines. ISO observations show, for example, that FIR fine structure lines can be used to separate starburst and AGN (Spinoglio et al. 2003). Star formation produces [O III]88 μm that originates from HII regions as well as strong [C II]158 μm emission from PDRs, whilst AGN activity produces strong emission of [O III] in the Narrow Line Regions (NLR), but also transforms a PDR into an X-ray dominated region (XDR), significantly enhancing the [O I]63 μm emission (Meijerink et al. 2007). The line ratio diagram of [C II]/[O I] vs. [O III]/[O I] can therefore be used to separate the two sources of energy production. Molecular emission lines can be used to probe the spatial extent and the geometry of the FIR continuum in the centres of galaxies (González-Alfonso et al. 2008). The high excitation lines of OH and $H_2O$ provide constraints on dust opacity models and OH and possibly the high-J lines of CO can be used to detect the molecular tori around AGN and study their physical conditions (González-Alfonso et al. 2008).

Herschel-PACS will observe complete 55 – 210 μm spectra of a relatively small number of local galaxies and confirm statistically the diagnostic power of the FIR molecular and atomic lines. In this way it will calibrate locally the tools with which to study distant galaxies which can only be seen with SPICA.

The rest-frame MIR waveband contains several emission lines and spectral features which measure the contributions from AGN and star formation to the overall energy budget. These features do not suffer the heavy extinction that affects the UV, optical and even the near-IR lines, and therefore provide an almost unique diagnostic in highly obscured regions (Spinoglio & Malkan 1992). ISO demonstrated that the ratios of emission lines tracing the hard UV field found in the narrow line region of AGN (e.g., [Ne V], [O IV], [Ne VI]) to those tracing stellar HII regions (e.g., [S III], [Ne II]) versus the strength of the PAH emission features - indicators of star formation - define a diagram which separates well the star formation and the AGN component in obscured galaxies (e.g., Genzel et al. 1998, Figure 1.13a). Spitzer observations have exploited these results and have demonstrated that MIR tracers can be used to quantify the properties of AGN (Tommasin et al. 2008, Tommasin et al. 2009, submitted), weigh the black holes driving the AGN activity (Dasyra et al. 2008), reveal previously unknown populations of highly obscured AGN (Goulding & Alexander 2009), pin down the physical properties of the local ULIRG population (Armus et al. 2007), trace directly the luminosity of the AGN (Meléndez et al. 2008; Rigby et al. 2009), and start to trace the lifecyle of local (U)LIRGs (Veilleux et al. 2009). The strong [Ne v] lines are unambiguous tracers of the presence of an AGN, as was shown by the detection of [Ne v]14.3 $\mu$m in NGC6240 (Armus et al. 2006), previously classified as a starburst galaxy before the detection of a buried AGN in the X-rays by Chandra (Komossa et al. 2003). Spitzer spectra have also demonstrated that the MIR hosts several new candidates for indicators of star formation in extragalactic objects: [Ne ii]12.8 $\mu$m, [S iii]34 $\mu$m, and [Si ii]35 $\mu$m lines, as well as the PAH features (e.g., 7.7 $\mu m$, 8.6 $\mu m$, 11.25 $\mu$m) and the $H_2$ pure rotational emission lines (e.g., 17.04 $\mu$m).

## 1.6 The role of SPICA-SAFARI

The SAFARI instrument on SPICA will exploit the power of MIR/FIR imaging spectroscopy and photometry to study the dust-enshrouded processes that drive galaxy evolution. With its broad instantaneous spectral coverage and high sensitivity, SAFARI will be able to cover the full 34 – 210 $\mu$m band in a few thousandths of the time that Herschel-PACS will take to obtain a deep spectrum of comparable resolution. SAFARI will be able to detect the key MIR and FIR lines from distant galaxies and use the MIR/FIR diagnostics that have, or will have, been revealed and calibrated by Spitzer and Herschel, respectively, in the local Universe. At a redshift of $z \sim 1$ the rest-frame 11 – 35 $\mu$m, which is very rich in ionic fine structure as well as $H_2$ rotational lines (see Figure 1.13b), moves to 22 – 70 $\mu$m and then out to 33 – 105 $\mu$m by $z \sim 2$ and will still be in the SAFARI spectral range out to a $z \sim 6$. We will see in the following sections that SAFARI will detect the brightest MIR lines to $z \sim 2$ in *intermediate luminosity* objects in a 1 hour integration (see also Figure 1.14b). This should be compared with Herschel-PACS, which only has the sensitivity to detect the very brightest MIR (and FIR) lines in several hours of integration *per line* on the most luminous objects at $z \sim 1$.

Photometrically, SAFARI will have the sensitivity to undertake very deep large-area, confusion-limited surveys at e.g., 70 $\mu$m (see Figure 1.16b), detecting all galaxies with $L_{IR} \geq 10^{11}\,L_\odot$ at a redshift of $z = 2$, all those with $L_{IR} \geq 5 \times 10^{11}\,L_\odot$ at a redshift of $z = 3$ and the most luminous IR galaxies ($L_{IR} \geq 10^{12}\,L_\odot$) out to $z = 4$, as well as the Milky Way-type populations ($L_{IR} \sim 10^{10}\,L_\odot$) out to $z \sim 1$.

The wide SAFARI FOV of $2' \times 2'$ gives an important spatial multiplexing advantage that will make it possible for the first time to collect *blind* spectroscopic surveys (Figure 1.14a) that are both wide and deep enough to measure the underlying physical processes driving galaxy evolution out to $z \sim 3$ and, in the most luminous/lensed objects, to even higher redshift. By comparing blind surveys with those targeted around known, high-$z$ objects, we will be able to determine the role of environment on galaxy evolution.

### 1.6.1 Spectroscopic cosmological surveys with SAFARI

The first question that we need to answer, before moving to more detailed modelling, is whether the local galaxies, for which we have detailed knowledge of their IR spectra, can be observed at high redshifts. In order to do so, we predicted the line intensities as a function of redshift (in the range $z = 0.1 - 5$) for three local template objects: NGC1068 (Alexander et al. 2000; Spinoglio et al. 2005) ($L_{IR} = 20 \times 10^{10}\,L_\odot$), a prototypical Seyfert 2 galaxy containing both an AGN and a starburst; NGC6240 (Lutz et al. 2003) ($L_{IR} = 50 \times 10^{10}\,L_\odot$), a bright starburst with an obscured AGN; and M82 (Förster Schreiber et al. 2001; Colbert et al. 1999) ($L_{IR} = 4 \times 10^{10}\,L_\odot$), the prototypical starburst galaxy.

Figure 1.14b shows the predicted intensities of a selection of fine-structure emission lines that trace AGN, stellar ionisation and PDR regimes. These have been plotted as a function of redshift for the three templates. It is clear from the figure that the Herschel-PACS spectrometer will be able to observe only the brightest object (NGC6240) up to $z \sim 2$ in the brightest line ([O i]63 $\mu$m). In contrast, the SAFARI *goal* sensitivity (5$\sigma$-1hr of $2 \times 10^{-19}$ W m$^{-2}$ at $R \sim 2000$) will enable us to detect, simultaneously - *in a single observation* - unambiguous

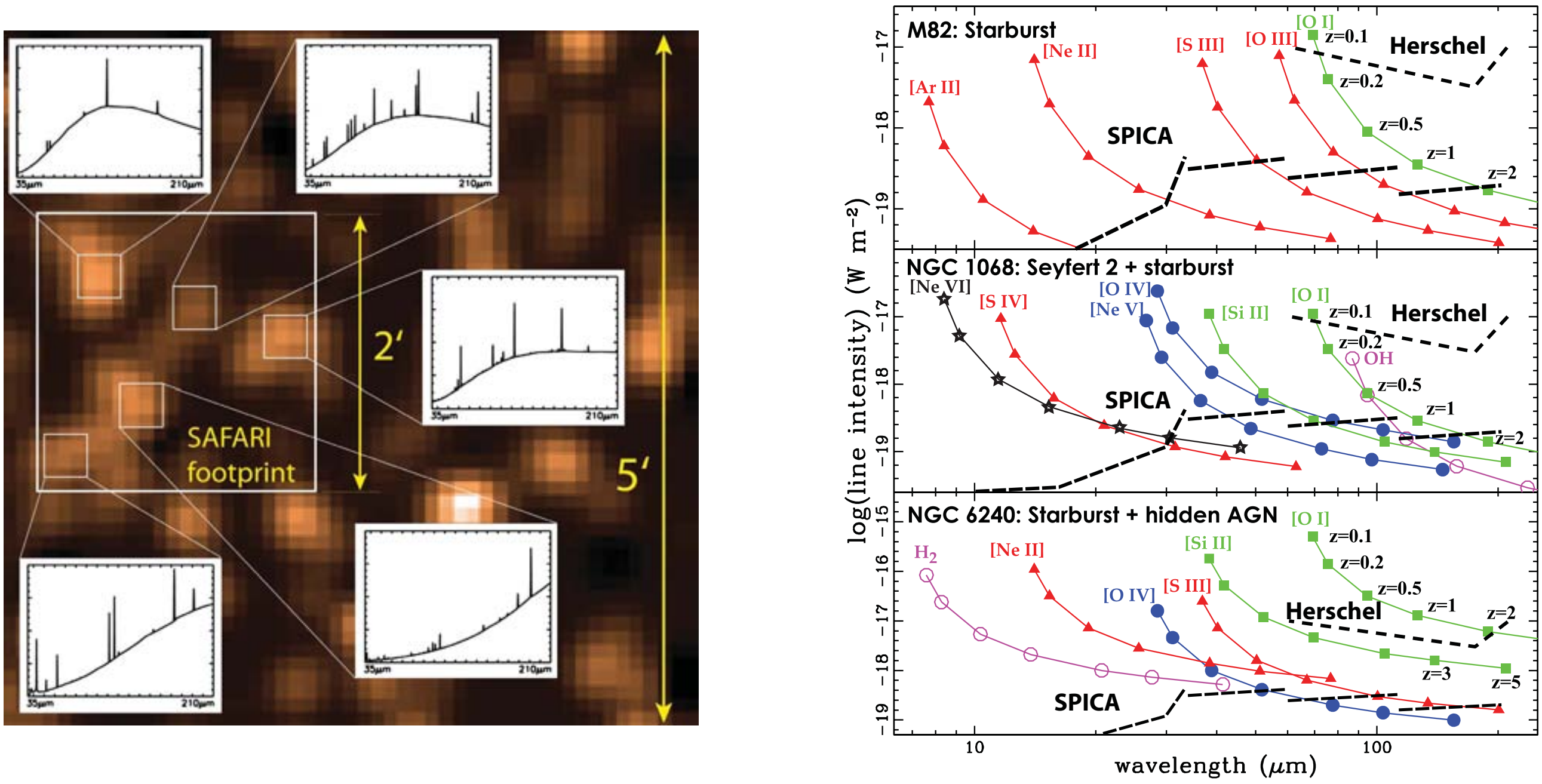

*Figure 1.14: **(a) Left panel**: A* $5' \times 5'$ *region taken from an off-source part of a 250* $\mu$*m ESA SPIRE publicity image, with the* $2' \times 2'$ *FOV of SAFARI outlined (credit: ESA and the SPIRE consortium). With its spectral imaging capability, SAFARI will be able to obtain spectral information covering the full* $34 - 210$ $\mu$*m range, in multiple sources, simultaneously – in a single pointing; **(b) Right panel**: SAFARI has the sensitivity to detect redshifted MIR/FIR emission out to high-z. Shown is a plot of line intensity vs. redshift of a selection of the key MIR/FIR emission lines visible with SPICA in three archetypical objects: M82 (*$L_{FIR} \sim 4 \times 10^{10}$ $L_\odot$*), NGC1068 (*$L_{FIR} \sim 2 \times 10^{11}$ $L_\odot$*) and NGC6240 (*$L_{FIR} \sim 5 \times 10^{11}$ $L_\odot$*) - in each panel the upper/lower dashed lines denote the* $5\sigma$*-1 hr sensitivity of Herschel-PACS and SAFARI, respectively.*

signatures of both AGN and star formation and to determine redshifts in these relatively low luminosity objects out to $z \sim 1-2$ for most lines, and even higher redshifts for the brightest lines.

Line diagnostic diagrams, such as the one shown in Figure 1.15a (using the [O IV]25.9 $\mu$m, [Fe II]26.0 $\mu$m and [S III]33.5 $\mu$m lines) can readily separate AGN-dominated and starburst-dominated galaxies. Critically, these redshifted MIR lines *all fall* into the SAFARI waveband at redshifts of $z \geq 0.4$.

The second task that we must fulfil to prove the feasibility of cosmological spectroscopic surveys, is to estimate of the number of galaxies that we expect to detect. We have used models constrained by observational data to predict the number of sources that we will be able to detect per SAFARI FOV at different redshifts in a blank-sky survey (see Spinoglio et al. 2009, for the details). Throughout, we assume the SAFARI goal sensitivity given above.

Our first simulation uses the galaxy evolution model developed by Gruppioni et al. (2009) and Gruppioni et al. (in prep.) that makes use of available IR data to determine luminosity functions from $z = 0$ to $z \sim 4$, with explicit contributions in the model from starbursts and AGN. To account for the population of spheroids (proto-ellipticals) which locally do not appear as substantial far-IR emitting sources, but are expected to be relevant at high redshifts (e.g., SMG galaxies), we integrated the model by Granato et al. (2004). This model uses an ab-initio approach for galaxy formation and evolution (see also Lapi et al. 2006) capable of predicting multi-wavelength luminosity functions and number counts. By following this approach, we calculate that we will detect around 7 sources over the redshift range $0 < z < 3$ in the [Si II]34.8 $\mu$m line. The lines of [O IV]26 $\mu$m (a tracer of AGN) and [S III]33 $\mu$m (a tracer of star formation) are, in the local galaxies, a factor $1.5 - 2$ fainter than the [Si II] line, but comparable on average in LIRGs and ULIRGs (Veilleux et al. 2009). Given this, we expect to detect all three lines in these sources.

In a second approach we adopted the model by Franceschini et al. (2009). This is a backward evolution model which fits all available data from Spitzer, ISO, COBE, SCUBA, etc. It includes direct determinations of multi-wavelength redshift-dependent luminosity functions from Spitzer and accounts in great detail not only for star forming galaxies, but also for type-1 and type-2 AGN. Using this model we predict that it will be possible to detect about 4 starburst galaxies, 2 type-2 AGN and 1 type-1 AGN - a total of 7 galaxies - in the redshift range of $0 < z < 3$ per SAFARI FOV assuming the standard $5\sigma$ 1hr sensitivity, and using as examples the [Si II]34.8 $\mu$m line to detect starburst galaxies and the [O IV]26 $\mu$m line to detect AGNs.

Using these two independent methods we predict that we will detect of order 5 to 10 objects per SAFARI FOV per hour within the redshift range $z \leq 3$. These estimates confirm that in a deep SAFARI survey of about 500 hours/2000 square arcmin we will detect a few thousand sources in the key MIR/FIR tracers that will identify the energy sources powering the detected galaxies. Deep spectroscopic surveys with SAFARI will therefore provide the **first statistical and unbiased determination of the co-evolution of star formation and mass accretion with cosmic time**. We have assumed a redshift-independent relationship between line luminosity and continuum/IR bolometric luminosity throughout: Were these to evolve with redshift, then the number of sources that will be detected by SAFARI in blind surveys would increase significantly, particularly at high-redshift. It is important to note that the increased abundance of ULIRGs at higher redshifts is accompanied by an evolution in their mode of star formation. Mid-IR spectroscopy with Spitzer-IRS (Farrah et al. 2008; Menéndez-Delmestre et al. 2009) and ground-based submillimetre spectroscopy (Hailey-Dunsheath et al. 2008) have demonstrated that the emission from ULIRGs at $z \sim 1-2$ is markedly different from that observed in local ULIRGs, instead resembling that seen in lower luminosity starburst galaxies. This result can be understood if the high-redshift ULIRGs represent vigorous star formation extended over a few kiloparsecs, rather than the sub-kpc bursts that power local ULIRGs, and may be a consequence of the higher gas mass fractions present at these earlier epochs.

Beside the rest-frame fine-structure MIR lines, other potential lines that can be measured by SAFARI include the oxygen lines of [O I]63 $\mu m$ and [O III]52 $\mu m$ and the nitrogen [N III]57 $\mu$m line, that can all be seen in the SAFARI spectral range out to redshifts of $z \sim 2$. The enhanced [C II] emission at high-$z$ (possibly due to low-metallicity effects, Nagao & Maiolino 2009), indicates that these FIR lines at high-$z$ might indeed be stronger than in the local Universe. Moreover, Spitzer observations of distant radio galaxies (Egami et al. 2006; Ogle et al. 2007), have unveiled a class of objects termed as "$H_2$ luminous galaxies" whose spectra are dominated by $H_2$ rotational emission lines. The lack of detection of PAH emission and hydrogen recombination lines (Egami 2009, priv. comm.), ruling out UV fluoresence as a possible excitation mechanism, indicates that $H_2$ is produced by shocks which deposit huge amounts of kinetic energy into the ISM. These $H_2$ luminous galaxies may represent a population at high redshifts where interactions and mergers are more common. Theoretical work (Mizusawa et al. 2005) predicts that in these galaxies the pure rotational lines S(1) (17 $\mu$m), S(2) (12.3 $\mu$m) and S(3) (9.6 $\mu$m) should have luminosities in excess of $10^{35}$ erg s$^{-1}$ and thus within the reach of deep SAFARI surveys. Based on detections of $H_2$ in nearby galaxies, we find that the power emitted in the two $H_2$ lines S(0) (28.2 $\mu$m) and S(1) is typically around 20–25% of that found in the MIR lines of [Si II] or [S III]. For the disc galaxies (Roussel et al. 2007), this corresponds to about $3 - 3.5 \times 10^{-4}$ of the total infrared power, a limit that should be attainable in deep SAFARI surveys. Additional observational evidence for the existence of large molecular $H_2$ reservoirs in high-redshift galaxies is provided by the tight correlation found between $H_2$ and PAH emission in local star-forming and disc galaxies (Rigopoulou et al. 2002; Roussel et al. 2007), as well as in galaxies out to $z \sim 1$ (Dasyra et al. 2009). PAHs are ubiquitous in high-$z$ galaxies (Valiante et al. 2007; Pope et al. 2008; Huang et al. 2009), and so if the PAH-$H_2$ correlation holds at higher redshifts, then the $L_{H2}/L_{FIR}$ calculated above should serve as a lower limit. Simple calculations suggest that we might see up to 7 sources per SAFARI field of view in the S(1) line, out to $z < 3$.

#### Beating spatial confusion

The third, spectral dimension in deep spectral line surveys provides an important way by which to overcome spatial confusion. Narrow-band line emission from a source at a given redshift appears only at very specific and discrete wavelengths, and so the high density of sources in an individual beam that causes confusion is drastically reduced relative to all sources emitting in the continuum (Figure 1.15b). This enables individual objects, whose continuum may be substantially below the continuum confusion limit, to be detected. Through simulations (Clements et al. 2007; Raymond et al. 2009) we have found that sources more than 10 times below the traditional (photometric) confusion limit at, for example, 120 $\mu$m can be isolated and detected. In this way, the majority of sources that contribute significantly to the CIRB can be detected and classified spectroscopically, as described above.

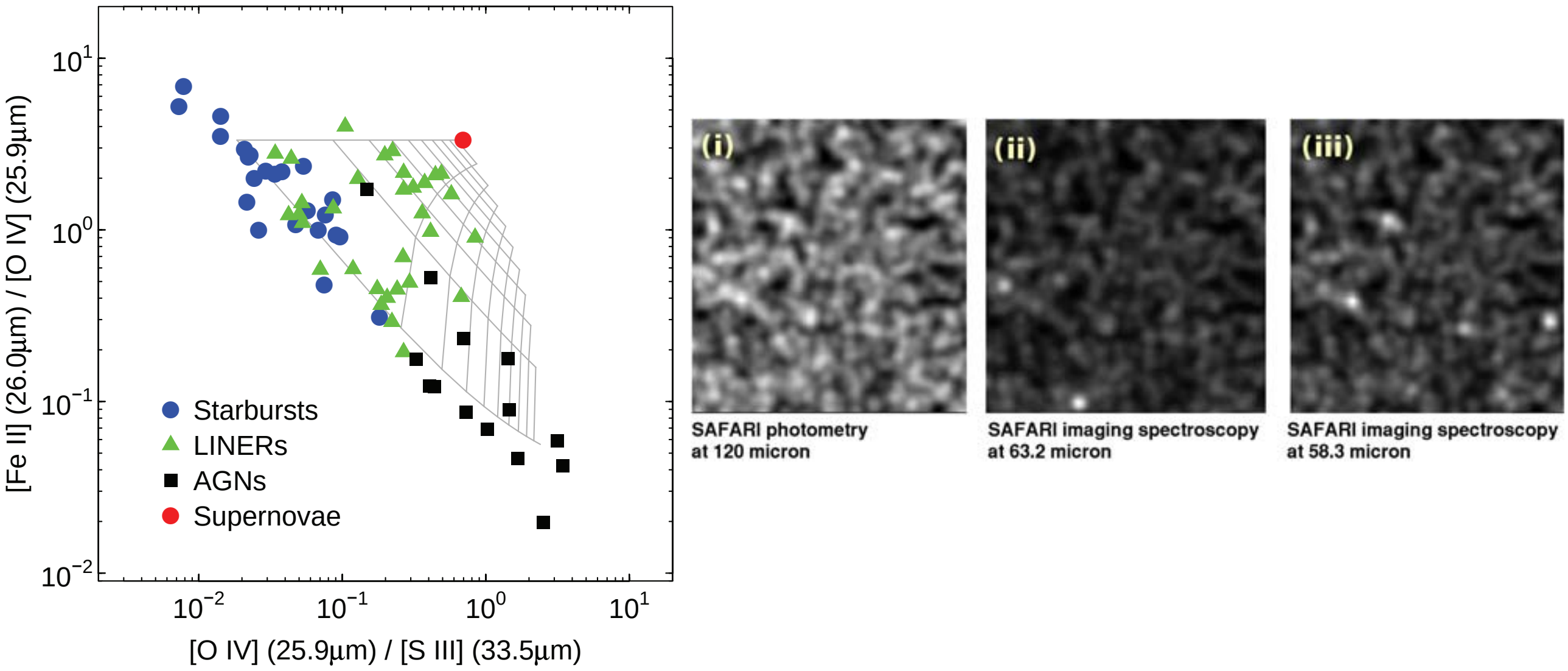

*Figure 1.15: **(a) Left panel**: MIR line ratios provide a powerful means by which to determine the dominant mechanism powering the prodigious IR output in the IR-luminous galaxy population. Plotted is the ratio of [Fe II]/[O IV] vs. [O IV]/[S III] for a selection of different types of galaxies. The overplotted grid represents a simple mixing by a linear superposition of starburst, supernovae (shocks) and AGN line spectra. The loci of the grid lines indicate where the contribution to the [S III] is 0%, 10%, 20% etc. for supernovae shocks and AGN (Sturm et al. 2000). All lines are accessible in the SAFARI band from a redshift of $z > 0.4$, providing a discriminator between star formation and mass accretion; **(b) Right panel**: Deep, spectroscopic surveys with SAFARI will enable us to break through the traditional confusion limit and will single out and characterise individual galaxies that would otherwise be indistinguishable from the background sky. This power of increased spectral resolution is illustrated in the three panels above: Shown in the panel (i) is an image covering a single SAFARI FOV, simulating photometric mode at 120 $\mu$m, at a spatial resolution of $9''$. In this particular realisation of the sky, no discrete sources can be distinguished against the background; shown in the panels (ii) and (iii) are the same FOVs made at the same angular resolution, but this time centred at 63.2 and 58.3 $\mu$m respectively, and with a spectral resolution of $R \sim 1000$. Several sources can be seen in each narrow band.*

### Clustering and first large scale structures

The source detection estimates presented in previous subsections for spectroscopic surveys assume random Poissonian distributions of galaxies all through the various cosmic epochs. However, extragalactic sources are known to be clustered. Significant clustering has been measured in high-redshift ($z \sim 1-3$) Spitzer galaxies (Farrah et al. 2006; Magliocchetti & Brüggen 2007; Magliocchetti et al. 2008; Farrah et al. 2006), with a strength which increases, as might be expected, with decreasing look-back time. Comparisons with theoretical models have provided a direct estimate of the dark matter mass of such sources, and the derived values ($M \sim 10^{13}$ $M_\odot$) indicate that luminous-IR galaxies at $z \sim 2$ are most likely the progenitors of the giant ellipticals which reside locally in rich clusters. This implies that studies of the infrared population at redshifts $z > 1-1.5$ can provide a unique tool with which to investigate the formation and evolution of super-structures such as proto-clusters, clusters and of the galaxies that belong to them from even before the peak of cosmic star formation activity. While the results from Spitzer necessarily suffered from some limitations due primarily to the lack of measured redshifts, SAFARI in its spectroscopic photometric mode will, for the first time, be able to overcome such problems and provide definite answers to a number of crucial issues:

**(i) Direct and unbiased investigation of the evolution of the Large Scale Structure in the Universe from z $\sim$ 3**. Thanks to the SAFARI capabilities, no previous selection on the observed population is needed as redshifts will be taken for all the objects brighter than some limiting flux which fall in the SAFARI FOV . With

our goal 5$\sigma$ sensitivity, such a task could be easily reached in about 500 hours by "blindly" surveying an area of approximately 0.5 square degree (see page 34). These results will not only be of invaluable importance *per se*, but will also complement those that will possible with NIR surveys which are limited for such studies to $z \sim 1.5$.

**(ii) Studies of galaxy formation and evolution as a function of environment.** Since high-redshift sources exhibit strong clustering, galaxies observed by SAFARI in a blind spectroscopic survey will either reside in overdense or underdense regions. This will give us the unprecedented possibility of investigating the impact of environment on galaxy formation and evolution as a function of redshift up - and possibly beyond - the epoch which marks the bulk of AGN/stellar activity. By doing this, it will be possible to provide answers to a number of questions such as whether there is any large-scale influence between surrounding environment and galactic AGN/stellar activity and, if so, if seeds for such a phenomenon were already in place by $z \sim 2-3$. The spectroscopic diagnostics will be essential to *measure* directly the star formation and AGN luminosities in each cluster component.

**(iii) Targeted studies of Proto-clusters**. As discussed above, most of high-redshift active galaxies reside in overdense regions or what we call proto-clusters. This was not only seen by Spitzer, as a number of other works report the discovery of $z \sim 2-3$ proto-cluster candidates in the NIR and for instance via the concentration of Lyman Break galaxies around powerful radio sources. With the imaging capabilities of SAFARI it will be possible for the first time to collect a large sample of high redshift proto-clusters in a relatively short time. When targeting regions centred on "known" high-redshift sources, we expect the number of sources detectable per SAFARI field to be boosted by as much as *a factor 10* above the predictions presented earlier in this section.

### Tracing the evolution of Milky Way-type galaxies

Tracing the cosmic evolution of (U)LIRGs is one of the key goals of SAFARI; however, whilst this population makes a very significant contribution to the CIRB, such objects only make up a small fraction of the total number of galaxies. With SAFARI we will have, for the first time, the sensitivity to characterise the ISM of the complete $z \sim 1$ population, and so to follow the evolution as a function of redshift of the much less IR-luminous, less massive but more ubiquitous galaxies. Our own Milky Way will be easily detectable in the continuum out to $z \sim 1$ as will be seen in section 1.6.2, whilst the MIR/FIR diagnostic lines will be visible out to $z \sim 0.5$ where the universal SFR has already increased by close to an order of magnitude as compared to the local Universe.

### The diagnostic power of the PAH and silicate features

The ability of the UV excited PAH bands to trace any star forming component has been demonstrated for local galaxies (Soifer et al. 2002; Peeters et al. 2004). The ratio of the PAH bands has been related to star formation activity for a wide variety of galaxies (Galliano et al. 2005, 2008), while the ratio of PAH emission line strength to underlying continuum has been established as a proxy for disentangling star-forming and AGN contributions (Brandl et al. 2006). In the deepest spectra obtained with Spitzer IRS (Menéndez-Delmestre et al. 2009; Siana et al. 2009) these features can be seen out to $z > 2$ in sources that are both optically and MIR-faint ($S(24\mu m) \sim 100\mu$Jy). The observed similarities between MIR spectra of LIRGS at the present epoch out to $z \sim 2.5$ suggest similar dust properties, and thus that significant enrichment and evolution has already taken place by $z \sim 2.5$ (e.g., Swinbank et al. 2004). When then, did the bulk of dust production and evolution take place? Delayed injection of heavy elements from AGB stars would, for example, weaken PAH emission, and increase the relative importance of other sources of dust such as supernovae or AGN winds (Maiolino 2007). Detecting changes in the PAH emission/silicate absorption feature profiles and strengths with epoch require modest-resolution spectrometers in both the MIR and FIR ($R \sim 25$) in order to follow the PAHs as they move from one waveband to another with redshift (Figure 1.16a). With SPICA instruments we will have the sensitivity to exploit the diagnostic power of the PAH/silicate features: By combining the low-resolution modes of MIRACLE and the spectrophotometric mode of SAFARI we will be able to make large-scale spectroscopic surveys of very faint galaxies, down to $\sim 20\ \mu$Jy and $\sim$ a few 100 $\mu$Jy, respectively with the two instruments.

A wide-field, spectrophotometric survey with SPICA over an area of 1/4 the size of the COSMOS field

(0.5 square degrees) would yield spectra of $\sim$ 5000 high-$z$ galaxies in 500 hours, a factor of 10 more than the total number of high-z galaxies observed with Spitzer IRS, and a significant increase on those that would have been observed by JWST by the launch date of SPICA. The bulk of these flux-limited catalogued objects will be found at moderate redshifts ($z < 2$). With deeper integrations (e.g., a 500 hr survey covering 200 sq. arcmin with resolution $R \sim 20$), we anticipate being able to detect and characterise the spectrum (emission features and continuum) of several hundreds of objects at $z > 2$. As mentioned above, these numbers are likely to be boosted significantly when targeting regions around "known" high-redshift sources. By measuring the PAH and silicate features on top of the dust continuum, we will not only be able to disentangle directly the relative importance of starburst and AGN emissions and to estimate the rates of star formation and gravitational accretion in the distant sources. With these spectral surveys we will also pin down redshifts in a completely unbiased way, with which to evaluate the co-moving space density of these dusty populations and to constrain the bolometric luminosity functions of luminous infrared galaxies.

**Spectroscopy of very distant galaxy populations through gravitational lensing**

The strong gravitational lensing of the distant Universe by intervening galaxy clusters has been exploited at many wavelengths in order to push the luminosity limits and redshifts out to the farthest limits. In the IR/sub-mm, lensing clusters act as transparent lenses, and cluster surveys at these wavelengths have been particularly powerful at pushing below the confusion limits of sub-mm surveys to constrain the faint end of the source counts (Smail et al. 1997; Blain et al. 1999) and opening up the study of not only some of the most distant objects, but also of those that are less extreme at slightly more modest redshifts (Knudsen et al. 2006). Deep Herschel lensing surveys will do this in the FIR, but only *photometrically*. With SAFARI we will be able to exploit the strong lensing power of massive clusters to push the same deep spectroscopic imaging studies described above from $0 < z < 3$ out to $z \sim 3$ and beyond and will be able to probe the properties of LIRGs in the epoch of cosmic reionisation for the first time in the MIR. In a more quantitative way, if we assume that (i) the IR galaxy luminosity function at $z \sim 2$ (Caputi et al. 2007) is representative of that at higher redshifts, (ii) the fractional IR luminosity in the 7.7 $\mu$m PAH feature is at least that observed in $z \sim 2$ submillimetre galaxies (Pope et al. 2008), (iii) that a typical cluster has a lensing potential similar to that of Abell 773, then in a survey of 50 massive clusters with SAFARI in spectrophometry mode we would expect to detect $\sim$23 LIRGs at $7 < z < 10$ using the 7.7 $\mu$m PAH feature. A blank-field survey with the same areal coverage will produce $\sim$ 8 detections in the same redshift range, so the lensing survey will offer a factor of $\sim$ 3 gain in number. The gain will be even larger if there is an abundance of low-mass/low-luminosity galaxies at such high redshift as for instance predicted by hierarchical structure formation theories. In fact, by accessing an epoch at which the space density of IR-luminous galaxies is virtually unknown, SAFARI will provide valuable observational data in the epoch of reionisation, constraining the cosmic histories of star formation, metallicity evolution (e.g., the 7.7 $\mu$m PAH strength is sensitive to metallicity), and structure formation. This is especially timely since in the coming decade, various new observatories such as ALMA and JWST will push the frontier of extragalactic IR/sub-mm astronomy beyond a redshift of 4 and possibly up to 10.

## 1.6.2 Deep cosmological surveys with SAFARI

The existence of a large number of distant sources radiating most of their energy in the IR implies that the critical phases of the star-formation and nuclear accretion history took place in heavily obscured systems, embedded within large amount of gas and dust. The substantial reddening affecting these dust-obscured objects makes their characterisation in the optical/UV severely biased and sometimes even impossible. Observations in the FIR/sub-mm to very faint flux densities (few 10s $\mu$Jy) are often the only means to detect star-formation and/or AGN activity in most of these heavily obscured sources at large cosmological distances. Deep photometric surveys with SAFARI will be sensitive to the important galaxy populations that will be missed both by spectroscopic surveys and by the current generation of FIR/sub-mm surveys. Both ground-based sub-mm surveys with existing facilities (SCUBA, MAMBO, LABOCA, SCUBA2) and FIR surveys from space (ISO, Spitzer) are limited by the confusion that results from modest angular resolution, and are sensitive only to the most luminous and distant galaxies, or to those that are very nearby. Such galaxies contribute less than half of the background at these wavelengths, and make up an even smaller fraction of the integrated CIRB. Herschel-SPIRE and PACS will open new windows in the FIR above 100 $\mu$m - a taster of this has been given by the BLAST balloon experiment (Devlin et al. 2009; Pascale et al. 2009). However our current knowledge of source densities at these wavelengths suggests that we will resolve only 50% of the CIRB. Our ability to identify sources against the CIRB is governed by confusion which dominates the performance of Herschel long-ward of 100 $\mu$m. At shorter wavelengths the sensitivity of Herschel-PACS is insufficient to reach the confusion limit and precludes our ability to resolve the CIRB. In contrast, SAFARI will be able to resolve the bulk (>90%) of the cosmological background over a wide wavelength interval from $\sim$ 30 to $\sim$ 100 $\mu$m and, as we show below, will be unrivalled by either ALMA or JWST.

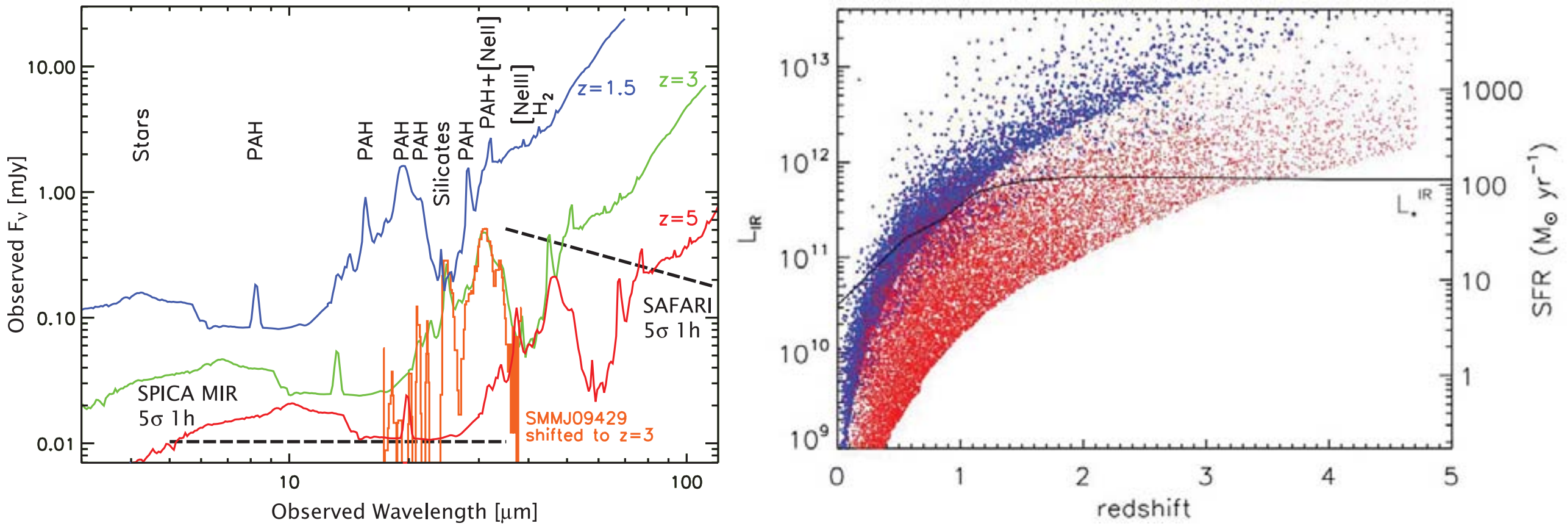

*Figure 1.16: **(a) Left panel**: Shown in green is a local ULIRG template that has been fitted to the observed spectrum SMM J09429 (red) (Valiante et al. 2007), with the SPICA spectrophotometric sensitivities overlaid (for SAFARI this is $R \sim 50$). Both spectra have been redshifted to $z \sim 3$ for illustrative purposes. SAFARI will have the sensitivity in low-resolution mode ($R \sim 50$) to detect PAH/silicate features in dusty distant galaxies out to $z \approx 3$ in 1 hour, and out to $z \sim 4$ in 10 hours. These features will provide important pointers to the physical processes powering distant dusty galaxies; **(b) Right panel**: Deep photometric surveys will identify many thousands of distant FIR luminous galaxies. Shown in blue is a plot of the luminosity-redshift distribution of FIR galaxies that will be detected by Herschel in its deep cosmological survey programmes. In red we show the galaxies that will be detected in a SAFARI 70 μm confusion-limited survey of 900 square arcmin. For a given IR luminosity, SAFARI will detect galaxies that are much more distant, whilst at a given redshift SAFARI will detect a much wider range of source luminosities thus identifying the bulk of the galaxy population. The bold black line traces the redshift evolution of an $L_*$ galaxy and represents the knee luminosity of the IR luminosity function derived using Spitzer-MIPS 24 and 70 μm (Magnelli et al. 2009).*

### Deep surveys at 70 $\mu$m – resolving the CIRB

According to current models, the 70 $\mu$m confusion limit for SPICA is around $\sim$ 50 $\mu$Jy. To reach such depths with Herschel-PACS would take an enormous and unrealistic amount of time (10 000 hours 5$\sigma$!), while with SAFARI this can be achieved in minutes. Using current assumptions for the intensity of the CIRB at this wavelength, a 70 $\mu$m confusion-limited survey (Figure 1.16b) would resolve more than 90% of the CIRB over 80% of the Hubble time ($z \sim 2$), detecting galaxies down to a star formation rate regime at which rest-frame UV observations meet the IR, i.e., 10 $M_\odot$/yr (Lagache et al. 2004) (more luminous galaxies emit the majority of their light in the IR). By observing at 70 $\mu$m we can avoid the effects of contamination by the strong MIR PAH features, and thus make very reliable determinations of the IR flux for sources with $z < 2$. Such a survey would increase significantly the range in IR luminosity and redshift over which galaxies will have been detected: at the more luminous end, LIRGs out to $z \sim 4$ and ULIRGs to beyond, and appreciable numbers of $L_*$ galaxies out to $z \sim 3$. With SAFARI it will be also be possible to detect galaxies as quiescent as our own ($L_{IR} < 10^{10}\ L_\odot$) out to $z \sim 1$ (Figure 1.16b), where the cosmic SFR peaks (see Figure 1.11b). Whilst this less IR-luminous population contributes only minimally to the CIRB, these galaxies are extremely abundant, and dominate the extragalactic background light at optical wavelengths. Confusion-limited surveys at 70 $\mu$m are unique territory

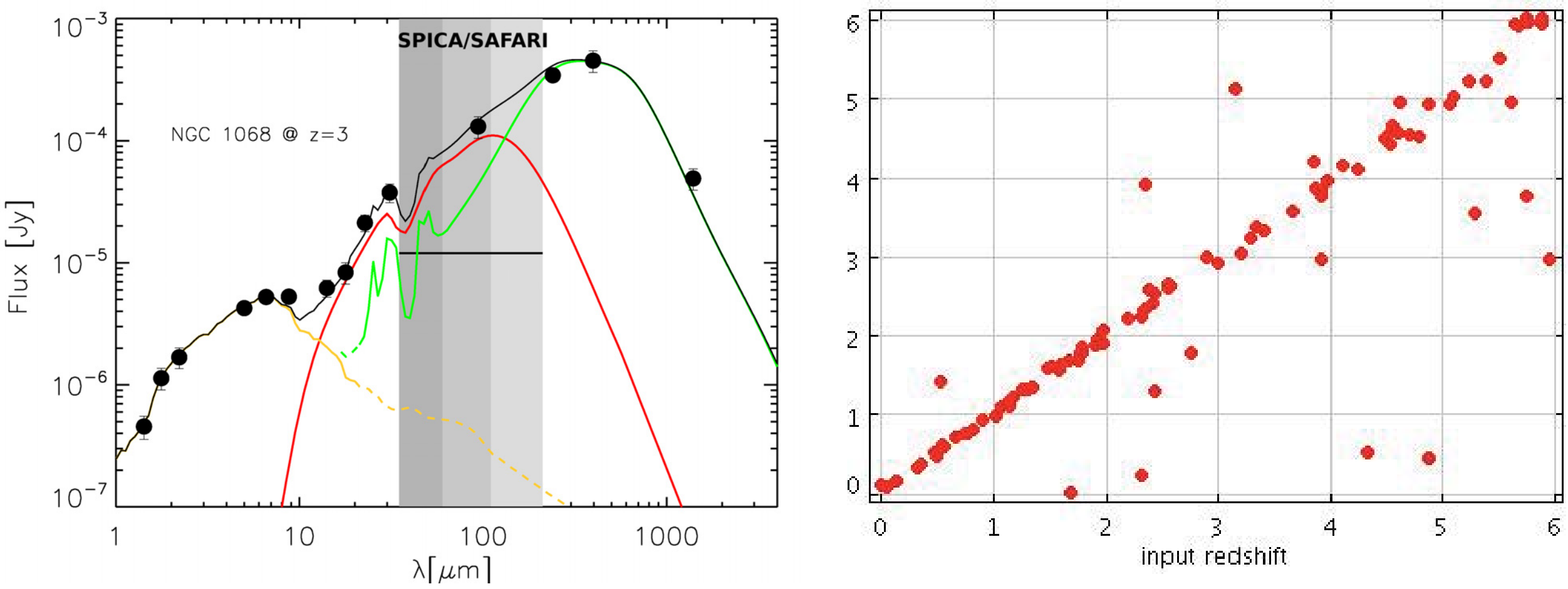

*Figure 1.17: **(a) Left panel**: The broadband SED of NGC1068 (in black) redshifted to z ∼ 3; overlaid are the different constituent emission components that are produced by a torus (red: model by Fritz et al. 2006), a starburst component based on the starburst galaxy NGC1482 and a stellar component (orange) based on the stellar evolution models. The vertical bars indicate the three bands of SAFARI, and highlight how the combination of 40 μm and 70 μm emission can be used to differentiate between AGN- and starburst-dominated galaxies; **(b) Right panel**: A plot to illustrate the accuracy with which a multi-filter photometric survey with SAFARI and MIRACLE will be able to recover redshifts. Note the outliers are typically elliptical galaxies and type-1 AGN, both of which have very little PAH emission.*

for SAFARI.

### Deep surveys at 40 $\mu$m – Broadband footprints of distant, obscured AGN

The surveys with SAFARI at 40 $\mu$m will not be confusion limited (estimated confusion limit $< 10$ $\mu$Jy). A survey at 40 $\mu$m down to a limit of a few 10s $\mu$Jy would be sensitive to moderate luminosity ($L_{IR} \sim 10^{11}$ $L_\odot$) type-2/obscured AGN out to $z \sim 5$, and particularly in the $3 \leq z \leq 5$ range, where the co-evolution of star-formation and accretion activity is expected to already be in place. Heavily dust-obscured, mostly Compton-thick, AGN are believed to be responsible for the unresolved peak of the X-ray background at 30 keV. These AGN are missed by the deepest current-generation X-ray surveys (the prototype Compton-thick AGN NGC 1068 would be missed above $z = 0.5$ even in 2 Ms with Chandra), yet reveal themselves in the MIR when stacking techniques are applied at X-ray wavelengths to strongly amplify their extremely dim fluxes. Obscured AGN can be identified through their infrared signature, which is reprocessed UV emission from the AGN itself. By selecting MIR sources with faint NIR and optical emission, both Daddi et al. (2007) and Fiore et al. (2008) suggest that the majority of the so called "IR excess" sources in the Chandra Deep Field-South are in fact highly obscured AGN at $1 < z < 3$. Spitzer MIR surveys have been key to complementing X-ray surveys, in identifying sizable samples of obscured AGN for $z < 3$ through the 9.7 $\mu$m silicate absorption (e.g., Weedman et al. 2006). The combination of the silicate absorption feature, the PAH features and the broadband contribution from the warm, dusty tori can be used to disentangle the AGN from starburst populations at redshifts $z > 3$ through photometry in the 40 $\mu$m (34 – 60 $\mu$m) band. This applies both at high and low infrared luminosities.The combination of both the 40 $\mu$m and 70 $\mu$m bands will provide a unique means in the SPICA era by which to disentangle co-evolving AGN and star-formation activity in the very distant universe (Figure 1.17a), and would complete the census of accreting supermassive black-holes (SMBHs) at $z > 3$.

SAFARI and IXO will provide a complementary perspective on distant AGN. IXO will characterise SMBHs out to $z \sim 7$ by measuring the X-ray spectral slopes and luminosities, which provide constraints on accretion rates, and so assess directly the growth of supermassive black holes at high redshift. SAFARI+IXO will be able to measure the bolometric output of AGN out to $z \sim 5$ at least, by combining contributions due to star formation and accretion, and together will provide the first direct measurement and study of the dust-to-gas ratio and its evolution from $z \sim 0$ to $z \sim 5$.

**Spectral Energy Distributions and photometric redshifts in the MIR/FIR**

The direct characterisation of the properties of dusty star-forming galaxies and AGN at MIR/FIR wavelengths is crucial for estimating the fraction of their bolometric energy output absorbed by dust and re-radiated at longer wavelengths. It also provides unique constraints on the nature of the different physical processes driving their ongoing activity (e.g., Smith et al. 2007; Spoon et al. 2007). Dust plays a dual role in extragalactic studies, simultaneously helping (emission by heated dust) and hindering (obscuration) our understanding of the physical processes driving galaxy evolution. The deep 40 and 70 $\mu$m surveys described above will provide unique insight into galaxy evolution at very high redshifts, and as a very important by-product will allow us to track changes in dust attenuation of stellar light in galaxies. GALEX has shown that dust attenuation in individual galaxies out to redshifts of $z \sim 1$ remains constant, with a hint of a decrease (Zheng et al. 2007; Buat et al. 2007a). This is in stark contrast to the *mean* dust attenuation in the Universe which has been shown by Takeuchi et al. (2005) to increase over the same epoch. This apparent contradiction can be attributed to the general brightening of the galaxy population with redshift, and the long-established correlation between dust obscuration and total luminosity of the galaxies (e.g., Hopkins et al. 2001; Buat et al. 2007b). At some point in cosmic history, obscured environments must become less common since it takes time for dust to not only be produced, but also to spread through the ISM of a young galaxy. The epoch at which this takes place can be pinpointed by comparing UV and rest-frame IR emission of galaxies which are responsible for the bulk of star formation at different epochs. The deepest Spitzer observations at 24 $\mu$m suggest that at $z > 2$ galaxies with bolometric luminosities $L_{bol}$ of $10^{11} < L_{bol} < 10^{11.5}$ $L_\odot$ contribute substantially (Caputi et al. 2007) or even dominate (Reddy et al. 2008) the star formation activity in the Universe. By detecting these galaxies *directly* in the FIR we can measure the emission from dust. SAFARI will detect galaxies ten times fainter than Herschel, corresponding to $L_{bol} = 10^{11.5}$ $L_\odot$ at $z = 2.5$ (Figure 1.16b). Whilst ALMA and JWST can both detect some fractions of this population, neither can make the crucial step of characterising the dust SED, since such observations either sample the redshifted stellar emission (JWST) or are restricted to the Rayleigh-Jeans dust emission (ALMA), and so cannot be used to establish dust temperates and source luminosities.

The same MIR PAH and silicate absorption features that contaminate and complicate the interpretation of deep surveys, on the other hand provide an excellent means by which to determine redshift using SPICA. To-date, photometric redshifts of SCUBA and Spitzer sources have been determined using optical/NIR data taken with many different surveys. With SPICA we will be able to determine redshifts using the combination of the NIR and MIR. Simulations suggest that with as few as 13 ($R \approx 8$) filters, evenly spaced over the 15 - 70 $\mu$m wavelength, it is possible to determine the redshift and type of a very wide range of galaxies, to an accuracy of a few % (Figure 1.17b). By including SAFARI bands we not only extend the redshift range of the technique, but also increase accuracy: with SAFARI bands we sample well the PAH and silicate absorption features which, in turn, break most of the degeneracy in the MIRACLE-only estimations. Sources for which the match between recovered and true redshift is poor are those lacking distinct PAH/silicate features: in the case of the AGN, these are intrinsically optically bright, and thus good candidates for spectroscopy.

## 1.7 SPICA in the context of ALMA and the JWST

SPICA will be the bridge between two other very large astronomical facilities that are due to become operational within the next five years - namely ALMA and JWST. The JWST will revolutionise near-IR and MIR astronomy through its impressive sensitivity, and ALMA will make a huge impact with its unprecedented sensitivity and high angular resolution in the sub-mm. SAFARI probes the crucial wavelength regime *between the JWST and ALMA regions*.

SPICA will cover critical spectral diagnostics (at their rest frame wavelengths) that are not accessible to ALMA. While the study of the oxygen chemistry with ALMA will be limited to observations of CO, HCO+ and other trace species, SPICA will observe the emission of the major oxygen reservoirs like atomic oxygen, water ice, $CO_2$ or the thermal line emission of water vapour. ALMA will detect the millimetre and sub-mm dust continuum emission, but only SPICA will be able to observe the specific grain and ice spectral features that are needed to determine the dust composition and its formation history along the life-cycle of interstellar matter.

ALMA and SPICA will be complementary to address the chemical complexity in the Universe. In particular, SPICA will allow us to observe molecules without permanent electric dipole or the mid-IR emission of PAHs, neither detectable with ALMA, which are major players in organic chemistry in space.

Since the discovery of the cosmic infrared background and the important populations of luminous dusty galaxies at high redshift, it has become clear that obscured star formation accounts for a large fraction of the stellar mass assembly budget. Enormous progress in infrared/(sub)mm/radio source count surveys will be made by Herschel, ALMA, EVLA and ground-based bolometer arrays such as SCUBA2. While ALMA will probe the molecular ISM in these objects and JWST the stellar component, the highly diagnostic mid-IR region is exclusive SAFARI territory. For instance, for an object at $z \sim 3$, JWST will be able to observe the rest-frame spectrum only out to 7 $\mu$m. The most luminous PAH features, hot dust continuum and key fine-structure lines will all be accessible only with SAFARI. This opens up the intermediate redshift Universe (at the peak of cosmic star formation activity) for scrutiny with the techniques that have been successfully used on low-z galaxies since ISO and Spitzer. The line sequence [Ne II]/[Ne III]/[Ne V] will characterise star formation as well as the importance of AGN activity. Many other similar probes are available for SAFARI, but neither ALMA nor JWST provides equivalent diagnostics. Thus the unique role of SAFARI is the characterisation of the sources of the cosmic infrared background. A further complementarity between ALMA and SAFARI is provided by SAFARI's ability to trace the energy budget of the star forming ISM out to $z \sim 2$ using the [O I]63 $\mu$m line and [C II]158 $\mu$m at lower redshift, while ALMA will access the [C II] line only at $z > 2$.

The relation between SMBH and spheroid mass is one of the most astonishing results of modern astrophysics, given the vastly different scales involved. The enormous difference between the Schwarzschild radius of the SMBH and the characteristic radius of the stellar population suggest that this relation is related to the coeval formation of both. While theoretical proposals to account for this relation exist, observational constraints are far and few between. To improve this situation, observations of galaxies at the peak of the cosmic star formation and black hole accretion rate are a key requirement. A fundamental diagnostic is provided by the [Ne V] line at 14.3 $\mu$m. This line is a very powerful probe of AGN activity in dusty galaxies, as shown by recent Spitzer work which has uncovered a large number of obscured AGN in the nearby Universe, including Compton-thick sources. Furthermore, as shown by Dasyra et al. (2008), the width of [Ne V] can be used to measure black hole mass even at the low spectral resolution of Spitzer. SAFARI will thus play a key role in establishing the inventory of AGN in obscured galaxies at intermediate redshifts, and trace the mass-buildup of the nuclear SMBHs. Again, equivalent diagnostics are not available with either ALMA or JWST.

Both JWST and ALMA will probe galaxies at the epoch of reionisation ($z > 6$). At these redshifts the most luminous CO lines shift out of the observing band, and the [C II] line becomes the key probe (Walter & Carilli 2008). In pristine (zero-metallicity) gas, however, this line is not available. The molecular medium in zero metallicity galaxies cools principally through $H_2$ rotational and vibrational lines. As shown by Mizusawa et al. (2004), a number of these lines may be detectable with SAFARI, over a wide range of conditions. Targets will be provided by JWST, which is expected to discover significant numbers of $z > 6$ galaxies, with the prime targets for SAFARI being those that are not detected in CO or [C II] by ALMA: the detection of the signatures of a metal-free molecular interstellar medium is again unique SAFARI territory.

## 1.8 Discovery Science with SPICA

The science cases we have presented illustrate the concrete and definitive progress in our understanding of the Universe that SPICA will allow. However, as with all strides in increased sensitivity or the opening of new wavelength bands, new discoveries are certain once SPICA starts observing. To give a flavour of what might be expected in the new discovery space opened by SPICA, we give here some examples of speculative research projects from both the nearby and distant Universe.

***The "end-point" of extra-Solar planetary systems:*** During the AGB phase, the radius of a star like our Sun will have expanded to a few AU (close to Jupiter's orbit) and it will be a few hundred times more luminous than during the main sequence stage. The associated increase of temperature in the outer Solar System may be sufficient to cause vapourisation of any icy body orbiting within a few hundred AU, i.e., the KBOs presented in section 1.2.2 (Stern et al. 1990). Indeed, the surprising detection of water vapour lines in the C-rich star IRC+10216 (Melnick et al. 2001) has been interpreted as the vapourisation of a collection of orbiting ice bodies caused by the luminosity increase in this evolved star (with an original mass similar to the Sun). SAFARI will simultaneously search for atomic oxygen, water vapour and ice features (i.e., the $\sim 44\,\mu$m feature) in many C-rich evolved stars where the prevailing carbon chemistry can not produce substantial amounts of water. If the previous scenario is right, SPICA/SAFARI not only will detect hundreds of planetary disk systems, and characterise tens of icy KBOs in our Solar System, but it will also trace the "end-point" of extra-solar systems around evolved stars; i.e., the sublimation of entire Kuiper belts at the end of the host star life.

***First identification of PAHs molecules in Space (the "GrandPAHs"):*** The ubiquitous MIR emission bands are attributed to the emission of a family of carbonaceous macromolecules: the PAHs. However, because these bands are due to the nearest neighbour vibrations of the C-C or C-H bonds, they are not specific to individual PAH species. Therefore, in spite of their relevance for astrophysics (as tracers of UV radiation fields or star forming regions in a broader extragalactic context), the identification and characterisation of a given PAH molecule in space has not been possible yet. The hope for such an identification lies in their FIR lower-energy bands (PAH skeleton modes) which are specific fingerprints of individual molecules. The energy that is emitted in these bands is expected to be very weak, typically a few tenths of the energy absorbed in the UV (Joblin et al. 2002; Mulas et al. 2006a,b). For example, it can be shown that if all the MIR emission observed in the ISM was due to a unique PAH such as ovalene, its FIR emission should be detectable even with a low ($\sim 10$) signal-to-noise ratio (Mulas et al. 2006a; Berné et al. 2009). Recent progress in the field suggests that there are in fact only a few (or at least limited number) of large and compact PAHs in space, that can resist the harsh interstellar conditions. These large PAHs are the so called "Grand-PAHs". The instantaneous broad band coverage of SAFARI will be especially adapted for deep searches of the Grand-PAH FIR bands, which is crucial since the positions of these bands are not known. Therefore, SAFARI can represent our first chance to identify specific PAHs in space.

***A new window in exoplanet research–far infrared observations with SAFARI:*** In the field of exoplanets (EPs), SAFARI will provide capabilities to complement SPICA studies in the MIR (either coronagraphic or transit studies). With very stable detectors, and efficient, high cadence observations, SAFARI will be used to perform transit photometry and, on some favourable candidates, spectrophotometry for the first time in the FIR domain. Therefore, SPICA could be the only planned mission able to study exoplanets in a completely new wavelength domain (neither covered by JWST nor by Herschel). This situation is often associated with unexpected discoveries. In the shortest wavelength photometric band ($\sim 48\,\mu$m), primary and secondary eclipses of Jupiter-size EPs around F to M stars will be detected in a single transit. FIR measurements will be specially suited for EPs around cool M stars. Since cooler EPs show much higher contrast in the FIR than in the NIR/MIR (e.g., Jupiter's effective temperature is $\sim 110$ K), if such EPs are found in the next 10 years, their transit studies with SAFARI will help to constrain their main properties, which are much more difficult to infer at shorter wavelengths. The SPICA longer wavelength range hosts a variety of interesting atmospheric molecular features (e.g., $H_2O$ at 39 $\mu$m, HD at 37 $\mu$m, $NH_3$ at 40 and 42 $\mu$m, etc.). Strong emission/absorption of these features was first detected by ISO in the atmospheres of Jupiter, Saturn, Titan, Uranus and Neptune (Feuchtgruber et al. 1999). Suitable EP candidates around M dwarfs for SPICA spectroscopic observations may be scarce, and a considerable amount of observing time will be needed to extract their main atmospheric composition. However, in a few optimal systems, SPICA observations may allow us to expand the variety of "characterisable" extrasolar planets. These are crucial steps to enter in a new era of quantitative characterisation of EPs.

***Searching for the first generation of star-formation and black hole growth:*** In regions of low metallicity, $\leq 10^{-2}$ solar, the pure rotational lines of $H_2$ (at 28, 17, etc. $\mu$m) are the primary coolant of UV and X-ray irradiated gas. Typical line intensities can be of the order of $0.1 - 1.0$ erg s$^{-1}$ cm$^{-2}$ sr$^{-1}$, exceeding those of the solar metallicity fine-structure and CO lines. Thus, molecular hydrogen emission provides a unique means by which to pinpoint the first episodes of star formation (population III stars) and early signatures of black hole growth at high redshift ($z \geq 6$). By definition, the nature of the search for such objects relies on the imaging capability of SAFARI, for which simple calculations suggest that between 1-10 of such halos might be detectable in a single SAFARI FOV. Whilst alternative cooling mechanisms will exist in more metal-rich systems, the $H_2$ rotational lines will still be valuable probes of the warm molecular gas reservoir.

***The diagnostic potential of the very high-J CO lines:*** The high-J ($J \geq 13$) CO lines provide a yet-to-be-demonstrated method of establishing the presence and physical properties of highly-obscured AGN (Krolik & Begelman 1988; Meijerink & Spaans 2005). The predicted differences between the expected CO spectral line energy distributions in X-ray (XDR) and photon-dominated regions are most significant (3 orders of magnitude) in the SAFARI band. By summing the CO, [C II]158 $\mu$m and [O I]63 $\mu$m emission one can evaluate the total cooling rate for gas within the locale of the AGN and, assuming thermal balance, derive the product of black hole mass and accretion efficiency. Thus, one can infer the properties of the black holes residing in obscured environments even in the absence of X-ray measurements. Herschel-PACS will make the first studies of these sources; the sensitivity of SAFARI will take us to the next stage and give access to a representative sample covering a wide range of mass and accretion luminosities, thus tracing the role and destiny of torus-covered black holes both locally and out to $z \sim 1$.

***Feeding hungry mouths–mass accretion from the IGM:*** The means by which primordial gas is accreted from the IGM to fuel on-going star formation in galaxies remains an unanswered question in galaxy evolution. Any mechanism for this direct fuelling requires a means by which to cool what will be hot gas. The cooling process is widely believed to be mediated by the injection of metals into the IGM in galactic winds, which themselves cool by inelastic collisions with dust grains at the IGM/wind interaction zone (Dwek & Werner 1981; Montier & Giard 2004). The recent discovery by Spitzer of a diffuse IR counterpart to the X-ray emitting intracluster medium of Stefan's Quintet supports this hypothesis, its luminosity exceeding that of the X-ray emission by two orders of magnitude. Herschel will search for the FIR signatures from hot plasmas in dense cluster and group environments; however, exploring this phenomenon in the bulk of the galaxy population will fall to SAFARI. SAFARI will be able to trace the FIR emission both photometrically and spectrophotometrically, thus enabling not only determinations of the cooling properties of the IGM to be made, but also characterisation of the dust grains.

**New frontiers in exoplanet research: Far infrared observations**

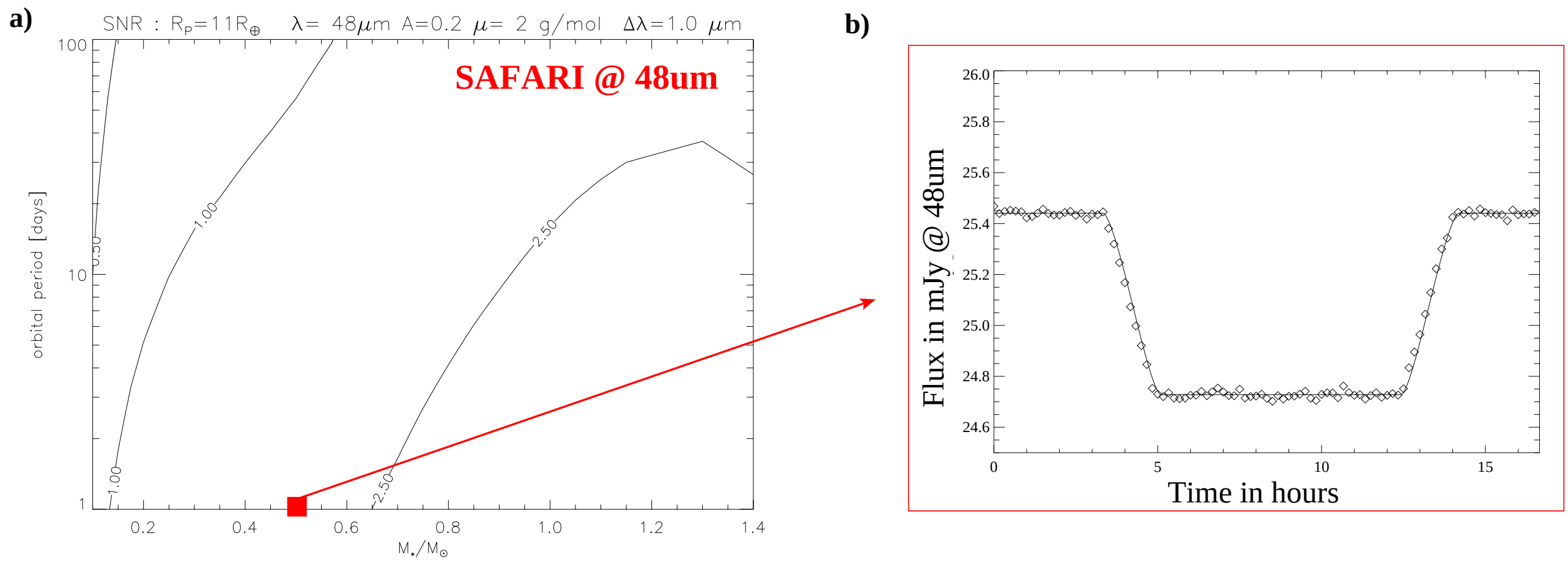

*Figure 1.18: **(a)** Signal-to-noise ratio for a primary transit of a Jupiter-mass EP at a distance of 1 AU from the host star with SAFARI (photometry at 48 $\mu$m). The star-mass vs. planet period parameter space is shown. A single spectral feature of 1 $\mu$m width is assumed in the bandpass. $R_p$: EP radius, $\mu$: mean atmospheric molecular weight, $\Delta\lambda$: spectral feature width (by A. Belu & F. Selsis). **(b)** Specific synthetic light-curve at 48 $\mu$m of a Jupiter-like EP around a cool M0 star at 10 pc (by S. Pezutto).*

## 1.9 Concluding remark

We have seen in the preceding sections that pushing forward our knowledge and understanding of the formation of galaxies and planets requires a leap in sensitivity in mid to far infrared spectroscopic and far infrared photometric capabilities. Only by placing a cold ($<$ 6 K) 3.5 m telescope in space with instruments sensitive enough to take advantage of the low photon background, can we achieve the detection limits required to fulfil the ambitions of the ESA Cosmic Vision and gain a true insight into the ***conditions for planet formation and the emergence of life, how planetary systems, including our own Solar System work and how galaxies originated and what the Universe is made of***. In other words, SPICA will play a central and vital part in ESA's Science Programme in the next decade.

# Chapter 2

# Scientific Requirements

This chapter describes the scientific requirements that should be fulfilled to achieve the scientific objectives described in chapter 1. The requirements are related to the following elements: i) Mission definition; ii) Payload: Telescope, mid infrared instruments, far infrared instrument, and mid infrared coronagraph instrument.

Requirements on the subsystems that are part of the European contribution (Telescope and SAFARI) have been explained in more detail and are given first. Requirements on the other satellite elements responsibility of JAXA are included here for completeness.

## 2.1 Telescope scientific requirements

The achievement of the scientific objectives is strongly dependent on the specification of the SPICA telescope. The telescope operational temperature, wavelength coverage and image quality are defined such that SPICA provides unique scientific capabilities with respect to other space missions like AKARI, Herschel, and JWST.

The "direct detection of exoplanets" scientific objective drives the scientific requirements on the mid infrared coronagraph instrument, which in turn impose the most stringent requirements on telescope image quality. However, since the mid infrared coronagraph is considered by JAXA as optional element of the payload, we have specified the requirements that it adds on the telescope separately. In this way it is possible to assess properly the delta on the telescope development related to the mid infrared coronagraph needs.

### 2.1.1 Wavelength coverage

The "design and operation" wavelength range of the telescope is defined as the interval 5 – 210 $\mu$m, which corresponds to the nominal baseline spectral coverage of SPICA core baseline focal plane instrumentation spectral range in the mid infrared and far infrared. A possible extension of the mid infrared instrument coverage (including the baseline operation range of the mid infrared coronagraph) down to 3.5 $\mu$m are considered in the definition of the goal "design and operation" wavelength range of 3.5 – 210 $\mu$m.

A "characterisation" wavelength range of 1 – 1000 $\mu$m has been defined over which knowledge of the telescope characteristics is required. This characterisation can be obtained by analysis, meaning that specific testing activities are not necessary. This knowledge is essential because the detector arrays of the core mid infrared and far infrared focal plane instruments will still have residual detectivity down to $\sim 1$ $\mu$m and up to $\sim 1$ mm respectively.

### 2.1.2 Telescope aperture

For diffraction limited performance, the angular resolution as provided by the telescope optics is equal to $1.22\,\lambda/D$. Many of the science cases described in chapter 1 require a 3-m class telescope to achieve the necessary collecting area and the spatial resolution. An aperture similar to Herschel's (3.5 m) is required for the science to go beyond Herschel's results. In particular, the current launcher selected for SPICA, the JAXA H-IIB, will allow a monolithic telescope of 3.5 m circular diameter.

The monolithic design of the SPICA telescope is particularly relevant for the mid infrared coronagraph. A clean and simple telescope Point Spread Function is essential to meet the required scientific performance of the coronagraph and for the quality of the planet image and spectra. In addition, the telescope size is determinant for the value of the inner working angle of the coronagraph ($= 3.3\lambda/D$), or minimum angular separation from a star at which an exoplanet can be detected.

### Obscuration and transmission

In order to maximise the efficiency of the telescope collecting area over the "design and operation" wavelength range, and minimise its possibility of self-emission, in particular in the far infrared, the total obscuration of the telescope shall be less than 12.5% (goal $< 10\%$) in terms of telescope pupil relative surface area, over the full telescope field of view (as defined in section 2.1.3).

The intensity transmission T of the unobstructed region(s) of the telescope pupil aperture shall be as high as possible to take advantage of the large collecting area, and more specifically higher than the piecewise linear transmission curve. Table 2.1 lists the values of the intensity transmission required over the "design and operation" wavelength range of the telescope.

*Table 2.1: Required and goal intensity transmission*

| **SPICA core wavelength ($\mu$m)** | 3.5 | 5 | 15 | 30 | 110 | 210 |
|---|---|---|---|---|---|---|
| $T_{\rm tel}$ **minimum requirement** | 85% | 90% | 95% | 97.5% | 99.0% | 99.0% |
| $T_{\rm tel}$ **goal** | 90% | 92.5% | 96% | 98% | 99.2% | 99.4% |

*Mid infrared coronagraph specific requirements*

The mid infrared coronagraph requires that, within the unobscured part of the telescope pupil, the telescope transmission is spatially uniform within to 1.5% rms or better over the spatial frequency range corresponding to the typical coronagraph high contrast region (i.e., 3.3 to 50 cycles/D), and in the mid infrared spectral range 3.5 – 27 $\mu$m. The non-uniformity of the transmitted flux across the pupil can reduce the coronagraph contrast. Although it can be partially compensated internally by the coronagraph, it is preferred to allocate the compensation budget for a wavefront error correction, assuming that the nominal high transmission of the telescope will nominally force possible variations of the transmission to be low.

The preferred shapes for the telescope pupil obscuration (if any) are, in the order of preference (i.e. best adapted to match maximal coronagraph performances):

1. 2 spider struts (e.g. along a complete pupil diameter)
2. 4 spider struts (e.g. in a centred orthogonal cross shape)
3. 3 spider struts

The exact shape and spatial distribution of the obscuration at the telescope pupil can affect the telescope PSF main lobe and (near and far-) sidelobes. A mid infrared coronagraph based on pupil apodisation by re-mapping or shaped mask is particularly sensitive to entrance pupil obscuration shape and geometry. No telescope pupil obscuration would be overall preferred, but may set overall unrealistic constraints (e.g., specific telescope design becoming an observatory-level design driver). For more than 4 spider struts, it is expected that the coronagraph rejection performance in its dark region will be too badly affected and therefore these cases are not considered.

### 2.1.3 Image quality and field of view

#### General

In order to achieve the maximum angular resolution for the given telescope diameter, the SPICA telescope shall have diffraction-limited imaging capability at $\lambda = 5\,\mu$m over a field of view of 5 arcmin radius (with a goal of 6 arcmin). In addition, the SPICA telescope shall keep diffraction-limited performances at 30 $\mu$m over a field of view of 10 arcmin radius (goal is 12 arcmin). The mapped telescope field of view shall be unvignetted over a radius of 12 arcmin (goal is 15 arcmin).

The mid infrared instruments will be located in the 5 arcmin radius area of the telescope field of view, where the image quality is expected to be higher. The far infrared instrument may be located in the periphery but still within the unvignetted field of view, not to damage the imaging efficiency nor to generate spurious spatial response from diffraction at a structure other than the one defined by obscuration in the telescope pupil.

#### Details of telescope aperture phase error distribution

Taking $\lambda_0 = 5\,\mu$m as the baseline shortest "design and operation" wavelength, and following the above, the main requirement is that the overall maximum SPICA telescope wavefront error, over a field o view of 5 arcmin radius (goal 6 arcmin), shall be 350 nm rms in order to be compatible with nominal diffraction-limited performances at $\lambda_0$.

Beyond this top-level requirement, imaging in certain modes and bands can be affected by phase or wavefront error at different pupil spatial scale and this leads to the following break-down of the overall telescope pupil phase error:

- Low spatial frequencies (i.e., below 3.3 cycles/D with D is the telescope pupil diameter): Overall figure errors to be limited for the SPICA telescope to $< 350$ nm rms. As an example, this can be spread equally between focus and higher order aberrations: $< 250$ nm rms for focus translates into a maximum defocus of $\pm\,0.185$ mm at the telescope focal plane. In addition, $< 250$ nm rms for higher but still low order aberrations (total telescope) translates into a maximum allocation of about one wave peak-to-valley at the standard visible test wavelength (e.g., 633 nm) or a surface figure error of $< \lambda_0/70$ per mirror in a 2-mirror telescope case.

- Medium spatial frequencies (i.e., from 3.3 to 1000 cycles/D): No specific requirements are defined for this interval. The residual mirror surface figure errors in these spatial frequencies range shall be limited only by the overall compatibility with the general 350 nm rms requirement. These errors arise from residual higher frequency surface error, mount induced stress, print-through effects or general mirror substrate medium-scale structure.

- High spatial frequencies (i.e., > to >> 1000 cycles/D): $< \lambda_0/285$ rms per mirror. These mostly arise from mirror surface micro-roughness, cracks and small local defects in coatings all of which induce large scale/wide angle scatter as well as generating small off-axis angle diffuse halo by near-specular scatter. The specification translates into a $< 17.5$ nm rms specification for each telescope mirror surface leading to a TIS $< 0.2\%$ at $< \lambda_0$.

*Mid infrared coronagraph specific requirements*

The maximum telescope wavefront peak-to-valley (PTV) error, within the spatial frequency range from 0 to 50 cycles/D and over the entire unobscured part of the telescope pupil, shall be no higher than 2 $\mu$m.

It should be noted that although baselined to operate down to 3.5 $\mu$m, the coronagraph does not require diffraction-limited performance at this wavelength as it is expected to perform its own wavefront compensation internally to recover the diffraction-limited performance at 3.5 $\mu$m. Nevertheless, the limitations of this wavefront correction require this extra PTV specification.

The mid infrared coronagraph imposes more stringent requirements on the telescope pupil phase and wavefront errors at the mid spatial frequencies (i.e., from 3.3 cycles/D to 1000 cycles/D):

1. From 3.3 cycles/D to 12 cycles/D: $< \lambda_0/45$ (= 111 nm) rms (with a goal $< \lambda_0/133$ (= 37 nm) rms) for the SPICA telescope in this spatial frequency range;

2. From 12 cycles/D to 50 cycles/D: $< \lambda_0/71$ (= 70 nm) rms (with goal $< \lambda_0/133$ (= 37 nm) rms) for the SPICA telescope in this spatial frequency range;

3. from 50 cycles/D to 1000 cycles/D: there is no specific requirement, residual error only limited by overall compatibility with the general wavefront requirement.

The limits of the mid spatial frequency range are defined to match the expected Inner Working Angle (~3.3 $\lambda_0/D$) and Outer Working Angle (OWA, potentially extending up to ~33 $\lambda_0$//D), including a 50% margin beyond the OWA for the aliasing effect from the surface error at spatial frequencies around the one equivalent to OWA. In this region, the goal telescope pupil phase (or wavefront) error is specified based on an equivalent maximum degradation on raw contrast of a factor 10 with respect to an assumed raw contrast of $10^{-7}$ (with quasi-perfect optics and/or active wavefront compensation), leading to the main aim of $10^{-6}$ raw contrast. In the nominal case of the baseline specification, it is assumed that only $10^{-6}$ is nominally achieved (i.e., 10 time less effective or absent active correction); with the telescope specification only $10^{-5}$ will be then obtained in final raw contrast.

Beyond 50 cycles/D, the coronagraph dark region is not expected to be affected and therefore the requirement is unconstrained (and known to be not easily accessible to verification by measurements) although the overall main requirement of diffraction-limited performances at $\lambda_0$ still applies.

### 2.1.4 Telescope temperature

The science objectives defined in Chapter 1 can only be achieved by increasing dramatically the sensitivity with respect to ISO, AKARI and Herschel. In particular, sections 2.3 and 2.4 give the sensitivity requirements on the mid infrared and far infrared instruments. To meet those, the development of more sensitive detectors needs to be complemented with the reduction to the maximum extent of the background photon noise generated by the telescope. In particular, it is a scientific requirement for SPICA that the observatory sensitivity is limited by the sky background.

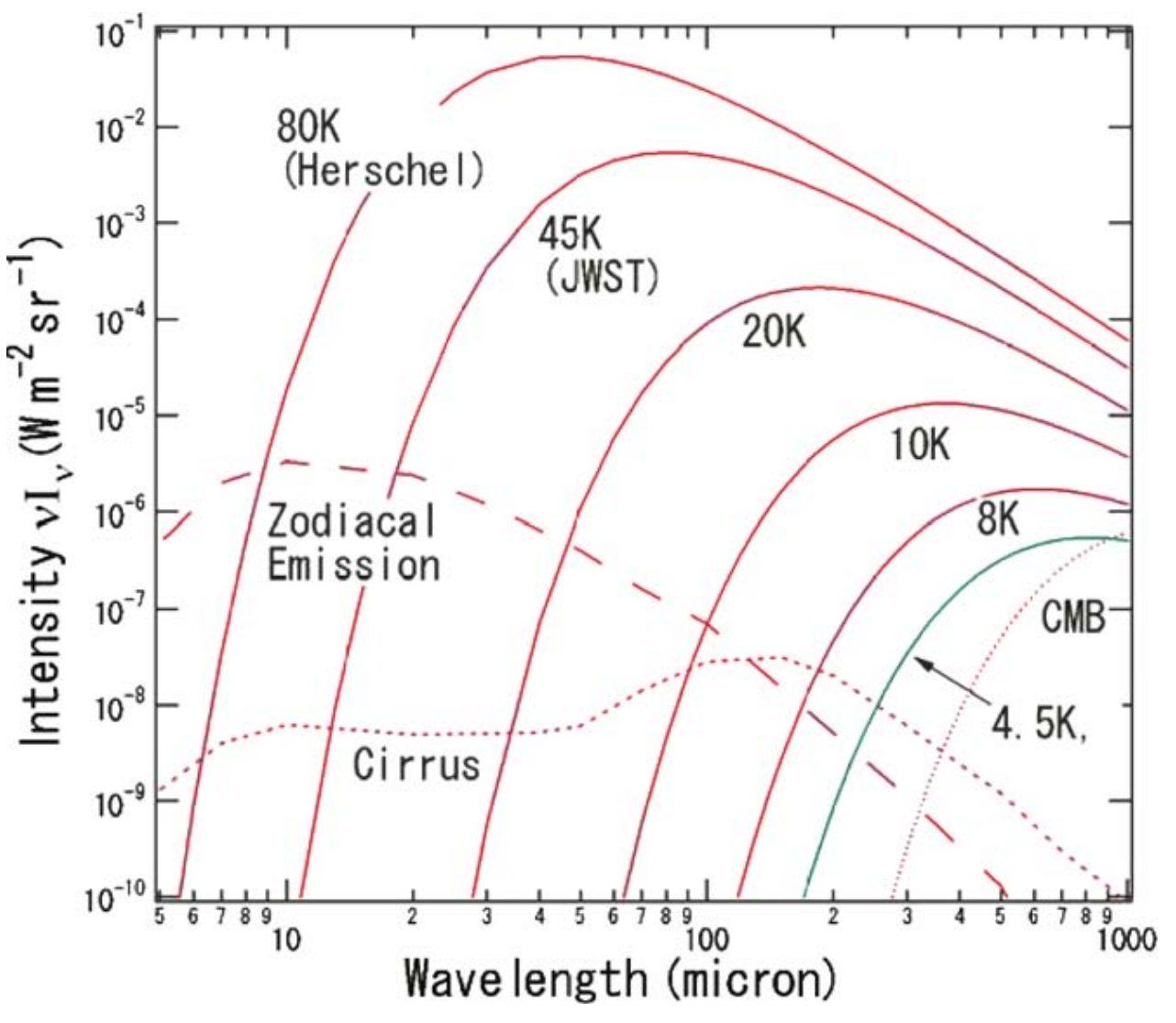

*Figure 2.1: Thermal emissions from telescopes at different temperatures and the natural sky background.*

Figure 2.1 shows the thermal emission for several telescope temperatures and the natural background defined by the zodiacal light, the cirrus and the CMB. The strong dependency of background on the telescope temperature is due to the locations of the mid infrared and far infrared wavebands in the Wien region of the telescope thermal emission. As can be seen in Figure 2.1, a telescope with a temperature $\lesssim 4.5$ K emits clearly below the sky background for wavelengths less than 200 $\mu$m. The figure also illustrates that the need for such low temperature is mainly related to the performance of the longer wavelength bands of the far infrared instrument.

From all the above, it is required that the temperature of the SPICA telescope is less than 6 K, with a goal of $< 5$ K. The limit of 6 K has been derived from calculations made with the sensitivity model of SAFARI, which show that, taking goal sensitivity detectors, this is the maximum telescope temperature that does not lead to a reduction of the instrument sensitivity.

Note that, in particular, the Herschel telescope emits several orders of magnitude higher than the required level for SPICA. The orders of magnitude lower photon noise, combined with the higher sensitivity of the SAFARI detectors, will lead to a breakthrough in sensitivity in the far infrared wavelength range. Regarding the mid infrared instruments, the low SPICA telescope temperature enables to achieve sensitivities similar to JWST in spite of the smaller telescope aperture. Actually, as can be seen in Figure 2.1, sky limited sensitivity in the mid infrared is already reached with a temperature of $\sim 25 \pm 5$ K, depending on spectral bandwidth.

**Spatial and temporal telescope temperature stability**

As explained above, the sensitivity to telescope elements temperature variations is very high for the far infrared instrument. For example, for wavelengths longer than 100 $\mu$m it becomes typically $\sim 300\%$/K. Telescope temperature gradients and spatial non-uniformities increase the instrument background associated noise, generate a field-dependent signal leading to a non-uniform background illumination across the far instrument array. While allowing the presence of "hotter spots" (typically up to 10 K) on the telescope aperture, their influence on the background needs to be minimised by being "weighted" through a smaller relative area and therefore through the view factor from the far infrared instrument pixel. Based on calculations of the contribution of higher temperature zones to the total emission, it is required that for every degree Kelvin above the maximum baseline telescope temperature of 5 K, the relative surface area (i.e., compared to the total telescope aperture) of this associated "higher" temperature zone shall be further reduced by at least a factor 5.

The strong dependence of the background noise on the telescope temperature imposes also a requirement on the temporal stability of the telescope operating temperature, which shall be stable to within 0.25 K over a typical observation duration, which is taken as one hour.

### 2.1.5 Straylight

The high sensitivity requirement of SPICA makes the avoidance of straylight essential. This needs to be tackled both in the telescope and instrument design. It is required that the system composed of the telescope and focal plane instruments shall provide adequate internal straylight control to enable SPICA observations with sky-limited sensitivity. In particular, the straylight rejection level shall be such that total background from out-of-field stray sources (artificial and natural) shall not increase the in-field optical background signal (from zodiacal light and telescope thermal self-emission) by more than 20% (goal 10%). The out-of-field sources, potentially seen by reflection, diffraction or scattering mechanisms, typically include the sky emission itself (zodiacal light) as well as particular "local" celestial objects such as Sun, Earth, Moon and planets, particular zones of the sky (e.g., galactic centre) as well as the SPICA observatory sub-systems self-emission.

In particular:

- Any emission from observatory structure at surface temperatures $> 10$ K (respectively $> 50$ K), that enters the far infrared instrument field of view (respectively the mid infrared instrument field of view) at telescope focal plane, shall be attenuated by the telescope system and associated baffles in order to reduce the view factor to these surfaces (including their in-band emissivity and from the focal plane instrument field of view) to less than 1 part in $10^4$.

- Scattered light from any natural source inside the telescope field of view (but outside the respective field of view of the focal plane instruments) shall contribute, into the respective aperture of any focal plane instrument, less than $10^{-3}$ of its peak irradiance when imaged by the telescope.
- Scattered light from any natural source outside the telescope field of view shall be attenuated to the maximum relative levels given in Table 2.2 defining the telescope PST (Point Source Transmission) in any azimuthal direction.

### 2.1.6 Surface cleanliness

Telescope cleanliness shall be optimised so that it does not affect the telescope collecting efficiency, in particular in the science bands where mid infrared and far infrared emission/absorption of particular ice compounds are intended to be studied with SPICA. For smooth infrared optical surfaces, surface scattering is dominated by contaminating dust particles which can raise the level of far-side lobes (i.e., wide field angle from boresight) of a diffraction-limited telescope PSF. The goal is that the telescope surface cleanliness level shall be kept within $\sim$ 2000 mm$^2$/m$^2$ or 2000 ppm for particulate and 3 mg/0.1 m$^2$ for molecular based on ECSS-Q-70-01A, or equivalently, (400±50)C in terms of MIL-STD-1246C classification.

## 2.2 Mission definition scientific requirements

### 2.2.1 Wavelength coverage

SPICA will cover the mid and far infrared wavelength region, which is one of the richest ranges of the electro-magnetic spectrum emitted by astrophysical objects. It is in this region where most energy is emitted during galaxy evolution. It plays host to a variety of atomic/ionic forbidden transitions, molecular lines and solid state features that provide unique diagnostics to understand physical and chemical processes in a multitude of astrophysical environments, as described in the Science Objectives (section 1.1.2).

The wavelength region that will be covered by SPICA is 5 – 210 $\mu$m. The goal is that the focal plane instruments provide continuous coverage over the full range, both for imaging and spectroscopy. A full spectral coverage has important advantages to investigate and model the physical conditions of the gas and dust in Ultraluminous Infrared Galaxies (ULIRGs), circumstellar discs, exoplanets and Solar System bodies. In particular, molecules like $H_2$, $H_2O$, CO and OH, and ions of Ne, S, Si, O, and C have essential diagnostic lines in the whole SPICA range. Broad spectral features associated with PAHs, ices, silicate and carbon grains can be much better interpreted with observations over both the mid and far infrared. Accurate measurements of their broad spectral shapes and of their spectral peaks along the mid infrared and far infrared region, enable to further constrain the models for an unambiguous identification of the makeup minerals or composites and for the determination the physical and chemical conditions. Furthermore, a continuous wavelength coverage from the mid to the far infrared allows the study of astronomical objects as a function of redshift, by analysing powerful diagnostic lines and spectral features as they shift from the mid infrared to longer wavelengths and into the far infrared.

*Table 2.2: Telescope Point Source Transmission, defined as the ratio of the signal level from the stray source reaching the telescope focal to its incoming level*

| **Off-axis angular region from telescope boresight** | **From 0.25° to 3°** | **From 3° to 30°** | **From 30° to 45°** | **Beyond 45°** |
|---|---|---|---|---|
| **Max PST level (base requirement)** | $< 5 \times 10^{-4}$ | $< 10^{-5}$ | $10^{-6}$ | Assumed dropping to $10^{-8}$ at 90° |
| **Max PST level (goal)** | $< 5 \times 10^{-5}$ | $10^{-6}$ | $10^{-7}$ | Assumed dropping to $10^{-9}$ at 90° |

However, constraints in the satellite resources for instrument allocation may prevent meeting the goal of full wavelength coverage both with imaging and spectroscopy. A review and selection of the wavelength regions and scientific capabilities to be covered will be carried out by JAXA in 2010, after an assessment of the mission requirements priorities and the scientific capability uniqueness with respect to other observatories.

### 2.2.2 Observational and environmental requirements

SPICA must be in an orbit that enables observation of all areas of the sky over the duration of the mission, and provides an instantaneous sky visibility of $> 30\%$ (TBC). SPICA's orbit shall provide:

- A benign and stable thermal environment, to facilitate the optimal performance of the cryogenic system and to minimise disturbances in the background photon noise.
- A relatively low-radiation environment, to avoid to the maximum extent radiation effects on the electronics and glitches on the detector signal that would compromise the required sensitivities.

### 2.2.3 Pointing requirements

The spatial resolution of SPICA, which corresponds to diffraction limited performance, combined with the need to reduce pointing instabilities that introduce additional noise in the signal, lead to the following pointing requirements:

Standard mode: Absolute Pointing Error $< 0.135$ arcsec ($3\sigma$), Relative Pointing Error $< 0.075$ arcsec in 200 s;

Coronagraph mode: Absolute Pointing Error $< 0.03$ arcsec ($3\sigma$), Relative Pointing Error $< 0.03$ arcsec in 20 min.

SPICA must be able to track fast moving targets in the Solar System, (e.g., comets), with a speed of 10 arcsec per minute during a period of 20 minutes. The pointing requirements specified above also apply during tracking.

SPICA shall provide a pointing mode to enable mapping of large areas of the sky. In this mode a stationary target star is measured by continuously shifting the spacecraft attitude with constant speed and direction according to the following specification:

Slow-scan speed range: 10 to 72 arcsec/sec for the far infrared instrument and from 0.054 to 2.28 arcsec/sec for the mid infrared camera and spectrometers.

Slow-scan speed accuracy: Less than 1% for the far infrared instrument and less than 10% for the mid infrared camera and spectrometers.

Slow-scan duration: Less than 600 sec for the far infrared instrument and from 5 to 50 sec for the mid infrared camera and spectrometers.

The absolute pointing at any point in time during scan execution shall be recoverable a posteriori (e.g., from housekeeping data) with the accuracy given for the standard pointing mode.

## 2.3 Far infrared instrument scientific requirements

The science case presented for the SPICA mission in chapter 1 includes the requirement for the provision of a flexible capability for imaging and spectroscopy working in the wavelength range from $\sim 34\ \mu$m, where the [Si II]34.8 $\mu$m line appears and the Si:Sb detectors stop operating, out to $\sim$ 210 $\mu$m, where the background from the telescope and baffles starts to dominate and the [N II]205.2 $\mu$m line is observed. This capability will be provided by the core far infrared instrument SAFARI. In Table 2.3 we present an analysis of the instrument requirements necessary to address the key science goals of the SPICA mission.

*Table 2.3: Requirements of the SAFARI instrument to match the scientific objectives of the SPICA mission*

| Science topic/section | Observing Requirement | SAFARI Requirement |
|---|---|---|
| **Planetary Formation** | | |
| Gas in protoplanetary discs | Spectroscopy of early stage gas discs in the main MIR/FIR gas cooling lines, light hydrides and organics to study physical conditions and oxygen/carbon chemistry as planetary systems form. | High sensitivity medium resolution ($R \sim$ few thousand) spectroscopy mode. Integral field spectroscopy over as large a field as possible is required for efficient surveys and to facilitate background subtraction. Uninterrupted wavelength coverage required from 34 to 210 $\mu$m. |
| Water in proto-planetary discs | Spectral survey of discs of a variety of ages to look for to the transition between gaseous water and water ice. In nearby systems spectral imaging of discs to resolve the "snow line". | High sensitivity medium resolution ($R \sim$ few thousand) mode for gas features and lower ($R \sim$ hundreds) resolution for solid state features. Spectral imaging mode required with field of view large enough to efficiently map nearby discs. Uninterrupted wavelength coverage required from 34 to 200 $\mu$m. |
| Occurrence and mass of debris discs | Unbiased survey of all stellar types out to a few hundred pc to determine frequency of dusty discs. | High sensitivity multiband (minimum 3) photometric mode with as large a field of view as practicable. |
| Dust mineralogy in debris discs | Spectral survey of dusty debris discs to look for differences in mineralogy indicating the chemical and thermal history of dust. In nearby systems spectral imaging of discs. | High sensitivity modest resolution ($R \sim$ few hundred) spectral imaging mode with a field of view large enough to efficiently map nearby discs. Uninterrupted wavelength coverage required from 34 to 100 $\mu$m. |

*Table 2.3: (continued)*

| **Science topic/section** | **Observing Requirement** | **SAFARI Requirement** |
|---|---|---|
| The composition of the Kuiper Belt | Unbiased survey of Kuiper belt objects. Mineralogy and physical properties of Kuiper belt objects and associated dust. | High sensitivity photometric mode with as large a field of view as practicable. High sensitivity modest resolution ($R$ ∼few hundred) spectroscopy mode for solid state features. Uninterrupted wavelength coverage required from 34 to 100 $\mu$m. |
| **Galaxy formation and Evolution** | | |
| The life cycle of gas and dust | Imaging low resolution spectroscopy and photometry of faint diffuse regions in our own and nearby galaxies to determine the mineralogical content of the dust in a variety of contexts and sources. Imaging spectroscopy of FIR cooling lines in faint diffuse regions in a variety of contexts and sources. | High sensitivity modest to moderate resolution ($R$ ∼few hundred to a few thousand) imaging spectroscopy mode required with as large a field of view as practicable giving uninterrupted wavelength coverage from 34 to 210 $\mu$m and Nyquist sampled imaging. |
| AGN/starburst connection over cosmic time | Spectroscopy of unresolved diagnostic features of ionic and atomic lines over wide wavelength range in individual sources with simultaneous characterisation of the background. Low resolution spectroscopy also required for detection of solid state PAH and silicate features.<br>Possible use of higher resolution spectroscopy to use line widths as a diagnostic of Black hole masses. | For line diagnostics we require high sensitivity medium resolution ($R$ ∼few thousand) spectroscopy mode. Integral field spectroscopy over as large a field as possible is required for efficient surveys and to facilitate background subtraction.<br>For PAHs and solid state dust features a spectro-photometric mode with $R \sim 25 - 50$ with uninterrupted wavelength coverage from 34 to 210 $\mu$m is required. |

*Table 2.3: (continued)*

| Science topic/section | Observing Requirement | SAFARI Requirement |
|---|---|---|
| The nature of the CIRB | Very large area photometric surveys to below the confusion limit at all bands (minimum 3). Very high sensitivity blind spectral surveys of as large an area as possible. | High sensitivity multiband (minimum 3) photometric mode with as large a field of view as practicable and with Nyquist sampled imaging capability. High sensitivity modest resolution ($R \sim$ few hundred) imaging spectroscopy mode with as large a field of view as practicable giving uninterrupted wavelength coverage required from 34 to 210 $\mu$m and Nyquist sampled imaging. |

### 2.3.1 Assessment of basic instrument concept

Inspection of Table 2.3 shows that the core far infrared instrument, i.e. SAFARI, needs to have a flexible imaging and spectroscopy capability with as high a sensitivity as possible given the constraints imposed by the presence of the photon background from the natural background and the telescope baffles. In principle the capabilities required could be provided by a mixture of a photometric camera and an integral field spectrometer such as employed in the PACS instrument of Herschel (Poglitsch et al. 2008) or the MIRI instrument on JWST (Wright et al. 2004). However, a study carried out in the pre-assessment study phase of instrument design showed that, given that megapixel detector arrays are not available in the FIR (unlike in the MIR), the spectroscopy mode of such an instrument would be prohibitively slow to cover the wavelength range required, a separate camera channel would be needed, the field of view of the integral field unit would be small and the requirements on the detector sensitivity to match the very low photon background would be extremely difficult to meet within the context of a space mission launching in 2017/2018.

We have therefore pursued an instrument concept based on an imaging Fourier Transform Spectrometer (iFTS) design such as used for the spectroscopy channel on the Herschel SPIRE instrument. This concept has the advantage of naturally providing both a photometric and spectroscopic imaging mode within the same instrument without the need for extra mechanisms such as switching mirrors etc. This instrument concept was also investigated in the ESA sponsored Concurrent Design Facility study carried out at the start of the present assessment phase (CDF Study Report, SPICA-SAFARI 2008). The naturally broad wavelength band of the instrument means that it will be limited in sensitivity by the natural photon background if the detector "dark" noise equivalent power can be kept to $<\text{few}\times10^{-19}$ W Hz$^{-1/2}$. Figure 2.2 illustrates the line sensitivity that can be achieved in an iFTS under a number of assumptions about the instrument throughput and photon background contributions as a function of the dark NEP of the detectors. We can see that the ultimate performance for a detector array that provides instantaneous Nyquist sampled images is reached when the detector NEP falls below $\sim 2-3\times10^{-19}$ W Hz$^{-1/2}$. We describe the detector technology necessary to achieve this performance in section 3.2.2 of the present report. For the purposes of setting the requirements on the instrument performance we do not invoke a given detector technology but rather give the requirement in terms of meeting the performance that could be achieved for the goal detector technology. In section 3.2.2 we discuss how the instrument design and capabilities would need to change in order to maintain the ultimate sensitivity if the detectors do not match the goal performance. Under any circumstances we aim to meet the sensitivity requirement for an unresolved spectral line on a point source.

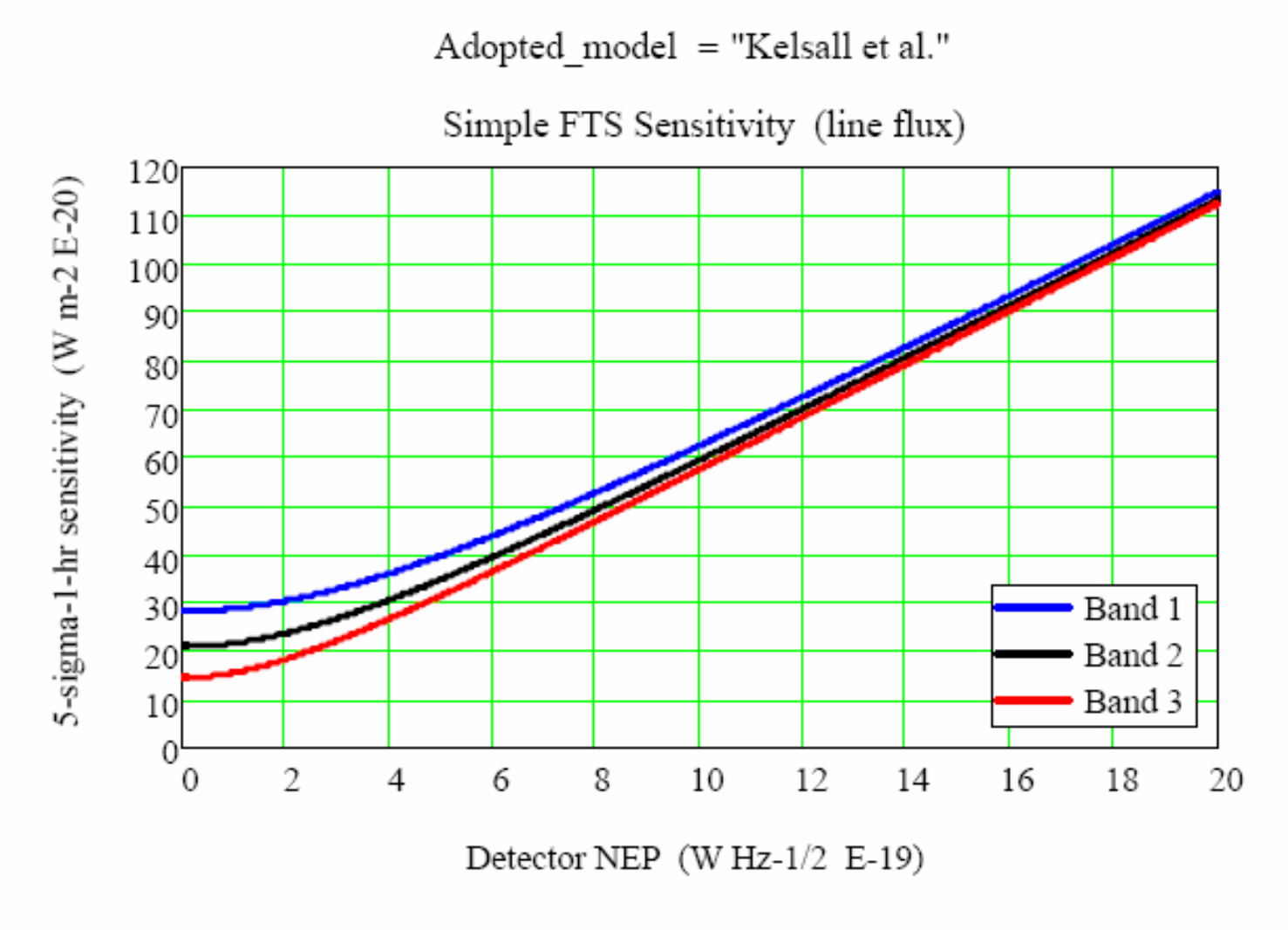

*Figure 2.2: Achieved unresolved line sensitivity on SPICA as a function of detector dark NEP for a broad band (R ∼ 3) Fourier Transform Spectrometer operating with three wavelength limited detector bands under the assumption of a photon background comprised of the Zodiacal light and the Cosmic IR background. The bands used were centred on 48, 85 and 160 µm and we assume the throughput of the instrument is that appropriate to instantaneous Nyquist sampling of the point spread function – i.e. 0.5 f/# λ pixels.*

### 2.3.2 Scientific requirements

The final scientific requirements for the far infrared instrument, SAFARI, are as follows:

1. Wavelength coverage over at least 34 to 210 $\mu$m with a design driver to achieve 28 – 210 $\mu$m (unless covered by other instrument). This wavelength range is chosen to match the Si:Sb (or Si:As for 28 $\mu$m coverage of the mid infrared instruments.
2. A photometric camera mode with $R \sim 2$ to 5.
3. Range of spectral resolution modes with a resolution of at least $R = 2000$ at 100 $\mu$m, $R \sim$ few hundred and a spectrophometric mode of $20 < R < 50$.
4. As large an instantaneous wavelength coverage as possible for spectroscopy but not at the expense of spectral resolution.
5. 3-D mapping in photometry and spectroscopy to be as fast as possible.
6. An instantaneous field of view of at least $2 \times 2$ arcmin$^2$.
7. Unresolved line sensitivity to be limited only by the natural background over as wide a wavelength range as possible. This equates to $\sim 3 \times 10^{-19}$ W m$^{-2}$ ($5\sigma$, 1 hour) at 48 $\mu$m and $< 2 \times 10^{-19}$ W m$^{-2}$ for a full $R \sim 2$ band.
8. For certain wavelengths higher sensitivity is desirable, which may be achieved only over a narrow spectral range through the use, for instance, of narrow band filters inserted into the optical path.
9. Continuum sensitivity of $< 50$ $\mu$Jy.
10. Unresolved line sensitivity to be maintained at the expense of spatial sampling in the event of less than goal performance from the detector.

## 2.4 Mid-infrared instruments scientific requirements

The scientific requirements for the mid infrared instruments have been defined by JAXA/ISAS in the SPICA Mission Requirements Document (2009). The achievement of the scientific objectives demands imaging and spectroscopic capabilities over the SPICA mid infrared range, which is defined as the interval from 5 to $\sim 34\,\mu$m.

### 2.4.1 Imaging and spectrophotometric capabilities

#### Spectral resolution

The spectral resolution for the imaging capabilities must be $R \sim 10$ and for the spectrophotometric capabilities, $R \sim 100$.

#### Sensitivity

The requirement for photometric sensitivity in the mid infrared is 5 $\mu$Jy, with a goal of 1 $\mu$Jy (5$\sigma$, 1 hour observation). A main driver for this specification is to perform imaging of a wide-area survey of galaxy clusters and of large scale structures in the early Universe where the star formation activities was at a peak, and whose redshifted emitting energy shifts into the mid infrared.

This sensitivity will also allow us to carry out imaging of debris discs whose amount of dust is comparable to our Solar System, and reveal the distribution and physical state of solid materials, particularly ice, in protoplanetary discs and dust discs in the main-sequence stars.

#### Field of view

The mid infrared camera must have a field of view greater than $5 \times 5$ arcmin$^2$, with a goal $> 6 \times 6$ arcmin$^2$. This large field of view will facilitate the execution of an efficient survey of galaxy clusters in the early Universe over an area corresponding to $\sim 300$ Mpc. The survey will allow us to trace the large scale structures and to reveal the star formation history in the early Universe (up to 9 Gyr ago) as well as the mass assembly history and its environmental effect on the galaxy evolution. The large field of view will be essential for the mid infrared imaging spectroscopy of 50 nearby galaxies of the AKARI sample, such that the ISM emission can be spatially resolved, to track galactic-scale material circulation from sources to sinks in the ISM in galaxies. Another important driver is the imaging and broad band spectroscopy of the galactic plane, to obtain a point source catalogue of $10^9$ sources, a low-resolution spectral catalogue of $10^7$ sources and diffuse line emission spectral maps.

#### Spatial resolution

The spatial resolution should correspond to diffraction limited performance as given by diameter of the SPICA telescope at 5 $\mu$m, which approximately corresponds to an angular resolution of 0.36″/pixel. This resolution is necessary to spatially resolve the ISM emission in nearby galaxies; to reveal the distribution and of solid materials in proto-planetary discs and discs around main sequence stars; to map in detail faint molecular and dust shells of evolved stars in the Milky Way and observe individual objects in the Magellanic cloud for the determination of the mass-loss histories and dust formation processes.

### 2.4.2 Spectroscopic capabilities

#### Sensitivity

The spectral sensitivity of the mid infrared spectrometers must be at least $5 \times 10^{-20}$ W m$^{-2}$ for a 5$\sigma$, 1 hour observation. This sensitivity is required to study the interstellar environment and dust emission characteristics of high-redshift galaxies out to $z \sim 3$, through PAH emission as well as atomic and molecular emission lines. These observations are essential to further understand the AGN/starburst connection at high redshift. It will also enable SPICA to make infrared imaging and spectroscopic observations of $\sim 1,000$ candidates in search for the forming super-massive black holes (SMBHs), that can not be observed easily in other methods due to the obscuration of dust, from the present to the early Universe. Supplementing these results with the results of observations for the galaxy formation history, we will understand the role of SMBHs in the galaxy evolution.

A second main driver for the sensitivity specification is to make surveys for emission lines which could be associated with warm gas (100 – 1000 K) in protoplanetary discs, and to correlate the amount and characteristics of gas with stellar masses and ages.

#### Spectral resolution

The spectral resolution ($R = \lambda/\Delta\lambda$) of the mid infrared spectrometers must be greater than 1000, which would satisfy most of the scientific objectives.

There is also the requirement for a high-resolution spectrometer with $R > 10\,000$, goal of $30\,000$. This higher resolution is crucial to elucidate the geometric, physical and chemical structure of proto-planetary discs by measuring the motion of gas, and therefore to witness the evolution of the disc structure due to planet formation. High-resolution spectroscopy will be particularly valuable when studying the atmosphere of gas giant planets through the transmission spectroscopy method.

Higher resolution spectroscopy is also required to examine the properties of molecules in the thick molecular layer surrounding the photosphere of red giants (discovered by ISO observations) and the properties of molecules in cold dense molecular clouds with embedded young stellar objects in the Milky Way. In general terms, the availability of a high resolution spectrometer on board SPICA will provide a unique scientific capability in the mid infrared that was neither available in Spitzer nor will be in JWST, whose spectral resolution is one order of magnitude lower.

#### Spatial resolution

The spatial resolution should correspond to diffraction limited performance at 10 $\mu$m. The drivers are the same as for mid infrared imaging.

## 2.5 Mid-infrared coronagraph scientific requirements

The main scientific driver for the mid infrared coronagraph is the capability to directly image and perform mid infrared spectroscopy of exoplanets, in particular young gas giant exoplanets that are within 1 Gyr of their formation. To accomplish this objective, the instrument needs to meet the following specification:

- The instrument wavelength range must be 5 – 27 $\mu$m to cover the atmospheric mid infrared spectral features from molecules expected in giant exoplanets atmospheres, like water, ammonia and methane, as described in section 1.3. As a goal, the minimum wavelength should become 3.5 $\mu$m to to reduce the inner working angle and to allow us to observe the 4 $\mu$m predicted spectral feature due to an opacity window in EGPs and cooler objects with temperatures between 100 and 1000 K.

- The planet/star contrast ratio must be $10^{-6}$ at all instrument wavelengths.

- An inner working angle (IWA) of $\sim 3.3\lambda/D$ ($D$ is the telescope diameter), which corresponds to the capability to detect a planet at a distance of the orbit of $\sim 9$ AU (Saturn's orbit) at 5 $\mu$m around a star at 10 pc. The IWA is the point where the required contrast ratio is achieved. The goal is to have an IWA of $\sim 1.5\lambda/D$, to detect at 5 $\mu$m a planet at $\sim 5$ AU from a star at 10 pc. The reduction of the minimum wavelength range to 3.5 $\mu$m would lower the minimum distance at which a planet could be detected.

- The spectral resolution must be $R = 20 - 200$.

- The outer working angle (OWA) should be large enough to allow the observation of planets at 5 $\mu$m at a distance of $\sim 50$ AU from a star at 10 pc, that is, $\sim 16\lambda/D$.
- The field of view must be $1 \times 1$ arcmin$^2$ to ensure the spatial coverage of close planetary systems and compatibility with the OWA.

The coronagraph instrument sets stringent requirements on the SPICA telescope as have been described in section 2.1.

# Chapter 3

# Payload

## 3.1 The SPICA telescope

In order to achieve unprecedented sensitivity levels in the mid and far infrared bands, the SPICA mission relies on a 3.5 m diameter cryogenic telescope operating at about 5 K. Such a cryogenic telescope (indicated hereafter as STA, SPICA Telescope Assembly) represents the main ESA contribution to the mission and it is described in detail in the paragraphs below.

After the selection of the SPICA proposal for the assessment phase, initial contacts between ESA and JAXA were established in November 2008 and preliminary interface specifications agreed on. Based on the work of the SPICA Telescope Science Study Team (STSST), science requirements applicable to the assessment activities were defined in SPICA Telescope Science Requirements Document (2009). Finally, a SPICA Telescope Requirements Document (2009), containing all engineering requirements applicable to the assessment study was compiled and maintained.

In order to obtain a preview on the main critical issues and properly define the scope of the planned industrial activities, an internal study was performed by ESA at ESTEC between March and April 2008, in close cooperation with JAXA (see SPICA Internal Assessment Study Report 2008).

Following the release of the Invitation to Tender, two parallel competitive contracts were awarded to EADS Astrium (F) and to Thales Alenia Space (F) for the industrial assessment study of the SPICA telescope. Both studies, with duration of 1 year, were kicked-off at the end of July 2008 and completed at the end of July 2009. The results of the study activities are summarised in this document.

The main objectives of the assessment study were: a) consolidation of all requirements applicable to the STA; b) further definition of the interfaces between the ESA provided telescope assembly and the JAXA provided mission elements; c) definition of a baseline design and demonstration of its compliance with the requirements and overall technical feasibility; d) definition of a complete development plan, comprehensive of any required technology validation; e) programmatic estimates (schedule and cost).

Based on the results of the industrial studies, very detailed for Phase 0/A level, we can conclude that all objectives have been met and that we have acquired a rather well defined picture of the challenges associated with the development of the STA. Based on the existing heritage and expertise, we claim that Europe is very well placed to deliver the SPICA telescope to JAXA, meeting the agreed requirements.

A Ritchey-Chrétien configuration is confirmed as the most suitable to meet the performance requirements. Both contractors have identified a design entirely based on ceramic material (SiC100 for Astrium and HB-Cesic for Thales), for the optical surfaces (M1 and M2) as well as the overall telescope structure. A quadripode configuration for the M2 support structure is selected as most appropriate from both a mechanical and optical point of view (PSF requirements induced by the Coronagraph instrument). Such a support structure would be connected to the Telescope Optical Bench (TOB) and not directly to M1, in order to minimise distortions of the mirror surface figure and meet the wavefront error requirements. The need for a separate Instrument Optical bench (IOB – to be provided by JAXA) is confirmed to host the Focal Plane Instruments (FPI's and properly manage their interfaces with the rest of the STA. A focus and tip/tilt mechanism acting on M2 will allow correcting for any residual deviation from the nominal alignment, while a set of heaters would allow decontamination during flight if required.

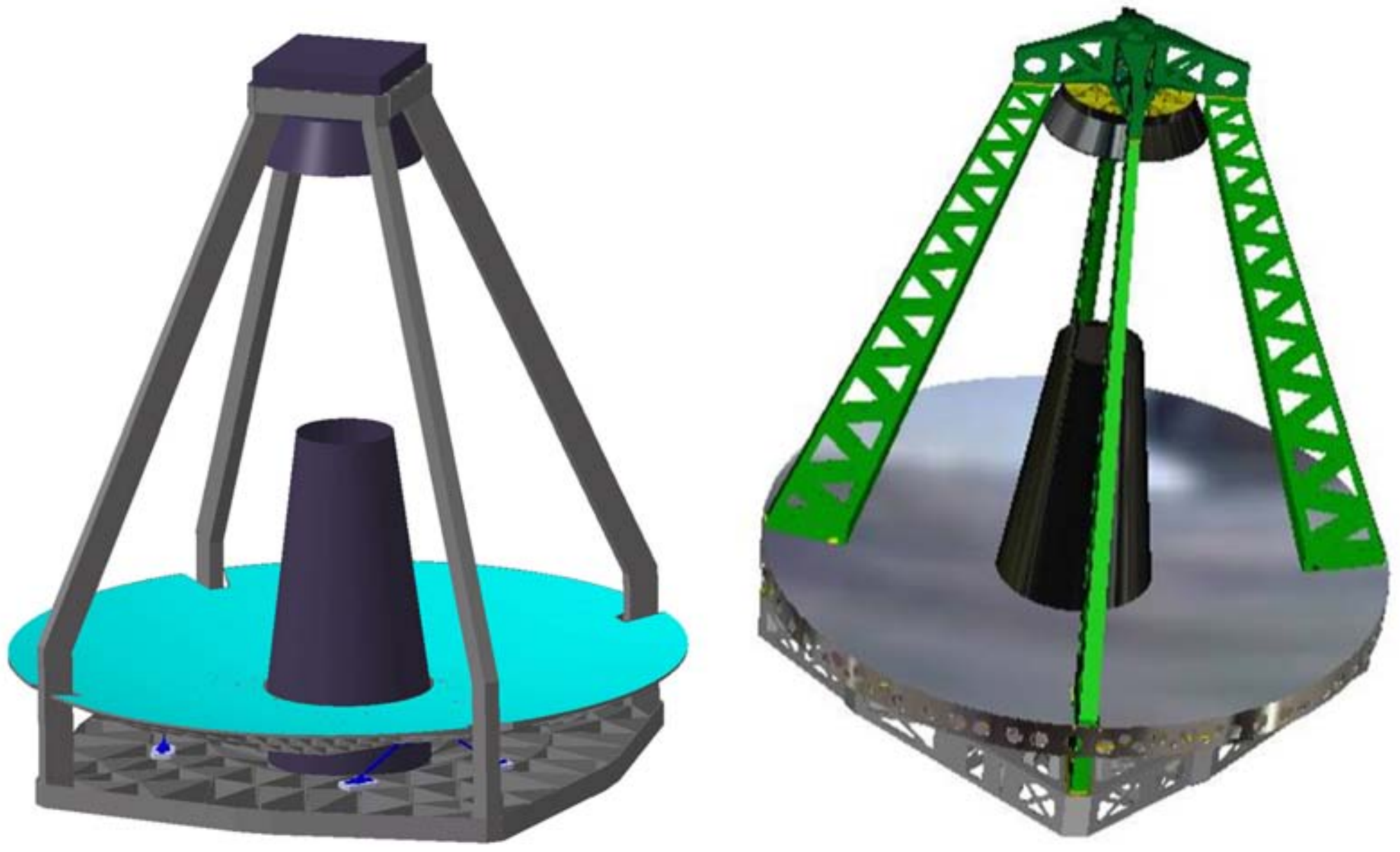

*Figure 3.1: Overview of the STA design from EADS Astrium (left) and Thales Alenia Space (right).*

### 3.1.1 STA requirements

The requirements applicable to the STA assessment study are described in the SPICA Telescope Requirements Document (2009) and are divided in different categories: a) general requirements (including a prescription for a R-C optical design); b) functional requirements; c) performance requirements; d) environmental requirements; e) interface requirements; f) qualification and AIV requirements. Given the collaborative nature of the SPICA project, specific attention has been paid to the interface requirements deriving from the JAXA provided elements of the spacecraft. Such requirements are based on the SPICA Telescope Interface Control Specifications (2009) delivered by JAXA.

Specific performance requirements (related to obscuration, transmission and image quality) are dictated by the inclusion of a coronagraph instrument in the SPICA payload complement. Such requirements have been applied to the telescope design activities, but, given their potential impact on the telescope development, industry has been also requested to identify the simplifications in case of a de-scoping (i.e. in case the coronagraph requirements were not considered).

A summary of the main telescope requirements is provided in Table 3.1. The additional requirements derived from the coronagraph instrument are listed in Table 3.2.

### 3.1.2 Description of interfaces and STA accommodation.

The SPICA Telescope Assembly (STA) includes the following parts under ESA responsibility:

- SPICA Telescope, including optical elements (primary and secondary mirror), M2 support structure, M2 refocusing mechanism with drive electronics, internal telescope baffles, adjustment shims, integration alignment devices.

- Telescope Optical Bench, as required to provide adequate mounting and mechanical support to the telescope elements and interface to the S/C structure and the Instrument Optical Bench (IOB).

*Table 3.1: Summary of main STA requirements.*

| | |
|---|---|
| Optical design: | Ritchey-Chrétien, axi-symmetric, EFL=20 m, M1 dia = 3.5 m |
| Operating temperature: | Nominal < 6 K, operating range 4.5 – 10 K |
| Wavelength range: | 5 to 210 $\mu$m (goal 3.5 to 210 $\mu$m) |
| Collecting aperture: | 3.5 m diameter (maximum allowed by HII-B fairing) |
| Total obscuration: | < 12.5 % on axis (goal < 10%) |
| Total transmission: | > 90% at 5 $\mu$m, > 95% at 15 $\mu$m, > 99% at 110 $\mu$m |
| Image quality: | Diffraction limited at 5 $\mu$m over a FOV of 5 arcmin radius |
| Field Of View: | > 12 arcmin (unvignetted) |
| Launch environment: | JAXA launcher H-IIB, warm launch. |
| STA mass allocation: | < 700 kg (including margins, excluding mass of IOB) |
| Stray-light rejection: | Total background from out-of-field stray sources (artificial and natural) < 20% of in-field background (zodiacal light and self emission from telescope). |
| Lifetime: | > 5 yr in orbit (and > 5 yr on ground) |
| Functional requirements: | • M2 refocusing at nominal operating temperature<br>• In-flight decontamination capability |

*Table 3.2: Summary of MIR coronagraph specific requirements.*

| | |
|---|---|
| Telescope transmission: | Spatially uniform < 1.5% rms over the range 3.5 to 27 $\mu$m. |
| Telescope pupil obscuration: | Preferred shape: 4 M2 spider legs |
| Image quality ($\lambda_0 = 5\ \mu$m): | • 0 to 50 cycle/D range: WFE < 2 $\mu$m (Peak-to-peak)<br>• 3.3 – 12 cycle/D range: WFE < $\lambda_0$/45 rms (goal: < $\lambda_0$/133 rms)<br>• 12 – 50 cycle/D range: WFE < $\lambda_0$/71 rms (goal: < $\lambda_0$/133 rms) |

- Thermal control hardware (MLI, electrical heaters, harness, temperature sensors) as required to monitor and control the telescope temperature, including decontamination procedures.
- Thermal interfaces to the S/C cooling chain (anchoring for thermal straps).
- Required transport, handling and ground support equipment.

The Instrument Optical Bench (IOB) and its interface mountings as well as the external baffle are not ESA provided. Model designs of both these elements have been used by industry to define in detail all STA elements and critical interfaces. In particular, the IOB (kept at 4.5 K via dedicated thermal strapping and supporting the cold unit of all instruments) has been modelled to have the following characteristics:

- Total mass of IOB and FPI's of 200 kg. The centre of mass of the FPI is assumed to be located on the optical axis, 250 mm below the STA interface plane.
- IOB of circular shape, 2m diameter, interfaced to TOB via 3 isostatic mounts.
- Global (IOB + FPI) stiffness requirements: $1^{st}$ axial (Z axis) natural frequency > 100 Hz, $1^{st}$ lateral (X,Y axis) natural frequency > 60 Hz.

The SPICA Telescope Assembly (STA) is a key element of the Payload Module (PLM) and is accommodated inside the telescope baffle (maintained at $\sim$11 K and also supported by the PLM truss) and shielded by the telescope tube, three thermal shields and the main Sun shield (see Figure 3.2). The STA is mounted onto the (PLM) support truss via 8 interface points on a diameter of 2700 mm (see Figure 3.2). A thermal strap connects the TOB to the 5 K mechanical cooler. Electrical heaters, thermistors and the M2 focusing mechanism are connected to their electronic units via dedicated harness. The reference frame of the SPICA Telescope Assembly is defined as follows:

- the origin ($O_{STA}$) is located on the telescope to S/C interface plane
- the Z-axis ($Z_{STA}$) coincides with the telescope bore sight
- the Y-axis ($Y_{STA}$) points in the direction of the Sun
- the X-axis ($X_{STA}$) completes the right-handed coordinate system

### 3.1.3 Optical design

The optical design of the SPICA telescope did not vary significantly during the study activities, which confirmed the axi-symmetric Ritchey-Chrétien approach, with the aperture stop located at the secondary mirror. During Phase 1 sensitivity analysis were performed on the main optical design parameters, coming to the following conclusions:

- Effective Focal Length optimised at 20 m to reduce size/mass of M2.
- Aperture stop confirmed on M2 to optimise the stray-light rejection.
- Back Focal Length defined at 828 mm (although adjustments still possible).
- M2 support structure based on four legs (driven by PSF characteristics).

During the design activities, effort concentrated on the definition of the polishing and coating specifications, on a preliminary stray-light analysis (with associated baffle design) and on the demonstration of the WFE feasibility (WFE budget). Additional sensitivity analyses have been performed, in particular on M2 de-center, tilt and de-focus. The main results are summarised in Table 3.3.

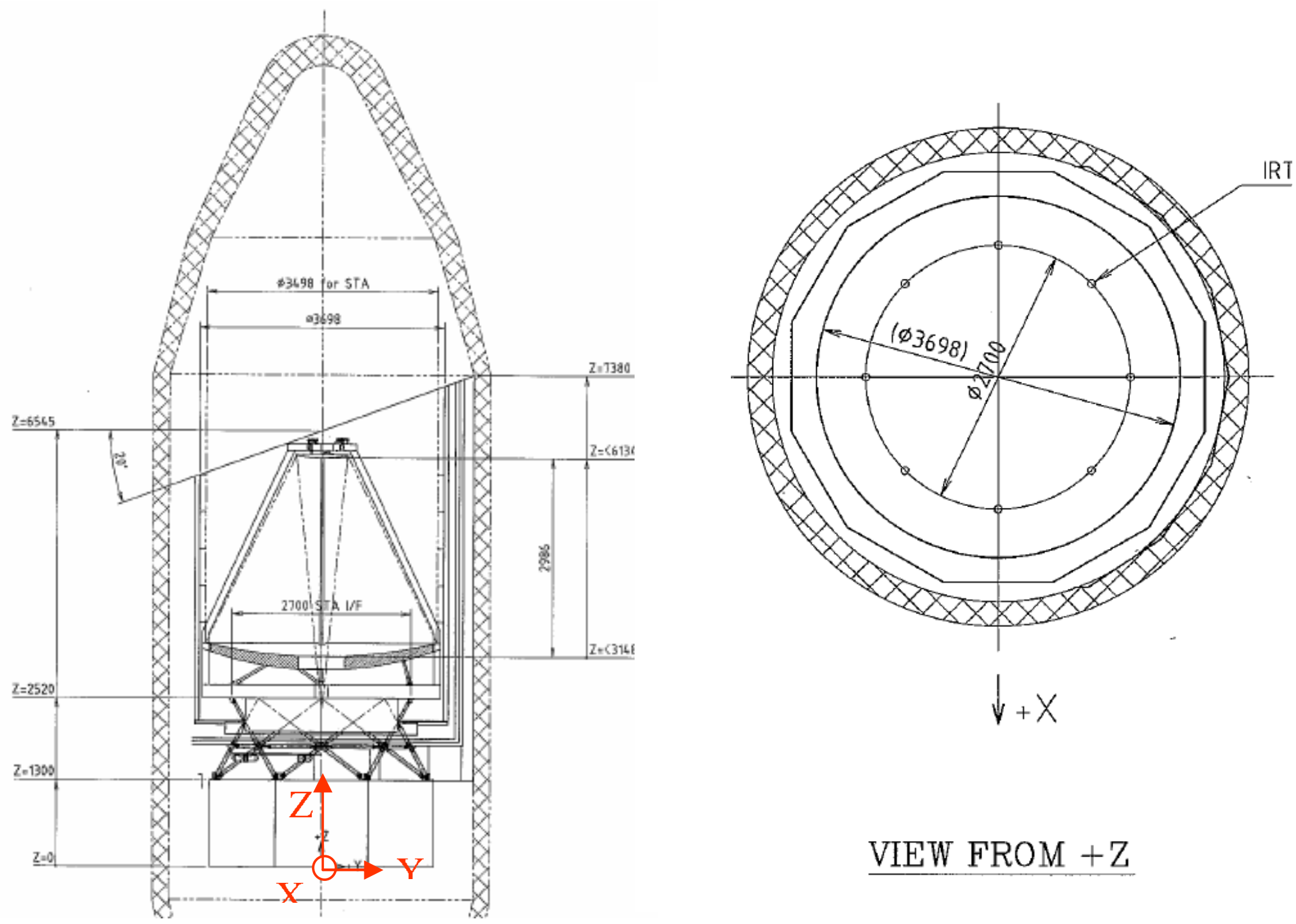

*Figure 3.2: **Right panel**: STA accommodation in the SPICA spacecraft. **Left panel**: STA-PLM interfaces.*

*Table 3.3: Summary of optical design specifications.*

| | |
|---|---|
| Configuration: | Ritchey-Chrétien, aperture stop on M2 |
| M1 diameter | Mechanical = 3500 mm, useful optical ~ 3470 mm |
| M2 diameter: | Mechanical ~ 660 mm, useful optical ~ 640 mm |
| M1 – M2 distance: | - 2986 mm |
| Focal Length, f-number: | 20000 mm, f = 6.015 |
| Back Focal Length: | 828 mm |
| Nominal design WFE: | ~ 350 nm rms at 5 $\mu$m, FOV radius of 5 arcmin |
| Total transmission: | > 0.90 at 3.5 $\mu$m, > 0.92 at 5 $\mu$m, > 0.95 between 15 and 210 $\mu$m |

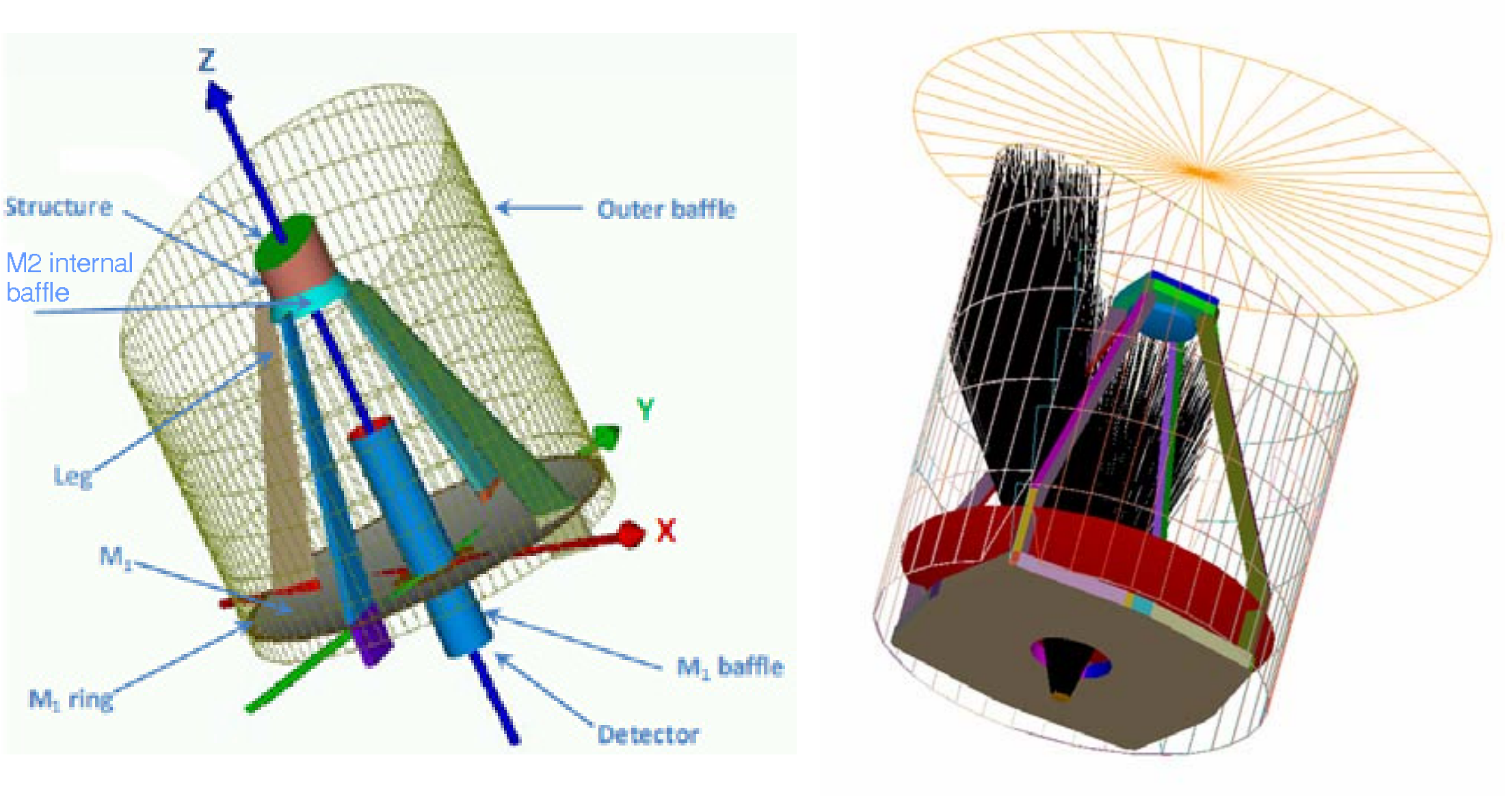

*Figure 3.3: Telescope internal baffling (Left: Thales Alenia. Right: EADS Astrium)*

The stray-light analysis has been conducted by both contractors at a level commensurate with phase-A, with the main objective of proving overall compatibility with the requirements. The approach has been based on a preliminary model of the external baffle provided by ESA (on JAXA behalf) and on the design of the inner M1 and M2 baffles by each contractor.

The preliminary stray-light activities have demonstrated the capability to meet the corresponding requirements by adopting the following measures: a) further optimisation of the external baffle thermo-optical properties (paint selection); b) leg profile and M2 support design minimising reflections; c) adoption of internal and external vanes in the M1 internal baffle; d) selection of paint for the internal baffles; e) adoption of cold stop for the instruments operating at $\lambda > 70\ \mu$m (i.e., SAFARI). In the range between 70 and 210 $\mu$m (dominated by thermal self-emission from the external baffle at $\sim 11$ K), compliance with the stray-light requirements depends on particulate contamination levels after launch, to be defined in the next project phases.

### 3.1.4 Mechanical design

The mechanical design of the STA design is based on a dedicated Telescope Optical Bench supporting M1 as well as the M2 support structure. This approach avoids transferring directly to M1 any loads due to M2, minimising distortions. All structural and optical elements are to be built in ceramic material (either HB-Cesic or SiC100), which ensures small and highly uniform Coefficients of Thermal Expansion (CTE), combined with high strength and low mass. Titanium alloy is used for the mirror fixation devices (bipods).

Given the different characteristics of SiC100 and HB-Cesic, a different manufacturing approach is baselined respectively for Astrium (M1 built by brazing of 12 segments and TOB in 8 segments) and Thales (M1 monolithic, TOB in two halves). In both cases M2 (of smaller size) is built in a single piece. Junctions between separate ceramic elements (e.g. M2 struts and TOB) are handled via Invar fittings. The JAXA provided IOB would be connected to the TOB via dedicated Ti bipods.

The mechanical design is driven by the quasi-static loads (particularly demanding at the M2 location) and the stiffness requirements provided by JAXA (see Tables 3.4 and 3.5). The quasi-static loads are subject to confirmation in the following phases, based on dedicated coupled load analysis (including spacecraft elements and STA behaviour). Detailed Finite Element Models have been constructed by both companies.

The mechanical analyses covered four main areas: a) modal analysis to demonstrate compliance with the stiffness requirements; b) thermo-elastic analysis to demonstrate proper behaviour during the cool-down from ambient to 5 K without affecting the telescope performance specifications; c) analysis of gravity effects; d)

*Table 3.4: Global STA (+ IOB) stiffness requirements.*

| | |
|---|---|
| $1^{st}$ natural freq. along X and Y axis: | > 30 Hz |
| $1^{st}$ natural freq. along Z axis: | > 60 Hz |
| $1^{st}$ natural freq. around Z axis (torsion): | > 30 Hz |

*Table 3.5: Design limits and quasi static loads.*

| | |
|---|---|
| M1 and M1 supporting structure (X, Y, Z non simultaneously): | 12.5 g |
| M2 and M2 supporting structure (X, Y, Z non simultaneously): | > 20 g |
| $1^{st}$ natural freq. around Z axis (torsion): | > 30 Hz |

analysis of quilting effects (induced by polishing).

The modal analysis has been conducted in different configurations: 1) Isolated STA and IOB system on rigid interfaces (representing the nominal test configuration in Europe); 2) STA and IOB on PLM support truss (closer to the flight configuration). Such analysis have assumed that the IOB (global behaviour including the FPI's) will have a first longitudinal mode (Z axis) at a frequency > 100 Hz and a first lateral mode (X, Y axis) at a frequency > 60 Hz. Both EADS and TAS design show compliance with the requirements of configuration 1) and 2). In case 2), the analysis of the truss behaviour has been based on a JAXA provided Finite Element Model.

The thermo-elastic analysis has been conducted combined the detailed telescope Finite Element Model with a corresponding temperature map provided by the thermal mathematical model and with materials data (CTE as a function of temperature). The main objective of this analysis was to demonstrate the capability of the truss supporting structure to absorb the differential thermal contractions during the cool-down (originated by the presence of different materials). Such a demonstration is of specific importance given the nature of the STA to PLM interface (based on 8 points). The main results of this analysis are:

- The supporting truss modelled by JAXA provides adequate radial "breathing" to absorb the differential contractions between STA and PLM elements, with a behaviour equivalent to isostatic mounting conditions.
- Optimising the TOB to IOB interfaces, it is possible to absorb any residual differential contraction without affecting the optical performance (even in case of large CTE differences as in the case of an Al based IOB).
- The telescope wave-front error, with proper shimming and corrective action from the M2 mechanism, remains within the allocated budget.

Industry performed the analysis of the gravity effects on both M1 and M2 WFE, thus being able to predict their behaviour during the polishing (with optical axis in vertical position) and telescope testing (with optical axis in horizontal position). The mechanical design of the mirrors and of the supporting structure has then been optimised to maintain the WFE within acceptable levels in 1g conditions (and different orientations), in agreement with the overall WFE budget allocations.

Quilting effects induced during polishing have been examined in order to establish the optimal back stiffening (primary and secondary) structure and front skin thickness (in particular for M1). Both contractors have identified designs ensuring a reduced impact on WFE (typical ~ 50 nm rms) as a result of the applied polishing pressure.

Finally the mechanical analysis has demonstrated compliance (in worst loading conditions) with the maximum stress levels allowed by each STA material, including required safety margins. This analysis included thermo-elastic effects induced by large temperature gradients (during cool down). The mechanical design showed compliance with the yield and ultimate load levels, including the required factors of safety for each different material used in the STA. Specific attention has been paid to the case of ceramic parts and to the definition of the related design margins, including latest lessons learnt from similar programs.

### 3.1.5 Thermal design

A first thermal analysis of the STA has been performed at ESA on the basis of a preliminary Thermal Mathematical Model and JAXA input. This model focused on determining the temperature distribution on the telescope given the boundary conditions imposed by the SPICA spacecraft (including thermal shields and mechanical coolers). During the industrial activities this model has been further refined and adapted to the specific STA designs by each contractor, including numerous sensitivity analyses. The analysis has assumed direct connection of the TOB to the 4.5 K cooler via a dedicated strap and thermal decoupling between TOB and IOB. The main results obtained with the Thermal Mathematical Model (also used to derive the T maps for the thermo-elastic analysis) are the following:

- In equilibrium conditions (after cool down and with nominal cooling power at 4.5 K and realistic interfaces) all telescope elements have a temperature below 7 K. In particular the optical surfaces have a temperature $< 6$ K.
- Temperature uniformity on the optical surfaces is very high (typically $< 0.2$ K).
- Adequate cooling of M2 can be achieved without complex thermal strapping, which could be limited to a few interface areas.
- The thermal behaviour of the STA is dominated by the external boundary conditions (SPICA spacecraft), in particular by the thermal shields, the cooling power of the 4.5 K cooler and heat conduction through the supporting truss.
- At equilibrium the heat load on the 4.5 K cooler is below 20 mW, compatible with the estimated performance of the JAXA cooler.
- The transient analysis showed that cool down of the PLM after launch takes approximately 160 days, and that the STA does not lag behind the rest of the system, while equilibrium conditions after operating the M2 mechanism are recovered in about 1-2 days.
- If required, decontamination of the telescope (up to 315 K) can be performed with a realistic heater power budget, compatible with the on-board resources.
- The heat load on the 4.5 K stage is sensitive to variations in the boundary conditions (e.g., temperature of heat shields and telescope shell).

Decontamination could be performed at beginning of life (out-gassing phase) and/or during the flight. Heaters would be implemented on the TOB. A summary of the thermal analysis results is provided in Table 3.6

### 3.1.6 Resource budgets

One of the main objectives of the assessment study was to consolidate the resource and performance budgets associated to the telescope design. The following budgets have been analysed in detail: a) mass budget; b) heat load budget; c) Wave Front Error. The results (all including conservative assumptions and maturity margins) are compliant with JAXA allocations and scientific requirements and summarised in Tables 3.7, 3.8 and 3.9.

*Table 3.6: Thermal analysis results (equilibrium conditions).*

| **Industry design** | **EADS Astrium** | **Thales Alenia Space** |
|---|---|---|
| Nominal cooler temperature: | 5 ± 0.5 K | |
| Telescope shell (boundary condition): | 24 K | |
| External baffle temperature: | 10.5 K | 11.3 K |
| M1 mirror temperature: | 5.3 K | 5.4 K |
| M2 mirror temperature: | 5.5 K | 6.0 K |
| M2 support structure temperature: | 5.4 K | 5.9 K |
| Telescope Optical Bench temperature: | 5.3 K | 5.7 K |
| Internal M1 baffle temperature: | 6.7 K | 6.9 K |
| Internal M2 baffle temperature: | 5.8 K | 6.0 K |

*Table 3.7: STA mass budget (including 20% maturity margins on all units)*

| **Industry design** | **EADS-Astrium [kg]** | **Thales Alenia [kg]** |
|---|---|---|
| M1 mirror and bipods: | 306 | 313 |
| Telescope Optical Bench: | 190 | 210 |
| M2 support structure: | 109 | 100 |
| M2 mirror and bipods: | 11.4 | 25.6 |
| M2 mechanism: | 12.0 | 7.0 |
| Refocusing mechanism electronics: | 3.0 | 4.8 |
| Thermal hardware: | 20.8 | 17.2 |
| Internal baffles (M1+M2): | 20.0 | 11.0 |
| Miscellaneous (screws, bolts, etc.): | 8.0 | 9.6 |
| **TOTAL (including maturity margins)**: | **680** | **698** |

### 3.1.7 Technology readiness

The technology readiness of all telescope elements has been reviewed during the assessment study; any remaining development risk areas have been identified and corresponding risk mitigation actions have been baselined.

Given the recent Herschel experience and other scientific and remote sensing space projects, Europe is very well placed to develop the SPICA telescope. The progress achieved in the manufacturing of large size ceramic mirrors is a main pre-requisite for the STA development, together with the experience acquired in the AIT of cryogenic telescopes and in the polishing of large mirrors for astronomical telescopes. No major technology developments are required for the SPICA telescope; the remaining critical areas and the corresponding risk mitigation actions are summarised in Table 3.10. The planned technology consolidation activities are expected to achieve TRL > 5 before the final M class selection (end 2011), thus fully meeting the Cosmic Vision 2015-25 needs.

*Table 3.8: STA heat load budget on 5 K cooler (telescope shell at 24 K).*

| Industry design | EADS-Astrium [mW] | Thales Alenia [mW] |
|---|---|---|
| Overall radiative load (from baffle): | 5.4 | 1.5 |
| STA interface (from truss): | 12.8 | 18 |
| **TOTAL heat load (5 K stage):** | **18.2** | **19.5** |

*Table 3.9: STA WFE RMS budget (including integration, polishing and cool down)*

| Industry design | Main contributor | Requirement [nm] | EADS Astrium [nm] | Thales Alenia [nm] |
|---|---|---|---|---|
| **WFE rms at $\lambda_0 = 5\ \mu$m over 5′ radius** | M1 cool down<br>Design aberration<br>M1 polishing | 350 | 342 | 258 |
| **WFE rms at $\lambda_0 = 30\ \mu$m over 10′ radius** | Design aberration<br>M1 cool down<br>M1 polishing | 2100 | 670 | 800 |
| **WFE rms at $\lambda_0 = 5\ \mu$m 3.3 to 12 cycle/D** | M1 polishing | 111 | 51 | 89 |
| **WFE rms at $\lambda_0 = 5\ \mu$m 12 to 50 cycle/D** | M1 polishing | 70 | 69 | 42 |
| **WFE PtV at $\lambda_0 = 5\ \mu$m 0 to 50 cycle/D** | M1 cool down | 2000 | 1768 | 1350 |

### 3.1.8 Assembly Integration & Verification plan

As part of the assessment study activities industry has produced a complete telescope AIV plan. The telescope development plan is based on the following elements:

- Proto-flight approach, with a Structural Thermal Model and spare units.
- Mechanism and drive electronics qualified at unit level prior integration.
- Risk mitigation via 2 main technology consolidation activities: 1) Light-weight ceramic mirror breadboard; 2) M2 cryogenic focusing mechanism.
- Additional risk mitigations included in the Definition Phase contract (e.g., materials characterisation measurements, metrology related aspects, etc.).
- Cryogenic testing limited to about 50 K (TBC), without major changes to test facilities existing in Europe. Final testing at 10 K in Japan.

The proposed model philosophy is the optimal solution to balance risk mitigation and cost / schedule constraints. The Structural Thermal Model would allow refining AIV issues, to address mechanical qualification aspects and to perform mechanical and thermal system level testing at JAXA. Static proof testing of all structural parts is also considered.

*Table 3.10: STA critical areas and identified risk mitigation actions.*

| **Critical area:** | **Technical driver:** | **Risk mitigation:** |
|---|---|---|
| M2 focusing mechanism | Qualification of linear actuator operating at cryogenic temperature (5 K). | Mechanism BB developed during Definition Phase. |
| Light-weight, large size ceramic mirrors. | Fabrication of 3.5 m diameter. Optimisation of polishing process. Gravity compensation. | Mirror BB and polishing tests during Definition Phase |
| Material properties at cryo temperature. | Detailed knowledge of CTE, Young Modulus, thermal conductivity, emissivity and BRDF properties of selected materials. | Perform measurements on representative samples during the Definition Phase |

The polishing of the primary mirror is, both from performance and schedule aspects, the single most critical activity of the programme. M1 will be designed to achieve a well controlled level of gravity induced distortion and its polishing will be performed in a dedicated facility, equipped with gravity compensation devices (vertical optical axis) and metrology equipment. The 0-g figure of M1 will be measured by moving the M1 optical axis in horizontal configuration and measuring the WFE at different orientations of the mirror around the optical axis.

Matching testing between M1 and M2 in (horizontal axis) is also foreseen, via dedicated Mechanical Ground Support Equipment and stitching of interferometric measurements performed on the full aperture of M1, via a rotating auto-collimating mirror. The same technique would also be used during the final telescope integration and performance verification, with M2 on its supporting structure and M1 on the TOB, including gravity compensation devices.

The test facilities existing in Europe are well suited for the characterisation of the telescope from ambient down to T ~ 50 K. Modest changes (adaptations to the STA size, inclusion of adequate metrology equipment and thermal shrouds) are required to perform the integrated STA testing (both vertical and horizontal configurations are possible). Preliminary analyses indicated that the extension of testing in Europe to lower temperatures (10 to 20 K) may be possible, subject to a more detailed analysis in the next project phases. Thermal vacuum testing in Europe will include a full characterisation of the optical performance of the telescope close to operating conditions before final shipment to Japan. Additional STA testing at 10 K is presently planned in Japan, before final integration on the SPICA spacecraft.

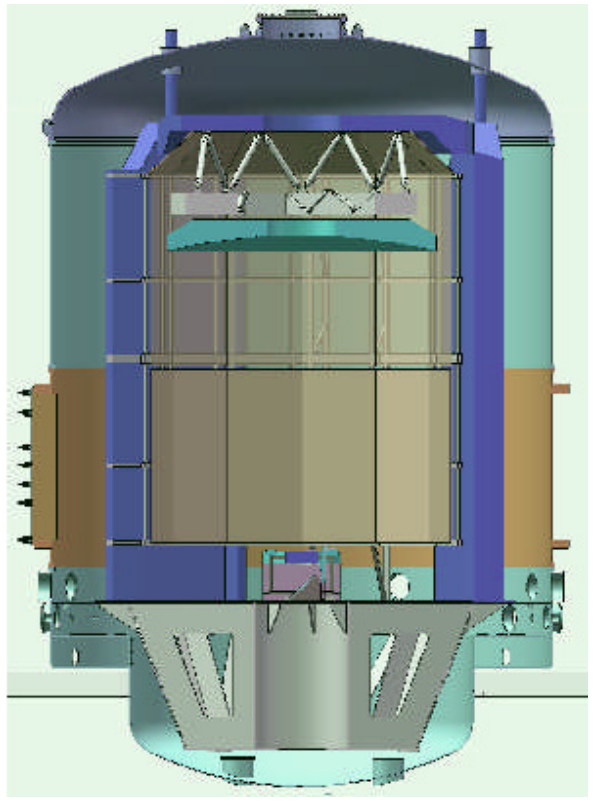

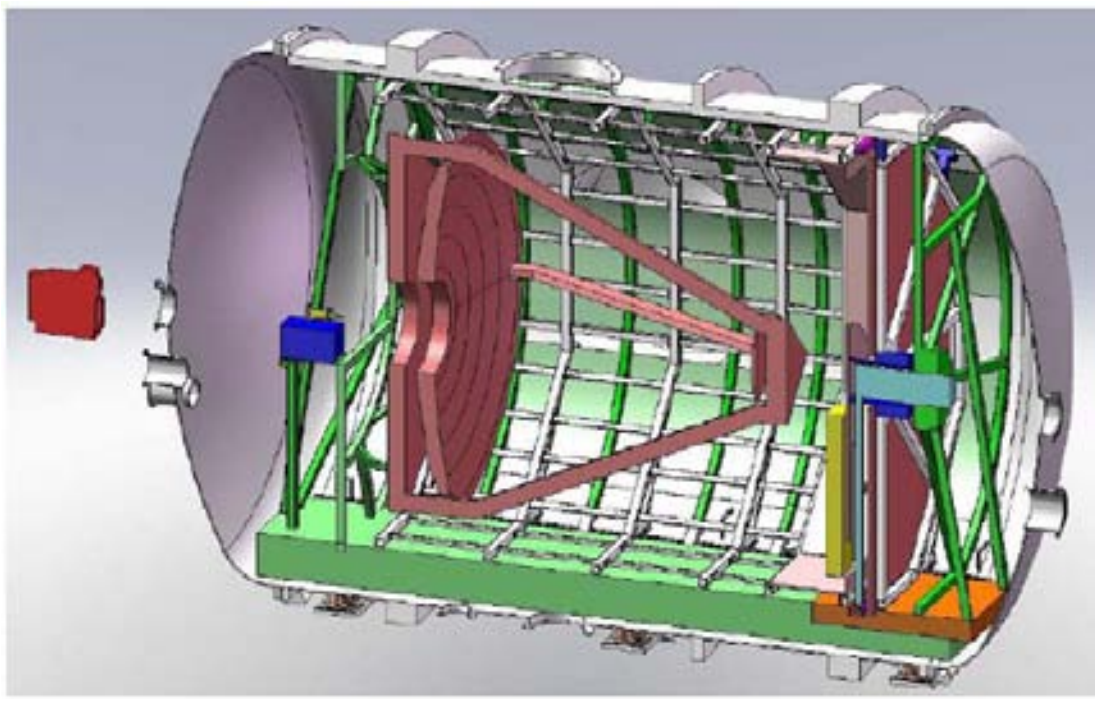

*Figure 3.4: **Left panel**: STA cryo testing in vertical configuration. **Right panel**: STA cryo testing in horizontal configuration.*

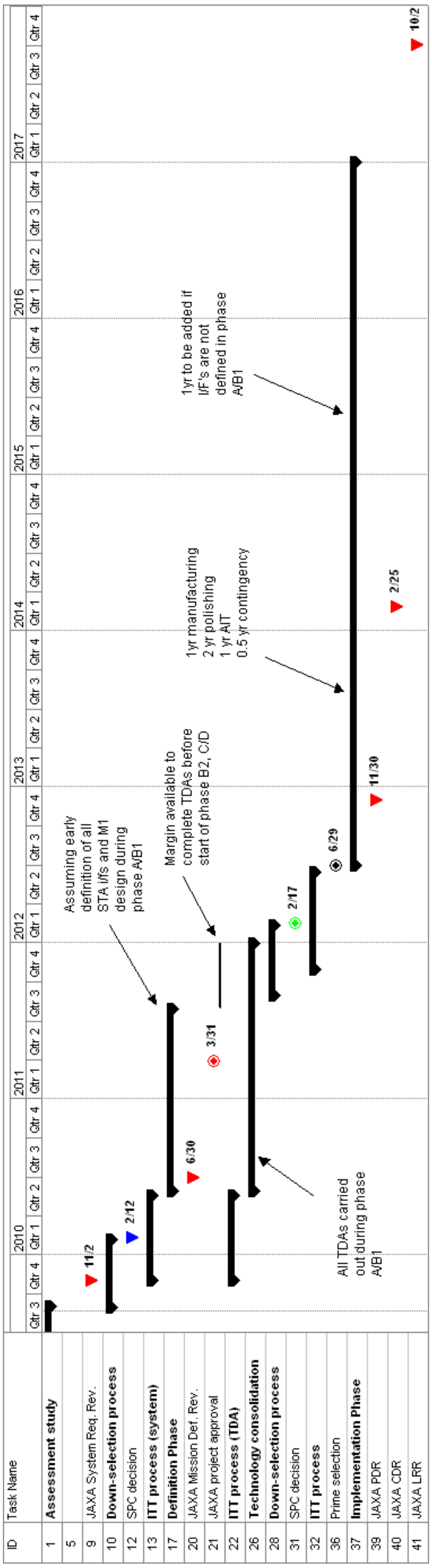

*Figure 3.5: SPICA Telescope Assembly – reference schedule (including M class down-selection milestones).*

### 3.1.9 Development plan

The SPICA development plan takes into consideration the boundary conditions given by the project approval process (at JAXA and at ESA) as well as all relevant technical issues. Key events to be accounted for (in case of down-selection) are: a) earliest start date of the Definition Phase at ESA (Q2/2010); b) JAXA final project approval in Q1/2011; c) earliest start date of the Implementation Phase at ESA (Q2/2012). Based on the assessment study, the tasks driving the telescope schedule are:

- Detailed design of all telescope elements (following freezing of requirements).
- Manufacturing of M1 blank (including related procurement actions).
- Polishing of primary mirror (estimated to take no less than 24 months; an additional 6 months would be required when including the coronagraph needs).
- Coating of primary mirror and final WFE characterisation.
- Telescope integration and test activities (at ambient and cryo temperature).

The reference development schedule of the SPICA Telescope is reported in Figure 3.5. The project plan takes into account all boundary conditions and assumes an early freezing of all design requirements during the ESA Definition Phase (A/B1), immediately after JAXA SDR (Q3/10), thus preparing key manufacturing drawings before the start of Phase B2-C/D. All technology consolidation and risk mitigation activities are also to be completed before starting the Implementation Phase. In this scenario, the delivery of the telescope to JAXA could take place by Q4/2016 to Q1/2017, assuming a minimum contingency of 6 months. Delays in the requirements definition would require and additional 8-12 months, while inclusion of the Coronagraph requirements would extend polishing an additional 6 months.

The SPICA reference schedule takes into account both the JAXA and ESA mission approval process as well as technical schedule drivers. The single most critical activity is the manufacturing ($\sim$ 1 yr) and polishing of the primary mirror ($\sim$ 2 yr). An early definition of the interfaces and optical requirements (to be agreed with JAXA) would allow anticipating the detailed design of M1, saving about 1 yr. An additional 6 months would be required for polishing in case the coronagraph requirements were enforced.

## 3.2 Focal Plane Instruments

### 3.2.1 Overview

The following instruments are candidates for the SPICA focal plane: MIRACLE (Mid-InfRAred Camera w/o Lens), MIRMES (Mid-IR Medium-resolution Echelle Spectrometer), MIRHES (Mid-IR High-resolution Echelle Spectrometer), SCI (SPICA Coronagraph Instrument), SAFARI (SPICA Far infrared Instrument), FPC (Focal Plane finding Camera; FPC-G is a guider camera for attitude control), BLISS (Background-Limited Infrared-Sub-millimetre Spectrograph).

Only one instrument at a time shall be in operation for astronomical observations. This excludes the FPC-G, which will be in operation for most of the observations. The Focal Plane Instruments (FPIs) onboard SPICA will be attached to the STA via the Instrument Optical Bench, which is thermally lifted at $\sim$4.5 K by a J-T cooler. A total mass of 150 kg is allocated for the FPIs. For SAFARI the allocation of 50 kg (cold units, including 20% margin) has been agreed. Cooling power constraints are tight: 15 mW total at the 4.5 K stage, and 5 mW at 1.7 K, lifted by the J-T coolers. At the 4.5 K stage, the routinely operated FPC-G plus the parasitic loads amount to approximately 2 mW. Since the resource allocations are rather stringent to allow the accommodation of all instrument candidates, an instrument review/selection process will be followed before the SPICA System Design Review (Q4/2010). In this process, BLISS is considered as an “optional” instrument.

Figure 3.6 shows a planned field-of-view (FOV) configuration (projected on the sky). Within the unvignetted field of STA (30 arcmin in diameter), all the FOVs, including two redundant FOVs of the FPC, shall be placed.

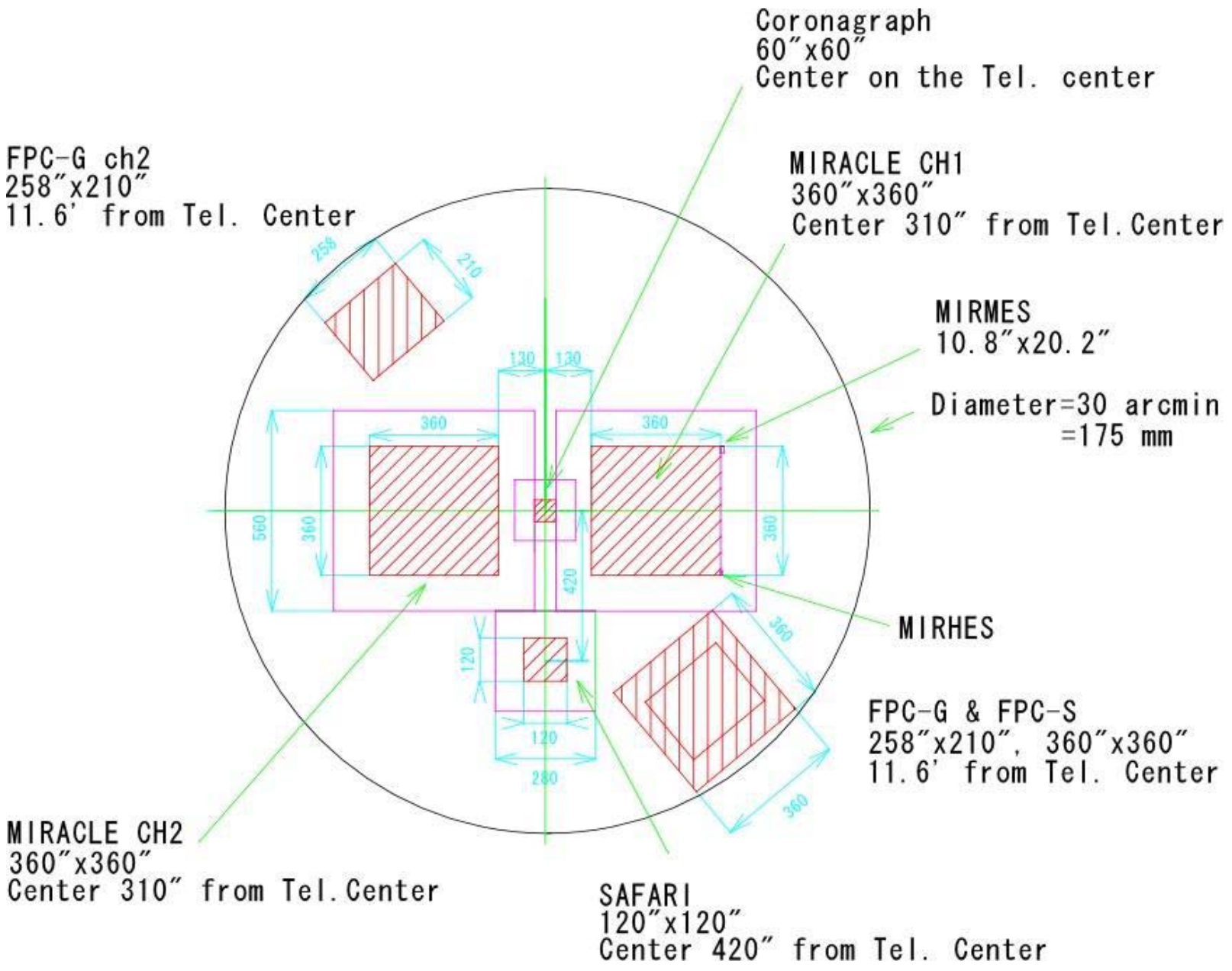

*Figure 3.6: Planned configuration of the field-of-views of the FPIs projected on the sky. Those of MIRMES, MIRHES, and BLISS are not shown.*

### 3.2.2 Far Infrared Instrument: SAFARI

The SpicA FAR infrared Instrument (SAFARI) is an imaging Fourier Transform Spectrometer (iFTS) designed to give continuous wavelength coverage in both photometric and spectroscopic modes from around 34 to 210 $\mu$m. The scientific performance requirements of SAFARI are outlined in section 2.3 “Far infrared instrument scientific requirements”, together with the rationale for choosing an iFTS concept for the overall instrument configuration. Figure 3.7 shows the current conceptual design of the SAFARI Focal Plane Instrument. It reflects much of the design optimisation carried out by the consortium during the Cosmic Vision Assessment study. For more information about the status of the instrument design, refer to the SAFARI Phase-A1 Study Report (2009).

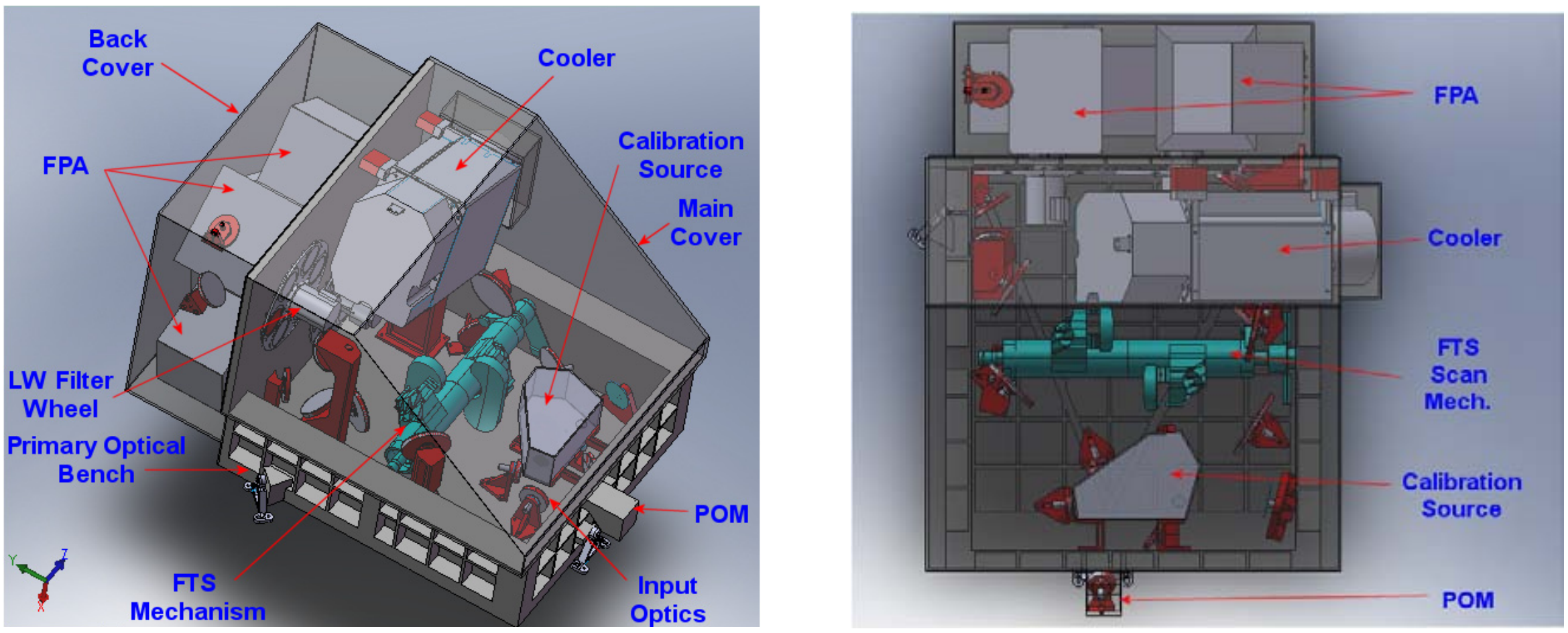

*Figure 3.7: SAFARI focal plane conceptual design at the conclusion of the Phase-A1 study.*

### Key design drivers

***Detector performance:*** The scientific performance of SAFARI is critically dependant on the sensitivity of the detector system. Given the importance of detector sensitivity to the ultimate impact of SAFARI science, the consortium has prioritised detector technology development activities so that the competing requirements of technological maturity / development risk vs. ultimate scientific performance can be optimised. There are currently four competing candidate detector technologies which are undergoing targeted development to meet the SAFARI requirements; TES, KID, Silicon bolometers and Ga:Ge Photoconductors. The four technologies are all credible candidates for SAFARI although they differ considerably in detail with respect to (1) likely in-flight sensitivity, (2) system impact on the overall instrument design and the attendant demands on the spacecraft resources and (3) technology readiness and development risk. JAXA have established a mission level requirement that the FPI detector technology is baselined prior to start of the SDR (Q4/2010) and therefore the consortium has planned to converge on a technology baseline in Q2/3 2010. The selection of the detector technology baseline will be carried out on the basis of the parameters of scientific performance, spacecraft/instrument interface compatibility, development risk and national programmatic support.

In the period leading up to the eventual selection of the detector technology baseline, the consortium has decided to adopt the TES detectors as the reference case. This is due partly to the fact that a prototype detector with a sensitivity meeting the SAFARI minimum requirement has successfully been optically characterised and behaves largely as predicted. It is also due to the fact that it is likely that the TES detectors will place the most demanding requirements on the cryogenic spacecraft resources, and can therefore act as a "design stressing" reference case.

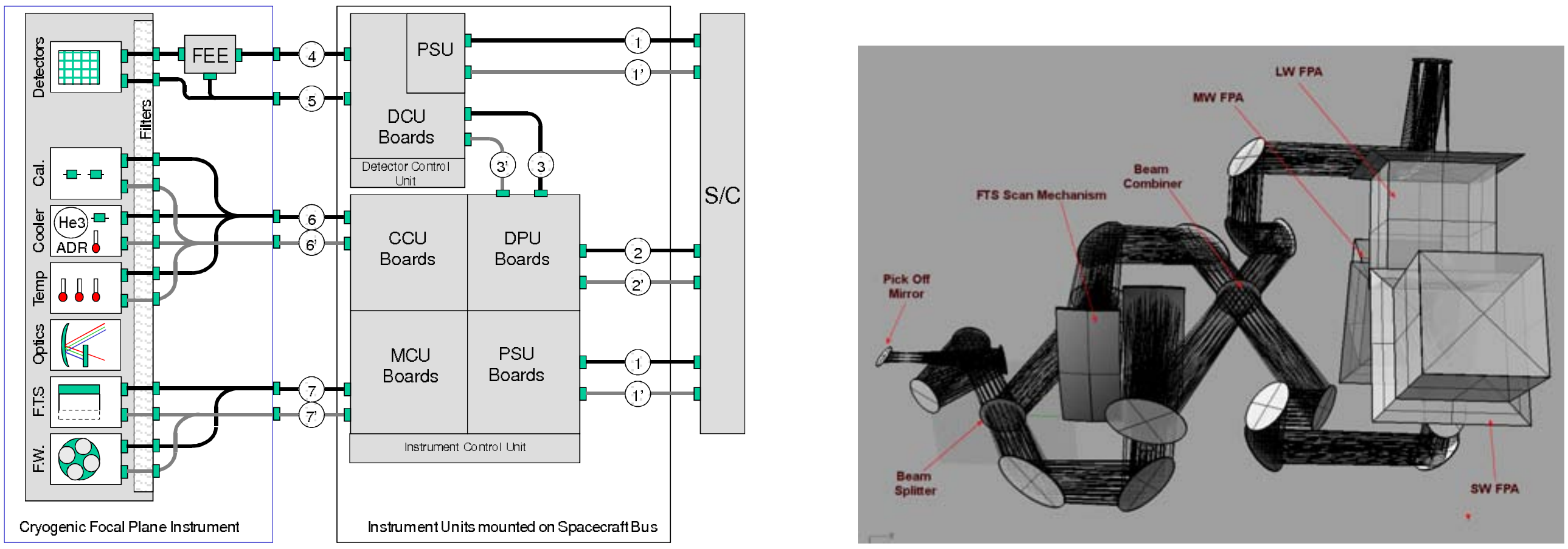

*Figure 3.8: **Left panel**: SAFARI flight hardware block diagram. **Right panel**: Optical elements of SAFARI.*

***Mass budget***: Due to the non-linear system level impact of added mass on the SPICA IOB to the thermal and structural design of the overall STA, the mass budget allocation of 200 kg for the IOB and FPI has very little flexibility. The JAXA agreed allocation of 50 kg (which includes 20% margin) for the SAFARI focal plane instrument is a significant instrument design driver. The instrument design produced during the ESTEC CDF study (CDF Study Report, SPICA-SAFARI 2008) was non-compliant on this basis. Recent evolution of the instrument design has reduced the current best estimate mass to 38 kg by (1) removing a mechanism used to stabilise the astronomical image on the instrument focal plane, (2) making the optical layout more compact and mass efficient and (3) by incorporating a Darwin Optical Delay Line heritage scan mechanism design into the instrument, which is lighter than the initial baseline. Although the current best estimate mass is compliant, strict mass growth and requirements creep discipline will have to be enforced during the definition phase to ensure that the instrument does not exceed the allocated mass budget. The main path to further optimisation of instrument mass is to reduce the volume of the Focal Plane Array, which would allow a more compact optical layout and a lighter instrument structure.

***Cryogenic thermal budget:*** The scientific promise of the SPICA mission is made possible thanks to the availability of the JAXA J-T and Stirling cryocoolers to cool the STA and the SPICA instrument suite. Despite the availability of these coolers, the available cryogenic heat lift places severe constraints on the design of the SAFARI focal plane instrument. Of particular concern is the heat loads on the spacecraft J-T coolers during instrument cooler recycle. The cryogenic design of the instrument is currently therefore one of the most important design drivers. Moreover, due to the highly coupled behaviour of the spacecraft / instrument thermal system, proper attention must be paid to the design optimisation of the overall integrated system.

***Electromagnetic design and analysis:*** The very demanding SAFARI detector sensitivity requirements imply that EMI coupling to the detector system must be stringently controlled. The susceptibility of bolometric detectors to EMI currents coupling into the detector system and dissipating power within the bolometers is a particular susceptibility of this class of detector technology. The detectors are also susceptible to DC and low frequency magnetic fields from the instrument cooler Adiabatic Demagnetisation Refrigerator (ADR) solenoids as well as the scan mechanism and other spacecraft systems. Provision for RF and DC/low frequency magnetic shielding has been incorporated into the instrument design, but further detailed design and analysis is required in order to confirm appropriate margins.

***Stray-light control:*** The SAFARI detector sensitivity requirements dictate that the level and modulation of the stray-light on the detector system is well controlled. This has been made apparent during SAFARI detector testing where standard stray-light control methods in test cryostats were found to provide inadequate shielding in the dark detector test cavity.

***Verification:*** The planning and design of the verification activities, particularly at spacecraft level, for most of these design activities (e.g., Electro- Magnetic Compatibility, telescope stray-light, cryogenic performance, detector sensitivity) is a challenge due to the difficulties of reproducing a flight-like environment in a clean room environment.

### SAFARI system design

The main flight hardware elements of the SAFARI instrument are shown in the block diagram (see Figure 3.8, left panel). The Focal Plane Instrument is accommodated on the SPICA Instrument Optical Bench while the Detector Control Unit and the Instrument Control Unit are mounted on the spacecraft bus. The primary sources of the high level requirements driving the instrument design are the SAFARI science requirements as contained in the SPICA Mission Requirements Document (2009) and the interface requirement with the spacecraft which are being controlled and documented in the SAFARI Interface Control Specification (2009).

#### *Optical design*

The main optical elements with overall dimensions are shown in Figure 3.8 (right panel). The Pick Off Mirror is centred 7′ from the telescope field of view slightly ahead of the focal plane. At this location, SAFARI has diffraction limited imaging at 40 $\mu$m over a $2' \times 2'$ square field of view taking into account instrument and spacecraft alignment tolerances. The input optics reimages the telescope pupil for stray-light control (Lyot stop) and expands and partially collimates the beam entering the spectrometer. The beam splitter creates two symmetric optical paths through the interferometer. The retro-reflectors in the scan mechanism are cat’s eye mirrors which is a configuration found to be well suited to both the simple design of the mechanism as well as achieving a compact and well controlled optical design. There is sufficient clearance between the elements of the interferometer to achieve an optical path difference of between -4 mm and +31.5 mm which (together with the 33 mm pupil diameter) is compatible with the spectral resolution requirement of the instrument of $R = 2000$ at 100 $\mu$m. The beam combiner is the final element of the interferometer and produces the interference beams in the two output arms. The camera optics focuses the image onto the three Focal Plane Arrays with f/20 beams. The overall fringe contrast and optical throughput of the instrument are compatible with the instrument sensitivity and spectral resolution requirements. The Focal Plane Arrays have an unvignetted field of view of $2' \times 2'$ and Nyquist sample the image (F $\lambda/2$) at the central wavelength of each band. The three science bands are defined by a series of edge filters and one dichroic filter. The nominal bandwidth can be tailored with two, six-position filter wheels accommodated within the overall camera optics. The approach for stray-light control is to enclose the instrument in a light tight enclosure, use a series of edge filters to attenuate out of band radiation and prevent it reaching the detectors, block off-axis light with pupil stops, locally apply black paint to critical locations within the instrument, enclose the Focal Plane Arrays in a light-tight 1.7 K box and make an appropriate design of the detector harness to prevent radiation "tunnelling" through the dielectric insulator. Considerable quantitative analysis and design will be required to ensure that best use is made of the low-background primary telescope.

***Thermal design***

A hybrid $^3$He sorption cooler coupled to a low temperature ADR will be used to cool the TES detectors[1] to around 50 mK and provide heat lift at 300-mK to intercept parasitic heat loads and cool detector front end electronics. The cooler will be operated in recycle mode with a duty cycle efficiency of 75% (10 hour recycle / 30 hours operation). The hybrid sorption/ADR cooler architecture was adopted due the fact that it is far lighter than a multi-stage ADR cooler and is therefore compatible with the overall SAFARI mass allocation. High purity, annealed Copper straps are used to provide a high conductivity link between the cooler and the detector arrays. The detectors will be mounted on light tight enclosures on the main chassis of the instrument which will be cooled to 1.7 K. The main chassis of the instrument will be isothermal with the IOB.

There will be three primary thermal interfaces with the spacecraft:

1. a 1.7-K interface to the $^3$He J-T cooler: used in the operation of the cooler, cooling the detector boxes and intercepting the electrical dissipation from the detector readout electronics,

2. a 4.5-K interface to the $^4$He J-T cooler: used in the operation of the cooler, intercepting cryoharness parasitics and the electrical dissipation in various subsystems mounted on the 4.5-K structure, and,

3. a 12-K interface to the Stirling cooler (2ST): used to recycle the cooler and possibly intercept dissipation from detector pre-amplifier (not part of the current baseline for TES detectors).

The main point of tension in the thermal interface with the spacecraft is the magnitude and duration of the peak loads from the instrument cooler during recycle; in particular the load on the 1.7-K interface from the enthalpy of the "hot" Helium de-absorbed from the sorption cooler pump and the enthalpy of condensation in the evaporator. The results of the thermal modelling (reported in detail in SAFARI Phase-A1 Study Report 2009) shows margins of over 30% for most thermal loads during instrument operations except the 4.5-K stage (−3.5%). Also, during cooler recycle, the peak load on the 4.5-K stage shows a margin of only 9%. These issues will be addressed during the design and development programme of the instrument and system level cooling systems by providing a breadboard cooler system to JAXA to test against the engineering models of the J-T coolers. In principle the instantaneous power from the cooler can always be reduced by taking more time to recycle the cooler, at the expense of some operational efficiency.

***Mechanical design***

The mechanical interface requirements of the FPI are contained in the ICS. For the purpose of the instrument assessment study, the two key requirements used to demonstrate overall compliance were (1) the mass (50 kg with 20% margin) and (2) Eigen frequency requirements (150 Hz axial / 100 Hz lateral). The conceptual structural design of the focal plane instrument is shown in Figure 3.7. The instrument chassis is fabricated from Aluminium alloy and mounted on three isostatic bipod mounts on the IOB to allow for the differential contraction across the interface (several mm). The optical components are either mounted on the main base optical bench or the vertical optical bench. Two covers are used to enclose the instrument and complete the stray-light/EMI tight enclosure as well as act as structural elements. As all elements (with the exception of the EMI filters) are mounted on the main chassis, subsystems can largely be integrated/de-integrated independently from each other. This approach also allows optical alignment to be carried out with the covers removed giving access to the individual components.

[1]Both the TES and the Silicon Bolometer detector options need to be cooled to around 50 mK. The KID detectors operate at a base temperature of around 150 mK whereas the photoconductors operate at 1.7 K and above and do not need a cooler.

The detailed breakdown of the various contributions to the mass budget are given in the SAFARI Phase-A1 Study Report (2009). In summary the best estimate for the mass of the cold FPI is 38.8 kg, giving a 22% margin over the allocation, and 23.6 kg for the warm electronics showing over 21% margin.

*Table 3.11: Status of critical subsystems for SAFARI*

| **Subsystem** | **Key Functional, Interface and Performance Requirements** | **Status** |
|---|---|---|
| Detector System | 5940 detector channels in three arrays, Sensitivity: $2 \times 10^{-19}$ W Hz$^{-1/2}$<br>Electrical and sensor 3 dB knee: 20 Hz<br>50 mK base temperature | Demonstrated $< 2\times10^{-18}$ W Hz$^{-1/2}$ optical NEP in a single detector.<br>Electrical and optical testing of a $5\times5$ sub-array with NEP $\sim 2 \times 10^{-19}$ W Hz$^{-1/2}$ on-going.<br>Baseband feedback readout electronics demonstrated in laboratory breadboard testing.<br>Conceptual design of Focal Plane Arrays completed.<br>Component level testing and prototyping ongoing (RF Notch Filters, SQUIDs, Focal Plane Optics, electrical interconnects). |
| Instrument Cooler | 1 $\mu$W heat lift at 50 mK<br>20 $\mu$W heat lift at 300 mK<br>Thermal interfaces at 1.7 K, 4.5 K and 12 K | Hybrid $^3$He sorption cooler / ADR architecture confirmed.<br>Breadboard cooler similar to the requirements of SAFARI tested at CEA under ESA Technology Research Programme study.<br>Mechanical CAD model of SAFARI cooler produced.<br>Critical issue is control of heat loads during recycle. |
| FTS Scan Mechanism | Total scan length $\sim$ 35 mm<br>Scan velocity noise: $< 3$ $\mu$m/s 0-20 Hz<br>Position measurement precision: 15 nm<br>Cryogenic zer0-g dissipation: $< 1$ mW | Baseline architecture based on TNO Darwin/Optical Delay Line mechanism.<br>Detailed feasibility study completed and no stopping issues identified. |
| Mirrors | $>$ 99% reflectivity in SAFARI band<br>$\lambda$/6 WFE per mirror | General design based on SPIRE technology and considered low-technical risk. |
| Filters, Dichroics and Beam splitters | Band definition of three arrays, out of band rejection and 50:50 amplitude splitting in interferometer. | Design based on well established design heritage as used on Herschel, Spitzer and many other programmes. |
| Filter Wheel | Two six position, $\sim$ 35 mm filter element, low tolerance mechanism | Filter wheel mounted on cryogenic stepper motor. |
| Structure | First mode $>$ 150 Hz and instrument mass $<$ 50 kg with margin | Solid model and Finite element model of instrument structure completed and is compliant with mass and Eigen mode requirements. |

*Table 3.11: (continued)*

| Subsystem | Key Functional, Interface and Performance Requirements | Status |
|---|---|---|
| Warm Electronics | Mass less than 30 kg and power less than 120 W. Bias and readout the detectors and focal plane subsystems and provide the electrical interface with the spacecraft. | Overall architecture defined. Some elements have been bread boarded. Digital ASIC development will be required for detector readout. |

***Critical subsystem development***

Table 3.11 gives an overview of the critical subsystems required for the SAFARI instrument and an assessment of their current development status. Three of these, the detectors, the sub-100-mK cooler and the scan mechanism were identified as critical development items during the initial assessment of the instrument design by ESTEC in June 2008. Active development programmes are ongoing to raise the technical readiness of these items and, as can be seen from the table, significant progress has already been made. The remaining technical challenges for the critical subsystems are well understood and Technical Development Plans have been submitted describing how these subsystems will demonstrate the required Technology Readiness Level ahead of the mission downselect in 2011 and entry into Phase B.

### 3.2.3 Mid Infrared Camera: MIRACLE

MIRACLE (Mid InfRAred Camera w/wo LEns) is the focal plane instrument for wide field imaging and low-resolution spectroscopic observations ($R = \lambda/\Delta\lambda \sim 5$ – 100) over a wide spectral range in the mid infrared (5 – 40 $\mu$m). MIRACLE consists of two channels (MIR-S and -L) with almost the same optical design. Each of them has fore-optics re-imaging the telescope focal plane. A field mask wheel is installed at the focal plane in order to provide optimal slits in the spectroscopic mode. The subsequent camera optics has a pupil position where filter wheels with band-pass filters and grisms are installed.

*Table 3.12: Specifications of MIRACLE*

| | MIR-S | MIR-L |
|---|---|---|
| Wavelength coverage | $5 - 26\ \mu$m | $20 - 38\ \mu$m |
| Detector | Si:As $1024 \times 1024$ | Si:Sb $1024 \times 1024$ |
| Pixel scale | 0.36″/pixel | 0.36″/pixel |
| FOV | $6 \times 6$ arcmin$^2$ | $6 \times 6$ arcmin$^2$ |
| Spectral resolution ($R = \lambda/\Delta\lambda$) | $\sim 10$ (imaging)/ $\sim 100$ (spec.)<br>Band pass filters/grism | $\sim 10$ (imaging)/ $\sim 100$ (spec.)<br>Band pass filters/grism |
| Size (mm) | 125(R) x 50(H) x 30 degree | 125(R) x 50(H) x 30 degree |

### 3.2.4 Mid-IR Spectrometer: MIRMES & MIRHES

**MIRMES (Mid-IR Medium-Resolution Echelle Spectrometer)**

MIRMES performs medium-resolution ($R = \lambda/\Delta\lambda \sim 1500$) spectroscopic observations over a wide spectral range in the mid infrared (10 – 40 $\mu$m) range. The instrument consists of two arms, ARM-S and ARM-L. They share the same field of view (FOV) area on the focal plane by means of the dichroic beam splitter, which separates transmission/reflectance at around 18.7 μm. Each arm has the field mask to limit the FOV and each arm has an image slicer as an integral field unit.

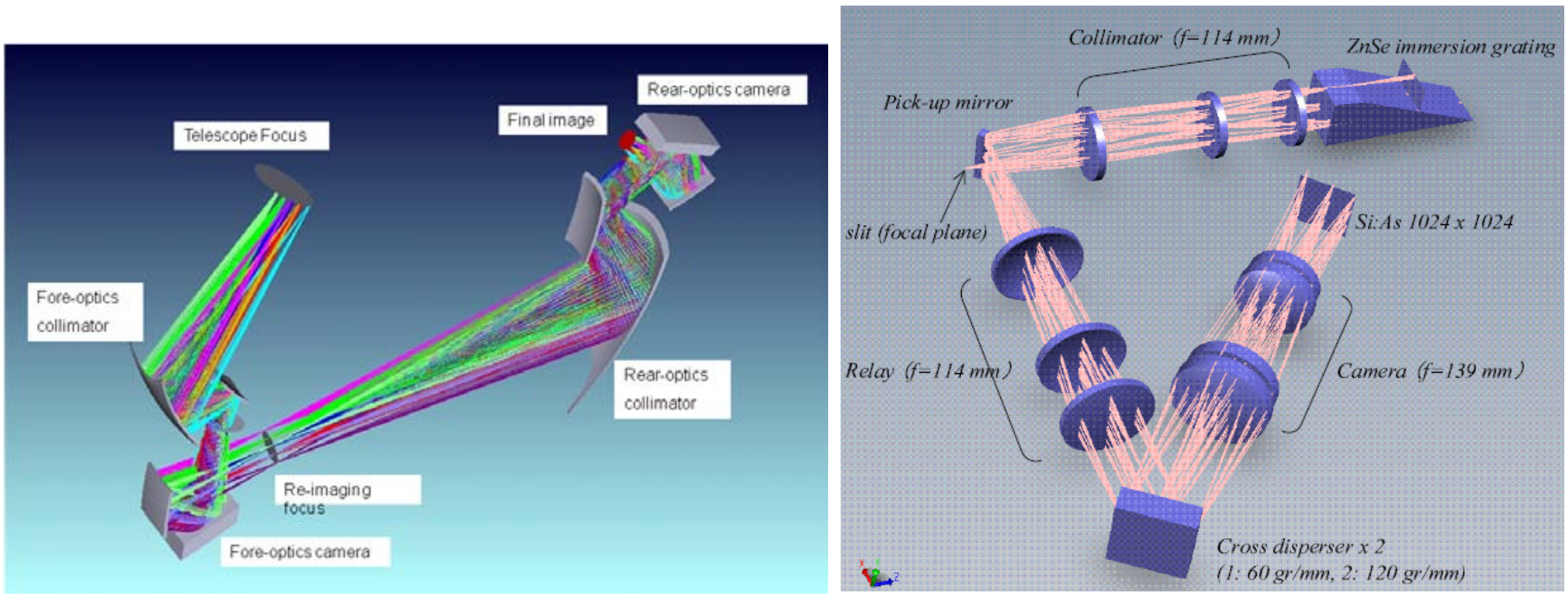

*Figure 3.9: **Left panel**: A successful optical design of MIRACLE. The folding mirror for beam pick-off at the STA focus is omitted for simplicity. The final F-number is set to fast value (3.15) in order to match the PSF size to small pixel size (18 μm) of the detector. **Right panel**: Optical layout of MIRHES S-mode.*

*Table 3.13: Specifications of the Mid-IR Medium-Resolution Echelle Spectrometer.*

| | **ARM-S** | **ARM-L** |
|---|---|---|
| Wavelength coverage | 10.32 – 19.35 μm | 19.22 – 36.04 μm |
| Spectral resolution ($R = \lambda/\Delta\lambda$) | ~ 1500 | ~ 900 |
| Pixel scale | 0.37″ | 0.72″ |
| Slit width | 1.11″ (3 pixel) | 3.6″ (5 pixel) |
| FOV | 12.95″ × 5.55″<br>(35 pixels × 3 pixels × 5 rows) | 25.2″ × 18.0″<br>(35 pixels × 5 pixels × 5 rows) |
| Size (mm) | 125(R) × 50(H) × 30 degree | 125(R) × 50(H) × 30 degree |
| Mass (kg) | 17.5 | 17.5 |

#### MIRHES (Mid-IR High-Resolution Echelle Spectrometer)

MIRHES performs high-resolution ($R = \lambda/\Delta\lambda \sim 20\,000 - 30\,000$) spectroscopic observations in two wavelength ranges, 4 – 8 μm (S-mode) and 12 – 18 μm (L-mode). Immersion gratings enable high spectral resolution with small mass and volume resources. Two independent spectrographs for S-mode and L-mode are considered (Figure 3.9, right panel).

### 3.2.5 SPICA Coronagraph Instrument: SCI

SCI (SPICA Coronagraph Instrument) is a high dynamic-range imager and spectrometer with coronagraph optics. The primary target of SCI is the direct observation (imaging and spectroscopy) of Jovian exoplanets in the infrared, whilst circum-stellar discs, other type of exoplanets, Active Galactic Nuclei, and any other compact systems with high contrast can be potential targets.

The specifications of SCI are summarised in Table 3.15. Though diffraction limited imaging at 5 μm wavelength is a specification for the SPICA telescope, SCI has as goal to observe at shorter wavelengths with wave front correction by a deformable mirror and a tip-tilt mirror. For the coronagraph method, binary-shaped pupil mask is the baseline solution because of the robustness against telescope pointing error, achromatic effects (except image size effects scaling with wavelength), and simplicity. A more challenging method, a hybrid solution of Phase Induced Amplitude Apodization (PIAA) and binary-shaped pupil mask, is considered as an option to obtain better Inner Working Angle (IWA) and throughput. With the InSb detector option, the observable range can be extended to shorter wavelengths, and simultaneous imaging will be possible with two detectors. Band-pass filters will be used for imaging, and transmissive dispersers (e.g., grism) will be used for spectroscopy.

*Table 3.14: Specifications of the Mid-IR High-Resolution Echelle Spectrometer.*

| | **Short(S)-mode** | **Long(L)-mode** |
|---|---|---|
| Wavelength coverage | 4 – 8 $\mu$m | 12 – 18 $\mu$m |
| Spectral resolution ($R = \lambda/\Delta\lambda$) | 30 000 | 20 000 – 30 000 |
| Slit width | 0.72″ | 1.20″ |
| Slit length | 3.5″ | 6.0″ |
| Dispersion element | ZeSe immersion grating | KRS5 immersion grating |
| Cross disperser | reflective | reflective |
| Size (mm) | 200(L) × 200(W) × 100(H) | 350(L) × 350(W) × 200(H) |
| Mass (kg) | 7.4 | 7.4 |

*Table 3.15: Specifications of SCI.*

| | |
|---|---|
| Wavelength coverage | 5 – 27 $\mu$m (goal: 3.5 – 27 $\mu$m)<br>(In 1 − 3.5 $\mu$m, SCI has sensitivity, though contrast is not guaranteed) |
| Observation mode | Coronagraphic/Non-coronagrahic,<br>Imaging/Spectroscopy |
| Coronagraph | Binary-shaped pupil mask |
| Contrast | $10^{-6}$ |
| Inner working angle (IWA) | $3.3\lambda/D$ |
| Outer working angle (OWA) | $16\lambda/D$ |
| Detector | Si:As 1k × 1k (long wavelength channel)<br>InSb 1k × 1k (short wavelength channel) |
| FOV | 1 × 1 arcmin$^2$ |
| Spectral resolution | $R = 20 - 200$ |

Thanks to a simple pupil shape and active optics, SCI provides the potential to perform coronagraphic observations with contrasts significantly higher than the coronagraph of JWST. This can be a unique capability to characterise atmospheric features of exoplanets in the 2010s decade. It should be noted that SCI can be useful for monitoring and observation of exoplanet transits.

### 3.2.6 Focal Plane Finding Camera

FPC (Focal Plane finding Camera) consists of two components. One is FPC-G that is used for the attitude control system to an accuracy of 0.05″. FPC-G is a system instrument and will be operated continuously during the whole SPICA mission. FPC-S is a near infrared camera for astronomical observations. FPC-S has a filter wheel with several filters and dispersive materials. FPC-G and FPC-S are integrated into one package with beam splitter. FPC specifications and optical design are shown in Table 3.16. The FPC mass is estimated at $< 10$ kg.

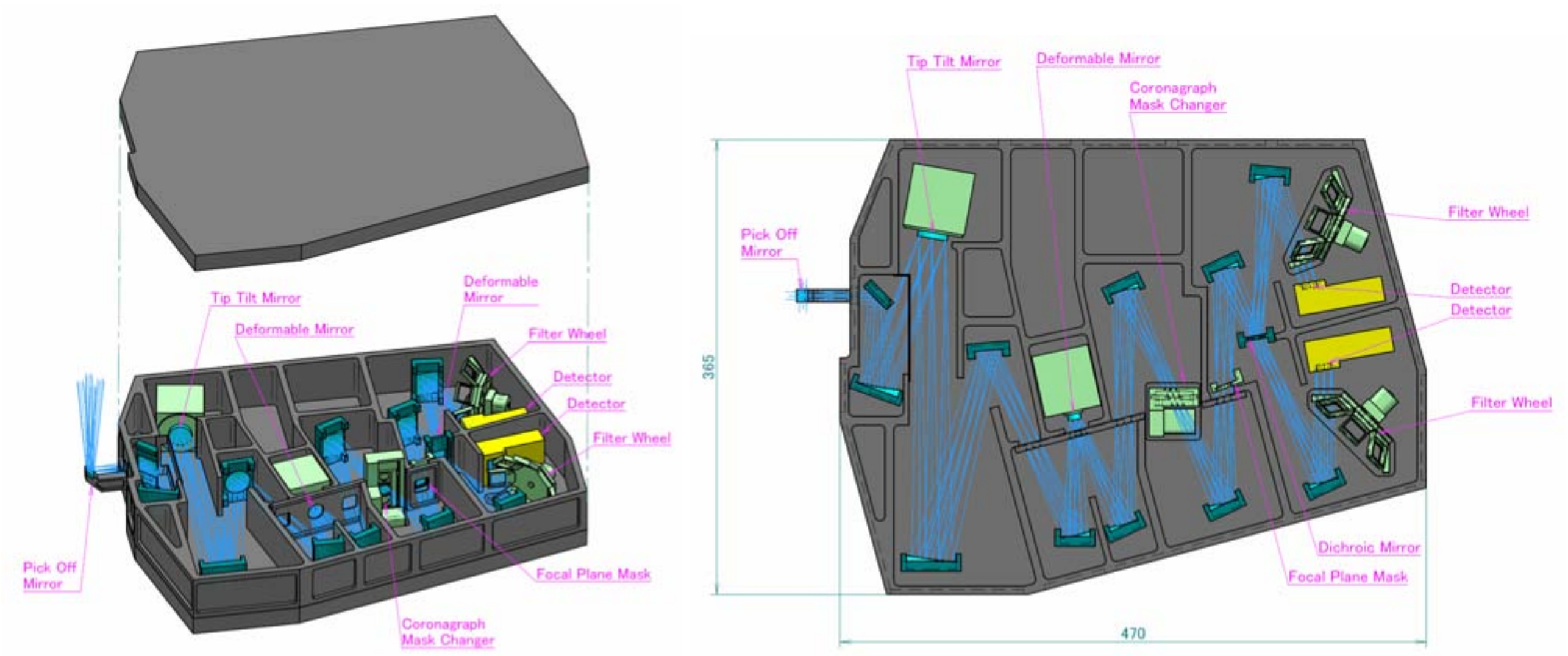

*Figure 3.10: Design of SCI. A tip-tilt mirror and a deformable mirror is used. All device before focal plane mask is made of mirror optics (i.e., no transmissive device). Mechanical changer is used to realise the coronagraphic mode and fine camera/spectroscopy mode without mask.*

*Table 3.16: FPC specifications*

| | **FPC-G** | **FPC-S** |
|---|---|---|
| Wavelength coverage | 1.6 $\mu$m (H band) | $2-5$ $\mu$m |
| Detector | InSb, $512 \times 412$ | HgCdTe, $2K \times 2K$ |
| Pixel scale | $0.5''$ | $0.18''$ |
| Slit width | $1.11''$ (3 pixel) | $2.16''$ (3 pixel) |
| FOV | $4.3 \times 3.5$ arcmin$^2$ | $6 \times 6$ arcmin$^2$ |
| Heat generation | $< 1$ mW | $< 2$ mW |

### 3.2.7 BLISS

BLISS (Background-Limited Infrared-Sub-mm Spectrograph) is an extremely sensitive broad-band far infrared & sub-mm spectrometer proposed mainly by US astronomers. BLISS covers the entire far infrared and sub-mm wavelengths (38 – 430 $\mu$m) with a spectral resolution $R = 700$, using 4200 superconductive bolometer arrays cooled down to 50 mK by a dedicated adiabatic demagnetisation refrigerator.

# Chapter 4

# Mission Design

The SPICA mission design is tailored to achieve the scientific objectives described in chapter 1. To this goal, the spacecraft design is based on a 3-m class telescope (the nominal diameter of the primary mirror is 3.5 m in the current baseline) operated at a temperature lower than 10 K (the nominal temperature of the telescope assembly is $< 6$ K in the current baseline). The telescope is designed to have diffraction limited performance at 5 $\mu$m (with a WFE $< 350$ nm rms). The main SPICA wavelength range is 5 – 210 $\mu$m, covered with five scientific instruments: MIRACLE (mid infrared camera and spectrophotometer), SAFARI (far infrared imaging spectrometer), MIRMES (mid infrared medium resolution spectrometer), MIRHES (mid infrared high resolution spectrometer), and SCI (mid infrared coronagraph) A sixth instrument, BLISS (FIR and sub-mm spectrometer) is considered as optional. The SPICA Telescope Assembly (STA) and the Focal Plane Instruments (FPI) are located in the cryo-module of the spacecraft, supported via a truss structure which is fixed to the Service Module (SVM).

The thermal environment required to operate the telescope and the instruments is obtained by a combination of passive cooling (via dedicated Sun and thermal shields combined to radiators) and active cooling (including a number of mechanical coolers, ensuring a base temperature of 4.5 K and 1.7 K). The total mass of the S/C is about 4 ton, in line with the capabilities of the baseline launcher vehicle, JAXA's H-IIB.

The Sun – Earth $L_2$ point is the optimum environment to obtain excellent sky visibility and a stable thermal environment to be able to cool the telescope. The onboard data generation rate is of order 4 Mbps, therefore requiring a high-speed downlink (compatible with X band). The guaranteed lifetime is 3 yr, with the goal of extended operations to 5 yr. The absence of cryogens onboard allows to extend the nominal lifetime beyond the nominal duration (finally limited by the AOCS propellant and any onboard failures). The main SPICA subsystems are illustrated in Figure 4.1.

## 4.1 Mission profile

The presently envisaged launcher vehicle for SPICA is JAXA's H-IIB, capable of delivering in excess of 4 ton to a SE-$L_2$ transfer orbit. The H-IIB is equipped with a 5S-H fairing with a useful diameter of 4.6 m, compatible with the SPICA needs. The launch site is JAXA's Tanegashima Space Center, while the launch is planned in 2018.

The orbit baselined for SPICA is a large amplitude halo orbit, with a period of about 180 days and semi-major axis amplitude of about 750 000 km (see Figure 4.3).

A first launcher dispersion manoeuvre is performed on the first day after launch, with the S/C already on a trajectory to SE-$L_2$. The spacecraft will be in proximity of the libration point after about 30 days from launch. The cool-down will start immediately after the out-gassing phase, during the transfer. The final orbit injection in the Halo orbit will take place approximately 120 days after launch. The guaranteed mission lifetime is 3 yr (nominal Operation Phase, driven by the guaranteed lifetime of the mechanical cooler), with the goal of 5 yr (including an Extension Phase of 2 yr).

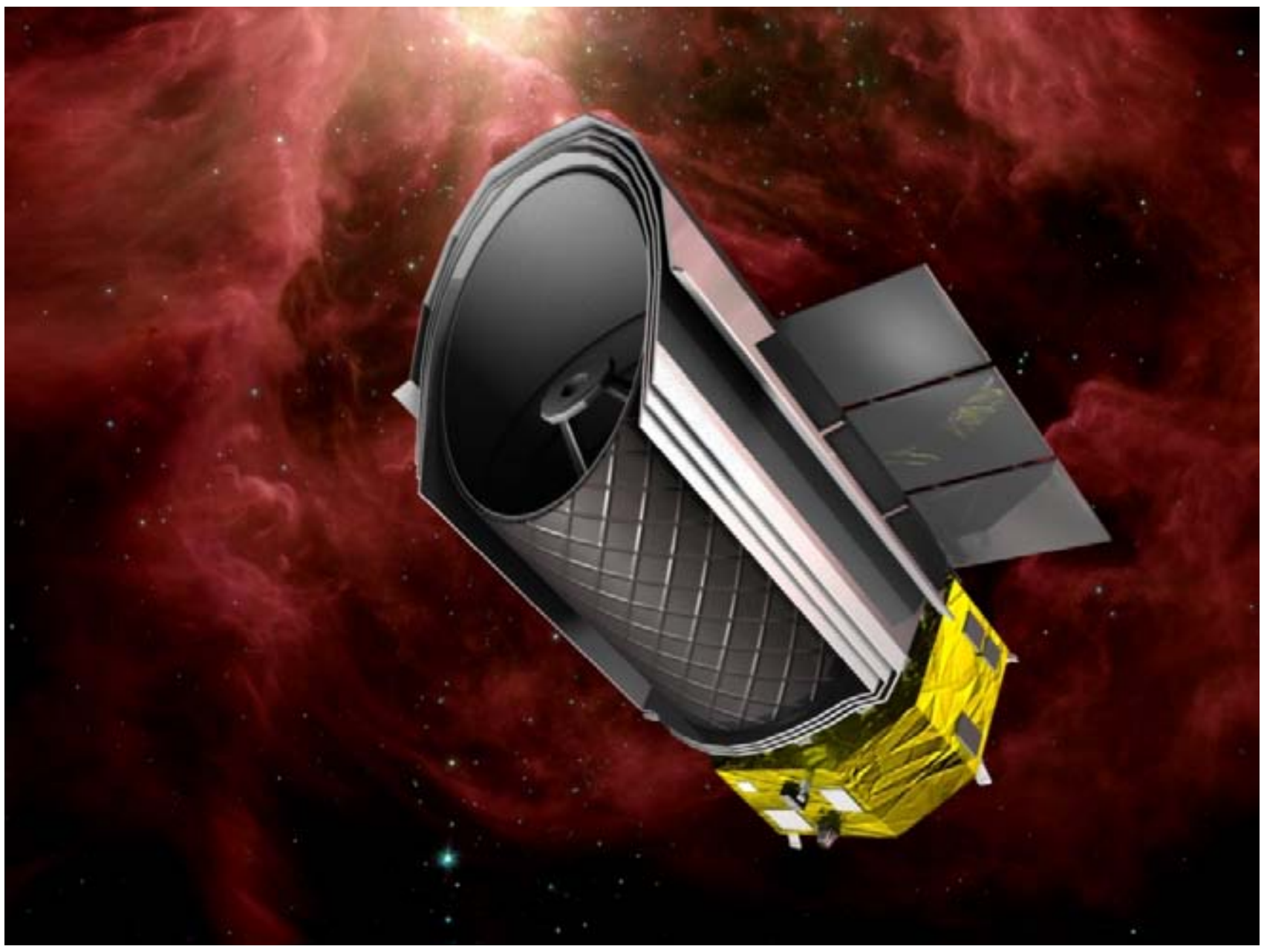

*Figure 4.1: Artist view of the SPICA spacecraft.*

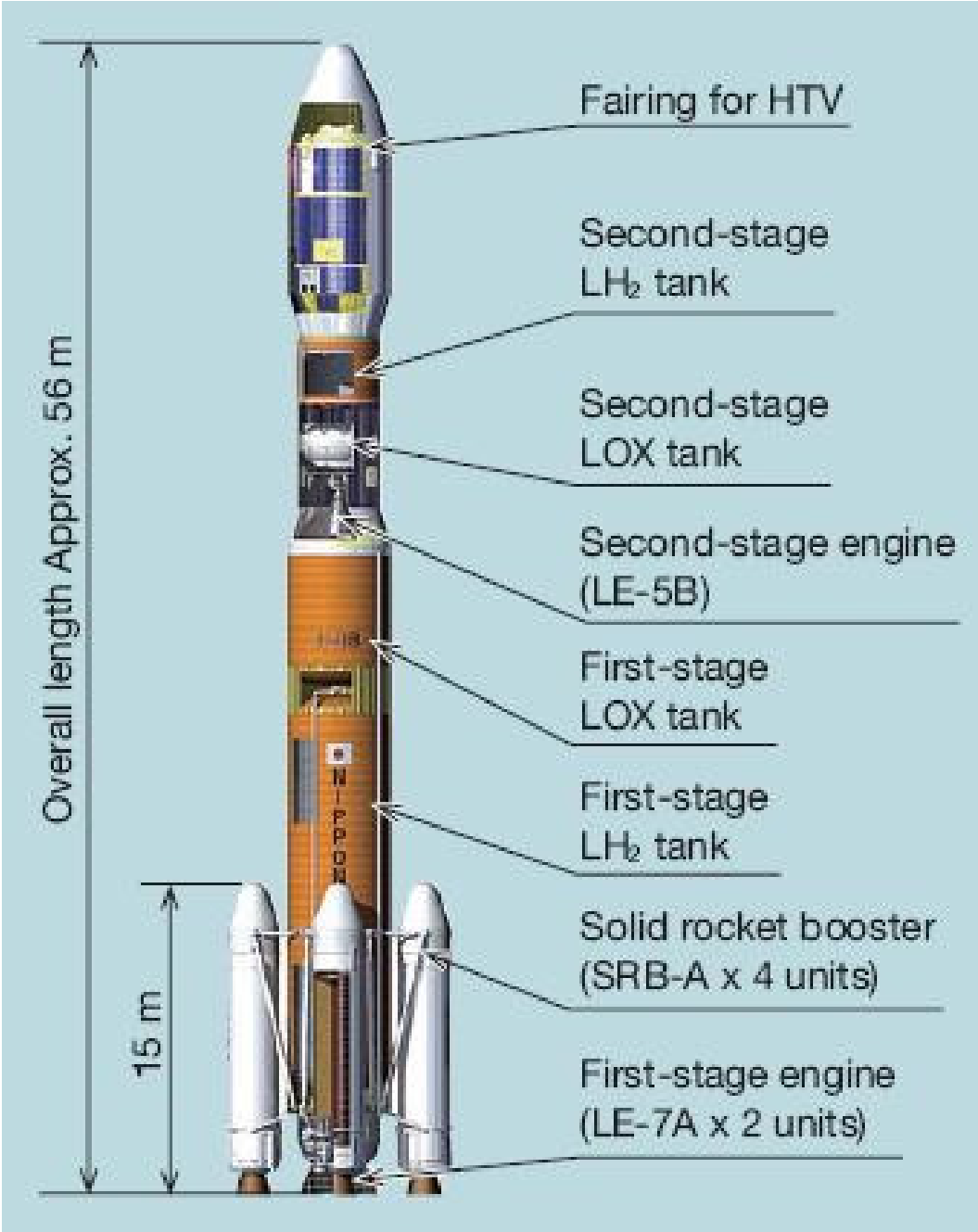

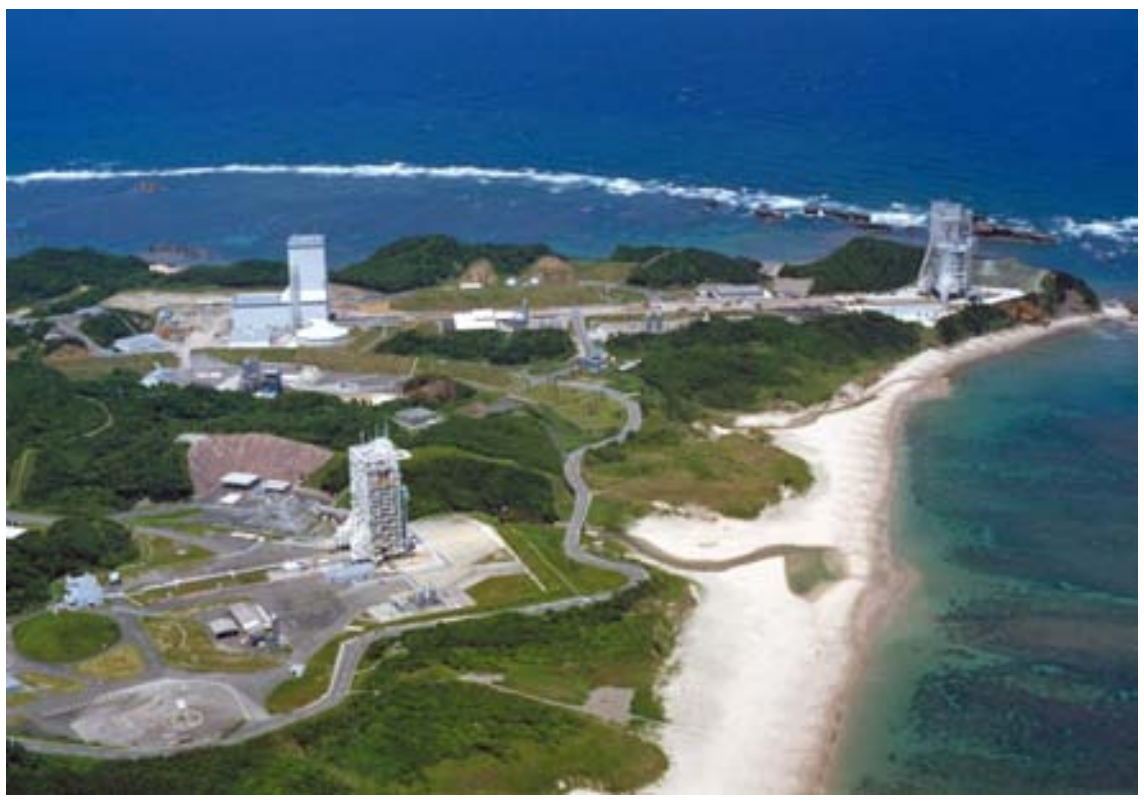

*Figure 4.2: H-IIB launcher vehicle and Tanegashima launch center.*

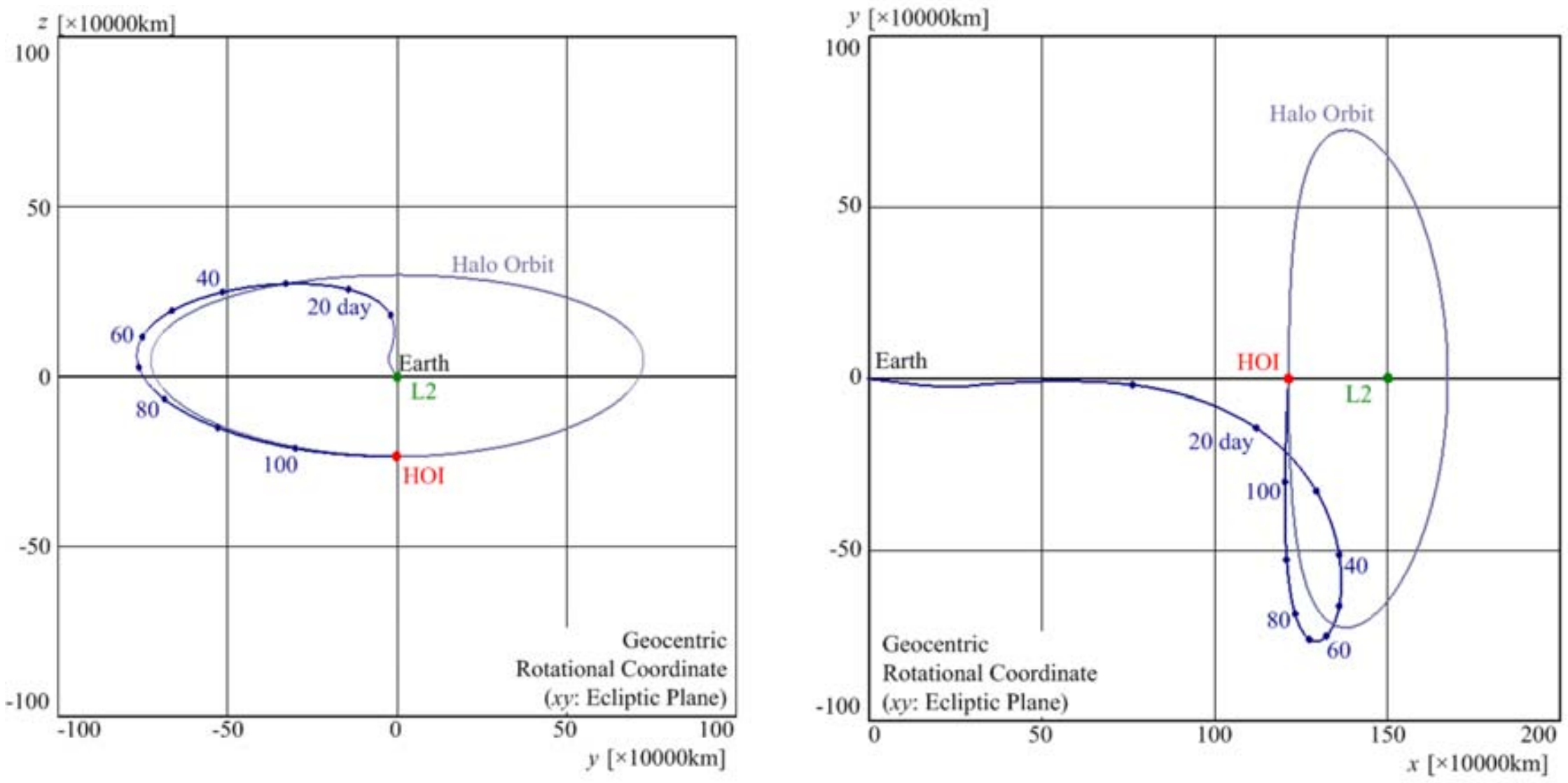

*Figure 4.3: $L_2$ transfer trajectory and Halo Orbit Injection.*

## 4.2 Spacecraft design

The SPICA spacecraft consists of the Cryo Payload Module (PLM) and the Bus (or Service) Module (SVM) as illustrated in Figure 4.4. The PLM includes the Cryogenic Assembly and the Scientific Instruments Assembly. The Cryogenic Assembly includes mechanical coolers and passive thermal shields, required to cool down and maintain at about 5 K the Scientific Instruments Assembly. The latter includes the STA and the Instrument Optical Bench (IOB). Additional coolers allow reaching the required focal plane detector temperature (down to ~ 100 mK). The overall PLM architecture is dominated by the need to have a large Sun shield, three coaxial thermal shields, the telescope shell and the additional telescope baffle. The PLM is connected to the Bus via a low thermal conductivity truss structure, which is also supporting the STA.

The Bus Module, of a conventional design, hosts most of the spacecraft subsystems (power, propulsion, attitude control, data handling, thermal control and telecommunications), as required to operate the observatory. A list of the SPICA subsystems can be found in Figure 4.5.

The total wet mass of the spacecraft (including system level margins) is estimated to be about 4000 kg (see Table 4.2), while the total electrical (peak) power required during operations is estimated to be about 2.3 kW (see Table 4.3). The spacecraft would have a maximum height of about 7.4 m, of which about 1.3 m for the Bus and 6.1 m for the PLM and support structure.

### 4.2.1 Bus module and spacecraft sub-systems

The SPICA bus module has an octagonal cross-section and a structure based on a combination of truss and panel elements. The octagonal geometry is reflected in the 8 interface points adopted between the launcher adaptor and bus module, between bus module and truss structure as well as between truss structure and telescope assembly, thus allowing a balanced transfer of the launch loads. The estimated dry mass of the bus module is 830 kg: this value takes into account all sub-systems and units, including the solar arrays.

The lateral panels allow easy access to the equipment and additional area to be used as radiating surface. The bus module is almost completely shielded from the direct Sun radiation thanks to the lower part of the PLM Sun shield, the Solar Arrays and the High Gain Antenna. The temperature of the upper panel of the bus and of the truss interface points determines the conductive loads to the cryo-module and needs to be minimised.

The Electrical Power Subsystem is based on triple-junction GaAs solar cells sized for a total maximum power of 2.4 kW (EOL) at an operating temperature of about 100 °C. The solar arrays, located outside the Sun shield, are arranged in two wings, each consisting of 3 elements, a deployment hinge, a holding mechanism and a yoke. The solar arrays are folded around the bus module and deployed immediately after separation from the launch vehicle. The Electrical Power Subsystem, designed for a mission duration of 5 yr, is based on an unregulated 50 V bus and a set of Li-ion batteries.

The Telemetry, Telecommand and Communication system includes a high speed X-band telemetry down-link (~ 10 Mbit/s), a low speed S-band TT&C data link and a low speed X-band TT&C data link. The high speed down-link includes steerable HGA and MGA. The low speed S band link is equipped with 2 Low Gain Antennas to ensure full coverage at different spacecraft attitudes.

The Data Handling System is based on an architecture already developed by JAXA for other science missions with the purpose of enabling re-use of existing equipment. The baseline configuration is illustrated in Figure 4.6. The onboard mass memory is sized for a minimum capacity of 48 Gbytes, corresponding to the data volume produced in slightly more than one day.

The architecture of the SPICA Attitude and Orbit Control Subsystem is represented in Figure 4.7. In order to achieve the required pointing stability performance, the information provided by the Inertial Reference Unit and by the Star Trackers is complemented by dedicated Focal Plane Cameras (FPC-G and C-FPC), located on the Instrument Optical Bench and used as Fine Guidance Sensors. FPC-G is used for observations not involving the Coronagraph instrument, while C-FPC is used in conjunction with the Coronagraph operations. Reaction wheels are baselined as actuators.

The AOCS performance during operations with the Coronagraph instrument and with the other FPI is provided in Table 4.1. The use of these two additional fine guidance sensors allows achieving rather good Absolute Pointing Error (APE) and Relative Pointing Error (RPE) performance.

The SPICA propulsion system is based on a blow-down, mono-propellant ($N_2H_4$ hydrazine) approach. The System includes four propellant tanks, four 23N thrusters, eight 3N thrusters and auxiliary equipment (such as piping, filters, valves). This simple configuration is well matched to the modest delta-V requirement ($\sim 100$ m s$^{-1}$) and the compensation of external torques (mainly related to solar radiation pressure).

The four tanks carry the propellant and the pressurant gas. The 23N thrusters are used for orbit control purposes and provide the thrust required to achieve the required delta-V (Z axis direction). The 3N thrusters are used mainly for attitude control purposes (Down-loading of the reaction wheels). The total mass of the propulsion system is estimated to be about 110 kg, with an additional 200 kg of propellant.

*Table 4.1: SPICA AOCS performance (APE and RPE)*

| **Observation mode** | **Absolute Pointing Accuracy** | **Pointing stability** |
|---|---|---|
| Except for coronagraph | 0.135 arcsec ($3\sigma$) | 0.075 arcsec (0-P) / 200 sec |
| Coronagraph | 0.03 arcsec ($3\sigma$) | 0.03 arcsec (0-P) / 20 min |

The mass and power spacecraft resource budgets are reported in Table 4.2 and 4.3. The unit or sub-system level mass values listed in the table are comprehensive of individual maturity level margins (depending on their design maturity). A system level margin of 20% is applied on top of the maturity margins in order to reflect the present level of design definition (commensurate to Phase 0/A).

Similarly the unit or sub-system level power values listed in the power budget table are comprehensive of individual margins and are representative of peak power demands.

### 4.2.2 Payload module – cryogenic system

The SPICA adopts a new concept of cryogenic system that uses no cryogen. The elements maintained at 4.5 K, including the STA, the IOB and some focal plane instruments (FPI), are refrigerated by the combined action of mechanical cooling and efficient radiative cooling in the stable thermal environment at the Sun-Earth $L_2$. In consideration of the cooling capacity of a 4 K-class mechanical cooler (4 K-MC) at the end of life (EOL) and on the assumption of Joule heating of the FPI being approximately 15 mW, the baseline design of the Thermal Insulation and Radiative Cooling System (TIRCS) was determined by the thermal and the structural analyses, so that the total parasitic heat flow from the higher temperature elements to the 4.5 K stage can remain below 25 mW.

*Table 4.2: Overall SPICA spacecraft mass budget.*

| Subsystem | Main units | Mass (kg) |
|---|---|---|
| Payload Module (PLM) | Cryogenic Assembly | 1200 |
| | STA | 700 |
| | FPI (including IOB) | 200 |
| | STA + FPI warm electronics | 100 |
| | Cryogenic electronics | 90 |
| | **Total PLM (with maturity margins)** | **2290** |
| Bus Module (BM) | Mechanical structure | 330 |
| | Thermal Control System | 66 |
| | Power control system | 32 |
| | Solar Array assembly | 94 |
| | TT&C | 115 |
| | Data Handling System | 22 |
| | AOCS | 106 |
| | Propulsion | 110 |
| | Harness | 39 |
| | **Total BM (with maturity margins)** | **914** |
| **Total Dry Mass** | **PLM + BM** | **3204** |
| System Level Margin | 20% on dry mass | 640 |
| Propellant | Hydrazine | 220 |
| **TOTAL** | **with maturity & system level margins** | **4064** |

The 4.5 K units, such as STA, IOB and FPI, are surrounded by the TIRCS, consisting of the external telescope baffle, the telescope shell, three shields and a sun shield to reject the heat flow from the outer environment, as depicted in Figure 4.8. Multi-Layer Insulation (MLI) is attached to the sun shield and to the thermal shield #3 to block thermal radiation, while a section of each thermal shield acts as a radiator to reject most of the absorbed heat from the sun and from the spacecraft service module (SVM) to deep space. The layout of these shields and of the solar array paddles is determined to optimise radiative cooling.

The STA, the telescope shell and the three shields are structurally supported by the SVM with main trusses between them. The trusses are made of the Carbon Fiber Reinforced Plastics (CFRP) and Alumina Fibre Reinforced Plastics (ALFRP) with low thermal conductivity, while the external baffle and the main part of the telescope shell are made of high thermal conductivity CFRP to dissipate the heat flow. The sun shield and three co-axial thermal shields are made of aluminium plates, and their structural frames are connected by means of dedicated trusses of ALFRP as well. The shield #3 and the external baffle are supported axially by the SVM and the telescope shell, respectively. Figure 4.9 shows a more detailed view of the truss structure separating the PLM from the SVM and supporting directly STA, FPI and thermal shields.

Wire harness between the FPI and their electronics equipments in the SVM is assumed to be made of Manganin (up to 1000 lines). The Heat Rejection System transports the heat exhausted from the mechanical

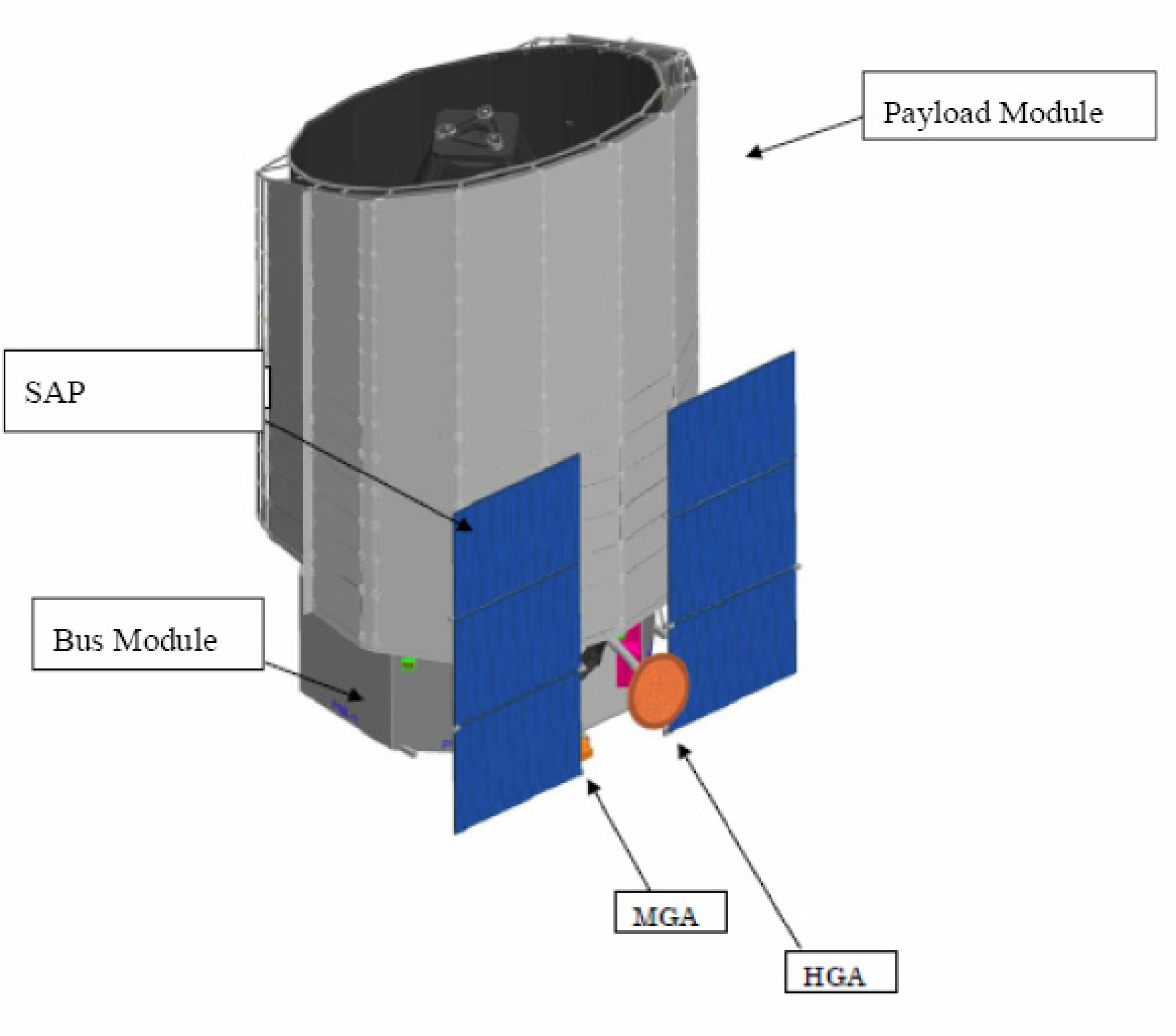

*Figure 4.4: Overview of the main elements of the SPICA spacecraft.*

*Table 4.3: Total peak power with margins*

| **Subsystem** | **Main units** | **Power (W)** |
|---|---|---|
| Payload Module (PLM) | Cryogenic Assembly | 900 |
| | STA | 0[(1)] |
| | FPI (including IOB) | 235 |
| Service Module (SVM) | Mechanical structure | 0 |
| | Thermal Control System | 160[(2)] |
| | Power control system | 35 |
| | TT&C | 185 |
| | Data Handling System | 100 |
| | AOCS | 130 |
| | Propulsion | 185 |
| **Total Power** | **PLM + SVM** | **1930** |
| Margin | 20% on total power | 385 |
| **TOTAL** | **Total power with margins** | **2315** |

[(1)]During normal operations the STA is passive. The STA is equipped with a refocusing mechanism used only during the observatory commissioning. The STA is equipped with heaters to allow for decontamination (with a power is of order several hundreds of W, available given the fact that the mechanical coolers would be stopped during decontamination).

[(2)]Not including heater power for the telescope decontamination.

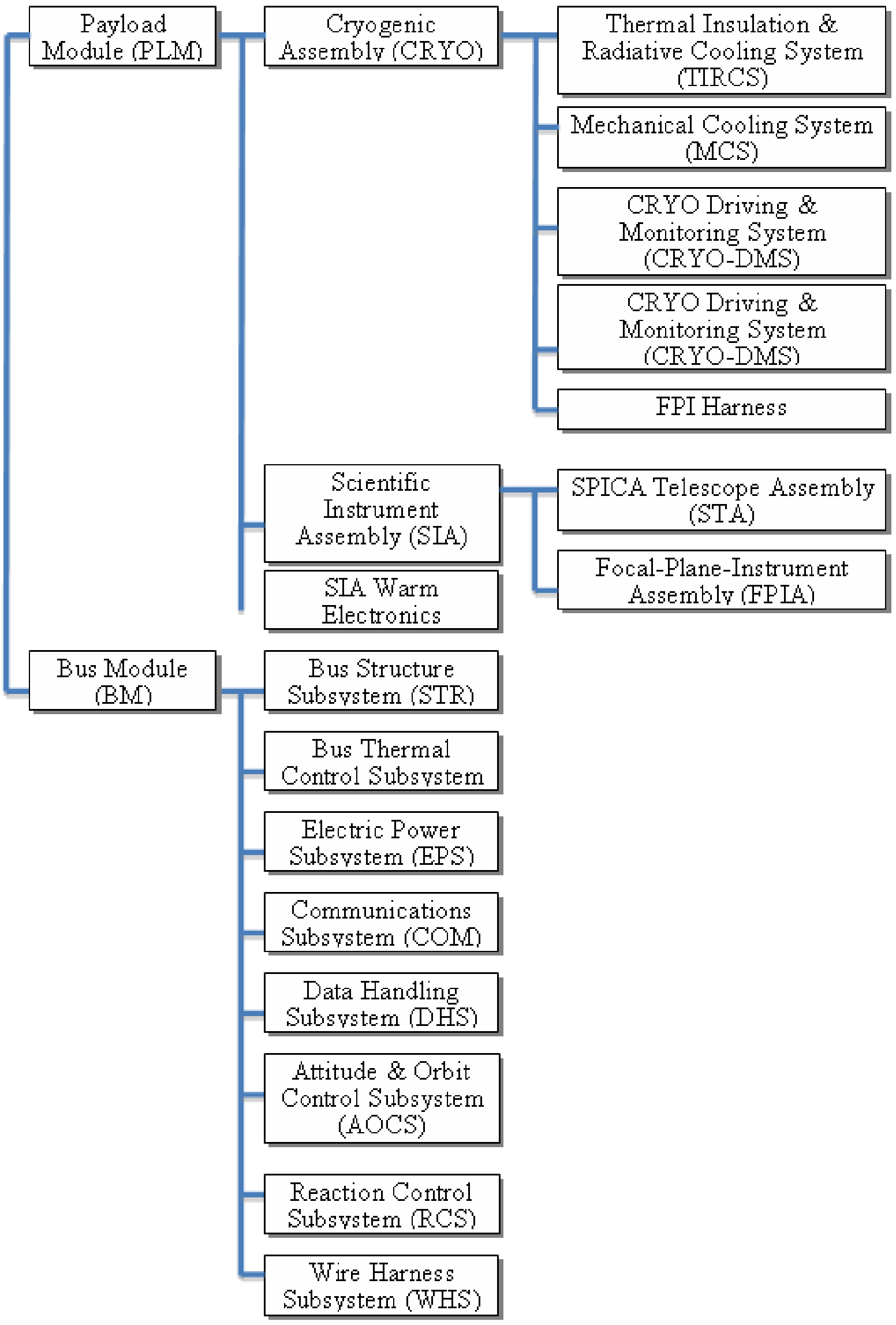

*Figure 4.5: List of the subsystems of the SPICA spacecraft.*

coolers on the cooler base plate to the cooler radiators. A loop heat pipe, which has some advantages such as flexibility in components arrangement and adaptability for long-distance transportation of large heat amount, is assumed to be used as heat transport device. The radiators dedicated to this purpose are located immediately above the service module, looking in the anti-sun direction.

A dedicated thermal analysis was carried out on the SPICA TIRCS configuration, showing that the heat load to the 4.5 K stage is dominated by conduction through structural supports (truss) and harness. The heat flow analysis for the steady state case shows that the total amount of parasitic heat and heat dissipation (from all elements of the PLM) is less than 40 mW at the 4.5 K stage, while the temperature of the external baffle and of the telescope shell are less than 14 K and 30 K, respectively. On the other hand, the bread board model of the upgraded 4 K Mechanical Cooler successfully demonstrated a cooling power larger than 50 mW, a cooling power compliant with the results of the TIRCS thermal analysis. Specific attention has been paid by the minimisation of vibrations transmission from the mechanical coolers to the PLM (and in particular to the FPI).

Concerning the transient case, a preliminary time-dependent analysis gives a cooling profile at the initial phase as shown in Figure 4.10, showing that a cooling time of about 170 days is required to reach less than 5 K after launch. The possibility to reduce the cool down time is presently being studied by JAXA.

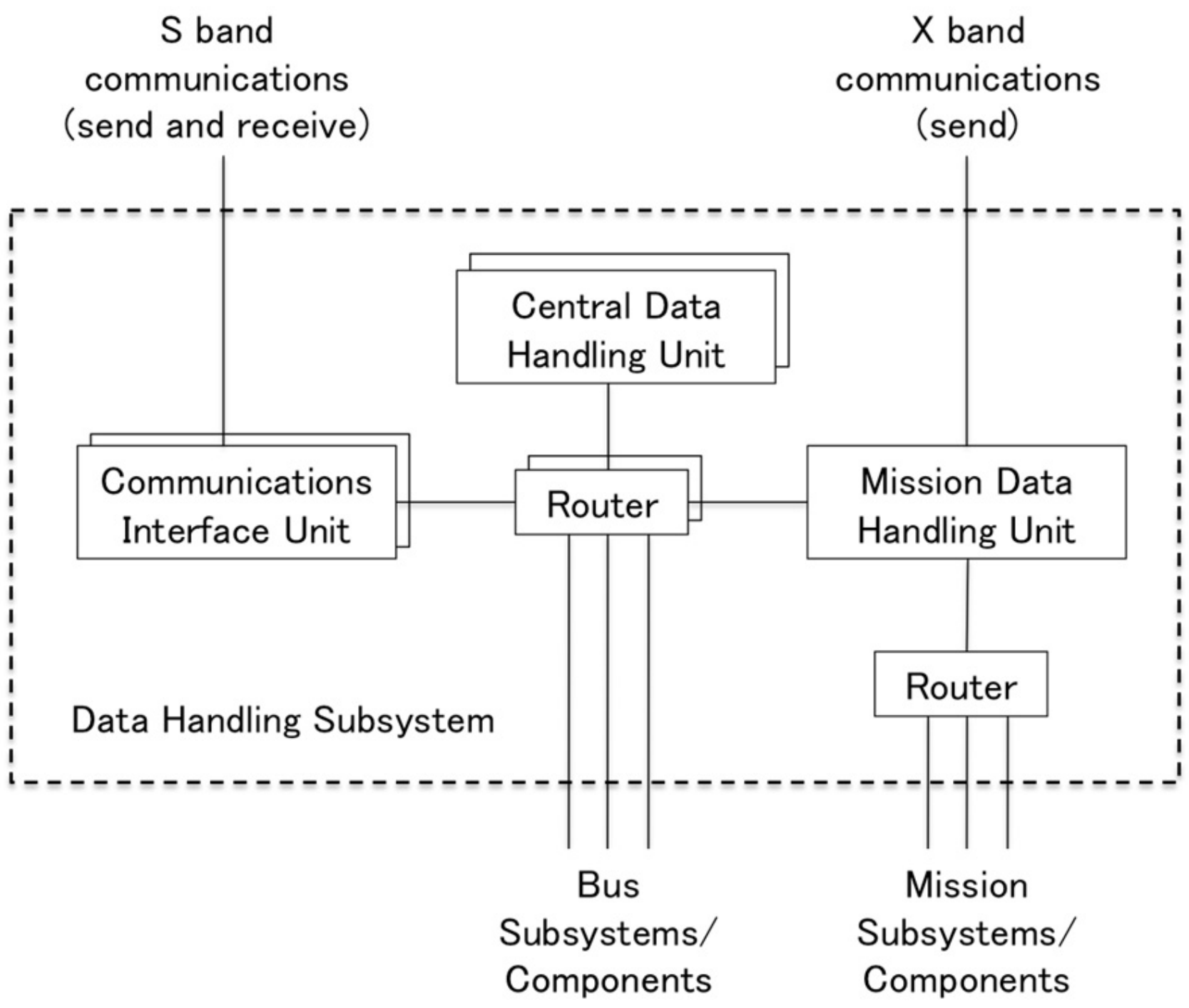

*Figure 4.6: Overall SPICA Data Handling System architecture.*

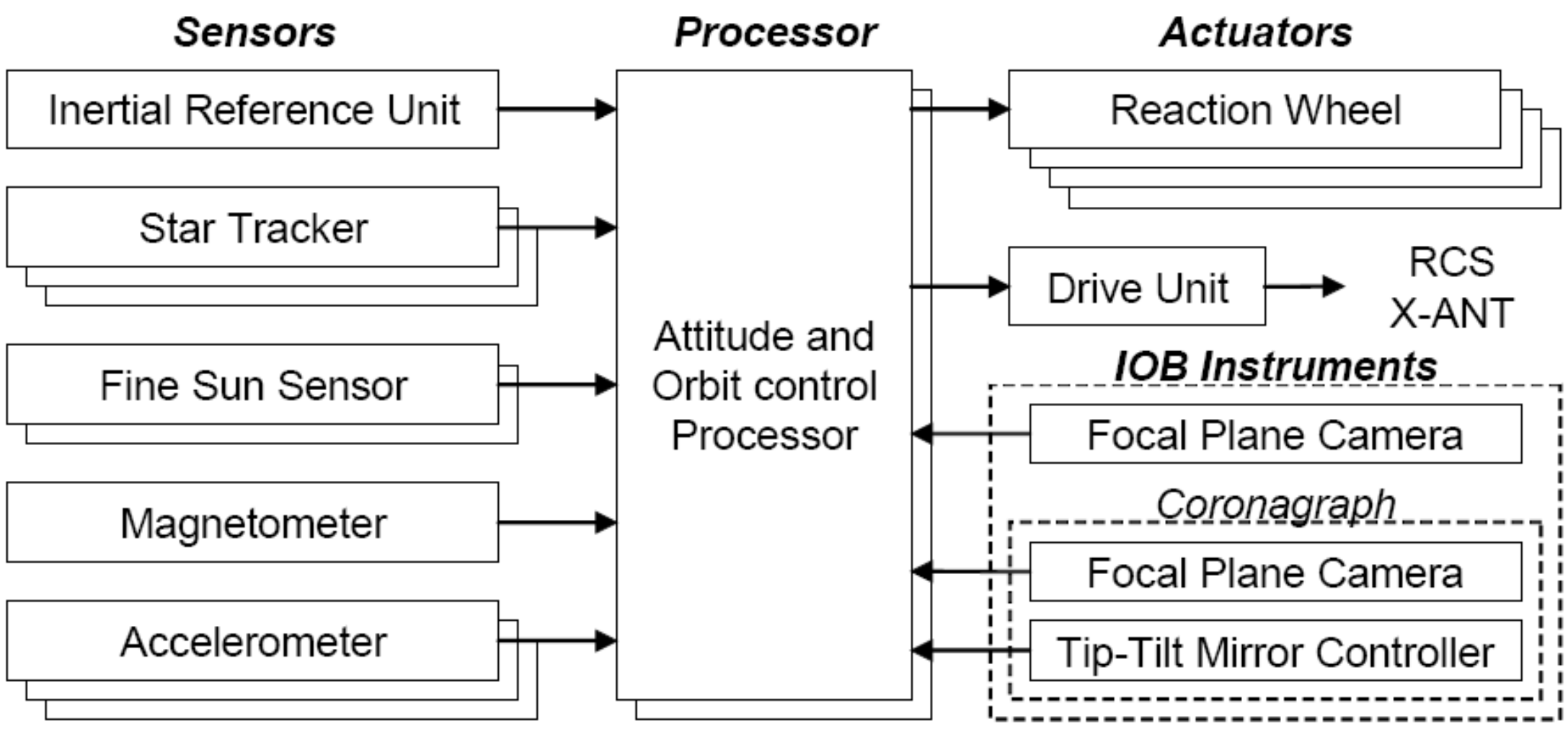

*Figure 4.7: Overall SPICA AOCS architecture (0-P = Zero to Peak).*

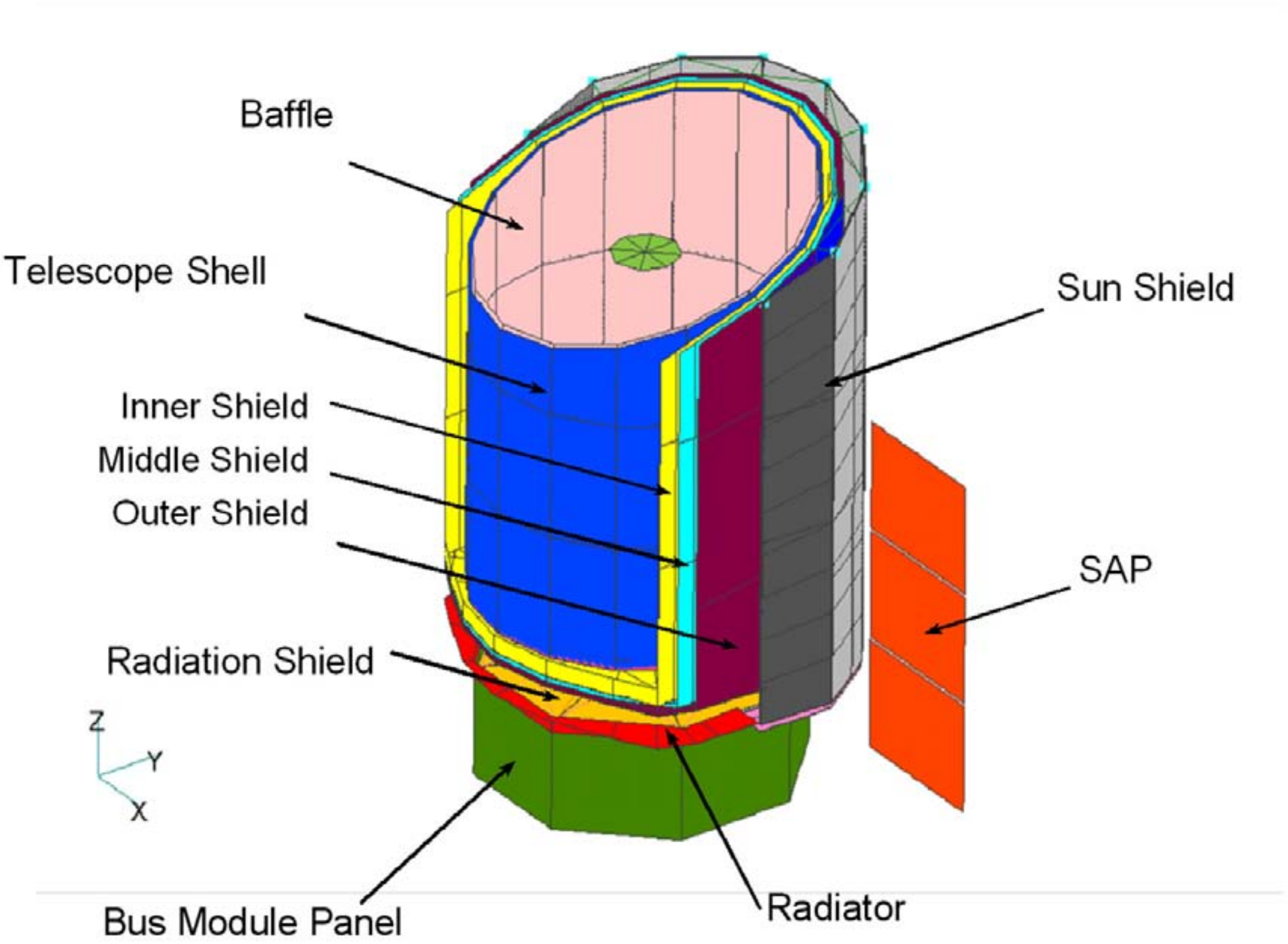

*Figure 4.8: SPICA Thermal Insulation and Radiative Cooling System (TIRCS)*

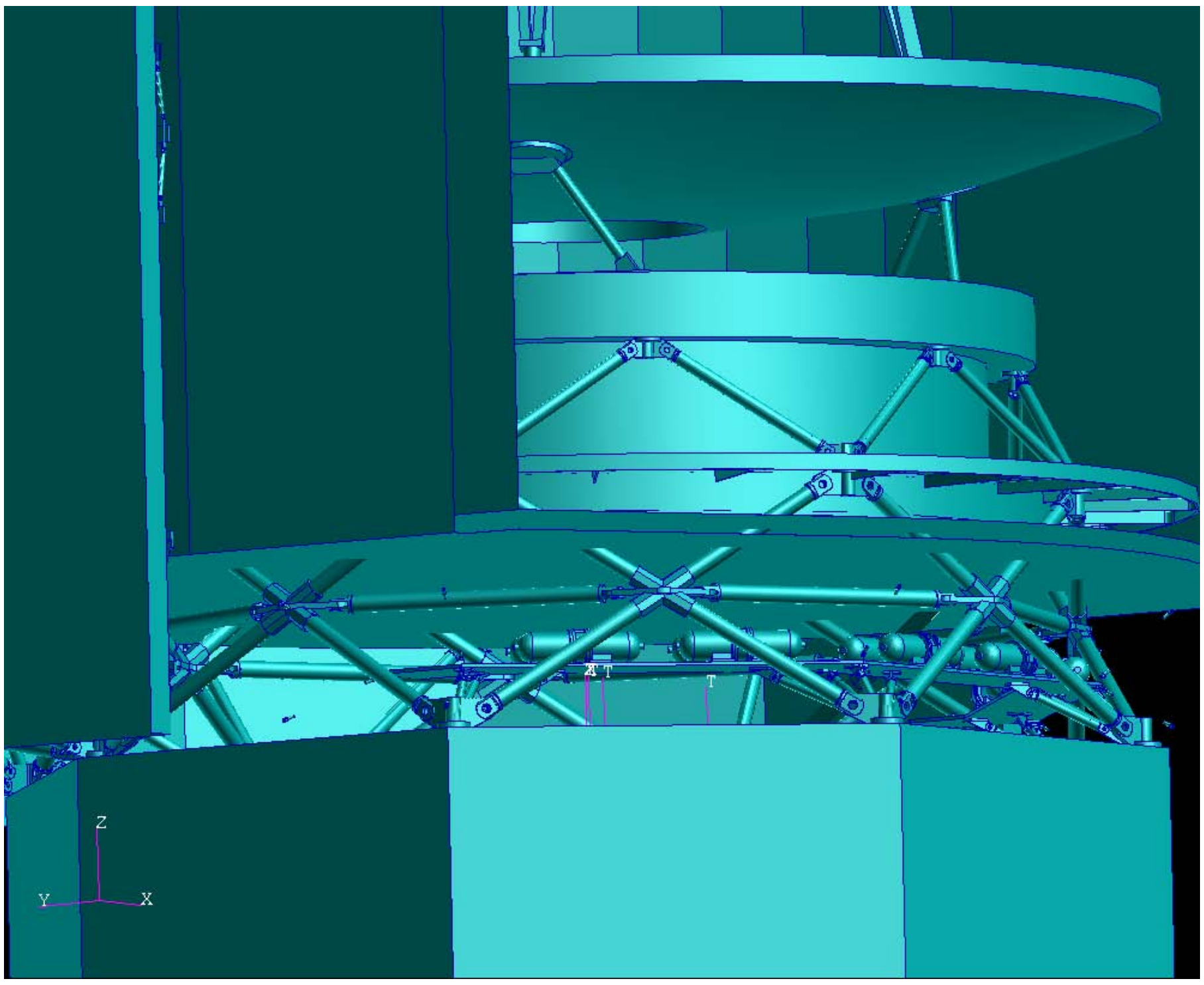

*Figure 4.9: Truss structure connecting the SPICA PLM to the SVM.*

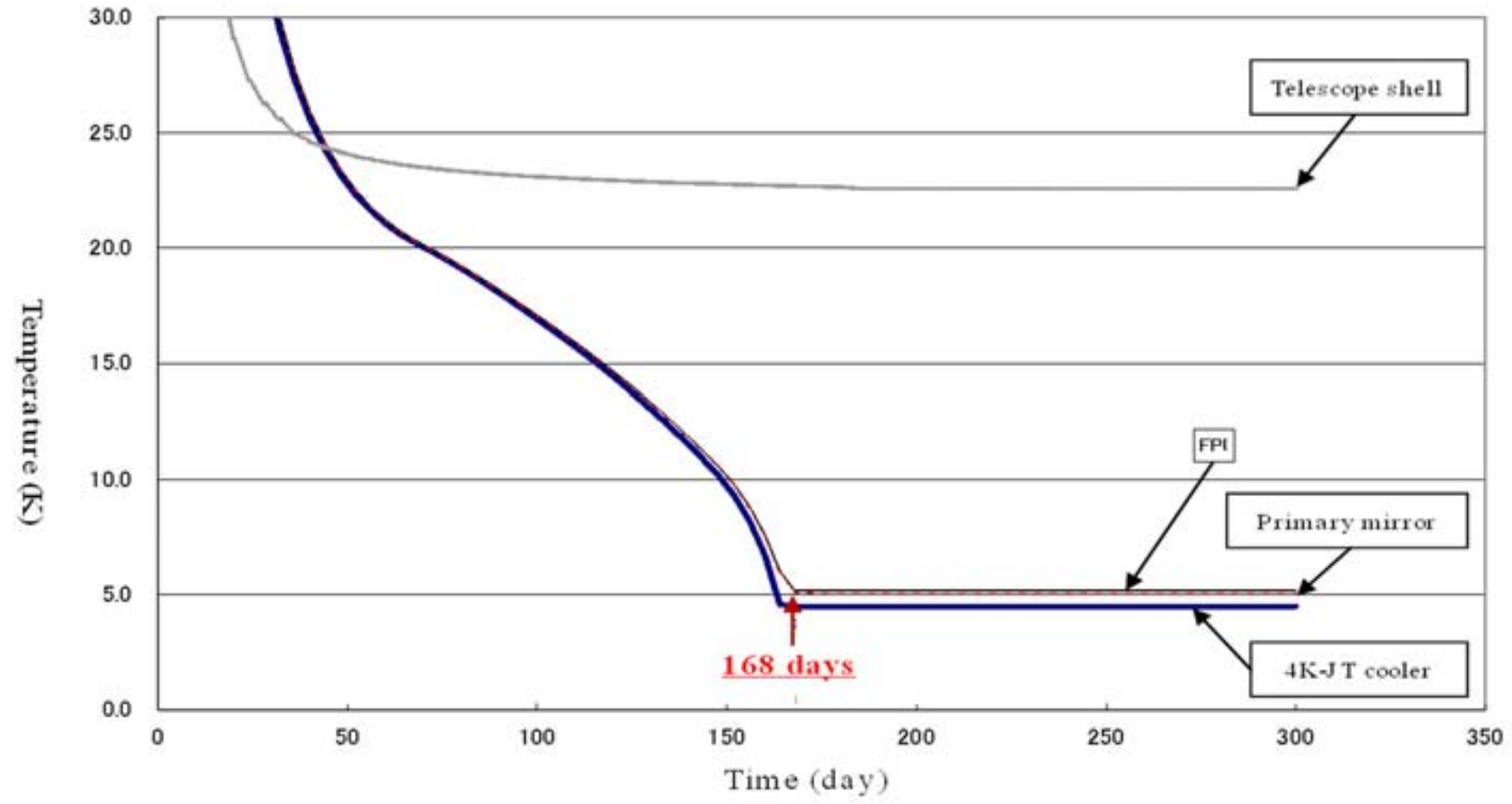

*Figure 4.10: SPICA cool-down profile.*

The SPICA Mechanical Cooling System is based on a set of advanced Stirling and Joule-Thompson coolers. The 4K Mechanical Cooler for the 4.5 K stage is a 4K-class Joule-Thomson cooler (4K-JT) connected with a pre-cooler of a 20 K-class two-stage Stirling cooler (2ST), which does not use any consumable cryogen. Far infrared instruments such as SAFARI on the IOB require further cooling to 1.7 K, reached thanks to a 1K-Mechanical Cooler consisting of a dedicated 1K-JT with $^3$He working gas and an additional 2ST pre-cooler. The overall architecture of the Mechanical Cooling System is illustrated in Figure 4.11. Specifications of mechanical coolers for the SPICA are listed in Table 4.4, also showing the good heritage existing, thanks to previous missions, such as AKARI and Astro-H.

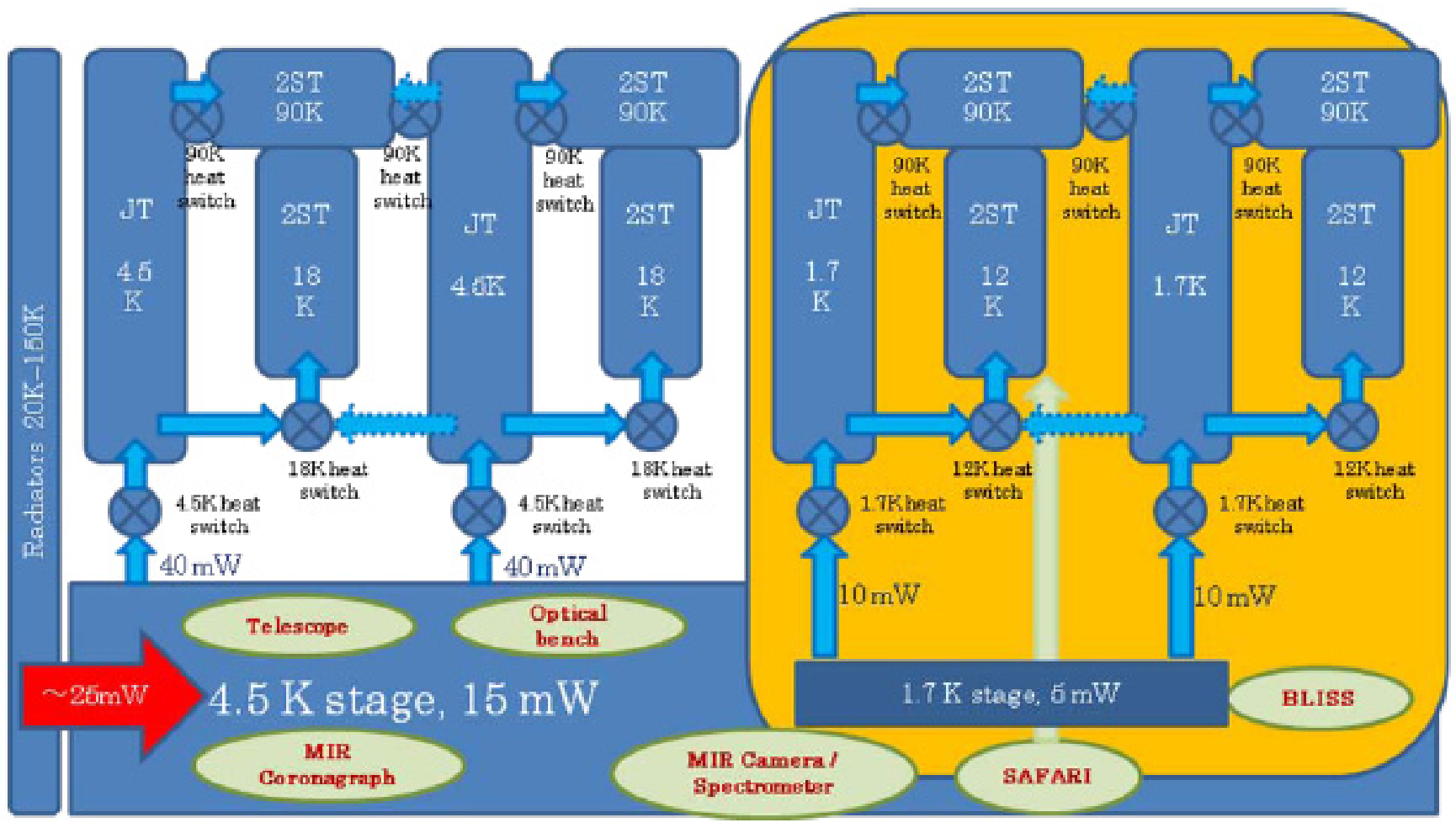

*Figure 4.11: Block-diagram of the SPICA Mechanical Cooling System.*

Although based on the heritage of the 2ST flown on board the successful AKARI, the 2ST for SPICA is required to have higher reliability, for continuous operation during more than 3 years to complete the guaranteed and extended mission phases. At the same time, the cooling capacity at 20 K has to be increased, as this largely contributes to increasing the cooling capacity of the 4K-JT and of the 1K-JT. The engineering model of the upgraded SPICA 2ST (with lower out-gassing materials) is shown in Figure 4.12. This engineering model was proved to provide higher cooling capacity, up to 325 mW at 20 K.

Based on heritage of the 4K-Mechanical Cooler developed for ISS/JEM/SMILES, the upgraded SPICA 4K-Mechanical Cooler was designed and fabricated by combining new JT heat exchangers (based on coaxial double tubes) with low pressure loss, coupled with the modified 2ST. Test results show that the maximum cooling power of 50.1 mW was efficiently obtained with an electric input power of 55.9 W (AC) for the JT compressors and of 89.2 W (AC) for the 2ST stage. The remarkable improvement of the cooling power at the 4.5 K stage is attributed to the increase of the mass flow rate; this was made possible by high-power pre-cooling at 18 K, thanks to the modified 2ST, with the extended 8 mm diameter second displacer.

Table 4.4: Specifications of the SPICA mechanical coolers.

| | 2ST | 4K-MC | 1K-MC |
|---|---|---|---|
| **Cooler type** | 2-stage Stirling | JT with 2ST | $^3$He-JT with 2ST |
| **Cooling object** | Exclusive for JT | STA, IOB and FPI | Far-IR Instrument |
| **Cooling requirement** | 200mW@20K (EOL) | 40mW@4.5K (EOL) | 10mW@1.7K (EOL) |
| **Driving power** | < 90 W | < 160 W | < 180W |
| **Heritage** | AKARI (2006-) | ISS/JEM/SMILES (2009-) | N/A |
| **R&D level** | EM for Astro-G (2013) Astro-H (2013) | BBM for SPICA | EM for Astro-H |

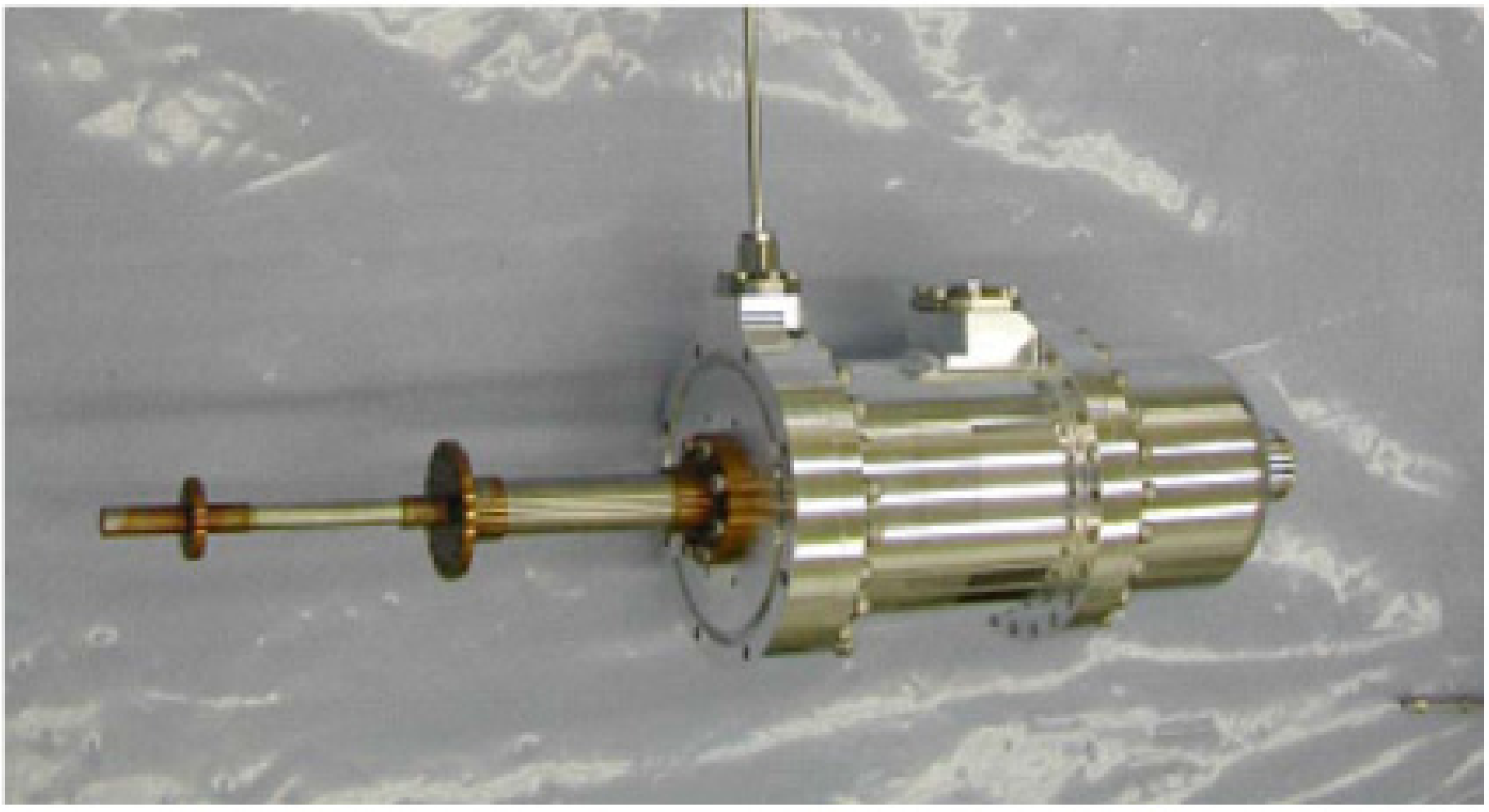

Figure 4.12: Bread board model of the upgraded SPICA two-stage Stirling cooler.

Test results for the 1K-MC indicate that the cooling power was successfully improved by 16.0 mW with an efficient input power of AC 76.6 W for JT compressors and AC 89.0 W for the powerful 2ST. Higher mass flow rate, obtained by cooling the $^{3}$He gas in the JT circuit at 12 K by the modified 2ST, drastically increases the heat lift capacity at 1.7 K.

It is notable that all the coolers satisfy the cooling requirements at the beginning of life (BOL), whereas long life tests for all the coolers are under preparation for the verification of reliability at EOL. In the SPICA Mechanical Cooling System baseline, two sets of 4K-Mechanical Coolers and 1K-Mechanical Coolers are employed for redundancy, respectively. During the development of the mechanical coolers, significant effort has been invested in minimising the generation of vibrations.

# Chapter 5

# Mission Operations

SPICA is a JAXA/ESA mission open to the general astronomical community as a multi-user observatory. The focal plane instruments are provided by consortia of scientific institutes. In particular, SAFARI is provided by a consortium of European, Canadian and Japanese institutes; MIRACLE, MIRHES and MIRMES and SCI are provided by Japanese consortia; FPC is provided by a consortium of Korean and Japanese institutes and BLISS has been proposed to NASA. The science management and the operations of the mission described here reflect the character of open observatory together with the distribution of responsibilities to the parties involved in the development of the spacecraft and the instruments.

## 5.1 Science management

The science management of the observatory accounts for the fact that SPICA is a JAXA led mission in which ESA is a partner. The scientific organisation and the allocation of observing time are defined to enable European astronomers to take maximum benefit of the observatory capabilities as a whole, and in correspondence with the relative European contribution.

### 5.1.1 Science teams

JAXA will establish a Science Advisory Committee that will consist of scientists representing all relevant fields in astronomy. It will be responsible to review the scientific objectives of the mission, monitor the development of the satellite in the context of the scientific performance, and give advice to the JAXA SPICA management team on all scientific aspects. The Science Advisory Committee will be chaired by the SPICA Project Scientist in Japan and will count with the participation of the ESA SPICA Project Scientist.

The ESA SPICA Project Scientist will be responsible for representing the interests of the astronomical European community in the Science Advisory Committee and in the interactions with the Time Allocation Committee (TAC). Together with the SPICA Project scientist, he/she will have the responsibility to maximise the scientific return of the mission, and to monitor that the development of the European elements (telescope, SAFARI, European SPICA Data Centre) is adequate to achieve the required scientific capabilities. The ESA SPICA Project Scientist will interface with JAXA, the ESA SPICA Project Team and the SAFARI PI. He/she will participate in the organisation of conferences, workshops or other activities to promote the mission and will be the interface with ESA corporate entities in support of Public Relations and outreach activities in collaboration with JAXA.

To execute these tasks, the ESA SPICA Project Scientist will be supported and advised by the European SPICA Science Team. The European SPICA Science Team will identify and advocate the interests of the European astronomers, provide input to the Science Advisory Committee, and ensure the maximum scientific return in relation with the elements of the European contribution, that is, the telescope, SAFARI and the European SPICA Data Centre. The membership will include European scientists representing the main fields associated with the SPICA science objectives, the SAFARI PI, telescope-optics experts, and a JAXA representative.

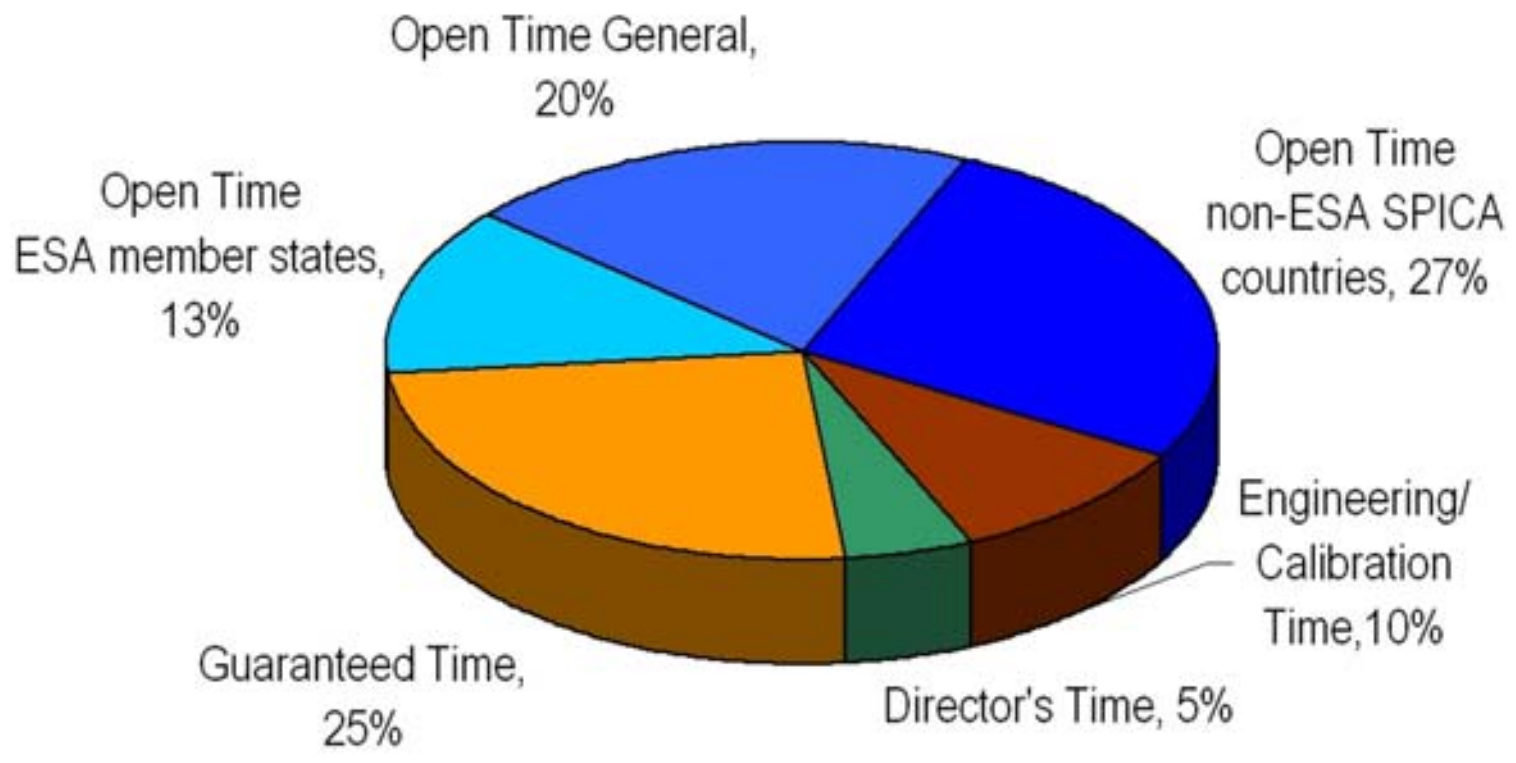

*Figure 5.1: SPICA operational time allocation.*

For community support activities, the ESA SPICA Project Scientist will be supported by the ESA SPICA Science Centre (described in section 5.3.2).

### 5.1.2 Observing time allocation

The SPICA observing time will be distributed following the principle of “guaranteed” and “open” time. Guaranteed time will be allocated to the scientific institutes directly involved in the building of the focal plane instruments. The instrument consortia will decide internally the distribution of their guaranteed time, under the final responsibility of the instrument PI. Open time will be allocated to proposals submitted by general astronomers, through a competitive process in which the TAC will provide the final recommendations.

The distribution of satellite time aims at keeping an adequate balance among the assignment of guaranteed time to the instruments scientific consortia, the availability of open time to astronomers from SPICA associated countries and the open time to the world community. In the following, the percentages of SPICA time for each category are specified (with respect to total satellite time in the Nominal Observation phase; see Figure 5.1):

- Engineering/Calibration Time: 10%
- Director’s time, including ToO observations: 5%
- Guaranteed time for SPICA team: 25%. It will be shared taking into account the relative contribution of each team to the project, and includes the guaranteed time for instrument teams, science teams members, and science operations centres.
- Open Time: 60%, which will be further subdivided as:
  - For astronomers from countries directly involved in the SPICA project: 40%
  - For the general community, without restrictions of nationality: 20%

The time allocation to European astronomers will be proportional to the European economic contribution to SPICA, which we assume will amount to one third of the total. Consequently, the fraction of European open plus guaranteed time should also be one third of the “allocated” open and guaranteed time. Within this scenario, time reserved to astronomers of the ESA member states is distributed in the following way:

- European guaranteed Time: 8.3% of total satellite time, of which 95% is reserved for the SAFARI consortium, 3.5% for the ESA SPICA Science Centre, and 1.5% for the European mission scientists.
- European open time: 13% of the total satellite time (or 22% of the total open time).

### 5.1.3 Observing programmes

Announcements of Opportunity will be made before and during the mission for SPICA Legacy Programmes and normal programmes. Legacy programmes are coherent and comprehensive observation proposals that aim to address key questions in Astronomy, and which require a large number of hours for their execution. The detailed rules for the SPICA Legacy programmes are under definition. Both Legacy programmes and normal programmes can be proposed through guaranteed and open time.

The observations in Legacy and normal programmes will have associated one year of proprietary time, counting from the moment that the observer receives the processed data. For Legacy programmes whose results are required for follow-up observations, the proprietary time may be reduced and proposers may be requested to make their results public as early as ∼ six months after receiving their observational data. Engineering and Director's time is not proprietary, that is, observations will immediately be made available.

The SPICA Science Operations Centre will distribute to the observers the raw observational data, ancillary data required for the processing (e.g., pointing) and calibrated products generated automatically by the instrument pipelines, in which the instrument effects have been removed. It is also the intention to provide a SPICA interactive analysis package for data reduction. Legacy programme teams will be responsible (together with the SOC) to make a coherent processing of their observational data and to provide highly processed products to the general community for observation follow-up and general scientific exploitation.

### 5.1.4 SPICA Observing Time Allocation Committee

A single international scientific Observing Time Allocation Committee will be established by JAXA, ESA and other potential mission partners sufficiently early before launch. The composition of the TAC will be arranged such that conflicts of interests with proposers will be avoided. It will be based on scientific excellence.

The function of the TAC will be to assess observing proposals made by the science community at large as part of the open time. In particular, the TAC will establish criteria for open time observing proposal selection, will review and categorise proposals on scientific merit, technical feasibility, and priority in light of the SPICA scientific objectives, and will recommend to JAXA and ESA the assignment of observing time. The TAC may decide to approve/reject full or partial parts of a proposal. The TAC will be supported by the SOC and by the ESA SPICA Science Centre, which will assess the proposals technical feasibility prior to the TAC meetings and provide statistics from the submitted/approved proposals on, e.g., distribution of requested time per instrument and per science area.

## 5.2 SPICA operational phases

The nominal lifetime of the SPICA mission is 3 years (guaranteed by the current design), with a goal of 5 years (expected from future developments in the cryogenic system). There are five operation phases from the time that the satellite is installed in the launch facility and the end of the mission:

- Launch Operations phase, which extends from the arrival of the satellite at the Tanegashima Space Centre until the end of launch operations.

- Initial Mission Operations phase, which starts at the end of launch operations. The satellite will be injected into the S-E $L_2$ Halo orbit. After 18 hours from the launch, the first orbit manoeuvre will be carried out. The following manoeuvres will take place on days 20 to 40. It will take the satellite approximately 120 days to be injected in the observational orbit. At this moment, we assume that it takes 168 days to cool the telescope by using 2 redundant 4K Mechanical Cooler at the same time. During the Initial Mission Operations phase, function checkout, performance verification and test observations are performed.

- Nominal Mission Operations phase, which is the “planned operation” period. During this phase, the routine scientific observations will be carried out. Nominal duration is 2-2.5 years, with the goal to achieve 4-4.5 years.
- Extended Mission Operations phase: Covers from the end of the Nominal Observation Phase to the time of termination of the satellite. This phase will start when one or more of the cryocoolers malfunction prevent nominal operations to continue. In this phase extended operation programmes will be executed with the instruments for which science operations in warmer conditions are possible.

## 5.3 SPICA operations centres

SPICA is a Japanese-led mission implemented on the basis of international cooperation. Following this principle, the mission operations will also be implemented as a collaborative effort from JAXA, ESA and the scientific institutes responsible for the provision of the SPICA instruments. The SPICA Ground Segment will consist of the following elements:

- Mission Operations Centre (MOC), located in Japan, and the associated ground stations, one of them supported from ESOC, Germany
- Science Operations Centre (SOC), located in Japan
- Japanese Instrument Control Centres
- European SPICA Data Centre, with two organisational units:
  - ESA SPICA Science Centre (ESSC), located at ESAC, Spain
  - SAFARI Instrument Control Centre (ICC). The SAFARI ICC will be geographically distributed in several European countries. The contact to the distributed ICC will be via the PI institute at SRON Groningen.

### 5.3.1 Mission Operations Centre and ground stations

The SPICA Mission Operations Centre (MOC) will be set up in JAXA and will be responsible for all spacecraft operations, including routine observations and contingency plans. The main functions of the MOC will be to generate and upload commands and receive telemetry data. The operations centre will control the conditions of the SPICA observatory by monitoring the housekeeping data and will forward to the Science Operations Centres the science data. The overall role of the SPICA MOC, its interfaces to the Science Operations Centres (located in Japan, Europe and possibly other countries depending on the consolidated cooperation scenario) is illustrated in Figure 5.2.

The prime ground station for the SPICA mission is the 64 m Usuda station. In the present spacecraft design, the telemetry downlink will be performed using X-band. A low speed data link in S band will also be available. The required downlink capacity is in the order of magnitude of 350 Gbit/day. This is more than the 316-237 Gbit/pass achievable with Usuda at 11 Mb/s and 8-6 hr dump duration (X-band). The JAXA downlink baseline is QPSK modulation and Reed Salomon coding with a maximum data rate of 11 Mbit/s. This would in fact lead to a daily downlink volume of about 316 Gbit/day in case of a 8 hr window.

The downlink, subject to final agreements, could be supported by an ESA ground station at a different longitude than Usuda, i.e., Cebreros and/or the new DS3 ground station in South America. A downlink rate of order of 9 Mb/s seems feasible for Cebreros (the same data rate can also be used for DS3). It is assumed that a respective variable coding is provided by the SPICA spacecraft. Given the envisaged loading of the ESA Deep Space Network at the time of the SPICA operations, it is realistic to assume a maximum dump duration of 4 hr/day. In this case the additional downlink volume given by the ESA ground station is about 130 Gbit/day. Thus the combined JAXA-ESA daily downlink volume would be about 446 to 367 Gbit/day, compatible with the mission needs. The actual need for an additional ground station needs to be confirmed, pending a final estimate of the actual payload demands and a more detailed evaluation of the SPICA telemetry data rate/volume needs.

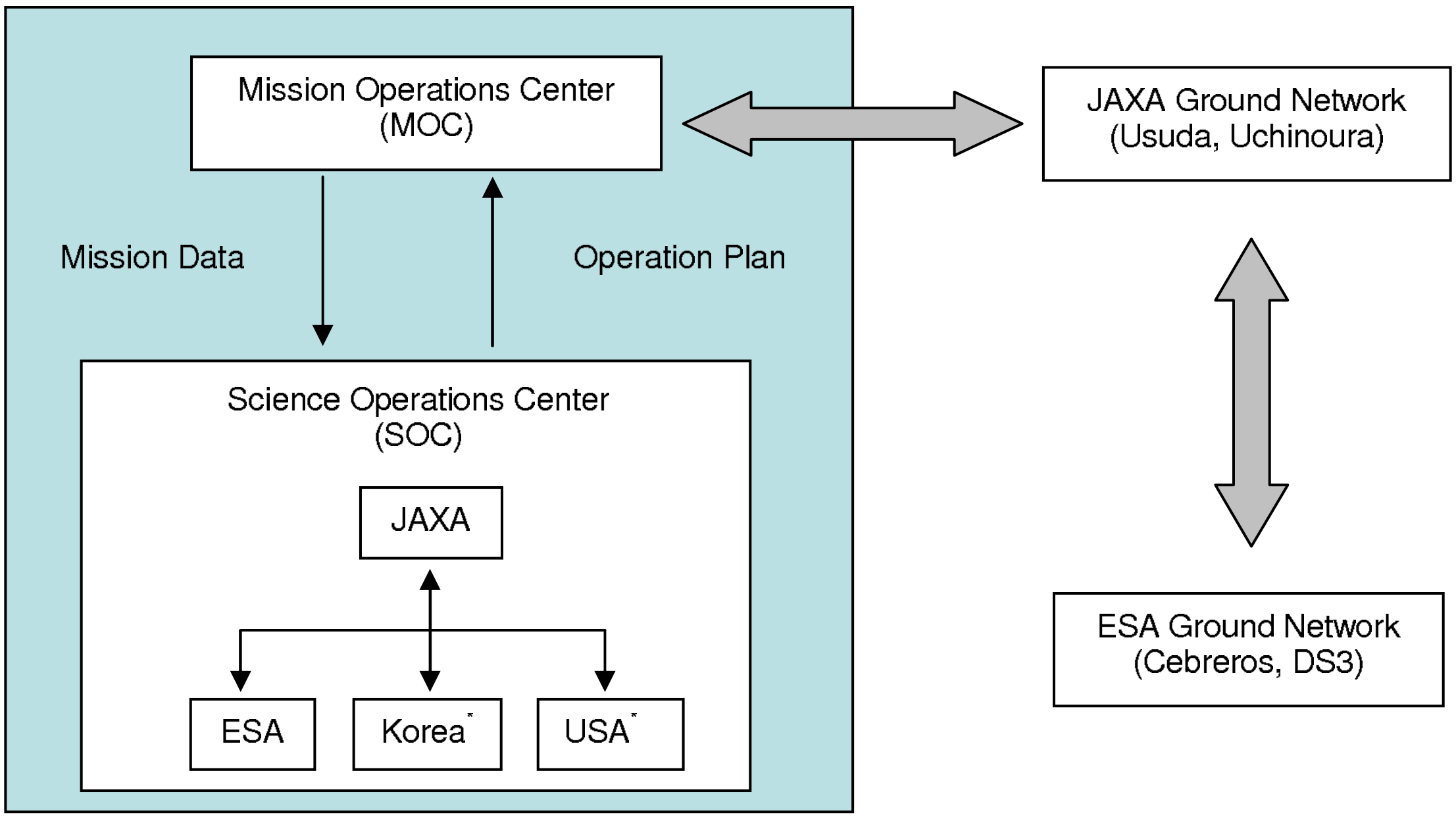

*Figure 5.2: Overall MOC and SOC interfaces (*: under discussion).*

It should be noted that the SPICA ranging standard is based on digital ranging different from the ECSS or CCSDS standards. Thus ranging could not be provided for SPICA by the ESA ground stations. The envisaged ESA involvement in the SPICA mission would not include any spacecraft or payload operations tasks (performed by JAXA). The ESOC involvement in the SPICA mission includes ground segment support in the form of ground station data relay services and data routing services to JAXA, as well as the necessary respective preparation and operations activities. The following high level requirements are presently considered for the ESA contribution to the SPICA ground segment:

- No ESA coverage during LEOP and transfer phase (ESA coverage starting during the commissioning phase).
- Science data reception at the Cebreros/DS3 ground station and off line data delivery in Japan to JAXA.
- 4h daily data dump duration at the Cebreros/DS3 ground.
- Firm ground segment availability and data return requirements: 95% (measured over any half year period).
- Timeliness: 24 hr for science data and real time (latency: TC 5 s TBC, TM 15 s TBC) for TC and HKTM.
- Planning to be compatible with long term planning (~ half a year in advance), minor changes can be processed up to 1 month ahead of the actual passes.
- Contingency services for the SPICA mission shall be made available on the basis of the general ESA – JAXA mutual support agreements.
- On line interface to be activated in case of acquisition problems.
- All facilities established for SPICA shall be based on the existing ground-segment infrastructure (only limited tailoring, such as software updates in relation to the final coding choice).

The ground segment preparations on the ESA side would start 3 years prior to launch. The readiness of the ground segment would be validated by a series of test campaigns, e.g., Radio Frequency Compatibility Test, Data Flow Test, System Validation Test. Besides software updates related to final coding choice, we are presently assuming that no additional infrastructure update is required.

As the spacecraft is built by and the mission is operated by JAXA, it must be assumed that the JAXA packet structure is applied, which is different from ESA's Packet Utilisation Standard (PUS). This is not a problem for the ESOC services, because they can be handled on CCSDS packet level, but instrument operations will be affected. Changes to existing instruments EGSE (e.g., if Herschel Ground Support Equipment was to be reused) may be required. The impact on command generation tools for the instruments needs to be investigated.

### 5.3.2 Science Operations and Instrument Control Centres

**Science operations concept**

SPICA is an observatory type mission open to the general astronomical community. Dedicated astronomical observations will be carried out following the specifications from the observers. Astronomers will be invited to submit proposals following Announcements of Opportunity. Observations will be specified following Astronomical Observation Templates (AOTs) based on pre-defined observing modes that will be optimised for each type of astronomical measurement to be performed with SPICA. Submitted proposals will be reviewed by the SPICA Time Allocation Committee. Approved programmes will be stored in the mission database with the detailed specification of the corresponding observations. It is assumed that the submission of proposals will be done electronically at the SOC, where the mission database will be administrated.

Scientific mission planning will be carried out at the SOC to generate the daily observation schedules. Because of the limited duration of the SPICA mission, it is essential that the planning is optimised and that maximum observation efficiency is obtained. Scientific schedules will include the sequence of scientific observations to be performed and the required associated instrument commands. Scientific schedules will be send to the MOC, where they will be completed and uploaded to the satellite.

The SOC will systematically process raw observational data through the pipelines to standard products and will add quality information to each observation. Standard products will contain calibrated data in which the instrument artefacts have been removed. Products should be processed to a level that will allow astronomers to perform scientific analysis and extract astronomical information, and such that they can be made available in multi-observatory environments like the Virtual Observatory.

In addition to standard processing, a dedicated interactive analysis package will be provided to the astronomical community for the interactive data reduction of the SPICA data. Observation standard products generated at the SOC will be made available to the observers and the astronomical community through the SPICA science archive, following the user rights and proprietary time policy specified in the SPICA Science Management Plan. The SPICA science archive will be located in Japan, with a mirror at ESAC. In addition, a European SPICA Data Centre Archive will be implemented to store SAFARI and related ancillary data produced from the beginning of the instrument level tests until the end of operations.

**Science Operations Centre**

The SOC is the main Science Operations Centre, and it is the interface to the MOC for scheduling, commanding and telemetry reception. In particular, responsibilities of the SOC include the of issue AOs with the required documentation, the administration of the observations database and the generation of observation schedules to be delivered to the MOC. The SOC will be the main responsible for the implementation of the data reduction pipelines and the systematic data processing (in collaboration with the ICCs and the ESA SPICA Science Centre) to generate the SPICA science products that will be distributed to the observers. The SOC will support non-European astronomers on the preparation of proposals and on data reduction.

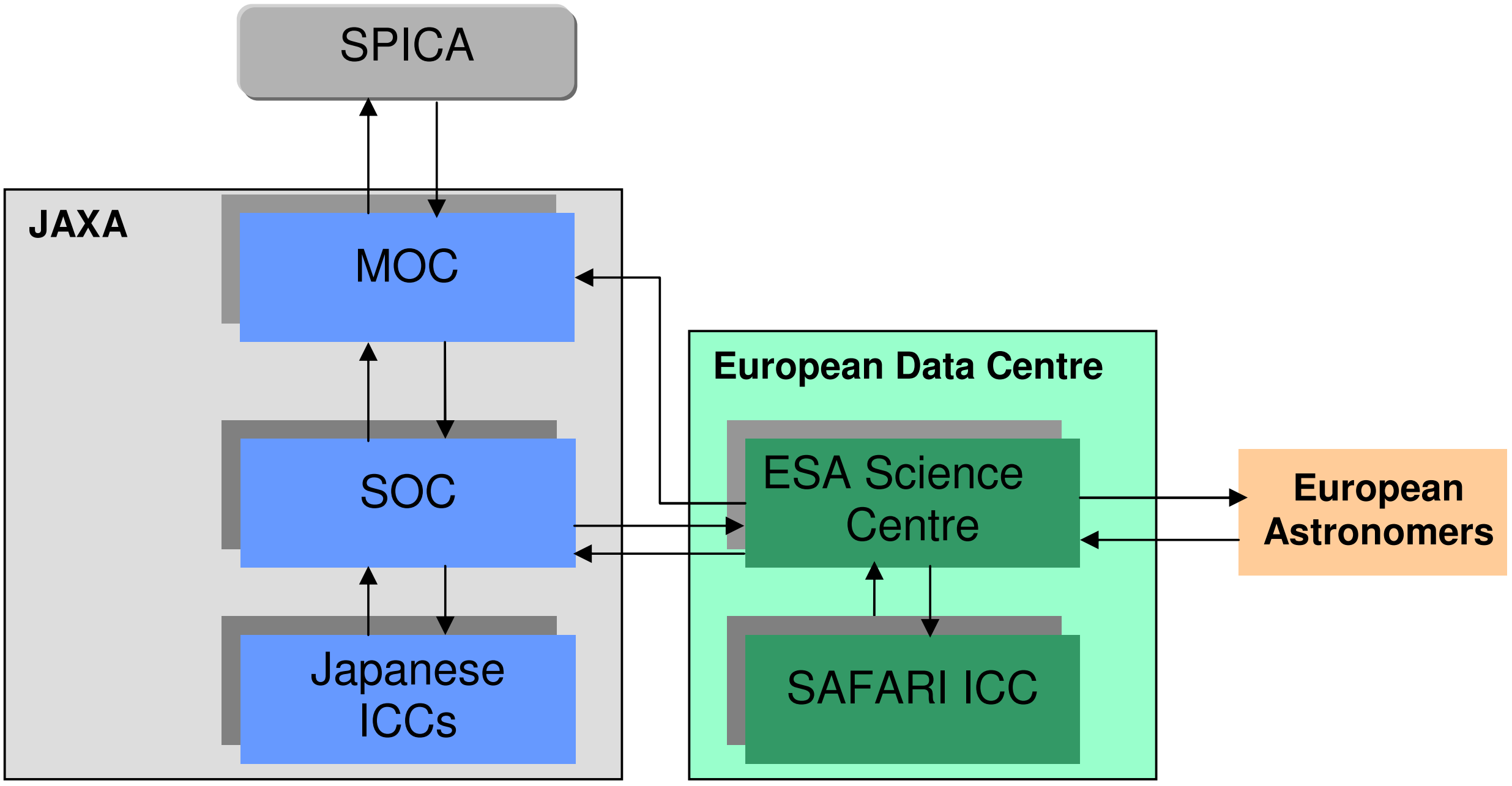

*Figure 5.3: The European SPICA Data Centre in the context of the SPICA ground segment*

### European SPICA Data Centre

The European SPICA Data Centre will consist of two elements: The ESA SPICA Science Centre and the SAFARI ICC. Figure 5.3 shows the European SPICA Data Centre in the context of SPICA ground segment and the general approach for interactions between centres.

The purpose of the ESA SPICA Science Centre (ESSC) is to act as interface for the European community to the SPICA observatory, and to represent the European users in the definition and testing of the science operations systems. In particular, the ESSC will support European observatory users in proposal submission, observation preparation and data reduction and analysis for all SPICA instruments.

The ESA SPICA Science Centre will also be the interface between the SAFARI ICC and the JAXA SOC and MOC for all SAFARI operational matters. This will involve the identification and definition of the ground segment interfaces, both operational and for exchange of data, and the definition with the SAFARI ICC of the instrument telecommands and instrument telemetry contents following the JAXA packet structure (which, as explained above, is different from ESA's Packet Utilisation Standard). The ESSC will support the ICC in the preparation of the SAFARI operational procedures following JAXA's operational requirements. With regard to software development, the ESSC will define the S/W interfaces to the JAXA system, carry out integration and testing and be responsible of the Quality Assurance for all S/W SAFARI items that will be delivered to the MOC/SOC (observing modes, pipeline, documentation). SAFARI documents intended for the general observers (Observer's Manual, Data User's Manual) will be prepared by ESSC personnel in collaboration with the SAFARI ICC. As an additional contribution to the observatory, the ESSC will implement the proposal submission system by re-using and adapting S/W implemented for the Hershel Space Observatory. Other possible contributions in terms of software recycling are under discussion.

The SAFARI Instrument Control Centre (ICC) will be set under the responsibility of the SAFARI PI, and will start its activities early on for the definition and execution of the instrument level tests. It will be geographically distributed and several national agencies have expressed an interest in participating in the ICC activities. The precise distribution of tasks and responsibilities has not yet been defined but in any event the PI institute will take overall management responsibility for the ICC and act as the point of contact between the consortium and the ESSC. The SAFARI ICC will be responsible for the calibration and characterisation of the instrument before and after launch, for which it will develop the required software, like e.g., the quick-look analysis and calibration analysis packages. The SAFARI ICC will provide the instrument command generation facilities, and define and validate the observing modes and the astronomical observation templates for the execution of the observations. It will also be responsible for the development of the software for the standard generation of observational products and for interactive data analysis, in collaboration with the ESSC. During operations, the SAFARI ICC will be in charge of the instrument on-board software maintenance, and will be the final responsible in relation with instrument health and operational procedures.

# Chapter 6

# Management

## 6.1 The SPICA Steering Group

SPICA is an international cooperation project led by JAXA, which is responsible for the overall project. If SPICA is selected as one of the Cosmic Vision programmes, ESA will be the prime project partner, responsible for development of the SPICA Telescope Assembly and for the delivery of the SAFARI instrument. Figure 6.1 shows the presently envisaged international cooperation scheme and the role of each country.

It is planned to establish a SPICA Steering Group early in the program prior to the formal agreement by both space agencies. This Steering Group, which will be chaired by a JAXA representative, will consist of members from ESA, JAXA and potentially other national funding agencies. During the System Definition Phase, the content of the formal agreement between the two agencies would be drafted by the Steering Group (i.e., agreement on the details of the ESA deliverables and overall joint SPICA program level agreements). The second role of the Steering Group, which will remain in place until the end of the SPICA operations, is to monitor the overall technical, financial and programmatic status of the mission. It will have the authority to make high level trade-off decisions when required (impacting on cost, schedule, risk, science performance etc.) and take executive actions to deal with problems arising throughout the programme. Additionally, a Joint Systems Engineering Team (JSET) will be set up with the aim to provide a forum for discussion and mutual debriefing on the mission development, thus avoiding any ”hidden gap” in the context of a collaborative project. JSET is expected to be co-chaired by JAXA and ESA and to consist of members from the project teams of both parties. In parallel a Science Advisory Committee will be established. This Advisory Committee, which will be chaired by the SPICA Project Scientist, will consist of a number of scientists representing important fields of astronomy and will count with the participation of the ESA Project Scientist. The main role of the Science Advisory Committee will be to review the project status and to advice the programme on science related matters.

## 6.2 ESA contribution to SPICA

The ESA contribution to the SPICA project includes: a) the SPICA Telescope Assembly; b) provision of ground segment support; c) interface management and responsibility for the delivery to JAXA of the nationally developed SAFARI instrument; d) provision of the ESA SPICA Science Centre. ESA will procure the Telescope Assembly from the European industry, following a parallel competitive definition phase (A/B1) and the selection of a single contractor at the beginning of the implementation phase (B2/C/D). The SPICA Telescope Assembly includes the following elements:

- SPICA Telescope, including optical elements (primary and secondary mirror), M2 support structure, M2 refocusing mechanism with drive electronics, internal telescope baffles, adjustment shims, integration alignment devices.
- Telescope Optical Bench, as required to provide adequate mounting and mechanical support to the telescope elements and interface to the S/C structure and the Instrument Optical Bench (IOB).

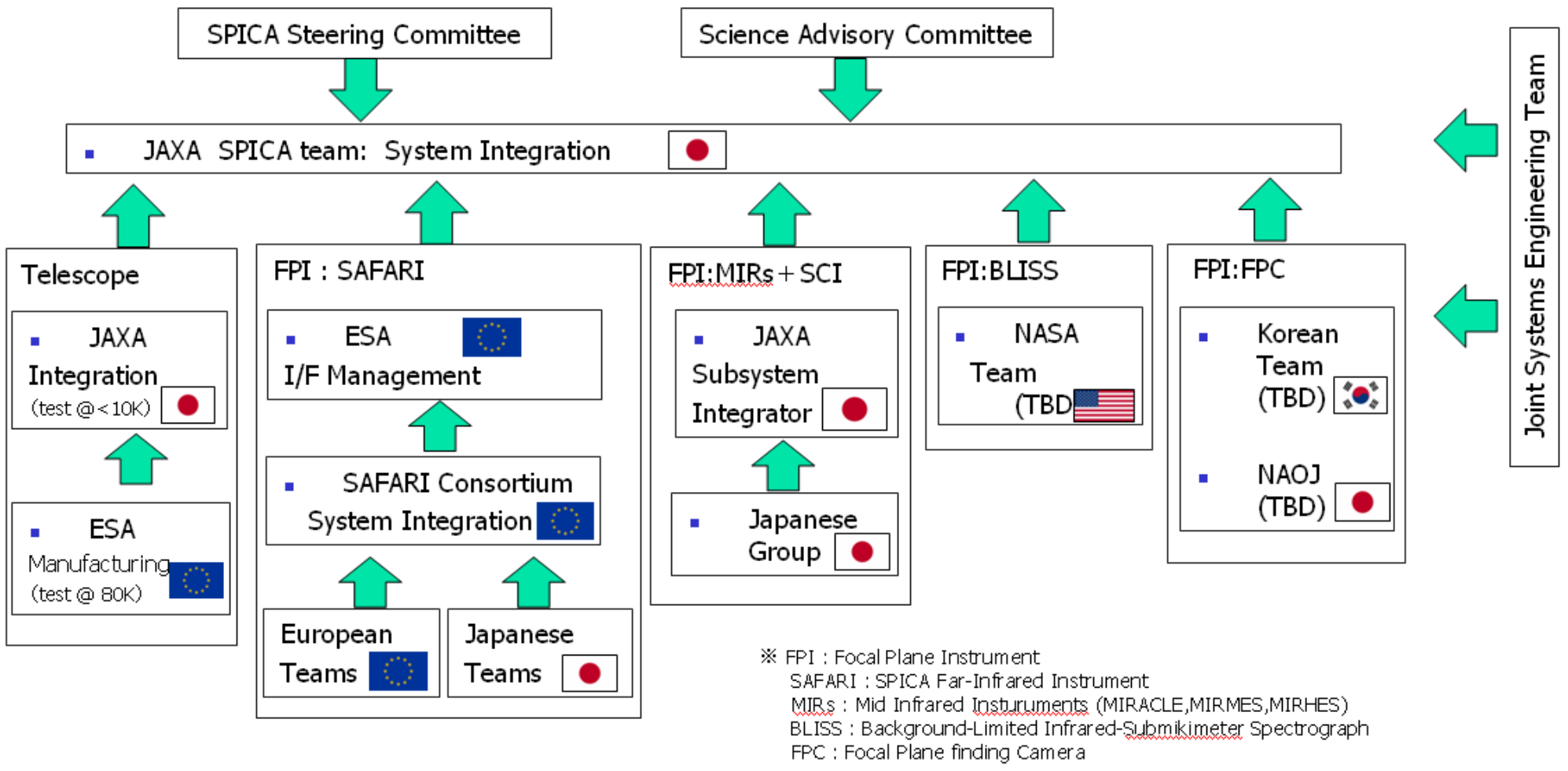

*Figure 6.1: SPICA international cooperation scheme.*

- Thermal control hardware (MLI, electrical heaters, harness, temperature sensors) as required to monitor and control the telescope temperature, including decontamination procedures.
- Thermal interfaces to the S/C cooling chain (anchoring for thermal straps).
- Required transport, handling and Ground Support Equipment.

The provision of ground segment support (in the form of data relay services from ESOC) is subordinated to the availability of a suitable ground station during the SPICA operations. Presently it is envisaged to baseline a 4 hr time window from Cebreros (or from the planned DS3 as an option).

ESA will take responsibility for the delivery to JAXA of SAFARI and will play the role of official interface between JAXA and the national consortium developing the instrument. In order to properly conduct this role, ESA will actively monitor the instrument development activities and the fulfilment of all interface requirements agreed with JAXA. In particular ESA will be responsible for conducting the review cycle, Flight Acceptance of the instrument, and co-ordination of system level activities with JAXA (integration and test at spacecraft level, launch campaign, LEOP and in orbit commissioning). It is presently planned that the European SPICA Science Centre will be based at ESAC and will facilitate the interface between the European astronomers and the SPICA project, providing adequate support to access the observatory capabilities.

## 6.3 SPICA project activities at JAXA

JAXA approved the SPICA project entering the so called pre-project phase (phase A) on 8 July 2008, with the nomination of a study team. The approved pre-project activities include a 2 year study phase, leading to the System Requirements Review (SRR - Q4/2009) and to the System Definition Review (SDR - Q4/2010). The next JAXA approval milestone is the formal Project Approval Review, planned by mid 2011 and enabling the project to enter the implementation phase (B/C/D). The JAXA SPICA pre-project team is presently working with three companies: Sumitomo Heavy Industry (SHI), NEC and MELCO. SHI has advanced space cryogenic technology and is studying the Payload Module. NEC and MELCO are satellite system experts, each of whom is working with JAXA individually. JAXA has contracts with these two companies and will define the system requirement before the SRR, also using elements provided by industry. After SRR, JAXA will submit the Request for Proposal (RFP) to industry and select a prime contractor for the SPICA project.

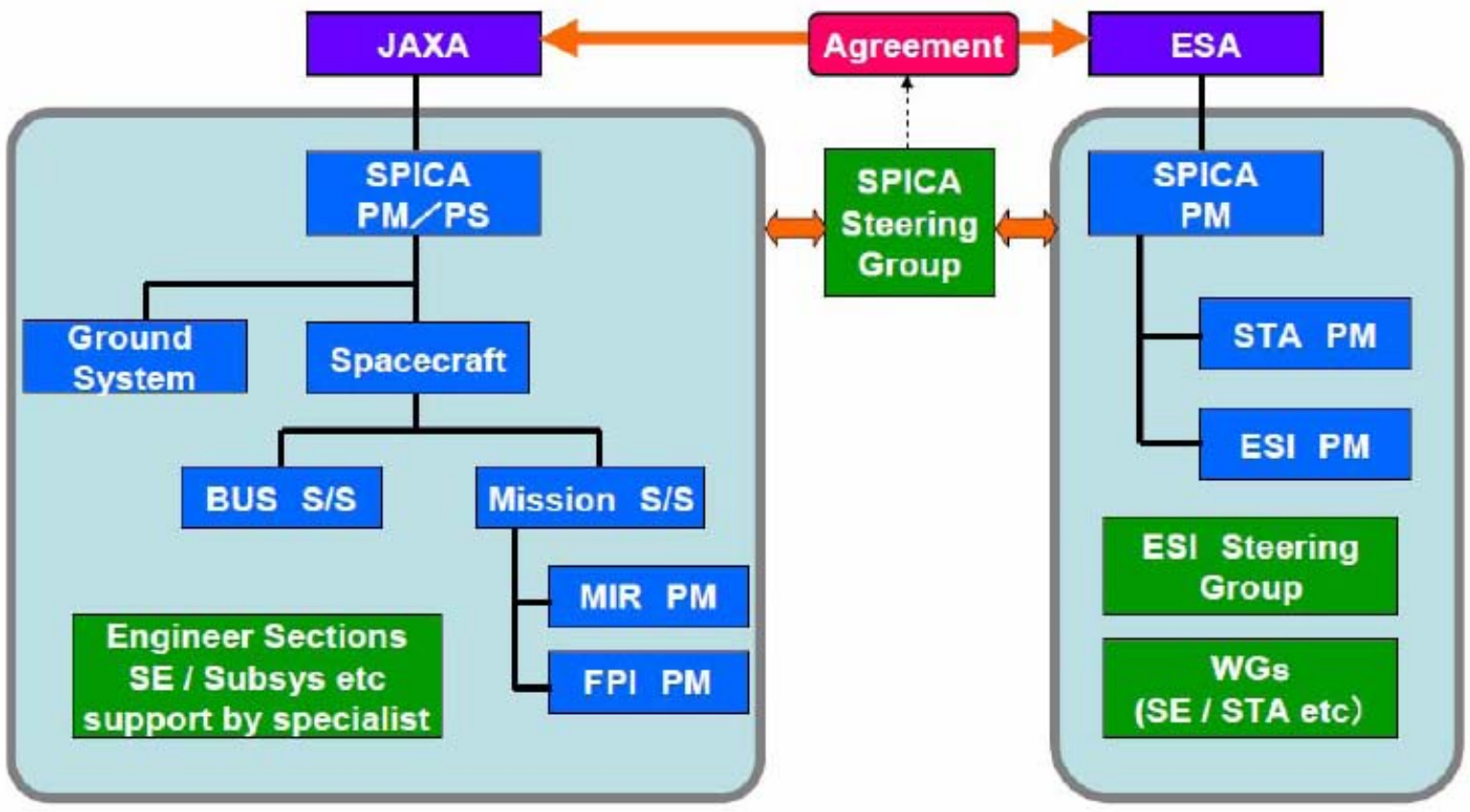

Figure 6.2: SPICA Management Structure

## 6.4 ESA/JAXA interfaces

A successful SPICA cooperation depends on the effective definition and management of the interfaces between JAXA's led activities and the elements provided by Europe. During the assessment phase, a continuous communication flow has been maintained between the JAXA and ESA study teams, leading to the definition of interface requirements and to the common understanding of the performance requirements applicable to the telescope assembly, to the ground segment and to the SAFARI instrument. This work, in addition to the consolidation of the telescope requirements document, has in fact resulted in the compilation of Interface Control Specifications applicable to both the STA and the SAFARI instrument. Under the assumption of SPICA entering the Definition Phase (A/B1), further progress will be made possible by the JAXA System Requirements Review (Q4/2009) and by the following System Definition Review (Q4/10). The assessment study has confirmed the critical role played by an early definition of the interface requirements applicable to the telescope in order to deliver the flight model in a timely manner: achieving such an early definition will be one of the challenges ahead of the ESA and JAXA teams.

For the definition phase an "exchange of letters" between ESA and JAXA is envisaged, to define the intended activities and to prepare the way to a formal Memorandum of Understanding (MoU), which can only be signed once the mission is approved for implementation. A single MoU is envisioned to cover all European contributions to the JAXA led mission (i.e., including the provision of the SAFARI instrument).

The issue of space standards applicable to the SPICA project has been only marginally addressed during the Assessment Study and will need further discussions in the next project phases. It is presently envisaged to follow an approach similar to the one already applied by ESA to the James Webb Space Telescope contributions, with each Agency applying its own standards to the respective contributions and a mutual recognition of the rules applied (without excluding the possibility of detailed analysis/discussions on specific and critical issues).

## 6.5 SAFARI procurement and management

In case of SPICA selection, the procurement of SAFARI will be the subject of an ESA Announcement of Opportunity to be issued, in early 2010. The SPICA far infrared instrument, SAFARI, is a nationally funded, Principal Investigator (PI) provided instrument. The development of the instrument will be led by a consortium of national institutes, under the leadership of an European PI and of the Instrument Project Manager. The PI will be responsible for guaranteeing the scientific performance of the instrument and its delivery within the overall schedule and financial constraints. The SAFARI consortium is led by SRON (Space Research Organisation of the Netherlands). The SAFARI consortium has been studying the scientific potential and technical requirements for over five years and has established that the technical capability exists within the consortium to deliver the instrument and its constituent subsystems.

Significant contributions to the instrument development are expected to come from Austria, Belgium, Canada, Denmark, France, Germany, Ireland, Italy, the Netherlands, Spain, Switzerland, the UK and, possibly, Japan. In all cases, these funding agencies are already aware of SAFARI and have been actively engaged in discussion during the assessment study. Each institution collaborating on SAFARI will have a nominated Co-I or Co-PI depending on the level of funding from a particular nation/institute to manage the work packages for that institution. An ESA sponsored Multi-Lateral Agreement (MLA) will be established to be signed by the entire consortium at funding agency level. Individual institutes will be asked to produce development plans and schedules for their deliverable items, in consultation and agreement between the PI, institute Co-I, SAFARI project manager and the institute project manager. Issues associated with resources within the consortium will be governed by the MLA and referred to a project steering committee attended by the PI, project manager and representatives of the nation funding agencies, ESA and JAXA.

The SAFARI consortium has performed a preliminary assessment study on the instrument (see chapters 2 and 3), provided a reference design that envelopes the required instrument resources in a "design stressing" (worst) case, and identified a road map, which shall lead to the selection of the focal plane detector technology and the instrument baseline design. As described in section 6.2, it is envisaged that ESA will be responsible for the delivery of SAFARI to JAXA and will act as official interface between the instrument consortium and the Japanese Agency.

## 6.6 SPICA schedule and coordination with JAXA

In case of SPICA selection, the Definition Phase (A/B1) study on the telescope is expected to start in July 2010 for a period of 16 months, with the objective to enable the project final adoption in early 2012. It will include two major reviews: the Preliminary Requirements Review (PRR), to be held by the mid-term of the study, and the System Requirements Review (SRR), which will close the Definition Phase. The Technology Development Activities (TDA's) will be initiated as soon as possible after the mission down-selection in February 2010. These activities will run in parallel with the Definition Phase and their intermediate results will be fed into the telescope study as necessary. Results from the TDA's which are critical to ensuring the project feasibility or its development schedule are expected to be available before the decision for the mission final adoption. At the PRR, the mission baseline should be well established and documented. It will be critically reviewed, with the aim of confirming the technical and programmatic feasibility of the space segment, and more generally of the overall mission concept. The System Requirements Review will close the Definition Phase by consolidating the overall mission concept for enabling an efficient start of the Implementation Phase, should the mission be finally adopted.

The coordination of JAXA's and ESA's schedules is subordinated to the respective project down-selection and approval processes. Despite of the different approval schemes, the coordination of the SPICA schedule is quite good, with a final go ahead from JAXA not before mid-2011 and a possible final approval from ESA by Feb 2012 (SPC). Given these two milestones and the parallel competitive approach used at ESA, the start of the telescope phase B2/C/D activities could not take place before Q2/2012, leading to a delivery of the flight model not before Q4/2016 (depending on specific options). This delivery date would allow a launch by the end of 2018 as from JAXA's schedule. Further improvements to the schedule synchronisation would be enabled by the delivery to JAXA of intermediate telescope models, such a structural model, allowing system level tests at an earlier stage.

## 6.7 Industrial organisation of ESA contribution

As for all other Cosmic Vision missions, the industrial Phase A/B1 will be opened for competition in early 2010, with the possibility of running two parallel industrial contracts. The final industrial organisation will be completed only in Phase B2, mostly through a process of competitive selection and according to the ESA Best Practises for subcontractor selection, by taking into account geographical distribution requirements.

# Acronyms

| | | | |
|---|---|---|---|
| **AD** | Applicable Document | **ADR** | Adiabatic Demagnetisation Refrigerators |
| **AGB** | Asymptotic Giant Branch | **AGN** | Active Galactic Nucleus/Nuclei |
| **AIV/T** | Assembly Integration Verification/Testing | **ALFRP** | Alumina Fibre Reinforced Plastics |
| **ALMA** | Atacama Large Millimeter Array | **AO** | Announcement of Opportunity |
| **AOCS** | Attitude and Orbit Control System | **AOT** | Astronomical Observation Template |
| **APE** | Aperture Pointing Error | **BFL** | Back Focal Length |
| **BHAR** | Black Hole Accretion Rate | **BLAST** | Balloon-borne Large-Aperture Sub-millimeter Telescope |
| **BLISS** | Background-Limited Infrared-Submillimeter Spectrograph | **BM** | Bus Module |
| **BOL** | Beginning Of Life | **BRDF** | Bi-directional Reflectance Distribution Function |
| **CAD** | Computer Aided Design | **CCSDS** | Consultative Committee for Space Data Systems |
| **CDF** | Concurrent Design Facility | **CDM** | Cold Dark Matter |
| **CEA** | Commissariat à l'Énergie Atomique | **CFRP** | Carbon Fiber Reinforced Plastics |
| **CIRB** | Cosmic InfraRed Background | **CMB** | Cosmic Microwave Background |
| **COB** | Cosmic Optical Background | **CSE** | CircumStellar Envelope |
| **CSO** | Caltech Sumbmillimeter Observatory | **CTE** | Coefficient of Thermal Expansion |
| **CXB** | Cosmic X-ray Background | **DC** | Digital Converter |
| **DPU** | Data Processing Unit | **DS3** | Deep Space Station 3 |
| **EADS** | European Aeronautic Defence and Space company | **ECSS** | European Cooperation for Space Standardisation |
| **EFL** | Effective Focal Length | **EGP** | Exo-Giant Planet |
| **EGSE** | Electrical Ground Segment Equipment | **EMI** | Electro-Magnetic Interference |
| **EOL** | End Of Life | **EP** | Exo-Planet |
| **ESA** | European Space Agency | **ESAC** | European Space Astronomy Centre |
| **ESOC** | European Space Operations Centre | **ESSC** | European Spica Science Centre |
| **ESTEC** | European Space research & TEchnology Centre | **EVLA** | Expanded Very Large Array |
| **FIR** | Far Infrared | **FM** | Flight Model |
| **FOV** | Field Of View | **FPC** | Focal Plane finding Camera |
| **FPI** | Focal Plane Instrument(s) | **FS** | Flight Spare |
| **iFTS** | imaging Fourier Transform Spectrometer | **GALEX** | GALaxy evolution EXplorer |
| **HII-A/B** | JAXA launcher vehicle | **HGA** | High Gain Antenna |
| **HKTM** | House Keeping Telemetry | **HW** | Hardware |
| **I/F** | Interface | **ICC** | Instrument Control Centre |
| **ICS** | Interface Control Specification | **IGM** | InterGalactic Medium |
| **IOB** | Instrument Optical bench | **ISAS** | Institute of Space and Astronautical Science |
| **ISO** | Infrared Space Observatory | **ISS** | International Space Station |
| **IWA** | Inner Working Angle | **JAXA** | Japan Aerospace Exploration Agency |
| **JEM** | Japanese Experiment Module | **JSET** | Joint Systems Engineering Team |
| **JWST** | James Webb Space Telescope | **KBO** | Kuiper Belt Object |
| **KID** | Kinetic Inductance Detector | **LEOP** | Launch and Early Orbit Phase |
| **LGA** | Low Gain Antenna | **LHB** | Late Heavy Bombardment |
| **M1, M2 ...** | Mirror 1 (primary mirror), Mirror 2 (secondary mirror), ... | **MGA** | Medium Gain Antenna |
| **MIR** | Mid-InfraRed | **MIRACLE** | Mid-InfRAred Camera w/o LEns |
| **MIRHES** | Mid-InfraRed High resolution Echelle Spectrometer | **MIRMES** | Mid-InfraRed Medium resolution Echelle Spectrometer |
| **MLA** | Multi Lateral Agreement | **MLI** | Multi Layer Insulation |
| **MOC** | Mission Operations Centre | **NEP** | Noise Equivalent Power |
| **NIR** | Near InfraRed | **NLR** | Narrow Line Region |
| **OWA** | Outer Working Angle | **QM** | Qualification Model |
| **QPSK** | Quadrature Phase Shift Keying | **PACS** | Photodetector Array Camera and Spectrometer |
| **PAH** | Polycyclic Aromatic Hydrocarbons | **PDR** | Photon Dominated Region |
| **PI** | Principal Investigator | **PIAA** | Phase Induced Amplitude Apodization |
| **PLM** | PayLoad Module | **PSF** | Point Spread Function |
| **PST** | Point Source Transmittance | **PTV** | Peak-To-Valley |
| **PUS** | Packet utilisation Standard | **RD** | Reference Document |
| **RF** | Radio Frequency | **rms** | root mean square |
| **RPE** | Random Pointing Error | **Rs** | Symbol Rate |
| **RT** | Room Temperature | **RW** | Reaction Wheel |
| **S/C** | Spacecraft | **SAFARI** | SpicA FAR-infrared Instrument |
| **SCI** | SPICA Coronagraph Instrument | **SCUBA** | Submillimetre Common-User Bolometer Array |
| **SDR** | System Design Review | **SE** | Sun-Earth |
| **SED** | Spectral Energy Distribution | **SF** | Star Formation |
| **SFR** | Star Formation Rate | **SKA** | Square Kilometre Array |
| **SMBH** | Super Massive Black Hole | **SMG** | Sub-Millimetre Galaxy |
| **SMILES** | Super conducting sub-mm Limb Emission Sounder | **SNe** | Super Novae |
| **SOC** | Science Operations Centre | **SPC** | Science Programme Committee |
| **SPICA** | SPace Infrared telescope for Cosmology and Astrophysics | **SPIRE** | Spectral and Photometric Imaging REceiver |
| **SRON** | Space Research Organisation of the Netherlands | **SRR** | System Requirements Review |
| **SSO** | Solar System Object | **STA** | SPICA Telescope Assembly |
| **STSST** | SPICA Telescope Science Study Team | **SVM** | Service Module |
| **TAC** | Time Allocation Committee | **TAS** | Thales Alenia Space |
| **TBC** | To Be Confirmed | **TBD** | To Be Defined |
| **TC** | Telecommand | **TDA** | Technology Development Activities |
| **TES** | Transition Edge Sensors | **TID** | Total Ionising Dose |
| **TIRCS** | Thermal Insulation and Radiative Cooling System | **TIS** | Total Integrated Scatter |
| **TM** | Telemetry | **TOB** | Telescope Optical Bench |
| **ToO** | Target of Opportunity | **TT&C** | Tracking Telemetry and Command |
| **ULIRG** | UltraLuminous InfraRed Galaxy | **UV** | UltraViolet |
| **WFE** | Wave Front Error | **XDR** | X-ray Dominated Region |